\documentclass[a4paper,openright,twoside,ngerman,UKenglish,cleardoublepage=plain]{scrbook}

\makeatletter%
	\renewcommand*{\@pnumwidth}{3em}
\makeatother%

\RedeclareSectionCommand[tocbeforeskip=1em plus 3pt]{chapter} %

\newcommand*{\ORIGchapterheadendvskip}{}%
\let\ORIGchapterheadendvskip=\chapterheadendvskip
\renewcommand*{\chapterheadendvskip}{%
	{%
		\setlength{\parskip}{12pt}%
		\noindent\hspace{-.5\marginparwidth}\rule[.3\baselineskip]{\linewidth+.5\marginparwidth}{1pt}\par
		\vspace{\baselineskip}%
	}%
	\ORIGchapterheadendvskip
}

 \def\thechapter{\Roman{chapter}}
 \def\thesection{\Roman{chapter}.\arabic{section}}
 \def\thesubsection{\Roman{chapter}.\arabic{section}.\arabic{subsection}}
 \usepackage{msor_thesis}
 \usepackage[utf8]{inputenc} %
 \usepackage{mathtools}
 \usepackage{braket,mathrsfs,units} %
 \usepackage{amsthm,amsfonts}
 \renewenvironment{subequations}{%
 	\refstepcounter{equation}%
 	\setcounter{parentequation}{\value{equation}}%
 	\setcounter{equation}{0}%
 	\ignorespaces
 }{%
 	\setcounter{equation}{\value{parentequation}}%
 	\ignorespacesafterend
 }

 \usepackage[locale = UK]{siunitx} 
 \DeclareSIUnit\annum{a} 
 \usepackage{dsfont} 
 \usepackage{amssymb,fixmath,upgreek} 
 \usepackage{lscape,rotating,placeins,graphicx,graphics,multirow,tabularx} 
 \usepackage{caption,subcaption}
 \usepackage{epigraph}
 \usepackage{slashed}
 \usepackage[nointegrals]{wasysym}

 \usepackage{hyperref}
 \usepackage{enumerate}
 \usepackage{tikz}
 \usepackage{appendix}
 \usepackage[newzealand]{babel}
 \usepackage[T1]{fontenc}
 \usepackage{csquotes}
 \usepackage{tcolorbox}
 \usepackage{nameref}
 \usepackage{cancel}
 \usepackage{microtype}
 
 \usepackage[sorting=anyt,doi=true,eprint=true,arxiv=abs,hyperref=true,style=nature-modified,citestyle=alphabetic,intitle=true,autolang=hyphen]{biblatex}
 \DeclareNameAlias{default}{first-last}
 
 \defbibenvironment{bibliography}
 {\epigraph{I have forgotten the books I have read and so I have the dinners I have eaten; but they both helped make me.}{Ralph Waldo Emerson (attributed)}
 \list
	{\printtext[labelalphawidth]{%
			\printfield{labelprefix}%
			\printfield{labelalpha}%
			\printfield{extraalpha}}}
	{\setlength{\labelwidth}{\labelalphawidth}%
		\setlength{\leftmargin}{\labelwidth}%
		\setlength{\labelsep}{\biblabelsep}%
		\addtolength{\leftmargin}{\labelsep}%
		\setlength{\itemsep}{\bibitemsep}%
		\setlength{\parsep}{\bibparsep}}%
	}
 {\endlist}
 {\item}
 \makeatletter
 \newcommand*{\currentname}{\@currentlabelname}
 \makeatother

 \setcapindent{0em}
 \allowdisplaybreaks
 
 \newcommand{\abs}[1]{\left\lvert #1 \right\rvert}
 \newcommand{\kl}[1]{\left( #1 \right)}
 \newcommand{\klg}[1]{\left\{ #1 \right\}}
 \newcommand{\kle}[1]{\left[ #1 \right]}
 \newcommand{\ed}[1]{\frac{1}{#1}}

 \newcommand{\defi}{\mathrel{\mathop:}=}
 \newcommand{\ifed}{=\mathrel{\mathop:}}

 \newcommand{\pdet}{\ensuremath{\operatorname{pdet}}}

 \newcommand{\dif}{\ensuremath{\mathrm{d}}}
 \newcommand{\scri}{\mathscr{I}}
 \newcommand{\eps}{\varepsilon}
 
 \newcommand{\Li}[1]{\ensuremath{\operatorname{Li}_{ #1 }}}

 \newcommand{\mi}{\ensuremath{{\left[\mu^{-1}\right]}}}
 \newcommand{\zt}{\kle{\zeta^\dagger}}
 \newcommand{\gi}{\kle{g_{\text{eff}}^{-1}}}
 \newcommand{\g}{g_{\text{eff}}}

 \newcommand{\kB}{k_\text{B}}

 \newtheorem{thm}{Theorem}

\begin{document}
 	\frontmatter
 	\title{Black Hole Evaporation: Sparsity in Analogue and General Relativistic Space-Times}
 	\author{Sebastian Schuster}
 	\date{\today}
 	\definecolor{darkgreen}{rgb}{.125,.5,.25} 
 	\definecolor{formulagreen}{rgb}{0.560181, 0.691569, 0.194885}
 	\definecolor{formulablue}{rgb}{0.368417, 0.506779, 0.709798}
 	\definecolor{formulared}{rgb}{0.922526, 0.385626, 0.209179}
 	 	
 	\subject{Mathematics}
 	\abstract{\addcontentsline{toc}{section}{Abstract}
 	Our understanding of black holes changed drastically, when Stephen Hawking discovered their evaporation due to quantum mechanical processes. One core feature of this effect, later named after him, is both its similarity and simultaneous dissimilarity to classical black body radiation as known from thermodynamics: A black hole's spectrum certainly looks like that of a black (or at least grey) body, yet the number of emitted particles per unit time differs greatly. However it is precisely this emission rate that determines --- together with the frequency of the emitted radiation --- whether the resulting radiation field behaves classical or non-classical. It has been known nearly since the Hawking effect's discovery that the radiation of a black hole is in this sense non-classical (unlike the radiation of a classical black or grey body). However, this has been an utterly underappreciated property. In order to give a more readily quantifiable picture of this, we introduced the notion of \enquote{sparsity}, which is easily evaluated, and interpreted, and agrees with more rigorous results despite a semi-classical, semi-analytical origin.
 		
 	Sadly, and much to relativists' chagrin, astrophysical black holes (and their Hawking evaporation) have a tendency to be observationally elusive entities. Luckily, Hawking's derivation lends itself to reformulations that survive outside its astrophysical origin --- all one needs, are three things: a universal speed limit (like the speed of sound, the speed of light, the speed of surface waves, \dots), a notion of a horizon (the \enquote{black hole}), and lastly a sprinkle of quantum dynamics on top. With these ingredients at hand, the last thirty-odd years have seen a lot of work to transfer Hawking radiation into the laboratory, using a range of physical models. These range from fluid mechanics, over electromagnetism, to Bose--Einstein condensates, and beyond. A large part of this thesis was then aimed at providing electromagnetic analogues to prepare an analysis of our notion of sparsity in this new paradigm. For this, we developed extensively a purely algebraic (kinematical) analogy based on covariant meta-material electrodynamics, but also an analytic (dynamical) analogy based on stratified refractive indices. After introducing these analogue space-time models, we explain why the notion of sparsity (among other things) is much more subtle and difficult to come by than in the original, astrophysical setting.}
 
 	\ack{\addcontentsline{toc}{section}{Acknowledgements}
 	First of all, I would like to thank Matt Visser. The list for what I want to thank you is long, and probably will not be complete in any case, so I will try to be brief: Thank you for having a lot of time for your students, and knowledge to share! Thank you for letting us go to conferences and give talks all across the globe! Thank you for making this PhD as fun as it was! Thank you for all the great feedback (and especially for the fast feedback on the thesis)!
 	
 	Equally invaluable was my family, in particular my parents Helga and Manfred: Thank you! Without your constant support, emotional, intellectual and otherwise, this would not have been possible. Thank you for all the help with moving, thank you for visiting me here at the other end of the world --- and for so much more!
 	
 	I can hardly overstate how much I value Benjamin Eltzner's general relativity tutorial in combination with our friendship: Without it, I doubt I would have been as enamoured with general relativity and quantum field theory on curved space-times! Thank you so much for introducing me to this! Likewise, I would like to thank Bernard Metsch for his great electrodynamics lectures, which first kindled my interest in relativity. Manuel Krämer, Prado Mart\'in-Moruno and Valentina Baccetti I want to thank for helping me meet Matt in Warsaw, and encouraging me to apply for a PhD under his supervision.
 	 	
 	During my PhD I had the chance to interact with a great many researchers who taught me with their talks, their insights, their discussions, and their help. For the sparsity, I am deeply grateful for the insight Don Page shared with us after our first version was uploaded to the arXiv. Natalie Deruelle, Chris Fewster, Dennis Rätzel, Friedrich Hehl, Jörg Frauendiener, Carlos Barcel\'o, and Eli Hawkins helped me a lot with their interest, questions, references, and ideas regarding the work on the electrodynamic analogue model, for which I am very grateful. I also enjoyed very much the discussions about physics in general with Philip Stamp, Yuichi Hirose, Roy Kerr, and David Wiltshire.
 	
 	I would also like to thank Claus Kiefer, Friedrich Hehl, Branislav Nikoli\'c, and the rest of the Cologne group for all the opportunities to join in on and contribute to great discussions about physics! I also express my gratitude to Shinji Mukohyama, for letting me speak at the Yukawa Institute in Ky\=oto --- I really enjoyed my stay there! My thank also goes to David Keitel, Sascha Husa, and Stefan Hild for the chance to give a talk in Palma de Mallorca, and in Glasgow!
 	
 	I am grateful Matt, Mark Srednicki, Uli Zülicke, and Michele Governale helped me realise my idea of a quantum field theory summer school here in Wellington. I enjoyed it a great deal, and am really imdebted to you for making this as successful as it was! Similarly, I want to thank all the other grad students for all the work they put in for the organisation of the NZMASP17 conference in Kaiteriteri! Both events were dear additions to my time as a PhD student here!
 	
 	Then, of course, all the great people I got to share my office with: Alex, Alex, Ana, Del, Finn, Jessica, thank you so much for the fun both physics-y, and out of academia, that we got to have. Also, thank you Del for helping with the proofreading --- and all the hour-long, heated, great discussions on life, the universe, and everything else! (And thank you to the rest for joining in, and tolerating these\dots Not to mention the fact that my books kept creeping onto more and more of other people's shelves\dots) Finn deserves a huge credit for letting me use his superbly annotated and documented numerical code! Thank you!
 	
 	I want to express my deep gratitude for all the help and the warm atmosphere in and from the School of Mathematics and Statistics (and beyond): Be it the team of the school office(s) (Kelsey, Simonette, Alex, Rebekka, Monoa, Ginny, Patricia), the people I tutored for or with (Steven, Matt, Dimitrios, Rod, Astrid, Noam, Petrik, Peter, Thuong, Mark), or the many, many other people who make this really a great place to work at. This is amazing! 
 	
 	A heartfelt \enquote{Cheers!} to the other grad students (in no particular order: Emma, Jessica, Del, Susan, Alex, Alex, Jasmine, Will, Andrew, Finn, Lei, Jayden, Sarah, Hamed, Steven, Ellis, Meenu, Gio, Sam, Ying, Lingyu, Yuki, Xander, Roy) for all the fun and games! Likewise I want to thank all my flatmates that gave (and give) me and my books such a great home here in Wellington: Fraser, Averill, Arianna, Stephen, Dawn, Ian, Miranda, Ela, Sage, and Danya. Very specifically for the thesis I have to thank Ela for lending me her laptop table --- my knees would not have survived this thesis without it!
 	
 	Next, I want to thank all the people who helped bring a bit of home to a place about $\SI{20000}{\kilo\metre}$ away from home (and/or welcomed me back with open arms, when I managed to make the trip): Vera, Benjamin, Carina, Christoph, Thorsten, Tobias, Aika, Paula thank you for all the Skype sessions! Thank you Tobias for helping with the soundtrack for writing a thesis! Again more gratitude towards my parents and my brother Torsten for making the trip through New Zealand with him possible --- without it, I hardly could have said to have been in New Zealand! Thank you Tim, Sonja, Johanna, Michael, Hannah and in particular Paula for your visits! Thank you Heger, Axel, Thomas (and Elke, Sandra, and Anke!) for making OhneHAST still last (in some form)! Thank you Kirsten and Norman for letting me be part of your big day, may the Force be with you! Thank you, Michael, Aika, Alex, and Julia, for making the trip to Japan so much greater and smoother! Anne, Mathias, Vera, Aika, Julia, David, Laura, also helped me out a lot by letting me stay while I was travelling back on the Northern hemisphere. Christine deserves credit for making an abstract understandable to non-physicists.
 	
 	Also thanks to the Fachschaft Physik/Astronomie Bonn, where I still feel every time like little to no time has passed; free coffee, great people, and of course the option to give talks on the really \emph{fun} things in physics! Kudos to Kevin Luckas for finding a typo instantaneously and at first glance without knowing what it was! Chapeau!
 	
 	I am certain, I forgot people who deserve my gratitude, and I am deeply sorry for that --- please forgive me! This thesis would not have been possible without the amazing people around me!
 	
 	I also want to thank the Victoria University of Wellington for the funding through both the PhD scholarship, as well as the thesis submission scholarship.
 	
 	\begin{center}
 		\vspace{\baselineskip}
 		\includegraphics[height=2cm]{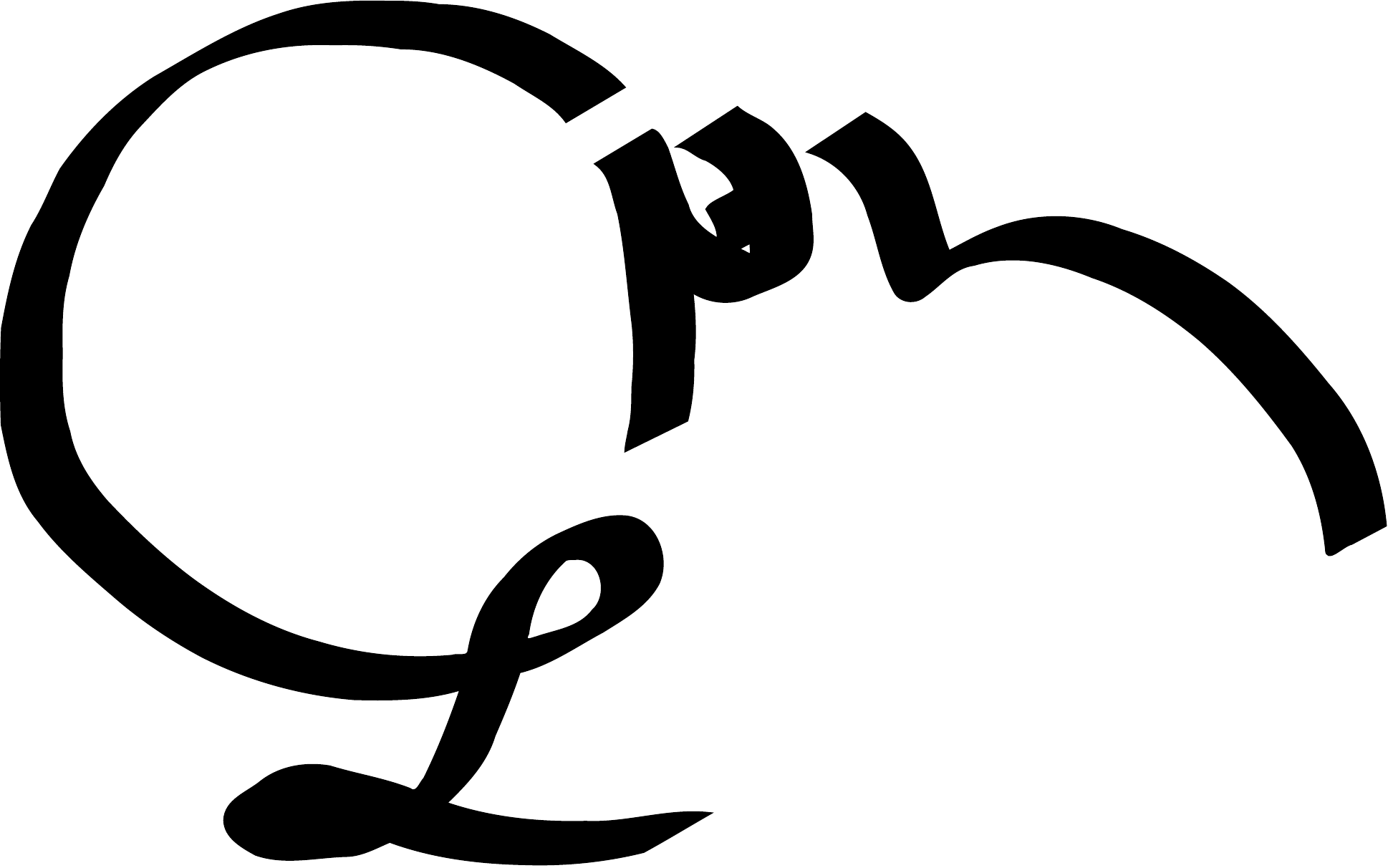}
 	\end{center}
 	} 
 	
 	\phd
 	\maketitle
 	\tableofcontents\addcontentsline{toc}{section}{Contents}
 	
 	\listoffigures\addcontentsline{toc}{section}{\listfigurename}
 	\listoftables\addcontentsline{toc}{section}{\listtablename}
 	
 	\mainmatter
 	
\chapter{Introduction}
\epigraph{\enquote{Strange and terrible books were drawn voluminously from the stack shelves and from secure places of storage; and diagrams and formulae were copied with feverish haste and in bewildering abundance. Of scepticism there was none.}}{H.P. Lovecraft, \emph{The Dunwich Horror}}

Contrary to their name, black holes have had a colourful history, and unlike their physical implications they gave rise to significant amounts of information. It is hoped that the present thesis can both add its own small strokes to the big picture, as well as shed some new light on previous brushwork. Imagery aside, before jumping right into the topics to be discussed, it is worthwhile to spend some time setting the stage. This is done most easily by first recapitulating the history of the field, in our case the field of quantum field theory in curved space-time, and with a special focus on the physics of black holes as (in this particular context) pioneered by Stephen Hawking. The aim of this first chapter is accordingly to give both a historical introduction, as well as a more elaborate introduction to the theme and goals of this thesis than the abstract does. This settled, we sketch the path we shall follow throughout the thesis towards our goals. Lastly, we will use the opportunity to also introduce and settle the relevant notation (including physical constants to appear quite soon) and conventions used.

\section{Some History} \label{sec:history}
 	The history of black holes starts already prior to the beginning of general relativity's tale: In 1783, John Michell conceived of the notion of black holes in a strictly Newtonian sense: Assuming light to be corpuscular, one can check at which radius for a given mass the escape velocity will equal the speed of light \cite{DarkStarsHistory}. This exactly corresponds to the later value found in a general relativistic setting --- but this discovery did not prevail for a long time, as it did not survive the discovery of light's wave nature. This only changed with Einstein's discovery of general relativity (GR), see \cite{GenGR1}, described by Einstein's field equations
 	\begin{equation}
	 	R_{ab} - \ed{2} g_{ab} R + \Lambda g_{ab} = \frac{8\pi G}{c^4}T_{ab},
 	\end{equation}
 	a set of quasi-linear partial differential equations for the metric tensor $g_{ab}$ of a Lorentzian manifold in terms of the Ricci curvature tensor $R_{ab}$ and Ricci scalar $R$ derived from $g_{ab}$ and its partial derivatives. $\Lambda$ is the so-called cosmological constant, and $T_{ab}$ the energy-momentum tensor of matter sourcing the metric. He himself suspected this to be a theory whose underlying differential equation would be too complicated to yield to exact methods. This suspicion was quickly overturned, when Karl Schwarzschild published the first exact solution (later to bear his name) to GR --- still using a precursor\footnote{The \enquote{November theory} as opposed to the now-used \enquote{Ricci theory}, see \cite{GenGR1}.} to it --- in 1916.\footnote{See \cite{SchwarzschildOuter}.} The most important issue for a long while was the presence of two singularities in this solution: One at the origin of the coordinates, one at the so-called Schwarzschild radius $r_\text{H}$. Mainly, this discussion was related to early problems in understanding coordinate (in)dependence of physical quantities. These issues were by no means limited to what was later to be known as \enquote{black holes} --- also gravitational waves' coordinate independent, physical content was only settled in 1957 at the famous Chapel Hill conference, involving many of general relativity's (and beyond) most eminent names of the time (including, but not limited to Weber, Wheeler, Feynman, and Bondi). This decade-long study of coordinate (in)dependence led to the conclusion that while the singularity at the origin of Schwarzschild's solution was not resolvable, the one at the Schwarzschild radius was of a more subtle nature. It turned out to be possible to find coordinate systems which were well-behaved at the corresponding 2-sphere, yet this 2-sphere \emph{does} have physical relevance.\footnote{For more on the historical development of general relativity, see for example \cite{BHTWarps}.} It is \emph{this} physical relevance that is at the very heart of this thesis. However, in order to fully introduce the thesis' context, it is worthwhile and necessary to dwell a bit more on the subsequent history of black holes. As the present text is, despite this, no thesis on the history of science, we shall restrict the references to the original (now \enquote{historical}) references to those of immediate concern to us. More historical references can be glanced either from the references cited earlier in this paragraph or the other references given in this chapter.
 	
 	Schwarzschild's solution turns out to describe a spherically symmetric space-time around a point-mass of mass $M$. Finding the solution corresponding to a \enquote{rotating point-mass} took considerably longer, and only in 1963 Roy Kerr succeeded in this endeavour \cite{Kerr}. (Granted, this mass is not quite \enquote{point-like}.) The structure of this solution proved to be much richer, just like in the case of already shortly thereafter found solutions for rotating and charged point-masses. Roughly around the same time, the first instances of the term \enquote{black hole} can be found, later taken up, popularized and canonized by John Wheeler. The first appearance in print of this term seems to be an article by Ann Ewing in the January~18, 1964, issue of the \emph{Science News Letter}, reporting on a January 1964 annual meeting of the American Association for the Advancement of Science \cite{BHOrigin}.
 	
 	The historically most important event for the purposes of this thesis, however, came with Stephen Hawking's eponymous discovery of the Hawking effect --- that is: Black hole evaporation. In order to properly set up its historical context, it might be worthwhile to quickly juxtapose the history of quantum theories, both of particles and fields, with that of general relativity, and how they cross-pollinated each other. 
 	
 	Quantum theory in general begun, similarly to (general) relativity, to prosper in the early twentieth century. Usually, this reckoning of quantum theory is told from the viewpoint of quantum mechanics --- the quantum theory of point particles; nevertheless, also the birth of the quantum theory of fields is to be found in these early days: As the limits of classical physics were first demonstrated in Planck's theory of black body radiation, and radiation being described by fields, already early versions of quantum mechanical studies of black body radiation contained bits of quantum field theory (QFT). For a retelling of this part of the history of quantum theory, see the first two chapters of \cite{Duncan}.
 	
 	Curved space-time quantum field theory (CSTQFT) developed --- to a degree --- at nearly the same time as quantum field theory, general relativity, and quantum mechanics --- their mutual interrelations occasionally shining through. However, much of this early history was closely associated with first attempts at quantum gravity.\footnote{For a historical account of the early history of quantum gravity, see for example \cite{Kiefer} and \cite{EarlyHistQG}.} But already in the 1930s, the Dirac field was investigated in a curved space-times context by Fock, Bargmann, and Schrödinger; and Schrödinger realized already in 1939 that cosmological expansion will bring about particle creation \cite{CosmPartCreat39}. However, these attempts were still \enquote{first quantised}. In the case of the Dirac field, and in modern phrasing, this means that the curved space-time version of the Dirac equation (\emph{e.g.}, as derived in \cite{SchrDiracCurved}) has to be understood as a classical field theory of a spinor bundle, rather than a fermionic quantum field theory in the modern sense. Nonetheless, much of CSTQFT has to rely on these early results as the underlying classical wave equation still is an ingredient of the quantised theory, and these classical developments already demonstrate the new feature of an ambiguity in this wave equation: One has to fix an additional, scalar parameter describing how curvature terms enter the original wave equation. 
 	
 	A truly early contribution (in some sense) was the Casimir effect, discovered 1948, and the only remotely CSTQFT-related effect that found quick experimental confirmation (in 1958), see \cite{ParkerToms}: The introduction of two conducting, parallel plates to the classical vacuum of flat space-time QFT changes the vacuum energy between the plates compared to that outside. This difference results in a force pushing the plates together. One has to admit at this point that the Casimir effect itself as such requires a bit of a stretch to be included in CSTQFT: The only difference from flat Minkowski space-time is the introduction of non-trivial boundaries. While this is now often considered a first step to moving beyond a flat space-time quantum field theory (and its Poincar\'e invariant vacuum state), the inclusion is still slightly tenuous. However, the effect is not a feature of flat space-times alone, and is also encountered in the more general context of CSTQFT.
 	
 	CSTQFT proper began in the 1960s and 1970s: Particular famous and lasting results were the particle creation during the expansion of the universe --- now the vast field of inflationary cosmology ---, and that seen by accelerating observers.\footnote{There certainly is room for speculation how well one can rephrase even cosmological particle creation in terms of some notion of acceleration. We shall, however, take the working principle to separate the field into the above-mentioned two broad categories.} Beyond the early prediction by Schrödinger, the former has been studied extensively by Chernikov, Fulling, Parker, Pitaevski, Starobinsky, Tagirov, Zel'dovich, and others. The latter case is of particular interest for this thesis. As accelerated observers as such are fairly general, it is natural to look for particularly interesting examples, the foremost being that of an accelerated observer in Minkowski space-time, and that of an observer at fixed position close to the horizon of a black hole. This particle creation close to a black hole was first discovered by Stephen Hawking in 1974 \cite{HawkingEffect1,HawkingEffect2}. Prior to this, the phenomenon of superradiance was observed in black hole space-times, and investigations of this in terms of quantum field theoretic notions (among others by Starobinsky and Unruh) formed part of the backdrop of Hawking's own research. In 1976, Unruh added the effect named for him which describes the particle creation related to accelerating observers in Minkowski space-time \cite{UnruhVacuum}. Around the time of these two results the field reached maturity, as exemplified by DeWitt's survey article \cite{DeWCSTQFT} from 1975. Additional developments of the field happening at the time were concerned with effects due to topology and boundary conditions (and their incarnation in the Casimir effect) on the one hand, and the renormalisation of the quantities involved in this program (like the energy-momentum tensor) on the other. Much of this historical aspect of CSTQFT can be glanced from \cite{BirrellDavies,Fulling,WaldBHTD,FabbriNS05,ParkerToms} and the original research referenced therein.
 	
 	While QFT in Minkowski space-time has Lorentz symmetry and an associated, preferred vacuum state, this notion is lost in the more general setting of CSTQFT. The resulting notion of an observer-dependent vacuum, and thus observer-dependent \emph{particle numbers}, is then the common denominator of the afore-mentioned effects and complications arising when transitioning from flat Minkowski space to a general, curved space-time.
 	
 	Already in the original research on CSTQFT it was noted that the new effects found were minuscule: Neither the Unruh effect of currently achievable accelerations nor the Hawking effect of black holes \enquote{at hand} can be observed with present technology --- possible surprises from so far unforeseen consequences of available physics not-withstanding. (A discoverer of such a direct and surprising experimental access would be guaranteed a lot of kudos from the community.) One way out is to look for analogies to these effects that are more readily testable. The key idea behind these analogies is that the physical origin of the effects we are interested in is not inherently astrophysical. While their general relativistic origin indeed seems to imply such a connection, the notion of Lorentzian geometry behind this is actually more general: \emph{Any} system exhibiting a \enquote{speed limit} (as the speed of light in relativity) confronted with a quantum theory may exhibit such effects, provided the \enquote{speed limit} becomes time- or position-dependent \cite{BHfromPDE,EssIness}.
 	
 	These features are very commonplace in physics. Not surprisingly with increasing understanding of this fact, the number of physical systems considered as analogies exploded over the years. This list now encompasses as diverse systems as surface waves, gravity waves, electromagnetism/light propagation in media, sound waves, superfluids, graphene, and many more. Following the review \cite{lrrAnalogue}, we shall distinguish between a historical period preceding 1981, and the modern history of this subfield post-1981. The transition year is related to Unruh's consideration and publication of an analogue space-time in a fluid mechanics setting. Further contributions to this line of inquiry then came from Jacobson, and Visser, soon followed by many more. The present thesis, however, is more concerned with one particular instance of an analogue space-times from the \enquote{historical period}: Those encountered in electromagnetic contexts. In 1923, Gordon \cite{GordonMetric} tried to use a gravitational field (and its resulting metric) to mimic light propagation in a medium. While most of the discussion employed (to varying degrees) the eikonal approximation, the idea can be extended beyond this. Similarly, his limitations from assuming a flat Minkowski background, always setting permittivity and permeability to be isotropic, and often assuming position independence for used quantities, can be lessened and the results generalised. His idea was then (in)famously taken up as an exercise question in \cite{LanLifII} and inverted: The medium is now used to mimic the gravitational field. This highlights that a good analogy should work both ways and occurs frequently in the context of analogue space-times. Not surprisingly, this raises the hope that they can be used to transmit expertise, understanding, and techniques between different fields of physics. Later, Pleba\'nski's and de~Felice's work to form an analogy between electromagnetism in dielectrics and general relativity formed the foundation of what now operates under the name of \enquote{transformation optics}. Here the goal is slightly different from that of the analogue space-time framework, however the link is close enough that we will have a short look at related work in section~\ref{sec:trafooptics}.
 	
 	Again, for details on and a more extensive exposition of (the rest of) the history of analogue space-times we refer the reader to \cite{lrrAnalogue}. It bears repeating that curved space-time quantum field theory as encountered in analogue space-times can be carried out completely independent of any astrophysical or special relativistic or general relativistic analogy. The philosophical implications of experimental observation of CSTQFT effects in an analogue are thus two-fold: (1) It shows the validity of CSTQFT \emph{of} the analogue system. (2) In a \emph{non-falsifiable} way, it raises researchers' confidence in applying CSTQFT methods to different space-times, be they analogue or not. The first implication is fully compatible with Popper's views on scientific progress, the second implication is more in line with Feyerabend --- if we slightly oversimplify the philosophy involved.
 	
 	\section{Goal}
 	As this historical overview shows, the Hawking effect highlights a lot of the issues arising from the mingling of quantum physics and general relativity. However, while studying the associated questions, a lot of folklore has accumulated. Probably the best example for this is the commonplace negligence of grey body factors in the Hawking spectrum. A result of this approach is that commonly the shorthand for describing the Hawking effect is the slogan: \enquote{A black hole radiates as a black body.} 
 	
 	The goal of this thesis, specifically, the goal of the term \enquote{sparsity}, is to have a closer look at the validity of this slogan. It turns out that --- while many of the predictions and conclusions borne of this slogan hold --- one certain(ly) intriguing, and important feature of this evaporation intrinsic to black holes is lost: The thermal wavelength of the black body's radiation usually considered in classical thermodynamics is smaller than the typical length scales of the black body itself, but in the case of black hole evaporation this particular ratio behaves exactly in the opposite way. The wavelength of emitted radiation being larger than the radiating body now can best be explained in terms of the notion of \enquote{sparsity}.
 	
 	While this distinction already in general relativistic space-times proves to be theoretically and physically stimulating, the inaccessibility of astrophysical Hawking evaporation still places all of this discussion outside the scope of today's or the near future's experimental verification. Nevertheless, as already mentioned in the historical outline, the principle behind black hole evaporation and similar effects, are a general property of quantum field theories on curved space-times. As these curved space quantum field theories do not need to arise only in the context of general relativity, sparsity can be analysed and looked for in the quantum field theories on analogue space-times.
 	
 	More specifically, the present thesis tries to look into the question of sparsity of \enquote{black hole} evaporation in electromagnetic analogue space-times. To provide the clearest framework to achieve this, we shall develop previous results in the literature into a covariant formalism of electromagnetic analogue space-times, including a (new) freedom to choose the background space-time, \emph{i.e.}, the lab in which this analogue is constructed. Strictly speaking, for example, this could be employed for using the plasma surrounding an astrophysical black hole as an electromagnetic analogue space-time.\footnote{This shall only serve as an example of the utility of the formalism demonstrated in this thesis; this particular example has not been worked out at the time of writing, but provides an avenue for future extensions of the work presented in this thesis.} However, as the traditional definition of an electromagnetic analogue is in purely algebraic terms, it is necessary (and instructive) to also examine carefully the differences in the corresponding wave equation for the electromagnetic four-potential $A^{a}$.
 	
 	\section{Outline} 
 	In chapter~\ref{ch:BH} we shall follow up the introductory remarks of the present chapter with the necessary background material for discussing black holes and their evaporation. This includes both a quick recapitulation of the necessary terminology and solutions from general relativity, as well as a short summary of the results from CSTQFT needed to derive the Hawking effect.
 	
 	Following this, chapter~\ref{ch:analogues} will discuss analogue space-times and the specific implementation of this framework in the context of electromagnetic analogue space-times. The chapter closes with an example from electrodynamics better described as \enquote{just} an analogy than as a case of an \emph{analogue space-time}. This chapter will constitute a significant part of the thesis, both in breadth, as well as depth.
 	
 	The penultimate chapter~\ref{ch:sparsity} shall discuss the overarching theme of sparsity, both in the context of general relativity --- where the concept was originally conceived ---, and in the context of the previously introduced electromagnetic analogue space-times.
 	
 	Finally, chapter~\ref{ch:finale} will summarise the gained results and outline possible, future avenues of research.
 	
 	The focus of the main part of the thesis shall lie on the physical application; the appendices provide more mathematical background material whose incorporation in the main chapters would detract from this focus on physics. In appendix~\ref{ch:DG} some notions of differential geometry are collected, while appendix~\ref{ch:Special} will list special functions applied in the main text together with properties used. More concretely, we shall collect in the first appendix some results on conformal equivalence of two metrics on the one hand, and on the other hand results related to orthogonal decompositions of (certain types of) tensors.
 	
 	Each separate chapter shall begin with a short outline of its subordinate sections.
 	
 	\section{Remarks on Notation and Prerequisites}\label{sec:notation}
 	Given that this thesis naturally encounters a plethora of physics' subfields --- through the notion of analogue space-times --- it is to be expected that clashes of notation between two subfields happen. To alleviate this problem before it occurs, here are collected the chosen conventions and the preferred terminology.
 	\begin{itemize}
	 	\item Sign conventions: This thesis follows the mostly-positive (also known as: space-like, relativity, East coast, Pauli) sign convention for the metric; the signature is chosen to be $(-+++)$. Similarly, for higher dimensions, as encountered in section~\ref{sec:ndim}, the signature will be $(-+\dots+)$. This also means that exponentials $\exp(-i\omega t)$ are the positive frequency modes.
	 	\item The previous choice also specifies the time coordinate to be the first coordinate, labelled by an index $0$.
	 	\item A vector at a point of a Lorentzian manifold is called \enquote{time-like}, \enquote{light-like} (\enquote{null}) or \enquote{space-like} w.r.t. the metric $g$ if $g(V,V)<0$, $g(V,V)=0$, $g(V,V)>0$, respectively. (This depends on the chosen sign convention!) Curves on a Lorentzian manifold are classified according to the behaviour of their tangent vectors $V$. If the tangent vector is always time-like, light-like or space-like w.r.t. the metric $g$, the same name is granted to the curve itself. A curve whose tangent vector is always either time-like or null is called \enquote{causal}.
	 	\item General space-time indices will be labelled with Latin letters starting from $a, b, \dots$. Where only index placement, but not explicit labelling is required, the corresponding space-time index will be denoted by the symbol $\bullet$.
	 	\item Purely spatial indices will be labelled with Latin letters starting from $i, j, \dots$. Where only index placement, but not explicit labelling is required, the corresponding space-time index will be denoted by the symbol $\circ$.
	 	\item If $n$ indices are enclosed in round brackets this indicates symmetrisation over these indices. Likewise, square brackets around indices indicate anti-symmetrisation. (Anti-)Symmetrisation of $n$ indices includes a factor of $1/n!$.
	 	\item Indices or quantities preceded by a comma indicate partial differentiation with respect to them, indices or quantities preceded by a semi-colon indicate covariant differentiation with respect to them.
	 	\item Levi-Civita (pseudo-)tensors are denoted by $\eps$, while Levi-Civita (pseudo-)tensor densities are written using $\tilde \eps$.
	 	\item While, technically speaking, a \emph{space-time} is the tuple $(M,g)$ of a manifold $M$ equipped with a Lorentzian metric tensor field $g$, occasionally the shorthand of identifying the metric itself with the space-time will be encountered. With two exceptions (clearly marked), we will refrain from being even more technically precise and not extend the meaning of space-time to be a triple $(M,g,\Gamma)$ also including the connection $\Gamma$.
	 	\item We will be using both abstract index notation and coordinate index notation. It is hoped that the difference is clear from context and thus no risk of confusion remains.
	 	\item Physical constants appearing frequently are: The gravitational constant $G$ (occasionally further adorned by an index \enquote{N} or \enquote{Newton}), Planck's (reduced) constant $\hbar$, the speed of light in vacuum $c$, Boltzmann's constant $\kB$, the vacuum permittivity $\epsilon_0$, and the vacuum permeability $\mu_0$.
	 	\item Normally, general relativistic calculations have a tendency to be done in natural units, where $G=\hbar=c=\kB=1$. However, the concept of sparsity, while unitless itself, is defined in terms of quantities that can (at least in principle) be observed. Thus it seems to be beneficial to keep physical constants present when discussing sparsity. However, natural units will often be employed without explicit mentioning. The introduction will aim to use SI units.
 	\end{itemize}
 	Less ubiquitous notation specific to certain sections will be introduced as and where required.
 	
 	While much of the required material will be introduced, that will happen more in the form of a reminder to settle remaining questions of notation. For example, most of the technical vocabulary of courses on general relativity and quantum field theory will be assumed. A selection of references for GR would be given by \cite{MTW,Wald,Carroll,Poisson,ChoBru,FrolovZelnikov2011}, while for QFT a short list would include \cite{PeskinSchroeder,ZeeQFT,Srednicki}. These lists are neither meant to be exhaustive nor are their notations in agreement. Where it seems appropriate, we shall refer to additional \enquote{standard texts}.
 	
\chapter{Black Holes and Their Evaporation}\label{ch:BH} 
\epigraph{\enquote{Ich erblickte das Alphabet der Sterne. Ein Firmament voller funkelnder Zeichen, eine unlesbare, aber wundervolle Schrift aus Licht, so alt wie das Universum.}\footnotemark}{Walter Moers, \emph{Das Labyrinth der träumenden Bücher}}
\footnotetext{\enquote{I beheld the stars' alphabet. A firmament full of flashing signs, an unreadable, but wonderful script made of light, as old as the universe.}}

\section{A Rapid Survey of Black Hole Space-Times}
 	As black holes are necessary for the Hawking effect proper, this section will collect some of the most useful space-times and notions for the remainder of the thesis. Given the context, this implies concentrating solely on black hole space-times --- we shall not delve into the plethora of space-times besides these: Gravitational wave solutions, cosmological space-times, numerical solutions of the Einstein equations, and (with one exception in section~\ref{sec:unphysical}) a range of pathological space-times of interest to either mathematics (\emph{e.g.}, for regularity questions) or physics (\emph{e.g.}, for causality questions). Several of the black hole space-times appearing in this section will re-appear later in section~\ref{sec:bespoke}. There, we will then introduce many of the following metrics in different coordinates than those presented here. Features which find more visibility in different coordinates (than the ones of this section) will then be explained when used in subsequent chapters. We will use mostly SI units, physical constants, and notation as described in the previous section~\ref{sec:notation}.
 	\subsection{The Schwarzschild Metric and Notions of Horizons}\label{sec:Schwarzschild}
 	Historically the first and as a toy model the most important example is the Schwarzschild space-time given (in spherical coordinates) by the line element (we also use the opportunity to define the line element of the two-sphere, $\dif\Omega^2$)
 	\begin{equation}\label{eq:Schwarzschild}
 		{\dif s}^{2} = -\left(1 - \frac{2GM}{c^2\,r} \right) c^2 \,\dif t^2 + \left(1-\frac{2GM}{c^2 \, r}\right)^{-1} \,\dif r^2 + r^2 \underbrace{\left(\dif\theta^2 + \sin^2\theta \, \dif\varphi^2\right)}_{\ifed \dif\Omega^2}.
 	\end{equation}
 	Here, the parameter $M$ has the physical interpretation of mass. This choice of spherical coordinates, the so-called Schwarzschild (or curvature) coordinates $(t, r, \theta, \varphi)$, has coordinate singularities both at the singularities of the standard spherical coordinates (\emph{i.e.}, the poles), and at the Schwarzschild radius $r_\text{H}\defi\nicefrac{2GM}{c^2}$. The range of the coordinates is: $t\in (-\infty,\infty)$, $r\in [0,\infty)$, $\theta \in [0,\pi]$, and $\varphi \in [0,2\pi]$. These are the standard ranges for spherical symmetry (and will prove useful even beyond that). Unless stated otherwise, the range of coordinates in the following sections will be assumed to follow this example when named similarly. We glossed in the statement of the coordinate ranges somewhat nonchalantly over the coordinate singularities at the poles, but that is easily remedied, and we will omit these technicalities. We abbreviated this by abuse of notation of allowing the coordinate ranges to be half-open or even closed. Nevertheless, a \emph{physical} singularity is at the origin, $r=0$, as can be seen by evaluating, for example, the Kretschmann scalar $R^{abcd}\, R_{abcd}$. Note that this is a statement about singularities of certain tensors at certain points on the manifold, not (necessarily) a statement about the coordinate range or which points have to be excluded from the manifold. Despite its appearance as a coordinate singularity in Schwarzschild coordinates, the Schwarzschild radius \emph{does} have physical relevance, as it describes the position of an \enquote{event horizon} in the Schwarzschild space-time.
 	
 	This then is the appropriate moment to discuss various notions of horizons. Given the mostly physical and heuristic approach taken in this thesis, we will attempt to forego the mathematical --- more specifically differential topological/geometrical --- definitions and focus on the physical meaning. However, a certain rigour is hard, if not impossible, to avoid in order to make the differences of the different notions of horizons apparent. For the required technical vocabulary, we refer to the second item of the list of notation in section~\ref{sec:notation}.
 	
 	The first notion is that of an event horizon. A past/future event horizon is defined as the boundary of the part of space-time which can be connected by causal curves to past/future null infinity. As this definition invokes global knowledge of a given Lorentzian manifold, this definition has little practical, experimental value \cite{VisserHorizonObs}. In contrast, an apparent horizon is quasi-local and characterised by being the boundary of a space-like hyper-surface such that outgoing, orthogonal (to this hyper-surface) null curves have non-positive expansion (as defined in the Raychaudhuri equation).\footnote{More precisely, this corresponds to a \enquote{MOTS}, a marginally outer trapped surface \cite{Wald}.} This is related to the pictorial interpretation of the horizon being \enquote{where lightcones tip over}. It has the obvious advantage over the notion of an event horizon, that it only requires knowledge of a subset of the Lorentzian manifold at hand. To jump ahead a bit, an apparent horizon also makes the notion of a black hole more palpable when confronted with Hawking radiation and its consequence of the black hole evaporating. 
 	
 	Lastly, we want to mention Cauchy horizons: A Cauchy horizon is the boundary of the domain of dependence of a given hyper-surface and thus measures that surface's failure to be a Cauchy surface; a Cauchy surface being in turn a hyper-surface on which initial data for an initial value problem can be propagated onto the whole of the manifold. Phrased differently, the presence of a Cauchy horizon implies unpredictability, a mathematical \enquote{here be dragons}. Many more (or more technically precise) definitions can be found in \cite{Wald,VisserWormholes,BoothBHBoundaries,VisserHorizonObs} and references therein.
 	
 	After this short, technical digression, let us come back to the Schwarzschild space-time and its properties: It is Ricci-flat ($R_{ab}=0$, meaning physically that it is a vacuum space-time), static (it permits a globally defined, and time-like Killing vector field, which permits a space-like hyper-surface orthogonal to it\footnote{\emph{Outside} the event horizon. The mathematical description of black holes is usually limited to the \enquote{domain of outer communication}, as the interior solutions tend to be mildly to mightily ill-behaved. While the Schwarzschild metric \enquote{only} has a physical --- seen from the interior even naked --- singularity, already the Kerr metric contains closed time-like curves in the interior. What was a time-like Killing vector fields outside the horizon ceases to be one in the interior. More on this will be elucidated below in section~\ref{sec:Kerr}.}), and is connected to Birkhoff's theorem: Every spherically symmetric vacuum solution to the Einstein equations must be static. \emph{This} remains true in higher-dimensions, where the generalisation of the Schwarzschild space-time is known as the Tangherlini space-time, to be further described in section~\ref{sec:ndim}. Many other similar uniqueness statements in 3+1 dimensions have been shown to not survive a transition to higher dimensions. Usually, one also considers a base manifold turning the Schwarzschild metric into something asymptotically flat. This is not the only option, see \cite{SchwarzschildManifolds}. Having introduced the above notions of horizons, it is also important to realise that in the Schwarzschild space-time, event, apparent and Killing horizon are the same and are located at $r_\text{H}$. Thus it makes sense to speak in this case of \enquote{the} horizon.
 	
 	In light of this, let us quantify it more than just by its location $r_\text{H}$. Fixing time $t$ and radius $r$ allows to talk about the surface of topological two-spheres; integrating the induced two-metric on these two-dimensional submanifolds then gives a surface area
 	\begin{equation}
	 	A_\text{H} = 4\pi r^2_\text{H}
 	\end{equation}
 	when evaluated on the horizon. This kind of integration will be performed each time we talk about the area of the horizon.
 	
 	For the purpose of Hawking radiation the key quantity is the \enquote{surface gravity}: The proper acceleration orthogonal to the horizon. Formally, it can be defined for any space-time possessing a time-like Killing vector field $t^a$. The surface gravity $\kappa$ can then be given by \cite{Poisson}
 	\begin{equation}
	 	\kappa^2 = -\ed{2} t_{a;b} t^{a;b}.
 	\end{equation}
 	In the case of spherical symmetry and staticity this reduces to the much simpler equation
 	\begin{equation}
	 	\kappa = \frac{c}{2} \frac{\partial_r g_{tt}}{\sqrt{g_{rr} g_{tt}}},
 	\end{equation}
 	or, in our even simpler case, to
 	\begin{equation}
	 	\kappa = \frac{1}{2} \partial_r g_{tt}.
 	\end{equation}
 	Thus,
 	\begin{equation}
	 	\kappa = \frac{c^4}{4GM}.
 	\end{equation}
 	This quantity is proportional to the Hawking temperature (to be discussed more fully below), as
 	\begin{equation}
	 	T_\text{H} = \frac{\hbar}{2\pi c \kB}\kappa.\label{eq:THawkingTooEarly}
 	\end{equation}
 	The next few examples of black hole space-times will further expand on this relation.
 	
\begin{figure}
	 	\centering
	 	\includegraphics[width=\textwidth]{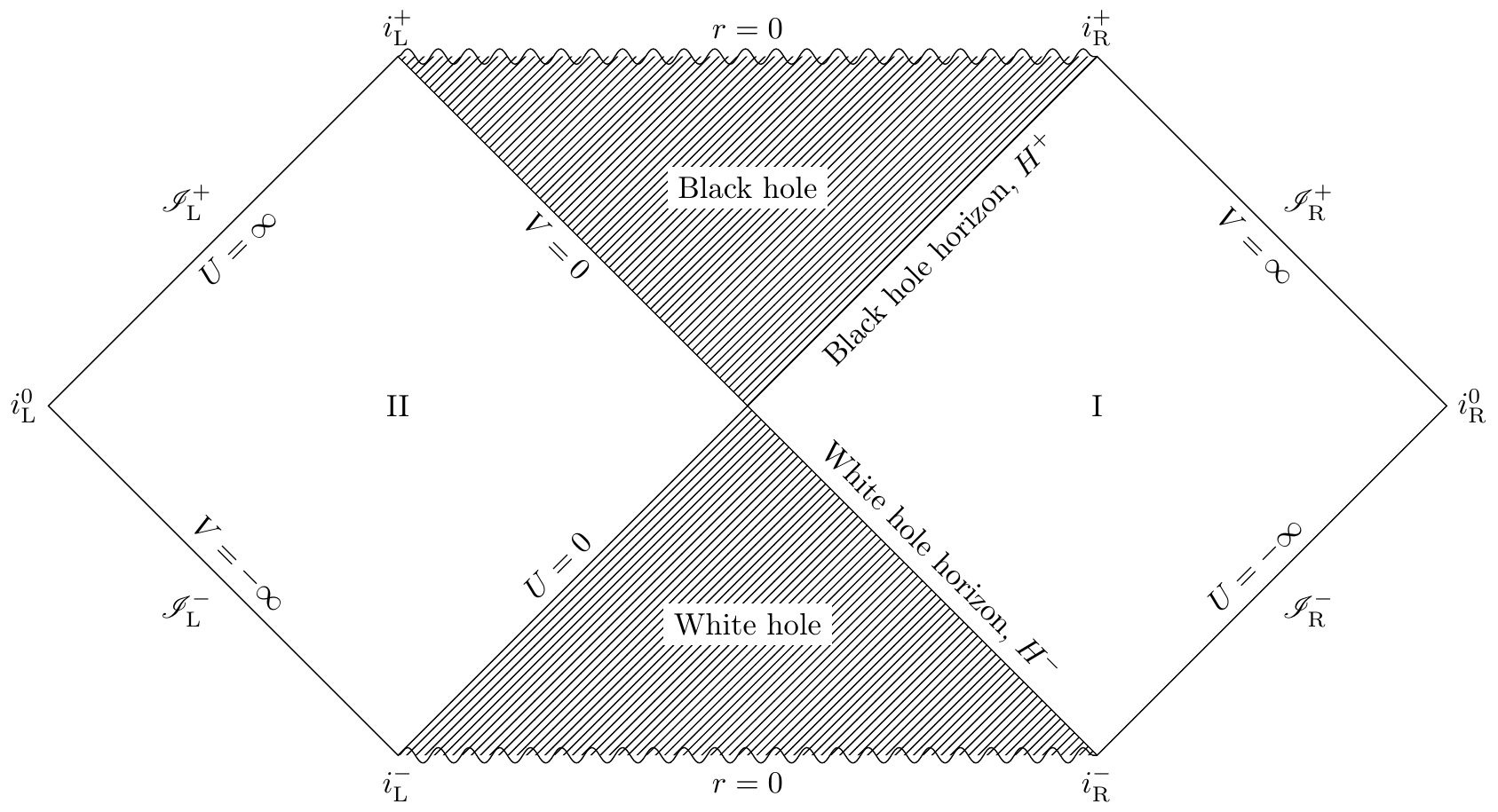}
	 	\caption[Carter--Penrose Diagram of Extended Schwarzschild Geometry]{Carter--Penrose diagram of maximally analytically extended Schwarzschild geometry.}
	 	\label{fig:Schwarzschild}
 	\end{figure}

 	In preparation for the technical part about the Hawking temperature~\ref{sec:Hawking}, we will now quickly summarise some additional coordinates that prove useful when discussing the Hawking effect. Classically, these are of similar importance as they help decide questions of coordinate artefacts or real singularities, just as they help extend the metric to a manifold larger than the one covered by the Schwarzschild curvature coordinates. First of all, let us introduce the so-called tortoise (or Regge--Wheeler) coordinate $r^*$:
 	\begin{equation}
	 	r^* \defi \frac{2GM}{c^2}\ln \kl{\frac{\abs{r-2GM}}{2GM}}.
 	\end{equation}
 	This turns the domain of outer communication into the range $r^*\in (-\infty,+\infty)$, with the right boundary corresponding to spatial infinity, and the left boundary corresponding to the event horizon. Using this, one defines the so-called retarded (or outgoing), and advanced (or ingoing) Eddington--Finkelstein null coordinates as
 	\begin{subequations}
 		\begin{align}
	 		u &\defi ct-r^*,\\
	 		v &\defi ct+r^*,
 		\end{align}
 		respectively. 
 	\end{subequations}
 	As their name suggest, their level sets are the future- (hence, outgoing) or past-directed (hence, ingoing) light cones, respectively, emanating from the singularity at $r=0$. Before discussing the part of the Schwarzschild geometry they cover, let us introduce one more coordinate system, which will greatly facilitate exactly this: The Kruskal--Szekeres coordinates. These come in both a version involving null coordinates, and a version involving time-like and space-like coordinates --- we will describe the former:
 	\begin{subequations}
	 	\begin{align}
		 	U &\defi -\frac{4GM}{c^2} e^{-uc^2/(4GM)},\\
		 	V &\defi \frac{4GM}{c^2} e^{vc^2/(4GM)}.
	 	\end{align}
 	\end{subequations}
 	They provide the maximal extension of the Schwarzschild geometry. This adds in addition to the future horizon a past horizon (a \enquote{white hole}), and a second region outside these two horizons; a mirror image of the original Schwarzschild space-time. From a physical point of view little of these features is relevant: To our current understanding, black holes form in gravitational collapse. This process does not possess a past horizon. As for the mirrored part, even in the maximally analytically extended Schwarzschild space-time this region is physically inaccessible as one would have to either pass through the singularity or exit a black hole region, neither physically possible. As the original Schwarzschild curvature coordinates, the coordinates exchanging $t$ for $v$ will cover the whole black hole and domain of outer communication, only this time the metric (determinant) will not be singular on the horizon. If one uses $u$ instead of $v$, a similar statement holds; but now the black hole part is not covered any more. Instead, the coordinates cover the white hole part. A conformal compactification of the coordinates $U,V$ is used to generate the Carter--Penrose diagram seen in figure~\ref{fig:Schwarzschild}. This boils the range $(-\infty,\infty)$ of $U$ and $V$ down to a bounded interval. We also can see in figure~\ref{fig:Schwarzschild} that now two domains of outer communication, I and II, exist --- their associated boundaries are labelled by the corresponding position in the Carter--Penrose diagram as either right (index \enquote{R}) or left (index \enquote{L}). Mostly, when referring to null infinity of the maximally extended Schwarzschild space-time, we will be talking about $\scri^\pm_\mathrm{R}$, and omit the index $\mathrm{R}$ for denoting the right half. In this vein: Future and past time-like infinity are at $i^\pm$, respectively. Spatial infinity is at $i^0$. Each exists once in the mirror universe II, and once in \enquote{our} universe I. Time increases from the bottom to the top.
 	
 	This quick (and not yet dirty, see below) introduction shall suffice for the time being, with more details provided along the way as needed.
 	
 	\subsection{The Kerr Metric}\label{sec:Kerr}
 	The next step beyond the Schwarzschild solution is to relax its demand of staticity. For this, let us call a region stationary, if this region has a time-like Killing vector field. As we look mostly at black hole space-times, we will make the slight abuse of notation of calling the space-time stationary, if its domain of outer communication is stationary. (A similar notation was already alluded to in the case of the Schwarzschild geometry.) Thus, our next space-time will lose the possibility to choose a space-like hyper-surface everywhere orthogonal to the time-like Killing vector field used in the definition of staticity. The resulting space-time is the Kerr solution. As demanded, it still is a vacuum/Ricci-flat solution, but now only axisymmetric, not spherically symmetric. This metric's line element reads:
 	\begin{subequations}
	 	\begin{equation}\label{eq:Kerr}
		 	\dif s^2 = - \frac{\Delta}{\rho^2} \kl{c \dif t - a \sin^2 \theta \dif \phi}^2 + \frac{\sin^2\theta}{\rho^2}\kl{(r^2+a^2)\dif \phi - a c \dif t}^2 + \frac{\rho^2}{\Delta} \dif r^2 + \rho^2\dif \theta^2,
	 	\end{equation}
 	where
 		\begin{gather}
	 		\Delta \defi r^2 - 2\frac{GMr}{c^2} +a^2,\\
	 		\rho^2 \defi r^2 + a^2 \cos^2 \theta,\\
	 		a \defi \nicefrac{L}{Mc},
 		\end{gather}
 	\end{subequations}
 	and $M$ is the mass of the black hole, while $L$ is its angular momentum. Despite the absence of spherical symmetry, the coordinates follow (traditionally) the naming scheme for spherical coordinates, $t,r,\theta,\varphi$.
 	
 	The Kerr space-time has a much richer structure than the Schwarzschild space-time: First off, it has not one, but two event horizons. They are determined by the zeroes of the $g_{tt}$-component w.r.t. $r$. This gives:
 	\begin{equation}
 		r_{\pm} = \frac{GM}{c^2} \pm \sqrt{\kl{\frac{GM}{c^2}}^2 -a^2}.
 	\end{equation}
 	As the inner horizon $r_-$ is hidden behind the outer one at $r=r_+$, just as any other physics behind the outer horizon, we shall refrain from looking into it.\footnote{One small comment is in order, though: The \emph{physical} singularity is, unlike in the Schwarzschild solution, not point-like but ring-like --- hence our putting quotes around \enquote{rotating point-mass} in section~\ref{sec:history}.} In section~\ref{sec:unphysical}, we shall see an example of a metric whose pathological or unphysical behaviour is not limited to the region behind the event horizon. Comparing the Kerr solution~\eqref{eq:Kerr} with the Schwarzschild solution~\eqref{eq:Schwarzschild}, one realises that while in the latter the $g_{tt}$-component was the inverse of the $g_{rr}$-component, this is not true any more in the Kerr solution. Thus, the zeroes of the $g_{rr}$-component will play a different role. They are at
 	\begin{equation}
 		r_{\text{E}}^\pm = \frac{GM}{c^2} \pm \sqrt{2\kl{\frac{GM}{c^2}}^2 -a^2\cos^2\theta}.
 	\end{equation}
 	Again, the lower value being hidden behind the outer event horizon, we shall only concern ourselves with $r_{\text{E}}^+$. The region between $r_{\text{E}}^+$ and $r_+$ is called the \enquote{ergo-sphere}. In this region, the frame-dragging of the black hole forces any time-like curve to be co-rotating with the black hole; moving withershins would involve moving outside the local light cone. The frame-dragging can also be captured by looking at the quantity
 	\begin{equation}
	 	\Omega_{\text{H}} = \frac{ac}{r_+^2 + a^2},
 	\end{equation}
 	which describes the angular velocity of the horizon.
 	
 	The area of the outer horizon can be evaluated to
 	\begin{equation}
	 	A_\text{H} = 4\pi(r_+^2 + a^2),\label{eq:KerrA}
 	\end{equation}
 	and on the horizon one has a surface gravity of
	\begin{equation}
		\kappa = \frac{c^2}{2}\frac{r_+ - r_-}{(r_+^2+a^2)},\label{eq:Kerrkappa}
	\end{equation}
	corresponding to a Hawking temperature of
 	\begin{equation}
	 	\kB T_\text{H} = \frac{\hbar}{2\pi c}\kappa = \frac{\hbar c}{4\pi}\frac{r_+ - r_-}{(r_+^2+a^2)}.\label{eq:KerrT}
 	\end{equation}
 	Looking at these quantities and the location of the horizon and its dependence on the angular momentum parameter $a$, one sees that as
 	\begin{equation}
	 	a\to \frac{GM}{c^2},
 	\end{equation}
 	outer ergo-surface and outer horizon merge, the surface gravity vanishes, and so does the Hawking temperature. This is the extremal Kerr solution. If one raises the value of $a$ even further, naked singularities occur. These solutions are (to our current understanding, experimental evidence, and belief in the cosmic censorship conjecture) not physical. This state of affairs is mirrored in the existence of the third law of black hole thermodynamics.
 	
 	\subsection{Interlude: Black Hole Thermodynamics}
 	This is an appropriate point to quickly recapitulate the four laws of black hole thermodynamics --- even though this will lead to some concepts being mentioned before they are introduced. The main reason for this is that the above space-times are astrophysically the most important ones and hence also the once most often encountered when black hole thermodynamics and the Hawking effect are discussed.
 	\begin{enumerate}
 		\setcounter{enumi}{-1}
 		\item The surface gravity is constant on the event horizon of stationary black holes. It is proportional to the black hole's temperature.
 		\item Perturbations of a stationary black hole are related to changes in energy:
 		\begin{equation}
	 		\dif E = \frac{\kappa}{8\pi}\dif A_\text{H} + \Omega_{\text{H}} \dif J + V_\text{H} \dif Q,
 		\end{equation}
 		where $\kappa$ is the surface gravity, $A_\text{H}$ is the area of the horizon, $\Omega_{\text{H}}$ is the angular velocity of the horizon, $J$ the black hole's angular momentum, $Q$ its charge, and finally $V_\text{H}$ the electric potential at the horizon. Charge will be introduced in the next subsection, subsection~\ref{sec:genBH}.
 		\item The horizon area is strictly increasing with time:
 		\begin{equation}
	 		\frac{\dif A_\text{H}}{\dif t} \geq 0.
 		\end{equation}
 		(This is subject to energy conditions being fulfilled.)
 		\item One cannot lower the temperature of a black hole in a physical process to $0$ (or below).
 	\end{enumerate}
 	The third law gets around the issue we encountered in the case of the Kerr metric, when a supercritical black hole started to exhibit naked singularities. Second and first law indicate the intimate relationship between black hole horizon area and the black hole's entropy, according to the Bekenstein entropy
 	\begin{equation}
	 	S_\text{B} = \frac{c^3\kB}{4 G\hbar} A_\text{H}.
 	\end{equation}
 	The first law is, as in traditional thermodynamics, a statement about energy conservation --- even though energy (and mass) in general relativity has a tendency to be an elusive and ambiguous concept. The zeroth law is, like the corresponding one in traditional thermodynamics basically a labelling of equivalence classes: The equivalence relation is \enquote{two systems are in thermal equilibrium with each other}, and each equivalence class is labelled by the temperature. This temperature is then the surface gravity at the horizon for black holes. If it was not constant all across the horizon it would not be fit to label these equivalence classes.
 	
 	More on the Hawking temperature will be covered briefly in section~\ref{sec:Hawking}.
 	 	
 	\subsection{General Black Holes I: The Kerr--Newman Family and Beyond}\label{sec:genBH}
 	There are other physically important black hole space-times. Since, however, we are mostly concerned with the two examples already provided --- Kerr and Schwarzschild --- we will discuss them only briefly.
 	
 	Already early in the history of general relativity, research was aimed at the interplay of the Einstein equations with Maxwellian electrodynamics. Starting from an action principle, the resulting system of partial differential equations (PDEs) is called the Einstein--Maxwell system or the Einstein--Maxwell equations. The simplest solution to these is the Reissner--Nordström solution,\footnote{Though, historically speaking, also Weyl's name could be associated with it, see its discussion in \cite{GriffithsPodolsky}. Jeffery then rediscovered it again in 1920.}
 	\begin{equation}
	 	\dif s^2 = -\kl{1-\frac{2GM}{c^2 r} + \frac{G Q^2}{4\pi \epsilon_0 c^4 r^2}}\dif t^2 + \kl{1-\frac{2GM}{c^2 r} + \frac{G Q^2}{4\pi \epsilon_0 c^4 r^2}}^{-1} \dif r^2 + r^2 \dif \Omega^2.
 	\end{equation}
 	It describes a static, spherically symmetric black hole, but now one that has besides a mass $M$ an electric charge $Q$, too. Obviously, its horizon structure is --- like in the Kerr geometry --- more complicated, even though this particular solution retains spherical symmetry and even staticity. Horizons are at
 	\begin{equation}
	 	r^\text{H}_\pm = \frac{GM}{c^2} \pm \sqrt{\frac{G^2 M^2}{c^4} - \frac{G Q^2}{4\pi \epsilon_0 c^4}}.
 	\end{equation}
 	Given the spherical symmetry, inner and outer horizon are necessarily concentric. Unlike the Kerr geometry, the causal behaviour is less pathological if the metric is taken seriously (or analytically completed) beyond the inner horizon. Between inner and outer horizon the metric is not static any more, as the $r$ coordinate becomes time-like. As this is of little relevance to an external observer, since this behaviour is hidden behind an event horizon, we can safely keep calling the Reissner--Nordström metric a static black hole. Behind the inner horizon the metric is again static. The Reissner--Nordström metric's Penrose diagram would be identical to that of the Kerr metric in the equatorial plane. Finally, as we would expect on physical grounds (or knowing the no-hair theorems), the metric turns into the Schwarzschild metric for $Q=0$. As one can see from the location of the horizons, if the charge exceeds
 	\begin{equation}
	 	Q_\text{crit} = \sqrt{4\pi \epsilon_0 G} M,
 	\end{equation}
 	the Reissner--Nordström solution becomes much more ill-behaved, just as the Kerr metric became for $a>GM/c^2$. Again, this relates to a vanishing surface gravity, and thus to the third law of black hole thermodynamics.
 	
 	The axisymmetry of the Kerr solution is a general feature of more general black hole space-times as encoded in the black hole uniqueness theorems, see \cite{Heusler}.\footnote{Even the Einstein--Yang--Mills system has (at least to our knowledge) still at least axisymmetry (and sometimes even spherical symmetry), despite having \enquote{hair}. This hair is often not captured in the uniqueness theorems which concern themselves (mostly) with the Einstein--Maxwell equations \cite{BHMenagerie}.} With this background, it should come as no surprise then that the metric of a rotating, charged black hole is of the axisymmetric variety. It reads:
 	\begin{subequations}
 		\begin{equation}
	 		\dif s^2 = - \frac{\Delta}{\rho^2} \kl{c \dif t - a \sin^2 \theta \dif \phi}^2 + \frac{\sin^2\theta}{\rho^2}\kl{(r^2+a^2)\dif \phi - a c \dif t}^2 + \frac{\rho^2}{\Delta} \dif r^2 + \rho^2\dif \theta^2,
 		\end{equation}
 		where
 		\begin{align}
	 		\Delta \defi r^2 - 2\frac{GMr}{c^2} +a^2 + \frac{G Q^2}{4\pi \epsilon_0 c^4},\\
	 		\rho^2 \defi r^2 + a^2 \cos^2 \theta,\\
	 		a \defi \nicefrac{L}{Mc}.
 		\end{align}
 	\end{subequations}
 	This is the so-called Kerr--Newman metric or family of metrics. The physical interpretation of the different quantities agrees with those in the Kerr and Schwarzschild geometry. As the Kerr metric, this metric has a ring singularity and two horizons, and just like the Kerr metric, the innermost regions are plagued with closed time-like curves. The behaviour in the domain of outer communication, however, is as well behaved as is needed for a physical interpretation, and applicability to astrophysical scenarios. It is important to appreciate that this changes once one considers microscopic objects: Charged, \enquote{rotating} (in the sense that they have spin) objects like electrons and protons can \emph{not} be interpreted as being represented by this metric. The two most obvious issues with treating microscopic objects with (macroscopic) metrics are: (1) Given spin and mass it would correspond to a ring singularity already of a size, $\approx \SI{e-13}{m}$. This scale is ruled out by experiments concerning themselves with much smaller length scales on which flat space quantum electrodynamics happens to be perfectly adequate. (2) Treating them either as only rotating or only charged objects results in horizon length scales below the Planck scale of $\ell_\text{Planck}\approx\SI{1.6e-35}{m}$. Without some quantum theory of gravity (or a very good reason why one would not need it) this similarly is nonsensical. The occurrence of these phenomena is quickly explained by a look at the location of inner and outer horizons, now located at:
 	\begin{equation}
	 	r_{\pm} = \frac{GM}{c^2} \pm \sqrt{\kl{\frac{GM}{c^2}}^2 -a^2 -\frac{G Q^2}{4\pi \epsilon_0 c^4}}.
 	\end{equation}
 	As before, there are critical values for charge and angular momentum above which the solution becomes unphysical, and again this relates to vanishing Hawking temperature.
 	
 	Like with the previous metrics, it is easy enough to give expressions for the area of the horizon:
 	\begin{equation}
 		A_\text{H} = 4\pi(r_+^2+a^2).
 	\end{equation}
 	The surface gravity is
 	\begin{equation}
	 	\kappa = \frac{c^2 r_+ - GM}{r_+^2+a^2}.
 	\end{equation}
 	
 	While it is possible to extend most black hole space-times to other \enquote{cosmological backgrounds} --- the present examples are usually considered asymptotically flat --- by the introduction of cosmological constants (and thus consistent with this phrasing changing the asymptotic behaviour), we shall not describe these possibilities as we will not encounter them in the main part of the thesis. For this we refer to \cite{ExactSolsEinstein} and \cite{GriffithsPodolsky}. An extension of the thesis along these lines would be easily possible.

 	\subsection{General Black Holes II: Dirty Black Holes}\label{sec:dirty}
 	If we focus on general, static, spherically symmetric black hole space-times, we arrive at the notion of \enquote{dirty black holes}. The reason for the introduction of this terminology was to provide a general framework encapsulating common features of models combining classical fields (of various kinds) with spherically symmetric and static space-times. For a list of space-times fitting into this scheme, see \cite{DirtyBHs}. These solutions now, unlike the previous examples of black holes (and even though some of these can be described as \enquote{dirty black holes}), need \emph{not} be solutions to the \emph{vacuum} Einstein equations. We will therefore encounter the energy-momentum tensor in this example. The form chosen to describe this family of black holes,
 	\begin{equation}
	 	\dif s^2 = -e^{-2\Phi(r)}\kl{1-\frac{b(r)}{r}}\dif t^2 + \frac{\dif r^2}{1-\frac{b(r)}{r}} + r^2(\dif \theta^2 + \sin^2\theta\dif\phi^2),
 	\end{equation}
 	is also encountered in the context of wormholes, see \cite{MorrisThorne}. While this reference specifically asks for no horizon to exist to procure a notion of traversable wormholes, in our case we want the exact opposite: We want horizons to exist such that we can use this family of metrics as a model of black holes (surrounded by fields or matter of some kind). First, let us fix some terminology: The function $\Phi(r)$ is called the anomalous redshift, while $b(r)$ is called the shape function. The domains of the coordinates are the usual ones for spherical symmetry. Fixpoints of the shape function,
 	\begin{equation}
	 	b(r_\text{H}) = r_\text{H},
 	\end{equation}
 	are possible locations of horizons.\footnote{More concretely these fixpoints correspond to so-called \enquote{putative horizons}, zeroes of the lapse function.\footnotemark ~~This corresponds to a stopping of the chosen time coordinate --- which may or may not be related to an actual horizon, thus the different nomenclature.} \footnotetext{Given the staticity and spherical symmetry of the metric under consideration, this will be a genuine horizon of the Killing, event, or apparent variety.} Since our analyses are (with the possible exception of the next section~\ref{sec:unphysical}) concerned with the domain of outer communication, we shall limit our discussion to the outermost such fixpoint, \emph{i.e.}, the one with the largest value of $r$. In the discussion of these black hole space-times, we shall mostly paraphrase \cite{DirtyBHs,VisserWormholes}.
 	
 	Given our interest in the Hawking effect and its properties, it is only natural to have a closer look at the surface gravity $\kappa$. At the putative, outermost horizon, one evaluates the surface gravity to be
 	\begin{subequations}
	 	\begin{align}
	 		\kappa &= \lim_{r\to r_\text{H}} \ed{2} \frac{\partial_r g_{tt}}{\sqrt{g_{rr} g_{tt}}},\\
	 		&= \frac{c^2}{2 r_\text{H}} e^{-\Phi(r_\text{H})} (1-b'(r_\text{H})).
	 	\end{align}
 	\end{subequations}
 	To gain more physical insight into this expression, we make use of the Einstein equations, which allow us to link anomalous redshift and shape function with energy density $\rho$, radial tension $\tau$, and transverse pressure $p$. We get:
 	\begin{subequations}
 		\begin{align}
	 		b(r) &= r_\text{H} + \frac{8\pi G}{c^4} \int_{r_\text{H}}^{r} \rho\, \tilde{r}^2 \dif \tilde{r},\\
	 		\Phi(r) &= \frac{4\pi G}{c^4} \int_{r}^{\infty} \frac{\rho - \tau}{1-b/\tilde{r}} \tilde{r} \dif \tilde{r},\\
	 		p(r) &= \frac{r}{4}\kl{\frac{\rho - \tau}{1-b/r} \cdot \frac{b-8\pi G\tau r^3/c^4}{r^2} - 2\tau'} -\tau.
 		\end{align}
 	\end{subequations}
 	This, finally, allows to give the surface gravity at the putative horizon in physically more relevant quantities as
 	\begin{equation}
	 	\kappa = \frac{\exp\kl{-\frac{4\pi G}{c^4}\int_{r_\text{H}}^{\infty} \frac{\rho - \tau}{1-b/\tilde{r}} \tilde{r} \dif \tilde{r}}}{2 r_\text{H}} \kl{1- \frac{8\pi G \rho(r_\text{H}) r_\text{H}^2}{c^4}}.
 	\end{equation}
 	Notice that if the weak energy condition (WEC) is fulfilled, it means that
 	\begin{subequations}
	 	\begin{align}
		 	\rho-\tau&\geq 0,\\
		 	\rho &\geq 0.
	 	\end{align}
 	\end{subequations}
 	Adding the validity of the null energy condition (NEC) along the radial direction allows to further constrain the behaviour of $\Phi(r_\text{H})$, since then
 	\begin{equation}
	 	\Phi(r_\text{H}) = \frac{4\pi G}{c^4} \int_{r_\text{H}}^{\infty} \frac{\rho - \tau}{1-b/\tilde{r}} \tilde{r} \dif \tilde{r} \geq 0.
 	\end{equation}
 	Hence, if both the NEC and the WEC are valid, we know that
 	\begin{subequations}
 		\begin{align}
		 	\kappa_\text{dirty} &= \kappa_\text{Schwarzschild} \times \exp\kl{-\frac{4\pi G}{c^4}\int_{r_\text{H}}^{\infty} \frac{\rho - \tau}{1-b/\tilde{r}} \tilde{r} \dif \tilde{r}} \kl{1- \frac{8\pi G \rho(r_\text{H}) r_\text{H}^2}{c^4}},\\
		 	&\leq \kappa_\text{Schwarzschild} = \frac{\hbar c}{4\pi \kB r_\text{H}},
	 	\end{align}
 	\end{subequations}
 	if we assume a Schwarzschild black hole of the same $r_\text{H}$. There remains the issue whether $1- 8\pi G \rho(r_\text{H}) r_\text{H}^2/c^4$ can become negative or not, and \emph{if} it happens what this means. For this, first of all notice that we can define $b(r)\ifed 2 G m(r)/c^2$, where $m(r)$ can be given the interpretation of a mass within a sphere of radius $r$, so
 	\begin{equation}
	 	1- \frac{8\pi G \rho(r_\text{H}) r_\text{H}^2}{c^4} = 1-\frac{2G m'(r_\text{H})}{c^2}.
 	\end{equation}
 	So if $1- 8\pi G \rho(r_\text{H}) r_\text{H}^2/c^4$ becomes negative or zero, it corresponds to
 	\begin{equation}
	 	m'(r_\text{H}) \geq \frac{c^2}{2 G}, \qquad \Longleftrightarrow\qquad b'(r_\text{H}) \geq 1.
 	\end{equation}
 	However, given our definition of the outermost horizon, we know that
 	\begin{equation}
	 	b'(r) \leq 1
 	\end{equation}
 	in the domain of outer communication. At worst, we can therefore reach equality and $1- 8\pi G \rho(r_\text{H}) r_\text{H}^2/c^4=0$; this situation happens, for example, in the case of an extremal Reissner--Nordström black hole. This factor vanishing hence indicates the presence of some sort of extremal horizon --- it being negative corresponds (in our sense) to some ill-defined notion of outermost horizon.
 	
 	\section{Unphysical Black Holes: The Taub--NUT Family}\label{sec:unphysical} 
 	While the above examples of black hole space-times all had concrete physical applications and interpretations, these are not the only possible space-times exhibiting properties of black holes. In particular, while many more physically relevant solutions can be imagined (for example, adding a cosmological constant or various dynamical scenarios, including, but not limited to multi-black hole solutions, and evaporating black holes [with or without remnants]), there is also a plethora of unphysical solutions. Depending on how conservative one chooses one's personal physics, this might already include $N+1$-dimensional space-times, as discussed further below in section~\ref{sec:ndim}, but we shall not dwell on the physicality of extra dimensions. Rather, in the current section, we will dissect one particular example of unphysical black hole solutions: A special case of the Taub--NUT family. This family is named after its discoverers Abraham Haskel Taub, who laid the groundwork in 1951, Ezra T. Newman, Louis A. Tamburino, and Theodore W. J. Unti, who together generalised the original Taub space-time to what will be discussed here, in 1963. Given its pathological nature, it is not quite as well-known as the previous examples of space-times. Correspondingly, this family is occasionally rediscovered. The rediscovery by Hongsheng Zhang in \cite{ZhangTwisted}, prompted at least three re-examinations \cite{ChenJingTwisted,OngTwisted,TwistedUnphysical} of this space-time, all concluding with the realisation of its previous discovery. In the following, we will follow our contribution to this discussion, \cite{TwistedUnphysical}. Given the nature of this solution, we will forego the use of SI units in this section.
 	
 	Much more on the whole class of the Taub--NUT solution family, including generalisations beyond it, can be found in \cite{GriffithsPodolsky}, as well as, though with less emphasis on the physical focus to be set here, in \cite{HawEll,ExactSolsEinstein}. \cite{GeodesicsTaubNUT} contains an exhaustive study of the geodesics in this space-time. Before heading into our own discussion, it is worthwhile to spend a few words on the presentation given in the first mentioned reference of this paragraph, \cite{GriffithsPodolsky}. The following, defining equation we shall use, is precisely the one given in \cite{GriffithsPodolsky} as equation~(12.1). It is algebraically equivalent to the rediscovery in \cite{ZhangTwisted}, there as equation (1). Further variants discussed in \cite{GriffithsPodolsky}, as equation~(12.3), are different interpretations involving a different base manifold (by identifying previously separate points in the original manifold), and also an additional coordinate transformation. Even further changing the base manifold and, correspondingly, the range of coordinates, discussed in section~12.2 of \cite{GriffithsPodolsky}, leads to a singularity-free interpretation due to Misner. We shall only be concerned with the variant defined below.
 	
 	Concretely, the line element we shall use to define our particular version of Taub--NUT is given by
 	\begin{align}
 		\dif s^2 =& - \left( \frac{r^2-2mr-a^2}{r^2+a^2}\right) (\dif t - 2 a \cos\theta\; \dif \phi)^2 
 		+ \left( \frac{r^2+a^2}{r^2-2mr-a^2}\right) \dif r^2 
 		\nonumber\\
 		&\qquad+ (r^2+a^2) (\dif\theta^2 +\sin^2\theta\; \dif \phi^2).\label{eq:TaubNUT}
 	\end{align}
 	The coordinate ranges are: $t \in (-\infty,\infty)$, $r\in(-\infty,\infty)$, $\theta\in[0,\pi]$, $\phi\in[0,2\pi]$ --- where, as before, we ignore issues of coordinate singularities at the poles ($\theta =0$ or $\theta=\pi$) to keep the discussion transparent. As this is only a special case of Taub--NUT, we shall refer to this as \enquote{twisted black holes}, as suggested in \cite{ZhangTwisted}. We demonstrate several things related to these twisted black holes:
 	\begin{itemize}
 		\item They are \emph{not} asymptotically flat.
 		\item They connect two asymptotic regions via a time-like wormhole which is hidden behind an event horizon.
 		\item Their region(s) outside the event horizon(s) is (are) riddled with closed time-like curves. In the terminology of \cite{CausalHierarchy}, the domain(s) of outer communication is (are) \enquote{totally viscous}.\footnote{A space-time $M$ is called totally vicious, if for all $p\in M: I^+(p)=I^-(p)$, \emph{i.e.} each point's chronological future equals its chronological past. \enquote{Chronological} can be read as \enquote{causal} further restricted to time-like curves.}
 	\end{itemize}
 
 	\subsection{Massless Case}
 	As a first step it is useful to look at the case of mass zero, that is setting the parameter $m=0$ (which indeed can be identified with a notion of mass, see equation~\eqref{eq:TaubNUTChristoffel}) in equation~\eqref{eq:TaubNUT}. This results in the following, simplified line element:
 	\begin{equation}
	 	\dif s^2 = - \left( \frac{r^2-a^2}{r^2+a^2}\right) (\dif t - 2 a \cos\theta\; \dif \phi)^2 
	 	+ \left( \frac{r^2+a^2}{r^2-a^2}\right) \dif r^2 
	 	+ (r^2+a^2) (\dif\theta^2 +\sin^2\theta\; \dif \phi^2).
 	\end{equation}
 	All of what we will show in the following generalises to the case of $m\neq 0$. But as the analysis is naturally simpler in the present case, we will use this case to ease into the problem.
 	
 	The corresponding Ricci tensor fulfils
 	\begin{equation}
 		R_{ab} = 0,
 	\end{equation}
 	thus this is indeed a vacuum solution to the Einstein equations, while the Riemann tensor itself is non-vanishing, seen in the most compact way by the likewise non-vanishing of the Kretschmann scalar (even though the non-vanishing components of the Riemann tensor are a necessary ingredient in this),
 	\begin{equation}
 		R^{abcd} R_{abcd} = 
 		- \frac{48 a^2(r^2-a^2)([r^2+a^2]^2-[4ar]^2)}{(r^2+a^2)^6}.
 	\end{equation}
 	Many of these calculations are verified the easiest way by resorting to computer algebra systems (CAS).
 	\subsubsection{A Wormhole Hidden Behind a Horizon}
 	The claim made in \cite{ZhangTwisted} was the discovery of a new black hole space-time. Thus, it is prudent to actually analyse the horizon structure first. This will also allow us a closer look at the coordinate ranges of the solution given above. 
 	
 	To look for horizons, we look at the $rr$-component of the inverse metric and its zeroes
 	\begin{equation}
 		g^{rr} = \kl{\frac{r^2-a^2}{r^2+a^2}} \stackrel{!}{=} 0.
 	\end{equation}
 	This is easily solved, giving two horizons, located at
 	\begin{equation}
 		r_\text{H}^\pm = \pm a.
  	\end{equation}
  	Following \cite{GriffithsPodolsky}, we shall call the regions outside these two horizons NUT regions, while the region between the horizons shall be referred to as a Taub cosmology. Before having a closer look at the Taub cosmology and its connection to the horizons, let us have a quick glance at the ergo-surfaces and ergo-regions:
  	
  	The ergo-regions are found by looking at the zeroes of
  	\begin{equation}
  		g_{tt} =  -\left( \frac{r^2-a^2}{r^2+a^2}\right) \stackrel{!}{=} 0,
  	\end{equation}
  	again giving $\pm a$ as solutions. Hence, we have an empty ergo-region, as the horizons agree with the ergo-surfaces --- as in the Schwarzschild solution modulo there being \emph{two} horizons.
  	
  	If we examine the $r$ coordinate more closely, we see that the metric is invariant under sign changes of $r$. Looking at the Kretschmann scalar at $r=0$, or the components of the Riemann tensor in a tetrad basis, one sees that neither vanishes or diverges at $r=0$. More concretely, $R^{abcd} R_{abcd}\lvert_{r=0} = {48/a^4}$. All three facts hint at the need to actually consider the full real line as coordinate range of $r$. The final point to convince us of this is a careful evaluation of the area of two-surfaces at constant $t$ and $r$. The induced two-metric evaluates to
  	\begin{equation}
  		\dif s^2_2 =  - \left( \frac{r^2-a^2}{r^2+a^2}\right) (4 a^2 \cos^2\theta\; \dif \phi^2 )+ (r^2+a^2) (\dif\theta^2 +\sin^2\theta\; \dif \phi^2).
  	\end{equation}
  	This expression both remains invariant under the previously mentioned change of sign of $r$, and evaluates to a perfectly well-behaved two-metric at $r=0$:
  	\begin{equation}
  		\left.\dif s^2_2\right\lvert_{r=0} =  4 a^2 \cos^2\theta\; \dif \phi^2  + a^2 (\dif\theta^2 +\sin^2\theta\; \dif \phi^2).
  	\end{equation}
  	We can evaluate the surface area $A_0$ of the topological two-sphere at $r=0$ exactly (in terms of the complete elliptic integral of second kind, see~\ref{sec:elliptic} for a \emph{very} short reminder; the subsection on the massive case will see more use of this):
  	\begin{subequations}
  		\begin{align}
	  		A_0 &\defi 2\pi a^2 \int_0^\pi\sqrt{ \sin^2\theta + 4 \cos^2\theta} \; \dif\theta,   \\
	  		&=  4\pi a^2 \int_0^\pi\sqrt{ 1 - {\textstyle{\frac{3}{4}}} \; \sin^2\theta} \; \dif\theta,\\
	  		&= 4\pi a^2 \times 2\; \mathrm{EllipticE}\left({\textstyle{\frac{\sqrt{3}}{2}}}\right).\label{eq:areaorigin}
  		\end{align}
  	\end{subequations}
  	We deduce from this that, indeed, the natural range of $r$ is, as claimed when introducing the metric, $(-\infty,\infty)$. Note that this was missed in \cite{ZhangTwisted,ChenJingTwisted}, while \cite{OngTwisted} refers to our work \cite{TwistedUnphysical} for these matters. Integrating the surface element at the location of the horizon yields an area of
  	\begin{equation}
	  	A_\text{H} = 8a^2\pi.
  	\end{equation}
  	This can be compared to the area of a zero-mass Kerr space-time --- which, as it turns out, is just Minkowski space-time in oblate spheroidal coordinates; As to be expected (and a quick look at the formula for the area of the horizon in Kerr proves it), the area is $0$, a marked contrast to Taub--NUT.
  	
  	The better way to interpret the region \emph{between} the horizons therefore seems to be as an inter-universal wormhole, following \cite{VisserWormholes}. As it is hidden behind horizons, it is necessarily non-traversable. Nonetheless, the terminology as a Taub wormhole might be more appropriately than \enquote{Taub cosmology}. However, also this terminology has its merits: It highlights a cosmological feature, best emphasised by a minor coordinate renaming. For $r\in(-a,a)$, the $r$-coordinate becomes time-like between the horizons, just as the $t$-coordinate becomes space-like. This suggests the relabelling $r\leftrightarrow t$, resulting in the new metric
  	\begin{equation}
  		\dif s^2 = - \left( \frac{a^2+t^2}{a^2-t^2}\right) \dif t^2  + \left(\frac{a^2- t^2}{a^2+t^2}\right) (\dif r - 2 a \cos\theta\; \dif \phi)^2 
  		+ (a^2+t^2) (\dif\theta^2 +\sin^2\theta\; \dif \phi^2),
  	\end{equation}
  	where $t\in(-a,a)$. This coordinate range is important as we are otherwise not between the horizons of the full space-time!
  	
  	This new form of the metric corresponds to an anisotropic cosmology, more specifically, to Bianchi type~IX. This was precisely Taub's original solution. It corresponds to a universe undergoing a bounce in the $\theta$-$\phi$ directions, with a moment of maximum expansion in the $r$ direction.
  	
  	This structure, then, can be used to construct a maximal analytic extension along the lines of that for Reissner--Nordström or Kerr space-times.
  	
 	\subsubsection{Absence of Asymptotic Flatness}
 	While this demonstrates surprising structure behind the horizons, this is yet of little reason to worry about the physical usefulness. This changes once one looks at the Taub--NUT space-time's asymptotic properties. Specifically, let us test the metric~\eqref{eq:TaubNUT} for asymptotic flatness, as this is one of the standard litmus tests for the interpretation of a solution of the Einstein equations as one describing an isolated black hole. By its very nature, testing for asymptotic flatness involves a careful look at the behaviour for large values of $r$ --- provided the $r$-coordinate can be combined with an appropriate notion and interpretation of radial distance. This is indeed the case in the present example. As $a$ is the length scale set by the metric (at least in the massless case considered here), we look at $r\gg a$, and then
 	\begin{equation}
 		\dif s^2 \approx -  (\dif t - 2 a \cos\theta\; \dif \phi)^2  + \dif r^2 + r^2 (\dif\theta^2 +\sin^2\theta\; \dif \phi^2).
 	\end{equation}
 	However, the term $(\dif t - 2 a \cos\theta\; \dif \phi)^2$ breaks asymptotic flatness.
 	
 	This becomes even more noticeable if one looks at infinitesimal circles around the axis of rotation and compares their circumference $\mathcal{C}$ with their radius $\mathcal{R}$. As we chose the circle to be infinitesimal, these can be easily approximated to be
 	\begin{equation}
 		\mathcal{C} = 2 \pi \; \sqrt{g_{\phi\phi}},
 	\end{equation}
 	and
 	\begin{equation}
 		\mathcal{R} = \sqrt{g_{\theta\theta}} \; \theta = \sqrt{r^2+a^2} \; \theta.
 	\end{equation}
 	From these two equations it is possible to calculate the angle surfeit or angle deficit $\Delta\Phi$. In our case, it turns out to be a drastic (diverging) angle surfeit:
 	\begin{subequations}
 		\begin{align}
 			\Delta \Phi &= \lim_{\theta\to0} \left(\frac{\mathcal{C}}{\mathcal{R}}\right) - 2\pi, \\
 			&= 2\pi \left\{\lim_{\theta\to 0}\left({\frac{\sqrt{g_{\phi\phi}}}{\sqrt{r^2+a^2} \; \theta}}\right) - 1 \right\}, \\
 			&\to \infty.
 		\end{align}
 	\end{subequations}
 	As this is true along the full axis of rotation, it remains true for $r\to\infty$, thus in the asymptotic regime, and asymptotic flatness is violated.
 	
 	Another, third proof of the failure of being asymptotically flat is provided by looking at the $g_{\phi\phi}$ component of the metric in this asymptotic region of $r\gg a$. Then
 	\begin{equation}
 		g_{\phi\phi} \approx r^2\sin^2\theta - 4 a^2 \cos^2 \theta,
 	\end{equation}
 	which is negative if
 	\begin{subequations}
 		\begin{equation}
 			\tan^2\theta \lesssim \frac{4a^2}{ r^2},
 		\end{equation}
 		\emph{i.e.}, if
 		\begin{equation}
 			r \;|\!\tan\theta| \lesssim 2a.
 		\end{equation}
 	\end{subequations}
 	As the rotational axis is broken in two parts by the event horizon (see picture~\ref{fig:Taub-NUT})\footnote{While the picture is modelled on the massive case, the qualitative behaviour is the same.}, we can separate the previous condition into two cases: Once again employing the condition $r\gg a$ this then leads to the conditions
 	\begin{subequations}
 		\begin{equation}
	 		r \; \theta \lesssim 2a,
	 	\end{equation}
	 	close to the \enquote{North pole} ($\theta\approx0$), and
	 	\begin{equation}
	 		r\; (\pi-\theta) \lesssim 2a,
	 	\end{equation}
 	\end{subequations}
 	when close to the \enquote{South pole} ($\theta\approx \pi$). If one now constrains oneself to small azimuthal circles of constant $(t,r,\theta)$ around the axis of rotation by going through the full range of $\phi \in [0,2\pi]$, with a radius $R\lesssim 2a$, the resulting curve is indeed a closed time-like curve.
 	
 	So far, the existence of closed time-like curves in the asymptotic region $r\gg a$ only served to illustrate the absence of asymptotic flatness. However, below we will see that this is a generic feature of points outside the event horizon.
 	
 	\subsubsection{Causal Structure}
 	As we have shown above, in the asymptotic region and close to the axis of rotation, closed time-like curves are possible. The obvious questions are: How much of this was related to the asymptotic limit $r\gg a$? How compatible is this condition to being close to the axis of rotation? For this, we shall investigate the exact formula for $g_{\phi\phi}$ more closely, which reads
 	\begin{equation}
 		g_{\phi\phi} = \frac{4a^2(r^2-a^2) + (r^4+6a^2r^2-3a^4)\sin^2\theta}{r^2+a^2}.
 	\end{equation}
 	If we want to look for closed time-like curves along $\phi$ direction, we will need to find whether $g_{\phi\phi}<0$ can be achieved. For this, let us again fix $(t,r,\theta)$. Looking for the boundary of a region where such closed time-like curves are possible corresponds to $g_{\phi\phi}=0$, which happens if
 	\begin{subequations}
 		\begin{equation}
 			\sin^2\theta = \frac{4a^2(r^2-a^2)}{r^4+6a^2r^2-3a^4}
 			= \frac{4a^2(r^2-a^2)}{(r^2+a^2)^2 + 4a^2(r^2-a^2)},
 		\end{equation}
 		or equivalently if
 		\begin{equation}
 			\tan^2\theta = \frac{4a^2(r^2-a^2)}{(r^2+a^2)^2}.
 		\end{equation}
 	\end{subequations}
 	Therefore, $\sin^2\theta\in (0,1)$ if $|r|>a$, up into the asymptotic region $r\to\infty$, for which we already showed the existence of closed time-like curves around the rotational axis. The shape of this boundary is cigar-like, if one allows for cigars of infinite length. There are two cigars around the horizon: One surrounding the North pole at $\theta=0$, and one around the South pole. As we still have the symmetry $r\to -r$ in the condition $|r|>a$, there are two more cigars in the \enquote{other} universe of negative $r$, giving a total of four cigar-shaped regions containing azimuthal closed time-like curves.
 	
 	In order to now prove our earlier statement, that the domain(s) of outer communication is (are) totally vicious, we need to show more than this. Let us temporarily fix $r$ and $\theta$. On the cylinder spanned by $t$ and $\phi$ we then have the induced two-metric
 	\begin{equation}
 		\dif s^2_2 = - \left( \frac{r^2-a^2}{r^2+a^2}\right) (\dif t - 2 a \cos\theta\; \dif \phi)^2 
 		+ (r^2+a^2) \sin^2\theta\; \dif \phi^2.
 	\end{equation}
 	If a curve $t(\phi)$ is time-like, this then means that
 	\begin{equation}
 		\left( \frac{r^2-a^2}{r^2+a^2}\right) \left(\frac{\dif t}{\dif \phi}- 2 a \cos\theta\right)^2  >  (r^2+a^2) \sin^2\theta.
 	\end{equation}
 	The range of the derivative $\dif t/\dif \phi$ can thus be given for time-like curves as
 	\begin{equation}
 		\left(\frac{\dif t}{\dif \phi}\right)\in 
 		\left(-\infty, 2a\cos\theta - \frac{(r^2+a^2)\sin\theta}{\sqrt{r^2-a^2}}\right) \cup 
 		\left(2a\cos\theta + \frac{(r^2+a^2)\sin\theta}{\sqrt{r^2-a^2}},+\infty\right).\label{eq:tphi-interval}
 	\end{equation}
 	Were we to fix $\theta\notin \{0,\pi\}$, we can then make $r$ large enough to leave the cylinder with closed time-like curves, as this range turns into
 	\begin{equation}
 		\left(\frac{\dif t}{\dif \phi}\right)\in 
 		\left(-\infty, - r\sin\theta\right) \cup 
 		\left(+r\sin\theta,+\infty\right),
 	\end{equation}
 	which is just the usual time-like cone expected for a flat space-time.
 	
 	However, if we go about it the other way, that is fixing $r$ and then investigating what freedom $\theta$ has, we come to realise the following issue: If we approach the axis by varying $\theta$ (that is, either $\theta\to0$ or $\theta\to\pi$), at some point one of the two intervals in equation~\eqref{eq:tphi-interval} will cross the origin. The meaning of this crossing is that beyond this tipping point $\dif t/\dif \phi$ will become a time-like curve. This tipping point can also be rephrased as
 	\begin{equation}
 		\tan\theta = \pm \frac{2a \sqrt{r^2-a^2}}{r^2+a^2},
 	\end{equation}
 	just a rephrasing of the condition that $g_{\phi\phi}=0$, which we used as the starting point of our discussion. This also links the cigar shapes introduced earlier to any general closed time-like curve, besides just those merely in $\phi$ direction.
 	
 	Suppose one starts from any freely chosen point with a future-directed velocity: With this initial condition one can then construct a trajectory leading to a tip-over point as constructed just now. At this point, the light-cone has tipped far enough that one can encircle the axis along a helical curve long enough to leave the \enquote{cigar} again and reconnect the trajectory through the starting point's past light cone. This is illustrated in figure~\ref{fig:Taub-NUT}. The described procedure works as long as one does not cross the horizon --- each of the two domains of outer communication considered separately in its own right as a Lorentzian manifold thus becomes totally vicious.
 	
 	\subsubsection{\enquote{Mass} and Geodesic Coordinate Acceleration}
 	To finish the discussion of the massless case, we can take a look at
 	\begin{equation}
 		g_{tt} = -1 + \frac{2a^2}{r^2} + O(1/r^4).
 	\end{equation}
 	Comparing this, for example, with the weak field approximation to identify a possible gravitational potential, we note that we can find no match to the behaviour of a traditional gravitational potential. The only meaningful way to assign a mass, would therefore be to indeed associate mass zero to this space-time. However, it is apparent that more is happening.
 	
 	Looking then at the component
 	\begin{align}
 		\Gamma^r_{tt} &= \frac{2a^2 r (r^2-a^2)}{(r^2+a^2)^3 }, \\&\approx \frac{2a^2}{r^3}
 	\end{align}
 	of the Christoffel symbol, we observe a coordinate acceleration of a test particle dropped at rest following an inverse cube law. This again contrasts with the inverse square law we expect of a mass. In either NUT region the acceleration is towards the horizon.
 	
 	\subsection{Massive Case}
 	To consider the massive case, let us repeat the full (massive) Taub--NUT metric:
 	\begin{align}
	 	\dif s^2 =& - \left( \frac{r^2-2mr-a^2}{r^2+a^2}\right) (\dif t - 2 a \cos\theta\; \dif \phi)^2 
	 	+ \left( \frac{r^2+a^2}{r^2-2mr-a^2}\right) \dif r^2 
	 	\nonumber\\
	 	&\qquad+ (r^2+a^2) (\dif\theta^2 +\sin^2\theta\; \dif \phi^2).
 	\end{align}
 	The statements about the massless case and its curvature invariants remain qualitatively the same: We have a Ricci-flat geometry, whose Kretschmann scalar now evaluates to
 	\begin{equation}
	 	R^{abcd} R_{abcd} =  \frac{a^2-m^2}{a^2} \; (R^{abcd} R_{abcd} )_{m=0} 
	 	+  \frac{192\; ma^2r (3r^2-a^2)(r^2-3a^2)}{(r^2+a^2)^6},
 	\end{equation}
 	again requiring the existence of non-vanishing components of the Riemann curvature tensor.
 	
 	The discussion of the massless case caries through completely, with minor complications due to more involved expressions. To illustrate the similarity: Figure~\ref{fig:Taub-NUT} was created based on the massive case with $m=a=1$. The coordinates of the horizons change slightly to
 	\begin{equation}
 		r_\text{H}^\pm = m \pm \sqrt{m^2+a^2}.
 	\end{equation}
 	Again, the second horizon cannot be ignored, together with the accompanying need to extend the range of the $r$ coordinate to $(-\infty,\infty)$. However, the \enquote{mirror symmetry} changes: Instead of invariance of the metric under $r\to -r$, we now have to simultaneously change $r\to -r$ and $m\to-m$.
 	
 	It turns out that one can integrate the surface element exactly (a) even with mass, and (b) at general $r$, including, of course, the horizon. The general expression for the surface element is
 	\begin{equation}
	 	\dif s_2 = \sqrt{(a^2+r^2)^2+(3a^4+2a^2r(4m-3r)-r^4)\cos^2\theta}\dif\theta\dif\phi,
 	\end{equation}
 	which can be integrated in the following way:
 	\begin{subequations}
	 	\begin{align}
		 	\int_{r=\mathrm{const}}\dif s &= \int_{0}^{2\pi}\int_{0}^{\pi}\sqrt{(a^2+r^2)^2+(3a^4+2a^2r(4m-3r)-r^4)\cos^2\theta} \dif\theta\dif\phi,\\
		 	&= 2\pi \sqrt{4a^4+4a^2r(2m-r)}\int_{0}^{\pi}\sqrt{1-\frac{3a^4+2a^2r(4m-3r)-r^4}{4a^4+4a^2r(2m-r)}\sin^2\theta}\dif\theta,\\
		 	&= 4\pi \sqrt{a^4+a^2r(2m-r)}\mathrm{EllipticE}\kl{\pi;\sqrt{1-\frac{(a^2+r^2)^2}{4a^4+4a^2r(2m-r)}}}.
	 	\end{align}
 	\end{subequations}
 	Making use of the relation~\eqref{eq:ellipticaddition}, we can further simplify this to
 	\begin{equation}
	 	8\pi \sqrt{a^4+a^2r(2m-r)}\mathrm{EllipticE}\kl{\sqrt{1-\frac{(a^2+r^2)^2}{4a^4+4a^2r(2m-r)}}}.
 	\end{equation}

 	The horizon area then evaluates to a reasonable
 	\begin{equation}
	 	A_\text{H} = 8\pi\sqrt{(m^2+a^2)(2mr_\text{H}+a^2)}.
 	\end{equation}
 	A look at the general formula reveals furthermore that at $r=0$ all mass-dependence drops out --- hence the surface area at the origin has still the same value as in equation~\eqref{eq:areaorigin}. We see that the addition of mass does not change the behaviour at the origin, nor will there be any new hindrance to extending the coordinate range of $r$ to the full real line.
 	
 	Re-examining the $g_{tt}$ component's expansion in search for a mass term, we are now able to identify $m$ (as promised) as a mass:
 	\begin{equation}
 		g_{tt} = -1 +\frac{2m}{r} +  \frac{2a^2}{r^2} + O(1/r^3).
 	\end{equation}
 	Likewise, we can now see more familiar behaviour in the Christoffel symbol, too:
 	\begin{align}
 		\Gamma^r_{tt} &= \frac{(m[r^2-a^2]+2a^2r)(r^2-2mr-a^2)}{(r^2+a^2)^3 }, \\&\approx \frac{m}{r^2} + O(1/r^3).\label{eq:TaubNUTChristoffel} 
 	\end{align}
 	However, there is a small caveat related to the range of $r$ in the interpretation of $m$ as mass: The mass on the side of positive $r$ is positive, but on the side of negative $r$, the mass is effectively $-m$. Inserting this in the Christoffel symbol~\eqref{eq:TaubNUTChristoffel}, we see that for positive $mr$ we get the familiar acceleration towards the nearest horizon. But if $mr<0$, the acceleration is away from the nearest horizon. More on the behaviour and implications of negative asymptotic mass can be found in \cite{VisserWormholes} and references therein.
 	
 	Lastly, let us describe how to derive the cigar shape as shown in figure~\ref{fig:Taub-NUT}: Solving the $g_{\phi\phi}=0$ for $\theta$ results in the expressions
 	\begin{subequations}
 		\begin{align}
		 	\theta_\text{N} &= \arccos\kl{\frac{a^2+r^2}{\sqrt{r^4+6a^2r^2-3a^4-8ma^2r}}},\text{ or}\\
		 	\theta_\text{S} &= \pi - \arccos\kl{\frac{a^2+r^2}{\sqrt{r^4+6a^2r^2-3a^4-8ma^2r}}}.
	 	\end{align}
 	\end{subequations}
 	This mirrors the fact that one has a cigar both surrounding the North and the South pole axis. The distance of this surface from this axis at radius $r$ is then given as
 	\begin{equation}
	 	d = r\sin\theta_\text{N/S}.
 	\end{equation}
 	This was used to plot the cigars.
 	
 	\subsection{Summary}
 	Let us collect some of the results we derived in the previous pages.
 	\begin{itemize}
 		\item The massive case contains a region of negative mass.
 		\item Both massless and massive case have two domains of outer communication that are each by itself \enquote{totally viscous}, that is completely infested with closed time-like curves.
 		\item The axis has infinite angle surfeit. This is a type of space-time singularity.
 	\end{itemize}
 	All these points are unphysical in astrophysical contexts.\footnote{At least according to the present state of knowledge. This state of affairs is incredibly unlikely to change.} Of less physical concern are the following two points:
 	\begin{itemize}
 		\item The solution is not asymptotically flat. While this precludes the interpretation as an isolated black hole, if it was just the lack of asymptotic flatness other physical interpretations might have been possible.
 		\item The solution contains a inter-universal wormhole hidden by an event horizon. As the wormhole is eternal, it does not violate no-go theorems on topology change. As the wormhole is inter-universal and hidden behind event horizons, the wormhole itself does not endanger the causality.
 	\end{itemize}
 	Both of these last two points' innocuousness is rendered entirely moot by the first three points. This is hard to overstress.
 	
 	\begin{figure}
 		\centering
 		\includegraphics[width=.75\textwidth]{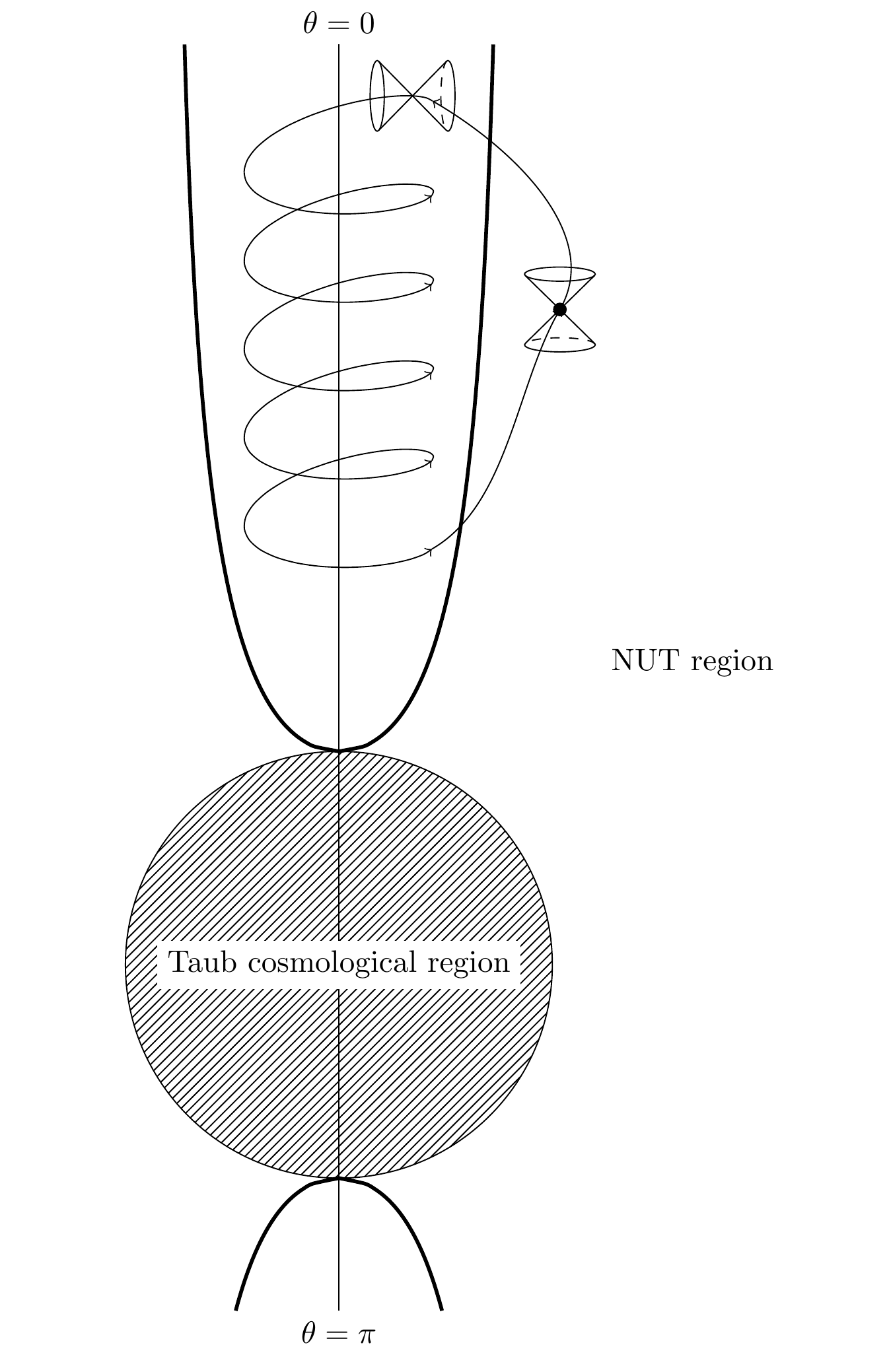}
 		\caption[Sketch of Acausal Behaviour of Taub--NUT]{A sketch of one half (\emph{i.e.}, either $r>0$ or $r<0$) of the massive Taub--NUT space-time with $m=a=1$. Outside the even horizon (shaded region) behind which lies the time-dependent, cosmological Taub region, are cigar-shaped regions surrounding the axes of rotation at the North and South pole of the horizon. Within these regions closed time-like curves are possible. An observer outside the cigar-shaped region (denoted by the black dot) can enter this region, circle around the axes, and exit the region in such a way that they can close their worldline to a closed time-like curve. Beware that the picture does mix spatial and time-like coordinates in a non-trivial way to illustrate this process.}
 		\label{fig:Taub-NUT}
 	\end{figure}

	\FloatBarrier
 	\section{Curved Space-Time Quantum Field Theory} 
 	
 	In this section and the following ones we shall limit ourselves to the study of free, real, massless scalar quantum fields. This is not only a technical restraint: Considering different spins, or including particle masses, let alone interactions, does change both the necessary assumptions, and also changes results, sometimes significantly. For example, if one wants to include fermions, the introduction of a spin bundle will lead to topological constraints.\footnote{The second Stiefel--Whitney class has to vanish for the existence of a spin bundle on a given Lorentzian manifold \cite{Wald,ExoticSmoothness,SpinorialGeo}.} Similarly, for a massless spin-1 vector field (\emph{i.e.}, Maxwellian electrodynamics) topological consideration may have to enter depending on whether one wants to quantise the field strength tensor directly, or an underlying four-potential. While most of these considerations will not enter our discussion of the Hawking effect in the context of sparsity, this latter consideration will be of some importance in the analysis of the electromagnetic analogue as developed in chapter~\ref{ch:analogues}: As the space-time itself is derived from electromagnetism, it seems much more natural to quantise a spin-1 field in this context. There exist both \enquote{traditional} CSTQFT approaches to this (as already a quick perusal of \cite{BirrellDavies} and its references will show\footnote{For a more recent, Dirac quantisation of linear electrodynamics, see \cite{DiracQuLinEdyn}.}), as well as \enquote{modern algebraic} treatments, see for example \cite{Spin1QEI,QuMaxwellNdim,MaxwellCSTQFT12,HadamardSpin1,GuptaBleulerSpin1CST,PreMetricAQFT,PreMetricQEI}. Already on the side of classical field equations, one can observe that curved space-time implementations of classical electrodynamics are non-trivial: As discussed in \cite{FriedlanderWaveEq}, the existence of constraints (in the form of charge conservation) makes most techniques developed in that reference in a mathematically rigorous way not applicable without further work. The resolution of this issue can be found, for example, in \cite{BaerGinouxPfaeffle,GreenHypOp,WaveDiracMfs}. For the Rarita--Schwinger field, the introduction of a curved space-time background is even much more restrictive and seems to place strong limitations on the curvature itself, as described in \cite{FrauendienerRaritaSchwinger} for the classical field and in \cite{NoGoRaritaSchwinger} for the quantum field. We shall mention the implications of quantising with our electromagnetic analogue space-time as background again, for example in section~\ref{sec:PDEana} and as a possible future development in the conclusion, section~\ref{sec:extensions}. Textbook/Pedagogical treatments of CSTQFT can be found in \cite{BirrellDavies,Fulling,WaldBHTD,FroNo98,FabbriNS05,BaerGinouxPfaeffle,CSTQFTSpringerLectureNotes,ParkerToms,FrolovZelnikov2011}, though this list makes no claims to be complete and is naturally far from it.
 	
 	After this rough summary of (part of) the available literature, it is time to start with the actual description of curved space-time quantum field theory.
 	
 	\subsection{Fixing the Notation in Flat Space-Time Quantum Field Theory}
 	Following more or less any other (non-algebraic) textbook presentation on curved space-time quantum field theory, it is useful to settle the notation for the quantum field theoretic aspects of the next discussion by a quick recapitulation of the flat space-time quantum field theory for the scalar field. The classical, real, and massless scalar field $\psi$ without interactions is described in Cartesian coordinates by the massless Klein--Gordon equation
 	\begin{equation}
 		\partial_a \partial^a \psi = 0.
 	\end{equation}
 	A solution $\psi_0$ to this wave equation now can be written using a Fourier decomposition as
 	\begin{equation}
	 	\psi_0(x) = \int \kl{a(\vec{k}) e^{-i\omega t +\vec{k}\cdot\vec{x}} + a^*(\vec{k}) e^{i\omega t -\vec{k}\cdot\vec{x}}} \frac{\dif^3 k}{(2\pi)^3 2\omega},
 	\end{equation}
 	where $x$ denotes the contravariant four-vector $(ct, x_1, x_2, x_3)^T$, while $\vec{x}$ refers to its spatial components. $\omega$ and $\vec{k}$ can then be collected into the covariant four-covector $(-\omega/c, k_1, k_2, k_3)$. Note that our demand of a \emph{real} scalar field forced a particular type of combination of the two independent, distributional solutions $\exp(\pm i k_a x^a)$ on us, where their coefficients need to be the complex conjugate of each other. The additional structure, namely the appearance of $1/(2\omega)$, has been arranged to guarantee Poincar\'e invariance of the resulting theory \cite{WeinbergQFT1,Srednicki}.
 	
 	This is, for the purposes of a heuristic introduction into CSTQFT enough for the beginning: The heuristic approach takes its power from appealing to the concept of Fock spaces; something that rigorously is far from simple. We shall follow this in this introductory section, and only mention the deficits of this approach when relevant later on.
 	
 	The underlying (flat) Minkowski space-time allows a unique and preferred choice of vacuum which is characterised as being invariant under the proper orthochronous Poincar\'e group $SO(3,1)^+ \ltimes \mathbb{R}^{3,1}$. This interpretation is achieved by following the heuristic, canonical quantisation procedure: Exchanging the complex-valued numbers $a(\vec{k}), a^*(\vec{k})$ in the classical Fourier decomposition of the classical field with operators on a Hilbert space $\hat{a}_{\vec{k}}, \hat{a}_{\vec{k}}^\dagger$, such that
 	\begin{subequations}
 		\begin{align}
 			[\hat{a}_{\vec{k}_1},\hat{a}_{\vec{k}_2}] &= 0,\\
 			[\hat{a}_{\vec{k}_1}^\dagger,\hat{a}_{\vec{k}_2}^\dagger] &= 0,\\
 			[\hat{a}_{\vec{k}_1},\hat{a}_{\vec{k}_2}^\dagger] &= (2\pi)^3 2 \hbar \omega \delta^{3}(\vec{k}_1 - \vec{k}_2),
 		\end{align}
 	\end{subequations}
 	where square brackets denote the commutator, provides the corresponding ladder operators to create states of a given momentum. $\hat{a}_{\vec{k}}^\dagger$ creates a state of momentum $k$, while $\hat{a}_{\vec{k}}$ annihilates such a state. The vacuum state is now the unique, Poincar\'e-invariant state $\ket{0}$ such that
 	\begin{equation}
	 	\forall \vec{k}: \qquad \hat{a}_{\vec{k}} \ket{0} = 0.
 	\end{equation}
 	This provides a unique meaning of \enquote{no particles present} in a quantum field theoretic picture --- as long as the underlying Poincar\'e invariance can be guaranteed. For improved readability we will from now on omit the hats on operators.
 	
 	\subsection{Moving on to Curved Space-Times}
 	
 	In curved space-times, however, the peculiarity of the Minkowski space-time's Poincar\'e symmetry becomes apparent: Its lack leads immediately to the highly non-trivial question of how to choose an appropriate vacuum state. The solution is to make the vacuum observer-dependent. Already at this level\footnote{Strictly speaking, even before introducing curved space-times: It is possible to analyse the vacuum of an observer in Minkowski space, who is not moving inertially. The comparison with the standard Minkowski vacuum would then show similar differences. This gives rise to the Unruh effect.} a new question arises: How to relate two vacua arising from two different observers? The key idea to achieve this is realising that the two vacua and their respective ladder operators should describe the same kind of particle. As such, they would have to obey the same commutation rules. This, then, means that only isomorphisms of the abstract spaces described by these commutation rules could interrelate these different vacua of the same space-time. These isomorphisms are Bogoliubov transformations.
 	\footnote{It is worth noting that from the rigorous point of view, the construction of Bogoliubov transformations is far from trivial. For this we refer to \cite{WaldBHTD}, and its theorems (4.4.1) and (4.5.2), and their discussion, particularly the comments regarding (4.5.2) and unitary \emph{in}equivalence on page~83. As a result, we feel justified in presenting the present, heuristic approach to CSTQFT --- even though we shall mention additional subtleties further below when discussing analogue space-times.}
 	
 	So, suppose one has two observers $A$ and $B$ with corresponding vacua $\ket{0}_A$ and $\ket{0}_B$, and ladder operators $a_A(\vec{k}), a_A^\dagger(\vec{k})$ and $a_B(\vec{k}), a_B^\dagger(\vec{k})$, respectively. Each set of ladder operators satisfies the canonical commutation relations separately and with respect to a corresponding set of orthonormal (with respect to a scalar product $(\cdot,\cdot)$), positive/negative frequency solutions $u^{\pm}_{A/B}(\vec{k})$ (also called \enquote{modes}) to the wave equation underlying the particles motion.\footnote{For the Minkowski vacuum, these are (usually) taken to be $\exp(\mp i k_a x^a)$, and understood in a distributional sense. These plane waves are solutions to the D'Alembert equation which appears to some degree in all PDEs behind quantum fields. (In the case of Dirac fields this latter statement is true only after squaring the differential operator, but the plan waves remain good distributional solutions of the Dirac equation as a classical field equation.)} Each observer must have a set of modes defined on a Cauchy surface of the (part of) space-time under consideration. Otherwise, the evolution of initial conditions for the wave equation underlying the modes is not well-defined. Then the Bogoliubov coefficients $\alpha_{BA}(\vec{k},\vec{k}'), \beta_{BA}(\vec{k},\vec{k}')$ are defined through the relation
 	\begin{subequations}
	 	\begin{align}
	 		u^+_B (\vec{k}) &=\int \alpha_{BA}(\vec{k},\vec{k}') u^+_A(\vec{k}') + \beta_{BA}(\vec{k},\vec{k}') u^-_A(\vec{k}')\dif \vec{k}',\\
	 		\Longleftrightarrow \qquad u^+_A (\vec{k}) &=\int \alpha_{BA}^\ast(\vec{k}',\vec{k}) u^+_B(\vec{k}') - \beta_{BA}(\vec{k}',\vec{k}) u^-_B(\vec{k}')\dif \vec{k}',
	 	\end{align}
	\end{subequations}
 	relating the positive frequency modes of observer $B$ to a mixture of both positive and negative frequency modes of observer $A$. Using this and the orthonormality of the given modes, one derives a range of properties for the Bogoliubov coefficients (and the other quantities involved). We shall collect these here, following \cite{FabbriNS05}, adapted to our notation:
 	\begin{subequations}
 		\begin{align}
 			\alpha_{BA}(\vec{k},\vec{k}') &= (u_B(\vec{k}),u_A(\vec{k}')), \quad \beta_{BA}(\vec{k},\vec{k}') = -(u_B(\vec{k}),u_A^\ast(\vec{k}')),\\
 			\delta(\vec{k}-\vec{k}') &= \int \alpha_{BA}(\vec{k},\vec{k}'') \alpha_{BA}^\ast (\vec{k}', \vec{k}'') - \beta_{BA} (\vec{k}, \vec{k}'') \beta_{BA}^\ast (\vec{k}', \vec{k}'') \dif \vec{k}'',\\
 			0 &= \int \alpha_{BA}(\vec{k},\vec{k}'') \beta_{BA}^\ast (\vec{k}', \vec{k}'') - \beta_{BA} (\vec{k}, \vec{k}'') \alpha_{BA} (\vec{k}', \vec{k}'') \dif \vec{k}'',\\
 			a_A(\vec{k}) &= \int \alpha_{BA}(\vec{k}',\vec{k}) a_B(\vec{k}') + \beta_{BA}^\ast(\vec{k}',\vec{k}) a_B^\dagger(\vec{k}')\dif \vec{k}',\\
 			a_B(\vec{k}) &= \int \alpha_{BA}^\ast(\vec{k},\vec{k}') a_A(\vec{k}') + \beta_{BA}^\ast(\vec{k},\vec{k}') a_A^\dagger(\vec{k}')\dif \vec{k}',\\
 			\ket{0}_A &= {\vphantom{\braket{0|0}}}_B\!\braket{0|0}_A \exp\kl{-\ed{2\hbar} \int \beta_{BA}^\ast(\vec{k},\vec{k}'') \alpha_{BA}^{-1}(\vec{k}'',\vec{k}') a_B^\dagger(\vec{k}) a_B^\dagger (\vec{k}') \dif\vec{k}\dif\vec{k}'\dif\vec{k}'' }\ket{0}_B.
 		\end{align}
 	\end{subequations}
 	The inverse Bogoliubov transformation can be guaranteed if the two sets of modes were orthonormal (as we demanded). The existence of a Bogoliubov transformation is equivalent to
 	\begin{equation}
 		\int \beta_{BA}(\vec{k},\vec{k}') \beta_{BA}^\ast(\vec{k},\vec{k}') \dif\vec{k}\dif\vec{k}' = \int |\beta_{BA}(\vec{k},\vec{k}')|^2 \dif \vec{k}\dif\vec{k}' < \infty,
 	\end{equation}
 	meaning that only normalisable states can be related to each other in this way. If one introduces a particle number operator
 	\begin{equation}
 		N_B(\vec{k}) \defi \ed{\hbar} a_B^\dagger(\vec{k}) a_B(\vec{k}),
 	\end{equation}
 	this can be rephrased as a requirement that only a finite number of particles is created when switching observers as described here, as
 	\begin{equation}
 		\vphantom{\braket{0|N_B(\vec{k})}}_A\!\braket{0|N_B(\vec{k})|0}_A = \int |\beta_{BA}(\vec{k},\vec{k}')|^2 \dif\vec{k}'.
 	\end{equation}
 	
 	This gives us the necessary tools to analyse the Hawking effect.
 	
 	\section{The Hawking Effect} \label{sec:Hawking}
 	
 	Given the above prelude, in deriving the Hawking effect, things become a matter of choosing the right Fock spaces and comparing their respective particle content. On the level of the ladder operators, little changes --- here all is in the name of the ladder operator, likewise it is for the vacua they belong to. The big task then is to find the right notions of positive frequency and calculating the correct Bogoliubov transformation relating them. A look at the space-time diagram~\ref{fig:Schwarzschild} helps here: The more or less \enquote{natural} question is: Given particles coming from $\scri^-$, what does an observer at $\scri^+$ see? This question does not change with the particular black hole model under consideration. In particular, a similar reasoning works for eternal black holes (be they static, rotating, uncharged, charged, \dots) just as well as for collapse models (names associated with these would be: Vaidya, and Oppenheimer--Snyder --- both corresponding metrics have many variants). As this section is only meant to provide enough background to make the rest of the thesis self-contained, we will be considering the much simpler static, eternal case, even though Hawking's original derivation already included a much more detailed calculation involving a general, nondescript collapse model. This certainly helped its acceptance, as an eternal black hole's teleological nature begs the question of how much dynamical results arising from it can even be trusted. Given, however, both the amount of work done since using either approach, and the fact that already early follow-up publications, like those of Boulware, DeWitt, and Unruh \cite{BoulwareVacuum,DeWCSTQFT,UnruhVacuum}, happily (also) used eternal black holes in their analysis, there is little cause for concern when we choose to adopt this practice. It is even possible to draw an analogy between the last null ray just missing the formation of the event horizon in a collapse model on the one hand, and the past horizon $H^-$ on the other hand, see p.~281 in \cite{BirrellDavies}. We shall occasionally allude to this analogy.
 	
 	The three canonical choices of vacua are those made by Boulware, by Hartle--Hawking, and by Unruh. Each choice comes with a different physical interpretation. Before explaining these, let us remember a good guiding principle: As modes valid both inside and outside of a black hole will have to be traced over for calculations pertaining to an observer (who has to be either inside or outside), the originally pure vacuum state will be measured as a mixed state (which in many applications can then be argued to be thermal). This also exemplifies the necessity of a horizon being present in order to produce this or similar effects. The Boulware vacuum corresponds to choosing the Eddington--Finkelstein coordinates with respect to which positive frequency modes are defined. As \emph{both} coordinates $u$, and $v$ will have to be used to capture both the physics at $\scri^-$ and $\scri^+$, on both the past horizon $H^-$, and the future horizon $H^+$ one of these will be ill-behaved. In the Hartle--Hawking vacuum, one chooses the Kruskal--Szekeres coordinates to define positive frequencies. They can cover the whole \emph{maximally extended} space-time, but at the expense of physical realisability: It turns out that this vacuum state corresponds to a black hole in thermal equilibrium \cite{FabbriNS05}. This means, among other things, that even at spatial infinity, the state will not approach the Minkowski vacuum.
 	
 	The wave equation for a scalar field can be separated on black holes of the Kerr--Newman family. Since the Schwarzschild solution is of this family, this is also possible here. The coming discussion of two sets of modes for observer $A$ and $B$ rests on this. The spherical symmetry can be separated just as it is done in quantum mechanics, and for our immediate concerns we can focus on the corresponding radial equation, the Regge--Wheeler equation:
 	\begin{equation}
	 	\kl{\frac{\dif^2}{(\dif r^*)^2} + \omega^2 -\underbrace{\kl{1-\frac{2GM}{r(r^*)}\kle{\frac{\ell(\ell+1)}{r^2(r^*)}+\frac{2GM}{r^3(r^*)}}}}_{\ifed V_\text{RW}}} R_\ell(r^*) = 0,
 	\end{equation}
 	where we also defined the Regge--Wheeler potential $V_\text{RW}$. The angular momentum potential is responsible for back scattering, which is encoded in the \enquote{grey body factors}. These factors we will define and identify in the sketch of the Hawking effect below. Note that this potential does not vanish even for a scalar particle. However, for both $r^*\to \pm\infty$, that is both at spatial infinity and at the horizon, it vanishes and we can expect mode solutions there to be asymptotically plane waves. Note that we can give $r$ as a function of $r^*$ using the Lambert-W function:
 	\begin{equation}
	 	r(r^*) = 2GM\kle{1+W\kl{\exp\kl{r^*/(2GM) - 1}}}.
 	\end{equation}
 	We shall encounter the more general Teukolsky equation, the wave equation to be separated on more general black hole space-times, in section~\ref{sec:refractive} as a starting point for an electromagnetic analogy to radial wave propagation in curved space-times.
 	
 	Observer $A$ shall be located at $\scri^-$. As $\scri^-$ is a Cauchy surface for black hole and domain of outer communication, it is thus simple to give appropriate mode functions there: We take
 	\begin{equation}
	 	\left.u^A_{\ell m}(\omega)\right|_{\scri^-} \sim \ed{\sqrt{4\pi\omega}} \frac{e^{-i\omega v}}{r} Y_{\ell m}(\theta,\phi),
 	\end{equation}
 	where $\ell, m$ are the usual angular momentum quantum numbers of spherical symmetry, just as $Y$ are the usual spherical harmonics.\footnote{The separation for Schwarzschild reduces to this simple, known case.} The symbol \enquote{$\sim$} stands for the asymptotic equivalence of left and right hand side at the given boundary or point. Usually the right hand side is the leading term of an asymptotic expansion of the left hand side; if the expansion point (here: $\scri^-$) is not given in the formula itself it will be clear from context (naming of quantity on left hand side, preceding or following text, \dots).
 	
 	For the other observer, $B$, located at $\scri^+$, the situation is less simple. It makes sense, following \cite{DeWCSTQFT,FabbriNS05,DeWittGlobal2} to separate the modes $\{u^B_{\ell m}(\omega)\}$ for $B$ into three sets $u^\text{out}_{\ell m}(\omega)$, $u^\text{down}_{\ell m}(\omega)$, and $u^\text{up}_{\ell m}(\omega)$:
 	\begin{subequations}
 		\begin{align}
	 		\left.u^\text{out}_{\ell m}(\omega)\right|_{\scri^+}  &\sim \ed{\sqrt{4\pi\omega}} \frac{e^{-i\omega U}}{r} Y_{\ell m}(\theta,\phi),\\
	 		\left.u^\text{down}_{\ell m}(\omega)\right|_{H^+} &\sim \ed{\sqrt{4\pi\omega}} \frac{e^{-i\omega v}}{r} Y_{\ell m}(\theta,\phi),\\
	 		\left.u^\text{up}_{\ell m}(\omega)\right|_{H^-} &\sim \ed{\sqrt{4\pi\omega}} \frac{e^{-i\omega U}}{r} Y_{\ell m}(\theta,\phi).
 		\end{align}
 	\end{subequations}
 	Their meaning is the following: $u^\text{out}$ reach the observer $B$ at $\scri^+$, and vanish on the horizon $H^+$, while they will not necessarily vanish on $I^-$. This is due to possible backscattering due to $V_\text{RW}$. This contrasts with the down modes, $u^\text{down}_{\ell m}(\omega)$: These are the modes captured by the black hole, thus experiencing enough blue shift that the influence of the potential on them can be neglected and no reflection to $\scri^+$ needs to be considered. As a result, they vanish on $\scri^+$. This is a case of the geometric optics approximation being valid in a quantum field theoretic context, see the discussion below in section~\ref{sec:geoquantum}. Lastly, we have $u^\text{up}_{\ell m}(\omega)$, which --- after introducing a reflection coefficient $r_\ell(\omega)$ and a transmission coefficient $t_\ell(\omega)$ --- can be written
 	\begin{equation}
	 	u^\text{up}_{\ell m}(\omega) = t_\ell(\omega) u^\text{out}_{\ell m} + r_\ell(\omega) u^\text{down}_{\ell m}.
 	\end{equation}
 	This means that part of these modes will be reflected down into the black hole, while some of them will reach $\scri^+$. It is important to note that they vanish on $\scri^-$, and hence they are important to form a complete set of modes \cite{DeWittGlobal2}.\footnote{In the context of collapse models it is more apparent that these are the modes responsible for the trans-Planckian problem: Hawking radiation modes reaching distant observers gain nearly arbitrarily large blueshifts when followed backwards in time close to the horizon. Depending on whether one wants to trust more in the mode picture or in the picture of renormalised stress-energy tensors, this is a more or less serious issue.} Note that they would be identical to $u^\text{out}_{\ell m}(\omega)$, if $r_\ell(\omega)$ was absent.
 	
 	Expanding a general field now in both sets of modes, it is possible to derive the Hawking radiation including grey body factors due to backscattering off $V_\text{RW}$. Since we will not use quantum field theory itself in the research presented in this thesis, we will refrain from a full derivation and limit ourselves to core ingredients of the derivation --- and its results. As we constructed our modes in such a way to provide us with two sets of \emph{complete} modes, the Wronskian can be employed to derive various identities useful in the calculations. Furthermore, by construction one has
 	\begin{equation}
	 	\abs{t_\ell(\omega)}^2 + \abs{r_\ell(\omega)}^2 = 1,
 	\end{equation}
 	and we now name $\abs{r_\ell(\omega)}^2$ the grey body factor $T_\ell(\omega)$.\footnote{We apologise for the minor clash in notation ($t_\ell(\omega)$ versus $T_\ell(\omega)$), but the later use of already two separate gammas ($\Upgamma, \Gamma$) made it more sensible to use $T$ for the grey body factor, as done in \cite{HawkFlux1}.} After some longer calculations \cite{FabbriNS05,DeWittGlobal2}, one finally arrives at
 	\begin{equation}
	 	\vphantom{\braket{0|N_{B}(\vec{k},\ell)}}_A\!\braket{0|N_B(\vec{k})|0}_A = \frac{T_\ell(\omega)}{e^{\frac{2\pi \omega}{c \kB\kappa}}-1},
 	\end{equation}
 	where the $-1$ is a result of the spin statistics theorem (or: the commutation relations) for our scalar field under consideration. We also recognise the appearance of the surface gravity $\kappa$ of $H^+$ in the right way to be interpreted as a temperature
 	\begin{equation}
	 	T_\text{H} = \frac{\hbar c}{2\pi}\kappa\label{eq:THawking},
 	\end{equation}
 	the famous Hawking temperature. While here, in the context of the Schwarzschild space-time (and withal an eternal version) only depending on the mass, we have seen earlier that $\kappa$ can depend on more physical quantities of interest. Also, the absence of rotation or charge (as opposed to, say, the Kerr(--Newman) solution) induces two simplifications: The non-existence of chemical potentials appearing in the exponential besides the frequency $\omega$, and the lack of additional indices on particle creation operators, number operators, modes, and grey body factors. If the former is changed, the latter will induce superradiance. More on superradiance will be presented in section~\ref{sec:superradiance}, including references regarding this topic.
 	
 	The grey body factors indeed behave as grey body factors in classical thermodynamics, and one can even regain the notion of Einstein $A$ and $B$ coefficients \cite{HawkingEinsteinCoeff}. It hence seems only natural to appeal as much as possible to our confidence in thermodynamics, and apply and transfer as many concepts from classical thermodynamics and thermal radiation (particularly of black and grey bodies). But not only does a standard course in thermodynamics often invoke necessary shorthand explanations \cite{SmerlakBlackbodybox}, gravity adds naturally to the confusion \cite{JessicaMattTolman2,JessicaMattTolman1}, and --- the purpose of this thesis --- it very often misses the severe non-classicality of the Hawking radiation \cite{Page1,BekensteinMukhanovQuantumBH,KieferDecoherenceHawkingRad,HawkFlux1,BHradNonclass}.
 	
 	Finally, note that the astrophysical framing (\enquote{black holes}, \enquote{collapse}, \enquote{space-time}, \dots) of this effect is hardly necessary. If one were to look closer at the Unruh effect, that is the corresponding effect in Minkowski space-time where an accelerated observer's vacuum is compared to that of an inertial one, this already becomes apparent. But more drastically, even this space-time context is still more than necessary: Even a merely effective space-time can show radiation effects of the type of Hawking or Unruh, provided similar notions of horizon and quantisation are available \cite{EssIness}. This horizon also is not restricted to an actual event horizon --- apparent horizons are enough, thanks to the kinematical nature of the derivation.

 	\subsection{Degrees of Approximation}
 	While the existence of the Hawking effect is reasonably easy to derive from general arguments, as already mentioned, quantifying the effect is much less straightforward. The geometric optics approximation turns out to be particularly important in the context of this thesis --- for one, most of the sparsity results make use of this approximation (see chapter~\ref{ch:sparsity}), for the other, the whole point of the exercise of deriving the most general consistency condition for an electromagnetic medium to correspond to an effective space-time is to go beyond the geometric optics approximation often used in previous derivations (for this, see section~\ref{sec:CovEMana}). But beyond this approximation, there are more approximations to be considered and mentioned:
 	\begin{itemize}
 		\item Adiabaticity, that is slow evolution: If either the evolution of space-time or the observer's acceleration changes too fast, the mode decomposition at the very heart of the definition of Bogoliubov coefficients will become impossible or at least of little relevance with regard to actual, physical measurements. This will constitute an infrared (IR) cut-off.
 		\item Neglecting ultraviolet (UV) effects: Any mode of energy larger than the mass-energy of a black hole will not be part of the actual spectrum, and hence all integrals involving wave numbers or frequencies as integration variable will have a UV cut-off.
 		\item Presence of grey body factors: As already the presence of inner most stable orbits (ISCO's) shows, the effective cross-section of a black body for photons will be greater than the area of a disc of area $\pi r_{\text{H}}^2$, even in the geometric optics approximation. Within this approximation, we shall derive the correction factor in section~\ref{sec:ndim}. But going beyond this results in the presence of grey body factors in the Hawking spectrum, as shown above. One obvious and common approximation is setting them equal to $1$, thus turning the spectrum exactly into that of a black body. Our sparsity results will discuss this in more detail.
 		\item Backreaction: Whereas grey body factors are related to backscattering, the issue of backreaction is much harder. Usually, backreaction is neglected. Given the low temperatures, corresponding low luminosities --- and our sparsity results of chapter~\ref{ch:sparsity} fit right into this narrative, too ---, this seems a good approximation. Should one want to gain more insight from the Hawking effect into possible hints at a quantum theory of gravity, however, backreaction becomes an issue of fundamental physics. Then one wants to have the field that was quantised on a given background also feed back into the metric. Instead of the standard Einstein equation the goal would then be to solve a semi-classical Einstein equation of the kind
 		\begin{equation}
	 		R_{ab} - \ed{2}g_{ab} R = \frac{8\pi G}{c^4}\langle T_{ab} \rangle.
 		\end{equation}
 		It turns out that already interpreting the meaning of the brackets (\emph{i.e.}, what kind of expectation value one is looking at) is highly non-trivial and has baffled generations of physicists. It does not help that this adds even more non-linearity to the already non-linear Einstein equations. However, toy models to this question where a Hartree--Fock method can be employed are known, see the last chapter of \cite{FabbriNS05} for a review of this approach.
 	\end{itemize} 
 	More on these further complications, including more references discussing various of their aspects, can be found in, for example, \cite{Thermality}.

 	\section{Geometric Optics Approximation}\label{sec:geoapp}
 	Given the nature of this thesis, there will be plenty of wave equations invoked. Therefore, we do not deem it useful to consider only particular special cases in the following. Rather, we will explain the geometric optics approximation for classical fields in general --- even if this results in fairly unspecific statements and few concrete insights. Rather, our goal is merely motivating the alternative name of geometric optics: Ray optics. After deriving this result, we then continue with a few words on how to relate these results for classical fields to the quantum fields one would actually be looking at in the context of the Hawking effect (or similar effects). 
 	
 	We conclude with the most important background for the thesis: Cross-sections in curved space-times for black holes \emph{in the geometric optics approximation}. We will either derive or reproduce results from references needed later. The only exception to this is the $N+1$-dimensional cross-section of spherically symmetric black holes, which will be relegated to section~\ref{sec:ndim}.
 	
 	\subsection{Classical Fields}
 	The basic idea behind the geometric optics approximation for classical fields is captured in the slogan \enquote{wave packets propagating as particles on curves is good enough}. This slogan is applicable both to massive and massless fields --- however, the present section shall only carefully treat the latter. This also provides an ease of notation: The very notion \enquote{geometric optics} seems slightly paradoxical when applied to a massive field. In the following we will follow a mix of \cite{MTW,ThorneBlandford2017} and \cite{BenOrs}, but modified from their respective presentation. The reason for this choice is that while \cite{BenOrs} is fairly general, their discussion is limited to ordinary differential equations, while our interest is mostly in field equations which are partial differential equations. On the other hand, \cite{MTW} is focussed too much on the electromagnetic case. While this case certainly is relevant to us, and we will summarise its results below, the main application of the geometric optics approximation will be encountered in chapter~\ref{ch:sparsity}, where we consider more fields in this approximation than just photons. However, we will try to keep the technical tools and methods to a minimum. The purpose of this part of the thesis is to facilitate and motivate a physical understanding of the approximations used in the main parts, not to prepare for formal exercises in asymptotic analysis (as these do not appear in the coming discussion). Finally, we make changes to the presentation as it is in \cite{ThorneBlandford2017} to be more in line with the one given in \cite{BenOrs} to elucidate the mathematical context a bit more. Perlick gives a geometrical, very formal, covariant discussion of ray optics in \cite{PerlickRay}.
 	
 	When looking at the field equations of a field, one can imagine three important length scales influencing the propagation. The first length scale is the wavelength(s)\footnote{\cite{MTW} and \cite{ThorneBlandford2017} use instead the reduced wavelength, $\bar{\lambda} \defi \lambda/(2\pi)$. Given the value of $2\pi$ and the requirement of the ensuing inequalities to be rather strong ($\ll$ instead of just $<$), we shall omit this factor in our presentation.} of the field solution --- the plural applies when looking at a wave packet decomposed into modes of different wavelengths. The physicist's notion of Fourier analysis suffices for this picture, but more rigorously, methods of microlocal analysis are needed. The \enquote{classical} Fourier decomposition in flat Minkowski space-time mixes coordinates and (co-)tangent vectors (and is allowed to do so thanks to its affine nature) in $\exp(-ik_a x^a)$, which is difficult to reconcile with a corresponding notion on curved space-times. For more on this, we refer to the literature, for example, \cite{BaerGinouxPfaeffle,CSTQFTSpringerLectureNotes} or the introductory sections of chapter~III in \cite{LawMichSpin}. The physical intuition is to perform a \enquote{local Fourier transform}. Hence we also have a notion of frequency and wavelength available, even if looked at from a very rigorous point of view. We will therefore confine ourselves to the physicist's point of view.
 	
 	The second length scale will also be given by the solution of the field equations itself: It describes the length scale on which amplitude, polarisation, or wavelength (\emph{i.e.}, the precise way the solution is Fourier decomposed) change. As this can be independent of each other, strictly speaking, there would be even more length scales involved --- we shall just take the minimum of this set and call it $L$.
 	
 	The third length scale is given by the curvature of the space-time itself. To define this length scale, one defines a tetrad (or vierbein) and looks at the values of the Riemann tensor in this frame. This will give something scaling as $\text{length}^{-2}$, thus the square root of its inverse will provide a length scale. Looking at each component, one then chooses the minimum value as the most conservative estimate.
 	
 	Were we to look at massive fields, too, a fourth length scale would appear, given by the Compton wavelength,\footnote{Given the appearance of the Compton wavelength in quantum mechanics (within, for example, the Klein--Gordon or Dirac equation), here the reduced wavelength can make more sense depending on the context.}
 	\begin{equation}
 		\lambda_{\text{Compton}} \defi \frac{2\pi \hbar}{mc}.
 	\end{equation}
 	However, if we were to carry out the following analysis in the massive case, a simple rearrangement of factors of $\hbar$ would reveal that this length scale does not appear in a fundamentally new way in the resulting geometric \enquote{optics}. Rather, one would just have an additional term involving the mass in the corresponding dispersion relation~\eqref{eq:dispersion}, see \cite{RoseGeomOpticsElectron}.
 	
 	Having introduced the length scales involved, we are now in a position to explain more clearly the scope of the geometric optics approximation. Here, one is interested in an approximation of a given field equation that is valid when two inequalities hold simultaneously:
 	\begin{subequations}
 		\begin{align}
 			\lambda &\ll \min \kl{R^{\hat{a}}{}_{\hat{b}\hat{c}\hat{d}}}^{-1/2},\\
 			\lambda &\ll L.
 		\end{align}
 	\end{subequations}
 	Again taking the minimum of either scale, calling it $\tilde{L}$, allows the introduction of a small parameter with respect to which we can then analyse the solutions of our field equations:
 	\begin{equation}
 		\epsilon \defi \frac{\lambda}{\tilde{L}}.
 	\end{equation}
 	
	With a small parameter at hand, we can now take a Wentzel--Kramers--Brillouin(--Jeffreys) (WKB) ansatz for our solution.\footnote{We will refrain from keeping the fourth name of Harold Jeffreys in the abbreviation, as there seems to be little agreement on how to include it. The abbreviation \enquote{WKB}, on the other hand, seems fairly universally understood.} For a rigorous treatment of the WKB method, we refer to \cite{BenOrs}, with the caveat that they consider only \emph{ordinary} differential equations (ODEs) (however, fairly general ones). We are much more interested in the behaviour of \emph{partial} differential equations, as the field equations we are most concerned with are archetypes of PDEs: Klein--Gordon equation, Dirac equation, Maxwell equation, Proca equation, Einstein's field equations, \dots. While there exists rigorous literature on the WKB method in the PDE context, this would lead too far afield.
	
	The idea is to write the solution $u$ of the equation under examination as
	\begin{equation}
		u = A e^{iS},
	\end{equation}
	where $A$ is an amplitude and $S$ a phase (called the \enquote{eikonal}), both possibly complex. Wave fronts are the isosurfaces of $S$ at fixed times $t$. Now we make the ansatz that
	\begin{subequations}
		\begin{align}
			A &\sim A_0 + \epsilon A_1 + \epsilon^2 A_2 + \dots,\\
			S &\sim \ed{\delta} \kl{S_0 + \delta S_1 + \delta^2 S_2 + \dots},
		\end{align}
		where $\delta$ is a second small parameter (which at the end of the analysis will ultimately be related to $\epsilon$). The symbol $\sim$ signifies that both expansions may only be valid as asymptotic expansions; that is, they may be formally divergent, but still delivering good approximations if truncated in the right way.
	\end{subequations}
	
	As a quick look at the ansatz reveals, we have two concepts here: One, we have a multi-scale analysis, as both parameters, $\epsilon$ and $\delta$, are considered small, \emph{i.e.}, we look at limits to zero for both of them. Put differently, we are looking at a slowly varying amplitude and a rapidly oscillating phase. Two, looking at the $S_0$ contribution, we see that it is furthermore a \enquote{singular perturbation theory}, as the $S_0/\delta$ part will lead to an essential singularity in this limit. Additional difficulties arise, if the solutions happen at critical points of $S$. Again, the mathematical depth is vast, but we shall leave it at a mere mention.
	
	Next, one inserts this two-pronged ansatz in all relevant equations. This may be more than the original PDE, as often one looks at a variant of it introducing the need to fix a gauge condition which also has to be subjected to this ansatz. Following \cite{MTW,ThorneBlandford2017}, we can form a scalar $\abs{A}$ out of $A$, and one can then give the \enquote{polarisation}, $p\defi A/\abs{A}$, as the normalised amplitude with respect to this scalar. The covariant derivative of the phase, $\nabla_a S$ is identified with the $N+1$-dimensional wave vector $k_a$. Equivalently, we can say the $-\partial_t S$ gives the frequency $\omega$, while the spatial derivative $\partial_i S$ gives the \enquote{traditional} wave vector $k_i$.
	
	Precisely how this scalar is formed will depend on the field equation. For example, a solution of the Klein--Gordon equation will not need this, being a scalar itself. The solution of Maxwell's equation, as a second example, would involve either a scalar invariant formed out of the excitation tensor $G^{ab}$ (see section~\ref{sec:CovEMana} for details) or an appropriate four-potential $A^a$ (whose existence, however, will invoke topological constraints) whose obvious scalar invariant would be $\sqrt{A_a \overline{A}^a}$. Finally, the equation in question might be the weak field Einstein equations, where the tensorial nature is even more apparent than in (not yet gauged) electromagnetism.
	
	Assuming, for the moment, that one only looks at the wave equation itself (and does not need to also check the behaviour of gauge conditions under this ansatz), we can then reduce the lowest order terms in $\epsilon$ and $\delta$ to the dispersion relation of the wave:
	\begin{equation}
		\kl{-\omega^2 + \Omega(x^a,k^i)^2} = 0,\label{eq:dispersion}
	\end{equation}
	where the precise form of $\Omega$ depends on the PDE in question. Note that the lowest order in $\delta$ will be of order $-2$, as the PDE we consider are second order. Note that in the case of massless fields in vacuum, this will be
	\begin{equation}
		\kl{-\omega^2 + k_i k^i} = 0.\label{eq:knull}
	\end{equation}
	The equation above is the first result: Massless fields have a null wave vector. Note that this already implies that they are following null geodesics, as $k_a$ is the gradient of the scalar $S$. Likewise, for massive fields we will have the result that the wave vector will be time-like.
	
	Should the PDE describe wave propagation \emph{in vacuo}, then the next higher orders include parallel transport of $A$ and $p$ along $k$, and orthogonality of $p$ and $k$ (and hence also of $A$ and $k$). This continues to hold if $\Omega(x^a,k^i) = \mathrm{const}\cdot \sqrt{k_ik^i}$; but cannot be expected to hold in general, nor if higher order terms are considered. This higher order terms would, however, be already by definition beyond geometric optics.
	
	Closely related is the eikonal equation
	\begin{equation}
		\kl{\nabla_i \Phi(x^i)}\kl{\nabla^i \Phi(x^i)} = n^2(x^i),\label{eq:eikonal}
	\end{equation}
	which, however, assumes that the time dependence is fully separated, \emph{i.e.}, that
	\begin{equation}
		u(x^i,t) = A(x^i) e^{-i\omega t + i\frac{\omega}{c}\Phi(x^i)},
	\end{equation}
	which is more restrictive than the analysis above. It is possible to also call the dispersion relation~\eqref{eq:dispersion} \enquote{eikonal equation} (or other equivalent or more general forms of it), but we shall refrain from this and only refer to equations of the form of equation~\eqref{eq:eikonal} as eikonal equations.
	
	Already at this point we end the discussion: The first result we actually derived --- that the wave vector is null and follows a geodesic --- is the only result we require for the context of this thesis. The insinuated results following equation~\eqref{eq:knull} are sufficient for our intended usage.

 	\subsection{Quantum Fields}\label{sec:geoquantum}
 	
 	Given that our ultimate goal is to quantify aspects of Hawking radiation, it is worthwhile to have a careful look at how much the ray optics picture helps in this case. The Hawking effect is only brought into existence by quantum field theory considerations on curved space-times. The geometric optics approximation, on the other hand, is intrinsically classical. This seems to indicate that the geometric optics approximation will not help analysing black hole evaporation. At the very least, one would have to be extremely careful in trusting any results gained from this combination. But before we throw the towel, it is worthwhile to have a closer look and assess our physical intuition more closely. This will also bring some interesting, though purely formal analogies to light.
 	
 	When considering the classical limit of quantum theories, the fastest way goes through the path integral formalism. In the path integral formalism of quantum field theory, time-ordered vacuum expectation values of an operator $O[\psi]$ are calculated as
 	\begin{equation}
		 \langle O \rangle = \frac{\int \mathcal{D}\psi\; O[\psi] \exp\kl{\frac{i}{\hbar} S[\psi]}}{\int \mathcal{D}\psi\; \exp\kl{\frac{i}{\hbar} S[\psi]}},
 	\end{equation}
 	where $\psi$ is a collective place-holder for any fields appearing in the action functional $S$. $\int\mathcal{D}\psi$ is the path integral over all states of $\psi$ (depending on $S$ the mathematical justification may be ongoing research\dots). As the appearance of the term \enquote{time-ordered} already alludes to, it is far from obvious how to generalise this framework to curved space-times, even in the intuitive, physicist's point of view. It was and is, however, even in this field used proliferously. We shall not concern ourselves with matters of validity and rather participate in this nonchalant approach.
 	
 	Doing this, one can see the appearance of $\hbar$ in full analogy to the earlier introduced WKB method as a singular perturbation theory for some $\delta \to 0$. Then, the so-called classical limit (formally setting $\hbar \to 0$) is to be seen as the lowest order of the corresponding WKB approximation. In lowest order in $\hbar$ we are confronted with finding the minimum of $S$ when varying over $\psi$ in the path integral --- which is precisely what is needed to find the classical field equations. In quantum field theoretic language, this corresponds to finding tree level amplitudes. Thus tree level processes \emph{roughly} correspond to classical field theory.
 	
 	Our emphasis on the word \enquote{roughly} in the previous sentence requires some explanation: Intuitively, it may seem plausible to assume that the previously given heuristic goes through in any field theory and action. However, this is not correct: Strictly speaking, this first approximation only gives \emph{tree-level results}. These are intimately related to classical physics, but not all tree-level physics \emph{is} classical physics. This can be seen more clearly from the canonical quantum field picture, where the uncertainty relation is more accessible. In order for a state to be reasonably classical, one wants small relative uncertainty. If a state, however, has low occupation numbers, this state cannot have a small relative uncertainty in the occupation number. Hence, in order for a quantum field to be \enquote{close to} a classical state, it needs to have a high occupation number. This also explains why fermion fields cannot have a classical \emph{field} limit. Here the classical limit cannot be achieved by high occupation numbers. The resulting (statistical) particle mechanics needs to be gained through a different limiting procedure. For more on this, see \cite{SchwablSM,Duncan}.
 	
 	At first glance this excursion into the classical limit of quantum field theories might seem overkill for the purpose at hand, but as we want to actually relate the quantum fields as they appear in black hole evaporation to the geometric optics approximation as introduced in the previous subsection, we are confronted with exactly this classical limit. This can also be seen from the explanation in \cite{MTW} and \cite{ThorneBlandford2017} for the propagation law for the scalar amplitude $\abs{A}$ (introduced above) derived from the geometric optics approximation: In the case of electromagnetism in vacuum, this can be rewritten as
 	\begin{equation}
	 	\nabla_a \kl{A_b A^b k^a} = 0,
 	\end{equation}
 	a conservation equation. Classically, it becomes rather cumbersome to interpret the corresponding conserved quantity\footnote{This would be a conserved \enquote{light ray density}.}, while quantum mechanically, this corresponds just to conserved particle number (of photons). This also reiterates the need of a quantum state close to a classical state to have low relative uncertainty in particle number --- otherwise this conservation law would cease to hold on a classical level and violate the geometric optics approximation.
 	
 	However, \emph{if} a quantum state can be interpreted classically, and \emph{if} that state can be described by the geometric optics approximation, one could see this as a successful, double stationary phase approximation of the physical situation: First, the path integral is approximated well by the lowest order term in a WKB approximation, which is the \enquote{classical} tree level. Second, the tree level is indeed well described by the classical solution of the classical action, \emph{and} this solution is well portrayed by the eikonal approximation and the resulting ray optics.
 	
	As it will turn out, the notion of \enquote{sparsity} --- to be introduced in chapter~\ref{ch:sparsity} and developed there --- will show that Hawking radiation most certainly does \emph{not} fit the bill of being described well classically. Hawking radiation has very low occupation numbers, hence cannot have low relative uncertainty in the occupation numbers. Surprisingly, the semi-classical concept of sparsity is nevertheless a good measure to highlight this. We will see that the presence of non-classicality can still be derived if one approximates all mode functions at late times using the geometric optics approximation, see section~3.6.2 of \cite{FabbriNS05}; however one loses the backscattering and thus access to grey body factors. As we will occasionally need to include grey body factors in the course of our analysis of sparsity, we deemed it necessary to include their origin in our introduction to the Hawking effect in the previous section. Nonetheless, as evidenced by the down modes $u^\text{down}_{\ell m}(\omega)$ introduced in section~\ref{sec:Hawking}, even in this derivation there is room for the geometric optics approximation. How this can be done will be mentioned at the very end of this chapter when introducing the notion of the DeWitt approximation. Nevertheless, depending on field or observer, one still has to be careful not putting too much weight on the analogy to \emph{classical} optics.

 	\subsection{Cross-Sections}\label{sec:cross}
 	
 	If one assumes that one can make use of some sort of geometric optics approximation (be it for massive or massless particle fields, be it classical or close to classical as described above), we will be able to describe the motion of the corresponding wave packets by looking at the corresponding geodesics. The wave packets then follow these geodesics (approximately) like point particles would.\footnote{There still is, even classically, some room for discussion: Due to general relativity's non-linearity, the question of how much one can ignore the backreaction of even idealised point particles is far from trivial. A long tradition of research has looked into this particular question --- the takeaway being that, yes, for our purposes the approximation of particles following geodesics works. More on this can be found in \cite{LRRPointParticles,SelfForceGR}.} If the black hole was only a hard sphere in flat Minkowski space-time, the question of what its (total) (capture) cross-section $\sigma_\text{tot}$ is, would reduce to
 	\begin{equation}
	 	\sigma_\text{tot} = \pi r_\text{H}^2 = \frac{A}{4}.
 	\end{equation}
 	Here, $A$ denotes the surface area of the event horizon, and we have assumed (for the time being) a Schwarzschild geometry. The factor
 	\begin{equation}
	 	c_\text{eff} = \ed{4} \label{eq:firstceff}
 	\end{equation}
 	can be seen as a first instance of what we will later call a \enquote{correction factor}. This is the proportionality factor linking the cross-section back to the event horizon surface area $A$, that is
 	\begin{equation}
	 	c_\text{eff} \defi \frac{\sigma_\text{tot}}{A}.
 	\end{equation}
 	Depending on the specific black hole space-time and observer's position relative to the black hole, this may or may not be a constant.
 	
 	Now the whole point of the black hole space-time is that we are \emph{not} in flat space-times. A first and naive look at this question in terms of a differential cross-sections runs into the problem that gravity, both Newtonian and relativistic, is a force of infinite range. As in the case of electromagnetism this will lead to diverging differential and total scattering cross-sections if the impact parameter $b$ is taken to infinity (equivalently, if one looks at small scattering angles). Even so, we will consider first and foremost the capture cross-section which \emph{is} well-defined. Later, in section~\ref{sec:3+1cross}, we will mention some considerations regarding more general black holes space-times, as well as other influences on the encountered cross-sections and their importance for our work. The present discussion of Schwarzschild black holes remains valid for all static, spherically symmetric black holes, as, for example, the dirty black holes described above.
 	
	As we are interested in the area that an observer would see as \enquote{the black hole}, we want to have an understanding of what this means concretely. The name already implies that it is less the visibility of something and rather the absence of anything to be seen what constitutes a black hole. While from a causal, or Lorentzian geometric, point of view the important surface will remain the surface area of the event horizon, this observability criterion is more concerned with the above mentioned capture cross-section. Any ray coming at the black hole with an impact parameter $b$ less than some critical impact parameter $b_\text{crit}$ will be captured. For a hard sphere in flat space-time (and assuming fully inelastic scattering) this will result precisely in the aforementioned $\pi r_\text{H}^2$.
	
	In order to proceed, one needs to have a closer look at geodesic motion in Schwarzschild space-times. This will allow us to introduce conserved quantities for the particle undergoing geodesic motion, which in turns allows to introduce an impact parameter to relate the discussion back to traditional ideas of cross-sections and scattering theory. We will follow the discussion presented in chapter~7 of \cite{FrolovZelnikov2011}.
	
	The geometry we consider being spherically symmetric, we can reduce the question of geodesic motion to one of geodesic motion in the equatorial plane. This corresponds to setting $\theta = \pi/2$, and the line element we need to consider in the geodesic equation simplifies from equation~\eqref{eq:Schwarzschild} to
	\begin{equation}
		\dif s^2 = -\kl{1-\frac{2GM}{r}}\dif t^2 + \frac{\dif r^2}{1-\frac{2GM}{r}} + r^2\dif \phi^2.
	\end{equation}
	Next, we write the four-momentum of the particle under consideration as
	\begin{equation}
		p^a = \frac{\dif x^a}{\dif \lambda},
	\end{equation}
	where we push the distinction between massless and massive particle into the affine parameter: For a massless particle, this is simply an affine parameter, while for the massive particle it is the proper time $\tau$ rescaled by the mass, giving $\lambda= \tau/m$. From the normalisation of the four-momentum, we then get
	\begin{equation}
		-\kl{1-\frac{2GM}{r}} \kl{\frac{\dif t}{\dif \lambda}}^2 + \kl{1-\frac{2GM}{r}}^{-1} \kl{\frac{\dif r}{\dif \lambda}}^2 + r^2 \kl{\frac{\dif \phi}{\dif \lambda}}^2 = -m^2.\label{eq:psquared}
	\end{equation}
	
	Invoking Noether's theorem for the two Killing vector fields $\partial_\phi$ and $\partial_t$ gives two conserved quantities. Note that the third Killing vector field associated to $\theta$ has been discarded by the reduction to motion in the equatorial plane. These conserved quantities are:
	\begin{subequations}
		\begin{alignat}{2}
			E &\defi -\kl{\partial_t}^a p_a &&= \kl{1-\frac{2GM}{r}}\frac{\dif t}{\dif \lambda},\\
			L &\defi \phantom{-}\kl{\partial_\phi}^ap_a &&= r^2\frac{\dif \phi}{\dif \lambda}.
		\end{alignat}
	\end{subequations}
	$E$ can be given the interpretation of the particle's energy, while $L$ corresponds to its angular momentum. This allows to rewrite equation~\eqref{eq:psquared} as the following three ODEs:
	\begin{subequations}
		\begin{align}
			\kl{\frac{\dif r}{\dif \lambda}}^2 &= E^2 - \kl{1-\frac{2GM}{r}}\kl{m^2+ \frac{L^2}{r^2}}, \label{eq:crossODEr}\\
			\frac{\dif \phi}{\dif \lambda} &= L/r^2,\\
			\frac{\dif t}{\dif \lambda} &=\kl{1-\frac{2GM}{r}}^{-1} E.
		\end{align}
	\end{subequations}
	At this stage it becomes sensible to distinguish the two cases, massive and massless particles, again. From now on, we will limit ourselves to the discussion of equation~\eqref{eq:crossODEr}, as this is the equation relevant for capture cross-sections.\footnote{This becomes wrong in the absence of spherical symmetry. Particle capture and the resulting shadow of the black hole then become dependent on the angle between the observer's line of sight and the black hole axis. We will mention some of these complications in section~\ref{sec:3+1cross}.}
	
	\subsubsection{Massive Particle Trajectories}
	In the case of massive particles, we can unwrap the affine parameter back to the proper time $\tau$. Equation~\eqref{eq:crossODEr} then becomes
	\begin{equation}
		\kl{\frac{\dif r}{\dif \tau}}^2 = \frac{E^2}{m^2} - \underbrace{\kl{1-\frac{2GM}{r}}\kl{1+\ed{r^2}\frac{L^2}{m^2}}}_{\ifed U}.
	\end{equation}
	The thus defined quantity $U$ constitutes an effective potential. We can simplify its appearance significantly by now switching to dimensionless variables using the available length scale $r_\text{H} = 2GM$. For this, define
	\begin{subequations}\label{eq:massivedimless}
		\begin{align}
			\rho &\defi \frac{r_\text{H}}{r},\\
			b &\defi \frac{L}{r_\text{H} m},\\
			\iota &\defi \frac{\tau}{r_\text{H}}.
		\end{align}
	\end{subequations}
	Spatial infinity in these variables corresponds to $\rho=0$, while the horizon lies at $\rho=1$. The ODE we need to solve for radial motion now reads
	\begin{equation}
		\rho^{-4} \kl{\frac{\dif \rho}{\dif \iota}}^2 = \kl{\frac{E}{m}}^2 - U,
	\end{equation}
	where the effective potential $U$ now reads
	\begin{equation}
		U = (1-\rho)(1+b^2 \rho^2).
	\end{equation}
	
	For this effective potential we now have to find the conditions when a given impact parameter will lead to gravitational capture. In order to achieve this, we will consider the problem as only depending on the impact parameter $b$ and the new variable $\rho$ --- remembering that if we increase $E$ while keeping $b$ constant, the particle's trajectory will have a closer encounter with the black hole. Classifying the trajectories for a given $b$ then becomes a question of understanding the effective potential's dependence on these parameters. Fixing $b$ and looking for extrema of $U(\rho,b)$ gives
	\begin{equation}
		\frac{\dif U}{\dif \rho} = -1-3b^2 \rho^2 + 2b^2\rho = 0.
	\end{equation}
	This gives two extrema $\rho_\pm$ at
	\begin{equation}
		\rho_\pm = \frac{1\pm \sqrt{1-\frac{3}{b^2}}}{3}.
	\end{equation}
	Checking whether these are minima, maxima or saddlepoints by looking at the second derivative,
	\begin{equation}
		\frac{\dif^2 U}{\dif \rho^2} = 2b^2(1-3\rho),
	\end{equation}
	tells us that $\zeta_+$ is a maximum, and $\zeta_-$ a minimum. The only exception to this is if the impact parameter happens to be $b=\sqrt{3}$, in which case $\zeta_+ = \zeta_-$, and we have a saddlepoint. The associated minimal and maximal values of $U$, $U_\pm$, are
	\begin{equation}
		U_\pm = \frac{2b(b^2+9)\pm 2(b^2-3)^{3/2}}{27b}.\label{eq:Upm}
	\end{equation}
	Now we are in the position to discuss the condition when the particle is captured by the black hole: This will happen if the squared specific energy, $(E/m)^2$, is larger than $U_+$ of the given impact parameter $b$. For more on this, we refer to both \cite{Carroll} and \cite{FrolovZelnikov2011}.
	
	The total cross-section is related to the maximal \emph{physical}, apparent impact parameter for which capture happens by
	\begin{equation}
		\sigma_\text{tot} = \pi b_\text{phys}^2.
	\end{equation}
	The reason to call it an \enquote{apparent} impact parameter lies in the fact that its physical interpretation is given in terms of properties of the trajectory at spatial infinity. The connection of physical and dimensionless impact parameter is
	\begin{subequations}
		\begin{equation}
			b_\text{phys} = \frac{m r_\text{H} b}{p} = \frac{L}{p},
		\end{equation}
		with
		\begin{equation}
			p = \frac{m\beta}{\sqrt{1-\beta^2}}
		\end{equation}
	being the norm of the particle's three-momentum. $\beta$ is the three-velocity as measured at spatial infinity.
	\end{subequations}
	
	As we care for the limiting case of particle capture, that is, when the particle is \emph{just} not captured, we look at
	\begin{equation}
		\frac{E^2}{m^2} = U_+,
	\end{equation}
	and using $E^2+p^2=m^2$ allows us to deduce that for the corresponding value of $b$
	\begin{equation}
		p = m\sqrt{U_+ -1}
	\end{equation}
	and
	\begin{equation}
		U_+ = \ed{1-\beta^2}.
	\end{equation}
	In the last step we made use of equation~\eqref{eq:Upm}, which can also be interpreted as connecting the dimensionless impact parameter $b$ with the three-velocity $\beta$, thus giving $b(\beta)$ --- though at this point only implicitly. Explicitly, this connection reads
	\begin{equation}
		b(\beta) = \ed{\sqrt{8(1-\beta^2)}}\sqrt{20+8\beta^2+\ed{\beta^2}\kl{\sqrt{(1+8\beta^2)^3}-1}}.
	\end{equation}	
	Three other solutions can be discarded as being either negative or purely imaginary. Finally, this results in
	\begin{equation}
		b_\text{phys} = r_\text{H} b(\beta) \frac{\sqrt{1-\beta^2}}{\beta},
	\end{equation}
	whence
	\begin{subequations}
	\begin{align}
		\sigma_\text{capture}(\beta) &= \pi r_\text{H}^2 \frac{(b(\beta))^2 (1-\beta^2)}{\beta^2},\\
		&= {\pi r_\text{H}^2}\underbrace{\kl{\frac{5}{2\beta^2}+1+\ed{8\beta^4}\kl{\sqrt{(1+8\beta^2)^3}-1}}}_{=4c_\text{eff}}.\label{eq:sigmamass}
	\end{align}
	\end{subequations}
	
	In figure~\ref{fig:ceff} we have plotted $c_\text{eff}$ for $\beta\in[0,1]$. Note that $c_\text{eff}$ is monotonically decreasing towards the limiting value $27/16$. In a last step --- still following \cite{FrolovZelnikov2011} --- it is worthwhile to have a look at the two limiting cases $\beta \to 0$ and $\beta \to \infty$ also analytically.
	
	\begin{figure}
		\centering
		\includegraphics[width=.66\textwidth]{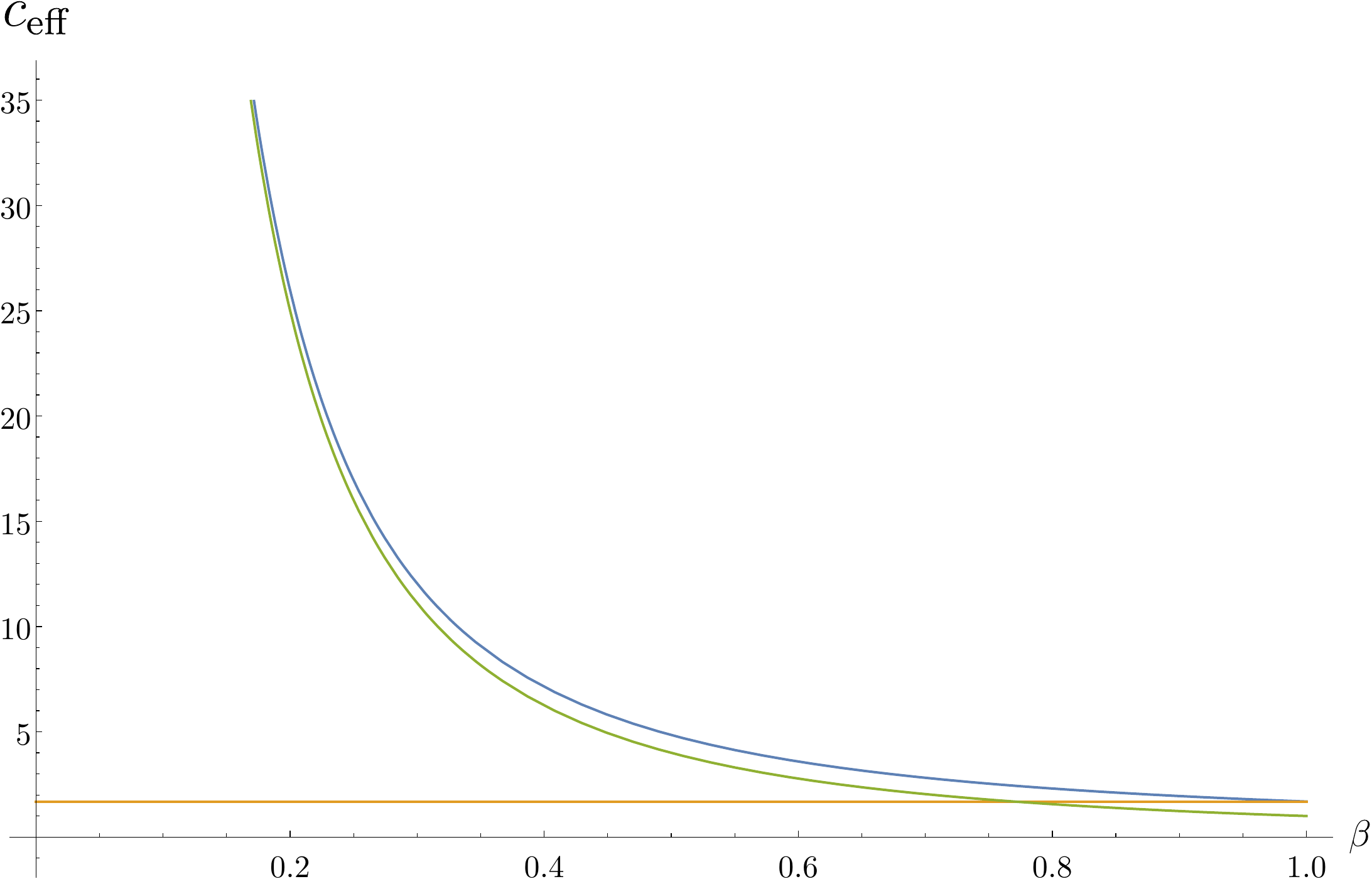}
		\caption[Plot of Correction Factor for Cross-Sectional Area of Massive Particles]{A plot of $c_\text{eff}$ for massive particles as a function of $\beta$. The orange horizontal line is the limiting value $27/16$ when $\beta\to 1$. The green line is the asymptotic behaviour $1/\beta^2$ for $\beta\to 0$.}
		\label{fig:ceff}
	\end{figure}
	
	For the first case, we have
	\begin{equation}
		b_\text{phys} \approx \frac{2 r_\text{H}}{\beta},
	\end{equation}
	and
	\begin{equation}
		\sigma_\text{capture} \approx \frac{4\pi r_\text{H}^2}{\beta^2}.\label{eq:capturemassive}
	\end{equation}
	While this can be rewritten in terms of the surface area $A$ of the horizon as
	\begin{equation}
		\sigma_\text{capture} \approx\frac{A}{\beta^2},\label{eq:sigmamassslow}
	\end{equation}
	the corresponding $c_\text{eff}$ as defined above diverges when $\beta\to 0$. This should come as no surprise: Gravity is a long-range potential, we \emph{should} expect diverging cross-sections. Also, if the three-velocity approaches $0$ at infinity, it means that the particle, given enough time, \emph{will} be captured by the black hole, simply following a purely radial geodesic. This corresponds just to the case of attraction of two massive bodies as seen already in Newtonian gravity.
	
	On the other end of the spectrum is the limit $\beta\to 1$. In this case:
	\begin{equation}
		b_\text{phys} \approx \frac{r_\text{H} 3\sqrt{3}}{2},
	\end{equation}
	leading to
	\begin{equation}
		\sigma_\text{capture} \approx \frac{27}{4} \pi r_\text{H}^2.\label{eq:massivelimitcrosssection}
	\end{equation}
	Here, 
	\begin{equation}
		c_\text{eff} = \frac{27}{16},
	\end{equation}
	a factor which will appear frequently in the discussion of (the yet undefined) sparsity in chapter~\ref{ch:sparsity}. However, there it arises straight from the consideration of massless particles --- therefore it makes sense to derive it without invoking a limiting procedure as presently.
	
	\subsubsection{Massless Particle Trajectories}
	If we want to calculate the impact parameter that separates capture of massless particles from other possible trajectories, we need a slightly different starting point compared to massive particles. With massive particles, we could frame the problem in terms of dimensionless variables involving the particle's mass. The way this was done in equations~\eqref{eq:massivedimless} can obviously not be extended to the massless case. A look at the resulting differential equation only reinforces this point. Nonetheless, with a different choice of parameters, we can still non-dimensionalise the system. We achieve this by setting
	\begin{subequations}
		\begin{align}
			\rho &\defi \frac{r_\text{H}}{r},\\
			b &\defi \frac{L}{E r_\text{H}},\\
			\iota &\defi \frac{E \lambda}{r_\text{H}}.
		\end{align}
	\end{subequations}
	At this point it is a good idea to remind ourselves that $\lambda$ stands here for a parametrisation (introduced above), \emph{not} a wavelength. Also, this time there is no need to distinguish carefully between the apparent impact parameter and the dimensionless one as in the case of massive particle trajectories. In the present case of massless particle trajectories, the only difference between the apparent, physical impact parameter $b_\text{phys}$ and $b$ is the normalisation by $r_\text{H}$.
	
	With this reformulation, the radial ODE takes the form
	\begin{equation}
		\kl{\frac{\dif \rho}{\dif \iota}}^2 = 1 - \underbrace{\kl{1-\frac{2GM}{r}}b^2 \rho^2}_{\ifed U_0}.\label{eq:Unull}
	\end{equation}
	We now look for zeroes of this equation. This either corresponds to a circular orbit, or (and this is what we aim for) a turning point. Turning points have all kinds of meanings --- they in principle could imply bound orbits, scattering, or an attempted, but failed escape. Many of these options do not exist for null geodesics, as a discussion slightly too far afield for our purposes would reveal, see \cite{Wald,Carroll}. Either way, setting equation~\eqref{eq:Unull} to zero allows us to derive a direct correspondence between $b$ and $\rho$:
	\begin{equation}
		b = \ed{\rho\sqrt{1-\rho}}.
	\end{equation}
	This function $b(\rho)$ has a minimum at $\rho=2/3$, corresponding to a minimal apparent impact parameter of $3\sqrt{3}\; r_\text{H}/2$. It immediately follows that for massless particles
	\begin{equation}
		\sigma_\text{capture} = \pi \kl{\frac{3\sqrt{3} \;r_\text{H}}{2}}^2 = \frac{27}{4}\pi r_\text{H}^2.
	\end{equation}
	This again results in $c_\text{eff}=27/16$, and it agrees with equation~\eqref{eq:massivelimitcrosssection}, as it should, and as we expected.
	
	\subsubsection{Further Complications}\label{sec:geocomplications}
	There are some caveats to be mentioned regarding the previous derivation of capture cross-sections. The discussion of these will be rather brief: For reasons discussed later in section~\ref{sec:3+1cross}, we have little need for technical precision in this particular point. Nevertheless, the existence of these additional effects itself warrants some discussion: So far, we reviewed only the capture cross-sections in the context of the Schwarzschild geometry. Obviously, this will not be enough for situations involving other black hole space-times. Take, for example, the case of a dirty black hole as described in section~\ref{sec:dirty}. These are still spherically symmetric, even still static, but their surface gravity changes, just as the position of the horizon changes. Most of these changes are either irrelevant for the calculation of the capture cross-section, or can easily be absorbed in the formalism presented above: We can generalise the analysis building on equation~\eqref{eq:crossODEr} by changing the factor $1-2GM/r$ appropriately to $b(r)$, while the anomalous redshift is picked up by the particle's energy $E$. This calculation cannot, however, be done in general --- unless one specifies anomalous redshift and shape function, it will not be possible to calculate the corresponding capture cross-section or its $c_\text{eff}$.
	
	Should one even deviate from spherical symmetry, the issue becomes more complicated. Taking the Kerr geometry as an example, the cross-section changes depending on the angle of the observer's line of sight to the black hole's axis. We refer to \cite{FrolovZelnikov2011} for more on how to calculate the form of the resulting shadow. The uptake of the analysis of the shadow area is that its size is usually comparable to that of the cross-section calculated above. The shape changes: If viewed from the equatorial plane, the shadow will not be centred on the line of sight, but rather shifted in the direction of rotation. The other side will appear flattened. Only if the observer is positioned on and looking along the line of symmetry (\emph{i.e.}, the axis of rotation) will the shadow be circular.
	
	Similarly, the discussion becomes more involved if one goes beyond ray optics: In this case, the shadow also becomes frequency dependent. This effect is well-known and can be traced back at least to the work of Don Page \cite{PageThesis,Page1,Page2,Page3}, but is still studied today, see, for example, \cite{SparsityNumerical}. It is also intimately related to the occurrence of superradiance, and this will be the context in which we will encounter this particular change to the capture cross-section, see section~\ref{sec:superradiance}. When doing so, the capture cross-section is more appropriately called \enquote{absorption cross-section}, to emphasise the difference between wave optics and ray optics.	To close on a lighter note (and not a complication), we mention that the previously mentioned Regge--Wheeler equation can reproduce the potential $U_0$, thus linking the wave optical approach to the geometric optics one, see \cite{FrolovZelnikov2011}, page~305. That wave optics produces similar $c_\text{eff}$ should thus not come as a surprise. This is most easily seen by taking note of two things: The geometric optics approximation means that we have a high occupation number, and it thus makes sense to approximate the angular momentum quantum number $\ell$ as being related to the physical angular momentum $L$ of the null geodesic. Neglecting $\ell$-independent terms in the angular momentum potential of the Regge--Wheeler equation (by noting approximating $\ell(\ell+1)\approx L$, $\ell \gg 1$), one can substitute in this approximation $\ell +\ed{2}$ with $E r_\text{H} b$ and gets $U_0$ back.
	
	This cross-pollination of the geometric optics approximation and the full wave optics (this time in the other direction) is called the DeWitt approximation. It can be rephrased as rephrasing the capture condition for null geodesics in terms of wave mode properties. The approximating assumption is then that all wave modes with $\ell < 3\sqrt{3}M\omega$ are absorbed. This corresponds to approximating the grey body factors by 
	\begin{equation}
		T_{\ell sm} \approx \theta\kl{3\sqrt{3}M\omega - \ell},
	\end{equation}
	where $\theta$ is the Heaviside function. Summing over all modes then gives a grey body factor depending only on frequency:
	\begin{equation}
		T(\omega) \approx \sum_{\ell=0}^{\infty} (2\ell+1)\theta\kl{3\sqrt{3}M\omega - \ell}\approx 27M^2\omega^2,
	\end{equation}
	which, rephrased in terms of $r_\text{H}$ reproduces, as claimed, $c_\text{eff}$. The big difference is that in the DeWitt approximation the analysis happens, basically, under the integral of the Bose--Einstein distribution, which explains the additional factor $\omega^2$ in the result. No particle physics (or statistical physics) input like this was needed in the original geometric optics approximation. Hence the DeWitt approximation is at least a (partial) explanation of why the geometric optics approach to \enquote{sparsity} (yet to be defined) will work as well as we will see in chapter~\ref{ch:sparsity}.

\chapter{Analogue Space-Times} \label{ch:analogues} 
\epigraph{\enquote{The snake spoke to her?\\No! no! It is a make-like.}}{Philip Pullman, \emph{The Amber Spyglass}}
This chapter will begin in section~\ref{sec:ast-general} with a general discussion of the aim of the analogue space(-time) framework --- for the moment allowing more general questions than the implicit mentioning of Lorentzian signature metrics would suggest. However, the section will finish with coming back to the specific context of this thesis, analogue space-times, reintroducing the Lorentzian signature. Afterwards, we will enter the main part of this chapter, section~\ref{sec:CovEMana}: A covariant, algebraic, electromagnetic space-time analogue. For this we will both quickly introduce the required notation and background from electromagnetism, as well as rederive the previously achieved answer to the question when a medium is describing an analogy to a space-time. (The other direction will turn out to be trivial.) These results have long been only looked at in explicitly or at least implicitly $3+1$-dimensional formalisms, occasionally even leading to expressions mixing tensors and tensor densities in incompatible ways in the results. One of the goals of our approach is to avoid this, both in the $3+1$-dimensional formalism, as well as in a full, four-dimensional and covariant approach. In section~\ref{sec:trafooptics}, we will give a very brief comparison to similar results out of the field of transformation optics. Section~\ref{sec:bespoke} then considers a list of examples of analogue space-times in our formalism, before section~\ref{sec:PDEana} discusses the difference between having an algebraic analogue and an analytical analogue. Finally, section~\ref{sec:refractive} will introduce an analytic analogy based on electromagnetism.

\section{General Remarks}\label{sec:ast-general} 

As already mentioned in the historical introduction, section~\ref{sec:history}, and independent of the necessities of the traditional realm of CSTQFT --- that is, actual astrophysical space-times ---, analogue space-times have a long history. In this section we want to make their meaning more concrete.

The principal idea behind analogue (or effective) space(-times) is explained in the easiest way as an exercise in differential geometry, specifically, (semi-)Riemannian geometry. Here the notion of a manifold together with a set of agreed-upon, geometric differential operators (Lie derivatives, exterior differentiation, connections, \dots) can give more geometric meaning to certain differential equations. Among the most traditional cases would be differential equations describing shortest paths on a manifold (geodesic equations), parallel transport, and notions of symmetry (via Lie derivatives). Common to these is, that the notion of a manifold arose when people first tried to make systematic sense out of the corresponding concepts as seen in Euclidean geometry: Shortest paths are simply straight lines, and both parallel transport along a given curve or recognizing symmetries at hand required comparatively few abstract concepts. With the advent of non-Euclidean geometry, however, these questions became substantially more subtle, and in turn created a need for more technical and abstract terms, which then in turn birthed the previously mentioned, different notions of differentiation on a manifold. In a similar vein, generalisations of the Laplace equation\footnote{If we allow for this to mean both homogeneous and inhomogeneous equations, as well as for varying metric signature, this will then encompass Laplace, D'Alembert and Poisson equations.} to general manifolds have to be viewed as coming originally from their \enquote{living in Euclidean (or Minkowskian) space}. The vanishing of curvature terms makes it far from obvious what exactly the generalisation of these equations to a general manifold would entail. As a result, there exist a variety of Laplacian operators on general manifolds whose difference depends characteristically on the curvature (and what kind of section they should act on: scalars, tensors, spin bundles, \dots), encoded in, for example, the Lichnerowicz formula or Weitzenböck identity, see \cite{GriHar,LawMichSpin}.

At first glance, this looks rather forbidding and unhelpful. However, the ubiquity of differing wave equations in physics means a similar plethora of structures. Should it be possible to reformulate any one of these in purely geometric terms, this would imply that geometric means can be employed in their study, too. What exactly these geometric terms \emph{are}, is \emph{a priori} not clear. What specific geometric structure\footnote{For example: Without any structure besides the tangent space construction (and dual spaces and tensor products), the only naturally arising derivative is the exterior one and, more or less, the Lie derivative (more because it needs no further structure, less because it involves the choice of a vector field). However, one can, as we shall, add a metric structure, which will add notions of curvature, and additional derivatives.} one wants to look for is then a matter of choice based on either what analogues one wants to look at (\emph{i.e.}, what field of physics one pursues) or what space(-time) one wants to look at (\emph{i.e.}, what geometric structure is sought).

Let us specialise, at this point, to analogue space-times, that is, let us fix the metric signature to be Lorentzian, thus adding a metric to the assumed geometric structures on the manifold. The most general way of phrasing the analogue space-time framework is then the following: Take any hyperbolic, second-order, partial differential equation and try to rewrite it in terms of a manifold with Lorentzian metric and its geometric derivatives and properties (curvature, torsion, topology, \dots). A mathematical perspective on this procedure can be found in \cite{BHfromPDE}. The most obvious, and straightforward example, that also highlights the naming of the analogue space-time framework, would be a PDE that could be rewritten as a certain scalar Laplace--Beltrami equation:
\begin{equation}\label{eq:Laplace-Beltrami}
	\nabla_a\nabla^a f = f_{;a}{}^{;a} = \ed{\sqrt{\abs{\det \g}}} \partial_a \kl{\sqrt{\abs{\det \g}}\;\partial^a f} = 0,
\end{equation}
where the index is raised with the effective metric $g_\text{eff}$.

However, we shall find that in our particular case --- the electromagnetic analogue --- the analogy is not necessarily of an analytic nature (\emph{i.e.}, through a PDE), but can also be given in a purely algebraic way. The resulting non-trivial interplay between the wave equation of the analogue and the wave equation in the background metric will be looked at in section~\ref{sec:PDEana}.

It is worthwhile to mention a philosophical point: As with \emph{any} analogy, \emph{any} statement achieved or experiment performed in the analogue framework, will necessarily be only non-empirical for the real counterpart of the analogy (in our case: astrophysical space-times). If, for example, Hawking radiation is observed in the laboratory, this will certainly improve theorists' confidence in the methods deriving Hawking radiation in any circumstance --- but the question of its existence in the original astrophysical context remains open until the explicit measurement of \emph{this} Hawking radiation. Unless revolutionary discoveries (either of technological nature or of entirely unforeseen effects (if not physics) already available to us) happen, the analogue framework can only improve psychological confidence, not empirical confidence for astrophysics. At the same time it means that CSTQFT has a meaning independent of astrophysics: As it is the case with quantum field theory or differential geometry, so is CSTQFT a framework applicable equally well to very different fields of physics.

\section{Electromagnetic Analogues --- A Covariant Introduction} \label{sec:CovEMana}
In order to set the stage, it is useful to recapitulate the formulation of both macroscopic and microscopic electrodynamics in terms of tensor fields.\footnote{A slight variation would be to consider tensor densities. This will not be done. Rather, we shall repeatedly highlight the strictly tensorial nature of the formalism to be developed. The same preference of tensors over tensor densities can be observed in previous results within the field of transformation optics \cite{CovOptMet1,CovOptMet2,CovOptMet3,CovOptMet4}.} Differential forms (as a more constrained tensorial structure) as foundation are also found, for example, for natural reasons in the context of pre-metric electrodynamics: As the goal of that field of research is to \emph{derive} a metric structure from electrodynamic properties, any notion of differentiability must not depend on the existence of that metric structure. Likewise, the metric duality of tangent space and co-tangent space is lost, nor is Hodge duality available to relate forms of different rank to each other. For more on the pre-metric approach, see \cite{RubilarPremetric,HehlObukhov,HehlLaemmerzahl,Itin1,Itin3,Itin2,Itin4,PreMetricAQFT,PreMetricQEI} and references therein.

However, we do believe that the use of forms will obfuscate the appearance of two different metrics in this analogue: For example, while varying the action it is easy to miss that the Hodge star will necessarily have to rely on the background metric of the laboratory, not the effective metric. We will discuss this in more detail at the end of subsection~\ref{sec:AnalougeAndMacro}.

\subsection{Maxwell's Equations and a Constitutive Tensor} 
Regarding the foundations of microscopic electromagnetism we shall take a pragmatist approach: For us, electrodynamics is described through the equations of motion derived from the variation of the action
\begin{equation}
	S = -\int \sqrt{-g} \ed{4}F_{ab} F^{ab} + A_a J^a \dif^4 x,
\end{equation}
where
\begin{equation}
	F_{ab} = \nabla_a A_b - \nabla_b A_a\label{eq:defA1}
\end{equation}
is the field strength tensor, derived from the four-potential $A_a$, $J^a$ is the four-current, and $g$ the determinant of the space-time metric $g_{ab}$, which in turn underlies the (metric compatible and symmetric) covariant derivative $\nabla$. The appearance of the covariant derivative is superfluous, and more a matter of taste,\footnote{This will change to a degree in what is to come, see section~\ref{sec:PDEana}!} as
\begin{equation}
	(\dif A)_{ab} = \partial_a A_b - \partial_b A_a = \nabla_a A_b - \nabla_b A_a.\label{eq:defA2}
\end{equation}
Less a matter of taste is the introduction of the four-potential, as this results in the simultaneous introduction of topological constraints on the manifold if this potential should be globally defined. Locally, its existence can be guaranteed by the Poincaré lemma, but this does not guarantee its extension to global existence. For the immediate context of this thesis these topological points are of little relevance --- however, it is easy to imagine extensions of the results presented in this chapter to be garnished with matters topological. This would provide new experiments on analogue space-times to be examined in the laboratory.

The resulting Euler--Lagrange equations are the microscopic Maxwell equations:
\begin{subequations}
	\begin{align}
		\nabla_a F^{ab} &= J^b,\label{eq:MicroInHomMax}\\
		\nabla_{[a}F_{bc]} &=0.\label{eq:MicroHomMax}
	\end{align}
\end{subequations}
The first set are the inhomogeneous Maxwell equations, also known under the name of Gauss--Ampère law, while the second set, the homogeneous ones, also goes by the name of Gauss--Faraday law.

The physical interpretation of these equations is that they describe electromagnetic fields on a space-time subject to prescribed charge and current encoded in $J^a$. The motion of charged particles can also be captured by adding an appropriate term to the action. As we are not concerned with the motion of charged particles, we shall not look into this. If one is not in flat space, an additional term involving the Ricci curvature tensor can make its appearance when using a four-potential. Since we will, for the most part of this chapter, not be concerned with the differential equation, we will not dwell on this. The Ricci tensor will make its appearance once it is of importance and does not detract attention from the subject of our analysis.

As the dependent variables we would be solving for are the components of the field strength tensor, we have six linearly independent dependent variables. At first glance it seems that the two sets of Maxwell equations over-determine this. That this is not so is most easily seen by going to a 3+1-dimensional description and paying heed to the fact that $\nabla\cdot(\nabla\times \mathbf{E}) = \nabla\cdot(\nabla\times \mathbf{B}) = 0$. This demonstrates that not all of the eight Maxwell equations are linearly independent. 

In preparation for both macroscopic electrodynamics and its constitutive relation, as well as to motivate the type of analogue space-times we want to consider in this chapter, we will have a closer look at the left-hand-side of equation~\eqref{eq:MicroHomMax}. More precisely, we are interested in the twice contravariant field strength tensor $F^{ab}$.
\begin{subequations}\label{eq:Zvacuum}
	\begin{align}
		F^{ab} &= g^{ac}g^{bd} F_{cd},\\
			&= \ed{2}\kl{g^{ac} g^{bd} - g^{ad}g^{bc}} F_{cd},\\
			&\defi Z^{abcd}_\text{vacuum} F_{cd},
	\end{align}
\end{subequations}
where we made use of the antisymmetry of $F$ to introduce a new tensor $Z$ with the properties
\begin{subequations}
	\begin{align}
		Z^{abcd}_\text{vacuum} &= Z^{cdab}_\text{vacuum}, \\
		Z^{(ab)cd}_\text{vacuum} = Z^{ab(cd)}_\text{vacuum} &= 0,\\
		Z^{[abcd]}_\text{vacuum} &=0.
	\end{align}
\end{subequations}
While this seems somewhat moot, in the context of (the as of now not described) macroscopic electrodynamics $Z$ describes the constitutive relations of the vacuum of a given space-time. As this is the core idea behind (algebraic) analogue space-times out of dielectric media, this seemingly trivial point bears belabouring, repeating and remembering.

\subsection{From Microscopic to Macroscopic Electrodynamics} 
While microscopic electrodynamics treats every charged particle as a separate source, the idea behind macroscopic electrodynamics is to distinguish between bound charges and bound currents, and free charges and free currents. For references, see, for example, \cite{HehlObukhov,AbrahamMinkowskiContro} --- the details of this separation of bound and free charges shall not be our concern, especially given some long-standing controversies, like the Abraham--Minkowski one, about these details.

For our purposes, we will neglect possible polarisation of the medium. As a result, macroscopic electrodynamics in our context means a system of partial differential equations for either the four-vector potential $A$ or electromagnetic excitation tensor $G$ and electromagnetic field strength tensor $F$. The system is comprised out of the above-mentioned homogeneous Maxwell equations, the inhomogeneous Maxwell equations to follow below, and the constitutive relations linking the field strength tensor $F$ and the excitation tensor\footnote{The excitation tensor is also known as the displacement tensor. Had we chosen to include polarisation and magnetisation, we would have had to introduce a third second rank, antisymmetric tensor $M^{ab}$, changing the constitutive relations to $G^{ab} \defi Z^{abcd}\,F_{cd} - M^{ab}$. As the polarisation tensor $M^{ab}$ would normally break our framework to mimic a different vacuum space-time, it would hinder our analysis, if not precluding it completely. This might change if one wants to incorporate vacuum polarisation effects.} $G$ in the following way:
\begin{equation}\label{eq:constitutive}
	G^{ab} \defi Z^{abcd}\,F_{cd}.
\end{equation}
Here, $Z$ is a fourth rank tensor with the properties that
\begin{subequations}\label{eq:Zsymprops}
	\begin{align}
		Z^{abcd} &= Z^{cdab}, \label{eq:Zsym}\\
		Z^{(ab)cd} = Z^{ab(cd)} &= 0.\label{eq:Zasym}
	\end{align}
\end{subequations}
The first property of symmetry, equation~\eqref{eq:Zsym}, follows from the desire to be able to derive the Maxwell equations from an action principle by varying the new action
\begin{subequations}
	\begin{align}
	S &= -\int \dif^4 x \frac{\sqrt{-\det  g}}{4} F_{ab} G^{ab},\\
	&= -\int \dif^4 x \frac{\sqrt{-\det  g}}{4} F_{ab} Z^{abcd} F_{cd}. \label{eq:action}
	\end{align}
\end{subequations}

The antisymmetry properties, equation~\eqref{eq:Zasym}, are likewise a consequence of the defining equation~\eqref{eq:constitutive}, together with the previous symmetry property: The contraction with the (antisymmetric) field strength tensor renders any possible symmetric parts irrelevant.

For counting the degrees of freedom (d.o.f.) of $Z$, we make again use of these properties. Each index pair has only $\nicefrac{4(4-1)}{2}=6$ degrees of freedom, which gives rise to the possibility to rewrite $Z^{abcd}$ as $Z^{AB}$, where $A,B$ range from 1 to 6. A change to these six-dimensional indices occurs frequently in the literature, for example in \cite{Post,HehlObukhov,AreaMetrClass}. We devote a subsection below to this notation. The symmetry on switching the index pairs then gives the total degrees of freedom for $Z$ as $\nicefrac{6(6+1)}{2}=21$.

It is possible to consider less restrictive constitutive tensors (for example, this is often encountered in a pre-metric context as done in \cite{HehlObukhov}), or considering a minor variation of the constitutive relations which gives rise to slightly different magneto-electric tensors, see \cite{AniBiGuide,TrafoOpticsCartographDistort},\footnote{Concretely, this is the difference between Tellegen and Boys--Post constitutive relations. We will be using the latter. This is, again, a matter of convention rather than of physics.\label{fn:TellegenBoysPost}} or more restrictive constitutive tensors (by demanding the first (algebraic) Bianchi identity to hold, as done in \cite{Post}). Our later application of a constitutive tensor to describe an analogue space-time will actually result in the first Bianchi identity to be fulfilled,\footnote{The Bianchi identity is connected to the vanishing of the fully antisymmetric part of $Z$. This vanishes as in our case (yet to be described) the ingredients are symmetric second rank tensors, the inverse effective metrics.} but at this stage it is not required to demand it, and we shall therefore refrain from doing so. Depending on one's choice of definition of constitutive tensor, $G$ may have a similar property as the dual field strength tensor $\ast F$: The positioning of electric and magnetic parts may change compared to that in $F$ itself: While the electric fields $\mathbf{B}$ are the purely spatial part of the field strength tensor, in $\ast F$ (or possibly $G$) this role will be filled by $\mathbf{D}$, similarly for $\mathbf{E}$ and $\mathbf{H}$ in the spatio-temporal part of $F$ and $\ast F$ (or possibly $G$). Our definition of $G$ does not do this --- the spatio-temporal components are $\mathbf{D}$, the purely spatial ones are $\mathbf{H}$. More on this, in terms of our convention, can be found in the appendix, section~\ref{sec:twoform}.

Having described the constitutive relations, let us give the Maxwell equations as they follow from the action~\eqref{eq:action}:\footnote{Again, we do not yet include a four-potential. We therefore also do not need to worry about curvature terms at this point. This will be considered further below in section~\ref{sec:PDEana}.}
\begin{subequations}
	\begin{align}
		\nabla_a G^{ab} &= J^b, \label{eq:InHomMaxwell}\\
		\nabla_{[a} F_{bc]} &= 0.\label{eq:HomMaxwell}
	\end{align}
\end{subequations}

As the resulting four-dimensional, homogeneous Maxwell equation~\eqref{eq:HomMaxwell} for the field strength tensor $F$ reads the same in both macroscopic and microscopic electrodynamics, and plays the role of consistency conditions (and is, for this reason, often called \enquote{merely geometric} in nature\footnote{The homogeneous Maxwell equation in the language of 2-forms reads $\dif F=0$, thus \enquote{merely} stating that $F$ is a closed two-form.}), the inhomogeneous Maxwell equation is at the core of electromagnetic analogues. It is here that the difference between \emph{in vacuo} and electromagnetic media appears.

Luckily for us, we are not looking for solutions of the Maxwell equations (at this point of the development): Finding a decent gauge, for example, is far more complicated in the realm of macroscopic electrodynamics than it was in microscopic electrodynamics. Related to this is the fact that at the time of writing, the search for a Green's function for macroscopic electrodynamics is ongoing. While, for example, for a constant(!) $Z^{abcd}$ Itin provides a general Green's function \cite{Itin1}, the situation is not yet settled for non-constant constitutive relations, even though partial results are available, for example \cite{MelroseDispersion}. The resulting bad news is that a careful re-examination (without too many approximations) of Hawking radiation in the corresponding analogue space-times would rely on just this.

\subsubsection{\texorpdfstring{The $6\times6$ Representation of $Z$}{The 6x6 Representation of Z}}\label{sec:6dim}
It is instructive to have a closer look at the representation of $Z$ as a symmetric $6\times6$ matrix, as indicated above. Written out, this matrix is
\begin{equation}
	\kl{Z^{AB}}_{A,B\in\{1,\dots, 6\}} = \begin{pmatrix}
	\epsilon & \zeta \\ \zeta^\dagger & \mu^{-1}
	\end{pmatrix},\label{eq:Z6x6}
\end{equation}
where $\epsilon$ is the $3\times3$ permittivity matrix, $\mu^{-1}$ is the (inverse) $3\times3$ permeability matrix, and $\zeta$ is the $3\times3$ magneto-electric matrix. We shall refer to these collectively as the constitutive \emph{matrices} in contrast to the constitutive tensor $Z$. Equivalently, we can denote them as \enquote{susceptibility tensors} or \enquote{susceptibility matrices}, and --- when no confusion with the plural of the constitutive tensor $Z$ is possible --- also as \enquote{constitutive tensor\underline{s}}. Here, $\epsilon$ and $\mi$ are real and symmetric, while $\zeta$ is real, but in general asymmetric. These link $\mathbf{E}, \mathbf{B}$ with $\mathbf{D}, \mathbf{H}$ in the following way\footnote{Just as the use of Franklin's \enquote{inconvenient} choice of the sign of the electric current (opposite to that of the flow of electrons) is a historical accident, so is the use of $\mi$ instead of $\mu$. We shall have to mention this again later on, as it sadly makes some subsequent results rather cumbersome in appearance.\label{fn:historylesson} This is related footnote~\ref{fn:TellegenBoysPost}}
\begin{alignat}{4}
	\mathbf{D} &=& \epsilon\; \mathbf{E}\; &+& \zeta\; \mathbf{B},\nonumber\\
	\mathbf{H} &=& \zeta^\dagger \mathbf{E}\; &+&\; \mu^{-1} \mathbf{B}.\label{eq:EBtoDH}
\end{alignat}
In terms of the $6\times6$-version of $Z$ this could be rewritten as
\begin{equation}
	\begin{pmatrix}
	\mathbf{D}\\\mathbf{H}
	\end{pmatrix} =
	\begin{pmatrix}
	\epsilon & \zeta \\ \zeta^\dagger & \mu^{-1}
	\end{pmatrix}
	\begin{pmatrix}
	\mathbf{E}\\\mathbf{B}
\end{pmatrix}.
\end{equation}
This demonstrates the issue with this formalism for our purposes: All fields involved implicitly depend on the four-velocity $V^a$ of the observer, as electric and magnetic fields/\allowbreak ex\-ci\-ta\-tions are part of the corresponding orthogonal decomposition of the field/\allowbreak ex\-ci\-ta\-tion tensors as described in appendix~\ref{sec:twoform}. Already at this level one expects that the constitutive matrices mix in a quite messy way under Lorentz transformations (which are important in the flat space-time context of moving media, for example in the Fresnel--Fizeau effect, see \cite{Post}), and even more so under general coordinate transformations (which become important, if we want to view $\g$ as an effective metric on a general, possibly curved background with physical metric $ g$). In the appendix~\ref{sec:4tensor} we shall further investigate the relationship between $V^a$ and the constitutive matrices --- they will prove to be the elements of the Bel decomposition (also known as the orthogonal decomposition) with respect to given $V^a$. This appendix also then settles the anticipated mixing of constitutive tensors under transformations within our notation.

The Maxwell equations in this case reduce to the well-known ones. For the inhomogenous ones we have
\begin{subequations}
	\begin{align}
		\frac{\dif \mathbf{D}}{\dif t} &= \nabla \times \mathbf{H} - \mathbf{j},\\
		\nabla\cdot \mathbf{D} &= \rho.
	\end{align}
The homogeneous ones are
	\begin{align}
		\frac{\dif \mathbf{B}}{\dif t} &= - \nabla \times \mathbf{E},\\
		\nabla \cdot \mathbf{B} &=\hphantom{-} 0.
	\end{align}
\end{subequations}

\subsection{Analogue Space-Times and Macroscopic Electrodynamics}\label{sec:AnalougeAndMacro}
We now have assembled the necessary ingredients to fully introduce (rather than continuously hint at) the method to produce \emph{algebraic} analogue space-times using covariant, macroscopic electrodynamics. First, let us remind ourselves of the constitutive tensor of vacuum electrodynamics,
\begin{equation}
	Z^{abcd}_\text{vacuum} = \ed{2}\kl{g^{ac} g^{bd} - g^{ad}g^{bc}}.
\end{equation}
With this we can rewrite the action of microscopic electrodynamics. Note that we are interested in keeping the conformal invariance, as we shall exploit it frequently in the future discussion. We will from now on assume a vacuum, equivalent to setting $J^a=0$. This is not strictly necessary, but otherwise we would have to keep track of conformal factors in the four-current, as its conformal weight is $-4$, see the corresponding discussion in appendix~\ref{sec:conformal}.
\begin{subequations}
	\begin{align}
		S &= -\int \dif^4 x \frac{\sqrt{-\det  g}}{4} F_{ab} F^{ab},\\
		&= -\int \dif^4 x \frac{\sqrt{-\det  g}}{4} F_{ab} Z^{abcd}_\text{vacuum} F_{cd}.
	\end{align}
\end{subequations}
However, this assumes that the Lorentzian manifold $(M,g)$ is both an electromagnetic vacuum and the space on which we vary the action. If we want to study an analogue space-time $(M,\g)$, however, we are required to introduce a distinction between the laboratory space-time $(M,g)$ (on which the variation happens), and the analogue space-time providing the \enquote{electromagnetic vacuum} $(M,\g)$.\footnote{Strictly speaking, the underlying manifolds might even be different --- at the very least the effective space-time will be a space-time with boundary. An optical medium could also easily accommodate a topology different to that of the laboratory. Some of these subtleties can be taken care of by carefully choosing a subset of $M$ as new base manifold.} 

We are now confronted with the issue of how to regain the correct Maxwell equations for $(M,\g)$: The factor of $\sqrt{-\det g}$ in the action will prevent this without additional thought. Let us provide a spoiler: Even after taking care of this, the solution turns out to be mostly aesthetic in nature. As we want our quantities to remain being tensors (rather tensor densities), our \enquote{solution} is to define a new constitutive tensor:
\begin{equation}
	Z^{abcd} = \ed{2}\frac{\sqrt{\det \g}}{\sqrt{\det  g}} \kl{\gi^{ac}\gi^{bd} - \gi^{ad}\gi^{bc}}.\label{eq:Zeff}
\end{equation}
It is trivial to see that the action for macroscopic electrodynamics on $(M,g)$ with this particular constitutive tensor will automatically \emph{look} like the action of microscopic electrodynamics of the Lorentzian manifold $(M,\g)$:
\begin{equation}
	S \stackrel{!}{=} -\ed{8} \int \dif^4x \sqrt{-\det \g} \kl{\gi^{ac}\gi^{bd} - \gi^{ad}\gi^{bc}}F_{ab}F_{cd}.\label{eq:actioneff}
\end{equation}
This seems to mean that the equivalence holds at the \emph{full wave optics} level of the analysis. No geometric optics approximation was needed, no WKB methods, no approximations. Of course, any material will introduce a natural frequency cut-off at the energy scale given by its lattice size, but this has not been introduced at this stage. Sadly, this appearance will turn out to be deceptive.

Still easy to see, but slightly more involved, is that this $Z$ indeed transforms as a tensor: First, suppose one is given a tensor density $A$ of weight $s$ and rank $(m,n)$ and a tensor density $B$ of weight $s'$ and rank $(m',n')$. Multiplying these together (in the sense of index-wise multiplication or, equivalently, tensor multiplication) results in a new tensor density $AB$ of weight $s+s'$ and rank $(m+m',n+n')$. Second, ${\sqrt{\det \g}}/{\sqrt{\det  g}}$ is the corresponding product of two scalar densities of opposite weight, thus resulting in a true scalar. Third, this is multiplied with a regular tensor, again resulting in a new, true tensor.

One more warning is again related to our demand of handling tensors instead of tensor densities: Any Levi-Civita symbol appearing will have to be interpreted as the corresponding Levi-Civita \emph{tensor} which differs from the more traditional Levi-Civita density\footnote{As many definitions of \enquote{tensor densities} actually involve the absolute value of the determinant, it could be argued that \enquote{Levi-Civita pseudo-tensor density} is even better a term. This was already alluded to in section~\ref{sec:notation}, where we set down our notational choices.} by factors of $\sqrt{-g}$ --- the precise way depends on the index placement.

It is important to now mention another caveat at this point: While we had no need to distinguish between the metric and the inverse metric for $Z_\text{vacuum}$ and $g$, for $\g$ this distinction is crucial: As we are still raising and lowering indices with the background (laboratory) metric $g$, $\g^{ab}$ and $\gi^{ab}$ are now different tensors. It is only valid for the background metric $g$ to equate $g^{ab}$ and $[g^{-1}]^{ab}$. It is this caveat that also tells us of the issues with the above appeal to the action $S$.

The precise issue is, for once, easier seen in the context of electrodynamics written out in forms. Here the microscopic action is represented by
\begin{equation}
	S = \int_M F \wedge \star_g F,
\end{equation}
where we emphasized that the Hodge star determines which Lorentzian manifold is used for the variation --- the Lagrangian $F \wedge \star_{\g} F$ would obviously lead to different electrodynamics. Macroscopic electrodynamics is now achieved through the action
\begin{equation}
	S = \int_M F \wedge \star_g G,\label{eq:formsaction}
\end{equation}
where we have used an excitation two-form $G$. This two-form is formed from the excitation tensor (by lowering indices) in the following way:
\begin{equation}
	G_{ab} = g_{ac} g_{bd} G^{cd} = g_{ac} g_{bd} Z^{cdef} F_{ef}.
\end{equation}
Even if the constitutive tensor is given by equation~\eqref{eq:Zeff}, we see that equation~\eqref{eq:formsaction} still has two points where the background metric came in: The Hodge star, and the two-form $G$. 

A different way of seeing this would be to remember that electrodynamics is conformally invariant. Thus changing the pre-factor of $Z$ to ${\sqrt{\det \g}}/{\sqrt{\det  g}}$ cannot change the physics. If one were to introduce a four-potential $A_a$ and write $F_{ab}$ in terms of covariant derivatives, one can understand the action~\eqref{eq:actioneff} two-fold: Once as $S(A_a, \nabla_a A_b)$ and once as $S(A_a, \nabla^\text{eff}_a A_b)$, where the covariant derivative is either the Levi-Civita connection of $g$ or of $\g$. We will follow this line of thinking some more in section~\ref{sec:PDEana}.

Nonetheless, this is all an \emph{analytic} argument. We simply will not reproduce the microscopic Maxwell equations of $(M,\g)$ perfectly with the macroscopic Maxwell equations of $(M,g)$. But the analogy to space-times here is strictly \emph{algebraic}. All we really ask for is for $Z$ to raise indices as if done with $\gi$. This is not a very common way of employing the analogue space-time framework --- but ironically it is with Gordon's original contribution \cite{GordonMetric} one of the oldest. So let us postpone the analytic complications until section~\ref{sec:PDEana} and focus on two other questions: When is this algebraic analogue space-time possible, and, \emph{vice versa}, when is a given material with electromagnetic properties encoded in $Z$ equivalent to an algebraic analogue?

\subsection{Consistency Conditions}\label{sec:conscond}
We start by remembering our d.o.f. analysis of the constitutive tensor: Our result was that it has 21 degrees of freedom. If we then take a look at the specific form of the constitutive tensor for an analogue space-time as set down in equation~\eqref{eq:Zeff}, we see that we only have the 10 degrees of freedom of a symmetric rank-2 tensor. However, the underlying theory being electrodynamics, we also have to take into account its conformal freedom. More specifically, any conformal rescaling of the metric will result in a rescaling of $Z$ with weight $-4$, see appendix~\ref{sec:conformal}. Hence, the electromagnetic space-time mimic will be unable to distinguish a metric from a conformal rescaling of it; both result in the same electromagnetic theory. Thus, any analogue space-time is in more precise words in truth a \enquote{analogue space-time class}. This also teaches us again not to take the simplifications in the action~\eqref{eq:actioneff} by our choice of conformal factor in equation~\eqref{eq:Zeff} too seriously. Instead of the 10 d.o.f. of a metric, a conformal class of metrics has only 9 d.o.f. available, and therefore we have a mismatch of 12 d.o.f. between the constitutive tensor of an analogue space-time and the most general constitutive tensor imaginable. The need for consistency conditions becomes clear. It is worthwhile to appreciate however, that already at this stage it should be obvious that \emph{every} four-dimensional, Lorentzian metric \emph{can} be written as an analogue space-time, and with the four-dimensional generalisations of the 3+1-dimensional constitutive tensors in place, one can immediately write down the corresponding constitutive tensors. Whether or not the resulting tensors are experimentally achievable is a different matter altogether. It is not hard to see that most coordinate forms of a given metric are likely to result in impractical, if not practically impossible, material properties.

What is usually done to find the consistency conditions, is to use the properties of the constitutive tensor $Z$ (as described above) and switch from four space-time indices $a, b, c, \dots$ ranging from 0~to~3 to two \enquote{field indices} $A, B, C, \dots$ ranging from 1~to~6, as described in section~\ref{sec:6dim}.
The issue here is that one loses the full covariance and instead implicitly uses an observer-dependent 3+1 decomposition. In the context of pre-metric electrodynamics (see, for example, \cite{HehlObukhov} and references therein) this is not a bug, but a feature. Our current approach is orthogonal to the pre-metric one: Not only do we want to keep the physical background metric $g$, we will also look for our effective metric $\g$. As both metrics will be four-dimensional and general, we want to stick with space-time indices. This particular approach was usually accompanied by adding additional assumptions on either the medium or the space-time to be mimicked, such as isotropy, geometric symmetries, vanishing magneto-electric effects, \dots --- our goal is to move beyond these suppositions.

As a result, the strategy we used in our paper \cite{CovEffMetrics}, and will be employing in this chapter, was two-fold: First, we wanted to showcase this fully covariant formalism for electrodynamics of media using only space-time indices as it is done, for example, in \cite{PerlickRay}, and \cite{BalZim}. In particular, this formalism enables us to evade additional constraints or assumptions on analogue space-time or medium. Second, we also wanted to find the consistency conditions in terms of the constitutive matrices themselves. As this second point in turn is important when engineering materials for this purpose, we shall give these consistency conditions (and the corresponding, analogue metric) in terms of the familiar matrices $\epsilon, \mu^{-1}$, and $\zeta$ (or, rather, their four-dimensional generalisations).

While the derivation of the consistency conditions has been done before (in numerous and various contexts and formalisms), see for instance references~\cite{HehlLaemmerzahl,BalZim,BalNi,AreaMetrClass,FavaroBergamin}, it still remained to explicitly write down the resulting effective metric once the consistency conditions are satisfied \emph{in terms of the corresponding constitutive tensor}. (In the context of pre-metric electrodynamics this is quite naturally done as soon as the space-time metric is recovered \cite{RubilarPremetric,HehlObukhov,HehlLaemmerzahl}.)

We will go about the derivation in steps: First, we re-derive the consistency conditions in a $3$$+$$1$-dimensional, flat space-time context, both for vanishing, and for non-vanishing magneto-electric effects. Second, we repeat this derivation in a fully covariant way. We achieve this by a careful translation of the $3$$+$$1$-dimensional notions into four-dimensional ones by making use of the Moore--Penrose pseudo-inverse and introducing the related, though little-known notion of a pseudo-determinant. After these careful derivations, we provide a physical argument reducing the analysis to that of a medium with vanishing magneto-electric effects. Each time we derive a different version of the consistency condition\footnote{Four times in total: Twice for the Minkowski $3$$+$$1$-dimensional formalism (vanishing and non-vanishing magneto-electric tensor), twice in a fully covariant formalism, independent of the background space-time (again, once for vanishing, once for non-vanishing magneto-electric tensor).}, we will also invert this consistency condition to then give the effective metric (or rather, a representative of its conformal class of metrics) purely in terms of the constitutive tensors occurring in (and \emph{as} in) the newly derived consistency condition.

\subsubsection{Step 1: Setting Up the Flat Space-Time, 3+1-Dimensional Formalism}
As a first preparation, it helps to make use of the freedom in the conformal factors: As our counting of degrees of freedom showed (based on the conformal invariance of electromagnetism, see appendix~\ref{sec:conformal}), the \enquote{effective metric} is a conformal class of metrics rather than a metric as such. This in turn means that any representative of this class of metrics is equally valid, and thus we can simplify our analysis tremendously by focussing on the representative for which
\begin{equation}
	\det \g  = \det  g.\label{eq:ConfChoice}
\end{equation}
Our constitutive tensor now takes on the form
\begin{equation}
	Z^{abcd} = \ed{2}\kl{\gi^{ac}\gi^{bd} - \gi^{ad}\gi^{bc}}.\label{eq:Zeffconffixed}
\end{equation}
If we use, for the time being, the effective metric $\g$ to raise and lower indices, it is then easy to show that
\begin{subequations}
\begin{align}
	\kle{[\g]_{ae} \; [\g]_{bf}  Z^{efcd}}\; \kle{[\g]_{cm}\; [\g]_{dn} Z^{mnpq\vphantom{f}}} &=[\g]_{ae}\; [\g]_{bf}  Z^{efpq},\\
	&= \ed{2}\kl{\delta_a{}^p\delta_b{}^q -\delta_a{}^q\delta_b{}^p}.
\end{align}
\end{subequations}
This corresponds to the reciprocity or closure condition as found, for example, in \cite{HehlObukhov, RubilarPremetric}. Note that since we are not in a pre-metric setting it is unimportant to distinguish the two concepts.

Up until this point, we have not yet introduced the announced simplification of the calculations by resorting to a flat-space-time analysis. Let us do this now. Again, it is certainly possible to immediately jump into the fully covariant, four-dimensional analysis. However, it is much more educational to first look at a 3+1 decomposition, and one that is more explicit than the six-dimensional one for a constitutive tensor described in equation~\eqref{eq:Z6x6}. Our restriction to flat-times, where $g = \eta = \mathrm{diag}(-1,1,1,1)$, is what makes the $3+1$ decomposition educationally palatable. By doing so, the previous choice of a conformal factor turns to $\det \g = -1$.\footnote{Note that this differs from the choice made in \cite{lrrAnalogue}, where the conformal invariance was used to set $\gi^{00} = -1$. That reference also has some minor inconsistencies in the form of signs and factors of 2 in relating the constitutive tensor to the constitutive matrices.} In the context of the coming steps~4 and~7, this means that we consider going to Riemann normal coordinates. More specifically, we choose an observer with four-velocity $V = (1,0,0,0)^T$; spatial projection simply means limiting the range of an index to $\{1,2,3\}$, while time-projection is equivalent to setting the index equal to~0. This also means that all remaining indices are spatial and raised or lowered with a three-dimensional Kronecker symbol. Should we need four-dimensional indices, they will start from $a$, three-dimensional ones start from $i$. This also links this first step to tetrad methods, as they are mentioned in passing in appendix~\ref{sec:ortho}.

It is easy to see that the definitions (see for example Appendix A in \cite{TrafoOptics2}, or \cite{PerlickRay})
\begin{equation}
	\begin{gathered}
		\epsilon^{ij} =  -2 Z^{i0j0}, \qquad \mi^{ij} =  \ed{2}\;\eps^i{}_{kl}\,\eps^j{}_{mn}\,Z^{klmn},\\ 
		\zeta^{ij} = \eps^i{}_{kl}\,Z^{klj0}
	\end{gathered}\label{eq:defconstmatrices}
\end{equation}
satisfy equation~\eqref{eq:EBtoDH}. It is useful to remind ourselves at this point that \emph{any} appearing Levi-Civita symbol $\eps$ is a (pseudo-)tensor, not a tensor density.\footnote{From a technical point of view \enquote{pseudo-tensor} would be even more appropriate than tensor, as they will still pick up signs when a coordinate change is introducing a change in the orientation. As it appears twice in the definition of $\mi$, the permeability will again be a tensor, not a pseudo-tensor.}

\subsubsection{Step 2: 3+1-Dimensional Consistency Conditions, Vanishing Magneto-Electric Effects}
The second step of our program then is to see what consistency conditions can be extracted under the simplifying assumption of a vanishing magneto-electric $\zeta$. Inserting equation~\eqref{eq:Zeffconffixed} into equations~\eqref{eq:defconstmatrices}, we find that
\begin{equation}
	0 \stackrel{!}{=}\zeta^{ij} = -  (\eps^i{}_{kl} \gi^{l0}) \gi^{kj}.\label{eq:3Dzeta0}
\end{equation}
Multiplying from the right with $[\g]_{jm}$, raising one index, this reduces to
\begin{equation}
	\eps^{im}{}_{l} \gi^{l0} = 0.
\end{equation}
Hitting this with another $\eps^k{}_{im}$ gives
\begin{equation}
	2\delta^k{}_{l}\gi^{l0} = 0
\end{equation}
and thus
\begin{equation}
	\gi^{k0} = 0.
\end{equation}
Using this first result, we get for the other two constitutive matrices:
\begin{align}
	\epsilon^{ij} &= \gi^{ij} \gi^{00},\\
	\mi^{ij} &= -\ed{2}\,
	\eps_{ikl}\,\eps_{jmn}
	\left( \gi^{km} \gi^{ln} \right).
\end{align}
Thus, $\g^{-1}$ block-diagonalizes. Since we know that $\det \g = -1$, we therefore can write this block structure as
\begin{equation}
	\renewcommand{\arraystretch}{1.25}
	\kl{\gi^{ab}}_{a,b\in \{ 0,\dots,3 \}} \ifed \begin{pmatrix}
	- \ed{\det \kl{\gamma^{\circ\circ}}} & 0\\ 0 & \gamma^{ij}
	\end{pmatrix}.
\end{equation}
Combining this with the following variant of Cramer's rule for $3\times3$ matrices,
\begin{equation}
	\eps_{ikl}\,\eps_{jmn} \{ X^{km} X^{ln} \} = 2 \det(X) \; X^{-1}_{ij},\label{eq:Cramer}
\end{equation}
we can then reduce the equations for $\epsilon^{ij}$ and $\mi^{ij}$ to\footnote{Remember that spatial indices are raised and lowered with the three-dimensional Kronecker symbol.}
\begin{equation}
	\mu^{-1}_{ij}  =  \det(\gamma^{\circ\circ})\;  \gamma^{-1}_{ij},\qquad 	\Longleftrightarrow \qquad \mu^{ij} =  \frac{\gamma^{ij}}{\det(\gamma^{\circ\circ})},\label{eq:mu3+1}
\end{equation}
and
\begin{equation}
	\epsilon^{ij} = \frac{\gamma^{ij}}{\det(\gamma^{\circ\circ})} = \mu^{ij}.\label{eq:conscond3+1}
\end{equation}
This last equation, \eqref{eq:conscond3+1}, is exactly the consistency condition we were after. If it is fulfilled, we can write $\g^{-1}$ as
\begin{subequations}
\begin{align}
	\gi^{ab} &= 
	\begin{pmatrix}
	-\sqrt{\det(\epsilon^{\circ\circ})} & 0 \\ 0 &  \displaystyle{\frac{\epsilon^{ij}}{\sqrt{\det(\epsilon^{\circ\circ})}}}
	\end{pmatrix},
	\\&= 
	\begin{pmatrix}
	-\sqrt{\det(\mu^{\circ\circ})} & 0 \\ 0 &  \displaystyle{\frac{\mu^{ij}}{\sqrt{\det(\mu^{\circ\circ})}}}
	\end{pmatrix}.\label{eq:geff3+1zerozeta}
\end{align}
\end{subequations}
This particular result is well known and can, for example be found in \cite{TrafoOptics2,TrafoOptics,AnalogueSurvey}. Of course the matching condition $\epsilon^{ij}=\mu^{ij}$ does not hold for naturally occurring media.\footnote{Already a quick check on Wikipedia or in your favorite material data reference table will reveal this.} It is only with the development of modern meta-materials that the $\epsilon^{ij}=\mu^{ij}$ matching condition becomes plausible physics.

What the consistency condition~\eqref{eq:conscond3+1} also shows is our need for using only Levi-Civita tensors --- otherwise the equation would equate a tensor (the permittivity) with a density (the permeability).

To see what the effective metric (\emph{not} the inverse effective metric!) would be, one now needs to invert the matrix~\eqref{eq:geff3+1zerozeta}. Doing this, we arrive simply at our final results for zero magneto-electric effects:
\begin{equation}
	[g_\text{eff}]_{ab} = 
	\begin{pmatrix}
	-\det([\gamma]^{\circ\circ}) &  0\\ 
	0 & \gamma^{-1}_{ij}
	\end{pmatrix}.
\end{equation}	
This implies:	
\begin{subequations}
	\begin{align}
		[g_\text{eff}]_{ab}
		&= \begin{pmatrix}
		-\sqrt{\det(\mu^{\circ\circ})}^{-1} &  0\\ 
		0 & \sqrt{\det(\mu^{\circ\circ})}\;\mu^{-1}_{\,ij}
		\end{pmatrix},\\
		&=\begin{pmatrix}
		-\sqrt{\det(\epsilon^{\circ\circ})}^{-1} &  0\\ 
		0 & \sqrt{\det(\epsilon^{\circ\circ})}\;\epsilon^{-1}_{\,ij}
		\end{pmatrix}.
	\end{align}
\end{subequations}
\subsubsection{Step 3: 3+1-Dimensional Consistency Conditions, Non-Vanishing Magneto-Electric Effects}
The big difference, obviously, in our third step is that with non-vanishing magneto-electric effects equation~\eqref{eq:3Dzeta0} does not hold. This complicates the algebra --- but not in an impossible manner. Setting
\begin{equation}
	\beta^{i} \defi \gi^{0i},\label{eq:defbeta}
\end{equation}
and, again using the conformal freedom to keep $\det \gi^{ab} = -1$, we consider the following, Kaluza--Klein-inspired form\footnote{As for the distinction between Kaluza--Klein (KK) and Arnowitt--Deser--Misner (ADM) formulations, note that they are dual to each other: The same decomposition is applied either to the metric (ADM, see \cite{ADM}), or to the inverse metric (Kaluza--Klein, see \cite{Klein26}). For a modern textbook treatment, see chapter~X, appendices~6 through 9 of reference \cite{ZeeGR}. This ADM versus KK duality holds in the sense of the cotangent space being dual to the tangent space. This distinction is independent of additional considerations of dimensionality.} for $\gi$:
\begin{equation}
	\gi^{ab} =  \begin{pmatrix}
	-\det(\gamma^{-1}_{\circ\circ}) + \gamma^{-1}_{kl} \beta^k \beta^l & \beta^j\\ \beta^i & \gamma^{ij} 
	\end{pmatrix} .\label{eq:geffKaluzaKlein}
\end{equation}
Clearly, the result for $\mi$ from the previous calculation, equation~\eqref{eq:mu3+1}, remains the same. However, the equations for $\zeta$ and $\epsilon$ will change and become more difficult to deal with. It is useful to remind ourselves of the earlier mentioned two ways to look at the consistency conditions: In the first case, one wants to take a given metric $\gi^{ab}$ and see with what material this metric could be achieved. After a bit of algebra (such as inverting $\gamma^{ij}$ as defined in equation~\eqref{eq:geffKaluzaKlein}), this can easily be done by looking at the following rewritten defining equations for the constituent matrices:
\begin{subequations}\label{eq:zeta3+1}
	\begin{align}
		\epsilon^{ij} =&  
		\left(\gamma^{ij} \{\det(\gamma^{-1}_{\circ\circ}) - \gamma^{-1}_{kl} \beta^k \beta^l \}  + \beta^i \beta^j \right),\\
		\mu^{ij} =&  \frac{\gamma^{ij}}{\det(\gamma^{\circ\circ})},\\
		\zeta^{ij} =& -\ed{2}\;
		\left(  \eps^{i}{}_{kl}\beta^l  \gamma^{kj}  \right).
	\end{align}
\end{subequations}
This gives the material properties for a given metric to be mimicked by a material. Do they not hold simultaneously, this material \emph{cannot} be interpreted as an effective metric in macroscopic electrodynamics resembling \emph{this} model metric. (Whether material properties failing one given model metric can be interpreted as an effective metric at all, is an aspect of the second viewpoint.)

The other way of looking at the consistency conditions is more involved and requires actually finding a concrete form of this condition. For this, take equations~\eqref{eq:zeta3+1} and use them to rewrite $\epsilon$ as
\begin{equation}
	\epsilon^{ij} =  
	\mu^{ij}\, (1 - \mu^{-1}_{\,kl} \beta^k \beta^l )  + \beta^i \beta^j.\label{eq:conscond3+1zeta}
\end{equation}
This is the consistency condition we were looking for. In deriving it, we had to express $\beta$ --- thus far only appearing in equation~\eqref{eq:defbeta} in terms of $\gi$ --- in terms of $\mi$ and $\zeta$; more on this shortly. Note that it indeed is a tensor equation: While the Levi-Civita pseudo-tensor appears, it appears in such a way that possible signs from orientation changes appear twice, thus cancelling and resulting in a true tensorial quantity. Thus, if you are \emph{given} the optical properties ($\epsilon$, $\mu$, $\zeta$) --- \emph{and they fulfill this consistency condition} --- then you can calculate the inverse effective metric via the two following equations
\begin{align}
	\gamma^{ij}  &= \frac{\mu^{ij}}{\sqrt{\det(\mu^{\circ\circ})}},\\
	\beta^m &=   \sqrt{\det(\mu^{\circ\circ})} \; \eps^{mk}{}_i \; \mu^{-1}_{\,jk}\; \zeta^{ij},\label{eq:beta3+1}
\end{align}
and insert in equation~\eqref{eq:geffKaluzaKlein} to arrive at:
\begin{equation}
	[g_\text{eff}^{-1}]^{ab} = \begin{pmatrix}
	-\sqrt{\det(\mu^{\circ\circ})}\; (1- \mu^{-1}_{\,kl} \beta^k \beta^l )& \beta^j\\ \beta^i &\displaystyle {\frac{\mu^{ij}}{\sqrt{\det(\mu^{\circ\circ})}} }
	\end{pmatrix}.\label{eq:NonZeroKaluzaKlein}
\end{equation}
This can be turned into an equivalent formula involving $\epsilon^{ij}$. In either case, if the consistency condition~\eqref{eq:conscond3+1zeta} is not satisfied, then the medium is simply \emph{not equivalent} to an effective metric.

While rewriting the previous result in terms of $\epsilon$ is not particularly edifying for the inverse metric itself, this does become helpful when looking at the actual metric. For this reason let us explore how to transform an expression involving $\mu$ into one involving $\epsilon$:

By multiplying the consistency condition~\eqref{eq:conscond3+1zeta} by $\mi_{jl}\beta^l$, we find
\begin{equation}
	\epsilon^{ij} \mi_{jl} \beta^l = \beta^i.
\end{equation}
This can be turned into the statement
\begin{equation}
	\mi_{jl} \beta^l = [\epsilon^{-1}]_{jl} \beta^l,\label{eq:muepsdegen}
\end{equation}
and hence
\begin{equation}
	\mi_{jl} \beta^j \beta^l = [\epsilon^{-1}]_{jl} \beta^j \beta^l.
\end{equation}
With this, we can rewrite the consistency condition as
\begin{equation}
	\mu^{ij} = \frac{\epsilon^{ij} - \beta^i \beta^j}{1-\epsilon^{-1}_{kl}\beta^k \beta^l}.\label{eq:conscond3+1mu}
\end{equation}
As for any column vector $u$ linear algebra provides us with the identity
\begin{equation}
	\det(X+u u^T) = \det(X)(1+u^T X^{-1}u),
\end{equation}
we can derive from this new alternative consistency condition the following identity for the determinants of $\epsilon$ and $\mu$:
\begin{equation}
	\det(\mu^{\circ\circ}) = \frac{\det(\epsilon^{\circ\circ})}{(1-\epsilon^{-1}_{kl}\beta^k \beta^l)^2}.
\end{equation}
This allows us to fix the magneto-electric effect in terms of either $\mu$ or $\epsilon$:
\begin{equation}
	\zeta^{ij} = - \ed{2} \sqrt{\mu^{-1}_{\circ\circ}} \eps^{i}{}_{kl} \beta^l \epsilon^{kj} = \ed{2} \sqrt{\epsilon^{-1}_{\circ\circ}} \eps^{i}{}_{kl} \beta^l \epsilon^{kj}.
\end{equation}
Finally, from this we can deduce the parameter $\beta^i$ in the KK decomposition such that we can easily switch between making $\mu$ or $\epsilon$ the featuring material property:
\begin{eqnarray}
	\beta^m = \sqrt{\mu^{\circ\circ}}\eps^{mk}{}_i \mu^{-1}_{kn} \zeta^{ni} = \sqrt{\epsilon^{\circ\circ}}\eps^{mk}{}_i \epsilon^{-1}_{kn} \zeta^{ni}.
\end{eqnarray}
These identities will appear again as key ingredients of section~\ref{sec:bespoke}, where we will calculate material properties for specified metrics to be mimicked through clever usage of the consistency conditions.

Doing either the decomposition in terms of $\mu$ and $\zeta$, or $\epsilon$ and $\zeta$, we can then evaluate the effective metric $[g_{\text{eff}}]_{ab}$ itself. In general, the inversion of the Kaluza--Klein decomposition~\eqref{eq:NonZeroKaluzaKlein} reads:\footnote{Given the aforementioned duality of KK and ADM formalism, one can get this by inverting the ADM decomposition and then dualising.}
\begin{equation}
	\renewcommand{\arraystretch}{1.25}
	[g_\text{eff}]_{ab} = 
	\begin{pmatrix}
	-\det(\gamma^{\circ\circ}) &  \det(\gamma^{\circ\circ})\, \gamma^{-1}_{jk}\beta^k\\ 
	\det(\gamma^{\circ\circ})\,\gamma^{-1}_{ik}\beta^k 
	&\;\;\gamma^{-1}_{ij} - \det(\gamma^{\circ\circ})( \gamma^{-1}_{ik} \beta^k) ( \gamma^{-1}_{jl}\beta^l)
	\end{pmatrix}.
\end{equation}
Inserting the consistency condition~\eqref{eq:conscond3+1zeta} we arrive at
\begin{subequations}
	\begin{align}
		[g_\text{eff}]_{ab} &= 
		\begin{pmatrix}
		-\sqrt{\det(\mu^{\circ\circ})}^{-1} &  \mu^{-1}_{\,jk}\beta^k\\ 
		\mu^{-1}_{\,ik}\beta^k 
		&\sqrt{\det(\mu^{\circ\circ})}\kl{\mu^{-1}_{\,ij} - ( \mu^{-1}_{\,ik} \beta^k) ( \mu^{-1}_{\,jl}\beta^l)}
		\end{pmatrix},\label{eq:g3+1mu}\\
		&= \begin{pmatrix}
		- \sqrt{\det \epsilon^{\circ\circ}}^{-1}(1-\epsilon_{kl}^{-1}\beta^k\beta^l) & \epsilon_{jk}^{-1}\beta^k\\
		\epsilon^{-1}_{ik}\beta^k & \sqrt{\det\epsilon^{\circ\circ}}(\epsilon^{-1}_{ij})
		\end{pmatrix},\label{eq:g3+1eps}\\
		\text{with~} \beta^m &\mathrel{\mathop:}= \sqrt{\det(\epsilon^{\circ\circ})} \; \varepsilon^{mk}{}_i \; \epsilon^{-1}_{\,jk}\; \zeta^{ij}.\nonumber
	\end{align}
\end{subequations}
This completes our $3+1$-dimensional analysis. The goal is now to go beyond the Kaluza--Klein split and the associated orthogonal decomposition and turn the results, as much as possible, into fully covariant statements.

\subsubsection{Aside: Pseudo-Inverses and Pseudo-Determinants}
Before we can continue, we have to take note of an apparent doorstop towards a fully covariant, four-dimensional formalism: Looking back at our equations in the previous steps we note that we frequently encounter both the inverses of $3\times3$ matrices and their determinants. Both notions at first glance make no sense in a four-dimensional context --- as explained in appendix~\ref{sec:4tensor}, the constitutive matrices remain four-orthogonal to $V$ when transitioning to the covariant formalism. Thus, their rank is $3$ and neither their determinant nor they themselves can be inverted. This, however, misses the availability of the Moore--Penrose pseudo-inverse $A^\#$ (see, \emph{e.g.}, \cite{MooreInverse,PenroseInverse,GenInverses}\footnote{\cite{GenInverses} also contains some more historic references about other (re)discoveries of the pseudo-inverse.}) and the pseudo-determinant\footnote{Early notions of the pseudo-determinant can be found in \cite{KhatriPDET}, while more modern appearances include \cite{KnillPDET,etalPDET}. Written as $\det' (A)$, a similar notion for operators can be found in the quantum field theory literature in \cite{AspectsSym} and probably even earlier. This notation has been adopted, for example, in \cite{BivsBi}.} $\pdet (A)$, defined for a general square $n\times n$ matrix $A$ with eigenvalues $\lambda_i$ as follows:
\begin{equation}
	\pdet (A) = \prod_{\substack{i=1 \\\lambda_i \neq 0}}^{\mathrm{rank}(A)}\lambda_i.
\end{equation}
Furthermore, the following identities hold for the pseudo-determinant, with the last equality valid for (anti-)symmetric or (anti-)Hermitian matrices:\footnote{For general (asymmetric) matrices, this can be generalized to 
\begin{equation*}
	\pdet(A\;A^\dagger) = \det\kl{\kle{\mathds{1}-A \; A^\#} + A\;A^\dagger}
\end{equation*}
using the singular value decomposition of $A$.}
\begin{subequations}
	\begin{align}
		\det\kl{\mathds{1}+zA} &= \pdet (A)\; z^{\mathrm{rank}(A)} + \mathcal{O}\kl{z^{\mathrm{rank}(A)-1}},\\
		\pdet (A) &= \lim_{z\to 0} \frac{\det\kl{A+z\mathds{1}}}{z^{n-\mathrm{rank}(A)}} = \lim_{z\to 0} \frac{\det\kl{A+z\mathds{1}}}{z^{\mathrm{nullity}(A)}},\\
		&= \det \kl{\kle{\mathds{1} - A\; A^\#} + A}.		
	\end{align}
\end{subequations}

Again: While the generally covariant $\mi$ and $\epsilon$ remain symmetric, due to their orthogonality to $V^a$ they will not have full rank as $4\times4$ matrices. Put differently, their rank being $3$, the null-space of $\epsilon$ or $\mi$ is one-dimensional, any two projection operators onto this null-space are therefore proportional to each other. As $V^a V^b = - t^{ab}$ is a projector onto this null-space of $\mi$ and $\epsilon$, this has to be proportional to the corresponding $\kle{\mathds{1} - A\; A^\#}^{ab}$. Note that $\kle{t^{\bullet\bullet}}^\#_{ab} = t_{ab} = -V_a V_b$. Furthermore, as we want the $3+1$ case to drop out if we chose $V = (1,0,0,0)^T$, we can see that
\begin{equation}
	\kle{\mathds{1} - A\; A^\#}^{ab} = - t^{ab} = +V^a V^b.
\end{equation}
Put to use on the pseudo-determinant, we can then give it in terms of a perfectly well-behaved, standard determinant:
\begin{equation}
	\pdet (\epsilon^{ab}) = \det \kl{-t^{ab} + \epsilon^{ab}} = \det \kl{V^aV^b + \epsilon^{ab}}.
\end{equation}

\subsubsection{Step 4: Translating 3+1-Dimensional Quantities to Four-Dimensional Ones}
The general idea for upgrading the analysis to a fully covariant approach is that the analysis in Minkowski space-time can be seen as the special case of an arbitrary space-time in Riemann normal coordinates. This ties into the orthogonal decomposition as described in appendix~\ref{sec:ortho}. Then, effectively, in our earlier calculation spatial indices $i = 1, 2, 3$ correspond to spatially-projected indices and time-like indices (indices set to zero) correspond to a contraction with the given four-velocity $V$.\footnote{Strictly speaking, the index should be hit with the temporal projector $t^a{}_b$ --- but the actual information contained in these processes is the same. Again, see appendix~\ref{sec:ortho} for more on this particular technical detail.} Any three-dimensional Kronecker symbol $\eps^{ijk}$ corresponds then to a contraction of the four-dimensional one with this four-velocity. Summarizing, we get the following set of translation rules:
\begin{subequations}
	\begin{align}
		X^i &\longrightarrow h^a{}_b X^b,\\
		X^0 &\longrightarrow V_a X^a,\qquad \Longleftrightarrow\qquad X^0 \longrightarrow t^a{}_b X^b,\\
		\eps^{ijk} &\longrightarrow \eps^{abcd}V_a.
	\end{align}
\end{subequations}
A quick consistency check: If we were to use these translation rules on the definition of the constitutive matrices~\eqref{eq:defconstmatrices}, we arrive at just the equations~\eqref{eq:WandDuals} in terms of the constitutive matrices $\epsilon$, $\mi$ and $\zeta$:
\begin{subequations}\label{eq:CovConstMatrices}
	\begin{align}
		\epsilon^{ab} &\defi - 2 Z^{acbd} V_c V_d, \label{eq:defeps4}\\
		\mi^{ab} &\defi \frac{1}{2} \eps^{ca}{}_{ef}\eps^{db}{}_{gh} Z^{efgh} V_c V_d, \label{eq:defmu4}\\
		\zeta^{ab} &\defi \eps^{ca}{}_{ef} Z^{efbd} V_c V_d. \label{eq:defZeta4}
	\end{align}
\end{subequations}
This links the previously considered special case with the general orthogonal decomposition presented in the appendix. Inserting the mimicking conditions~\eqref{eq:Zeff}, we get:
\begin{subequations}\label{eq:defconst4matrices}
	\begin{align}
		\epsilon^{ab} = &-\left( \gi^{ab}\;\gi^{cd} - \gi^{ac}\; \gi^{bd} \right) V_c V_d,\\ 
		\mu^{-1}_{ab} = &\hphantom{-\big(} \eps_{aefc}\,\eps_{bmnd}\left( \gi^{em}\; \gi^{fn} \right)   V^c V^d, \\
		\zeta_a{}^b = &-  (\eps_{amnd} \gi^{mc})\;  \gi^{nb} \; V_c V^d .
	\end{align}
\end{subequations}

Having introduced the notion of both a pseudo-inverse and a pseudo-determinant, we are now in the position to actually generalize the $\det \epsilon^{ij}$ or $\det \mi^{ij}$ terms that appear in, for example, equation~\eqref{eq:geff3+1zerozeta} or \eqref{eq:beta3+1}, to a fully covariant formalism. As determinants of a tensor pick up determinants of the physical metric under general coordinate transformations, we need the following rules for promoting determinants to quantities that behave as scalars under general coordinate transformations:\footnote{Note that as we only take determinants of symmetric matrices, $S^{[ab]}=0$, the bullet notation we employ is sufficient. This means, in terms of translation rules, that
\begin{equation*}
	S_{\circ\circ}\quad \longrightarrow \quad S_{\bullet\bullet},\qquad 		S^{\circ\circ} \quad \longrightarrow \quad S^{\bullet\bullet}.
\end{equation*}}
\begin{equation}
	\det (A^{ij}) \qquad \longrightarrow \qquad \frac{\pdet (A^{ab})}{-\det (g^{ab})}.
\end{equation}

We summarized all important rules for translating $3+1$-dimensional quantities into covariant, four-dimensional ones in table~\ref{tab:translation}.

\begin{table}
	\centering
	\begin{tcolorbox}[width=.75\linewidth]
		\renewcommand{\arraystretch}{1.25}
		\centering
		\begin{tabular}{rcl}
			$X^i$ &$\qquad \longrightarrow\qquad $& $h^a{}_b X^b$\\
			$X^0$ & $\qquad\longrightarrow\qquad$ &$V_a X^a$\\
			or $X^0$ &$\qquad\longrightarrow\qquad$& $t^a{}_b X^b$\\
			$\eps^{ijk}$ &$\qquad\longrightarrow\qquad$  &$\eps^{abcd}V_a$\\
			$(A^{ij})^{-1}$ &$\qquad\longrightarrow\qquad$&$(A^{ab})^\#$\\
			$\det (S^{\circ\circ})$ &$\qquad \longrightarrow \qquad$& $\frac{\pdet (S^{\bullet\bullet})}{-\det (g^{\bullet\bullet})}$
		\end{tabular}
	\end{tcolorbox}
	\caption[Table for Translating 3+1-Dimensional Quantities to Covariant Ones]{Translating 3+1 terms to fully covariant terms.}
	\label{tab:translation}
\end{table}

\subsubsection{Step 5: Four-Dimensional Consistency Conditions, Vanishing Magneto-Electric Effects}
After these preparatory steps, we are again in the position to tackle the consistency conditions, only now in a fully covariant, four-dimensional framework. Vanishing magneto-electric effects will, just as in the previous, 3+1-dimensional discussion, greatly expedite the calculation. And as we shall see later in step~7, this now is more than just a pedagogical introduction --- it actually has a connection to the final form of the consistency condition. This justifies our inclusion of the current step, as the question of mixing of constitutive matrices under coordinate transformations is much more obtrusive. With our translation rules in place, we can immediately proceed and get for the expression for the inverse of the effective metric
\begin{equation}
	\gi^{ab} = - \sqrt{\frac{\pdet(\epsilon^{\bullet\bullet})}{-\det(g^{\bullet\bullet})}}\; V^a V^b +  \sqrt{\frac{-\det(g^{\bullet\bullet})}{\pdet(\epsilon^{\bullet\bullet})}}\; \epsilon^{ab},
\end{equation}
while our consistency condition is turned into
\begin{equation}
	\epsilon^{ab} = \kle{\mi_{\bullet\bullet}^\#}^{ab}.
\end{equation}
If we then define
\begin{eqnarray}
	\mu^{ab} \defi \kle{\kle{\mu^{-1}_{\bullet\bullet}}^\#}^{ab},
\end{eqnarray}
we can simplify this to the familiar
\begin{equation}
	\epsilon^{ab} = \mu^{ab}.
\end{equation}
However, the hidden mix of inverse (from the traditional notation $\mi$ to link magnetic field to excitation, unlike for the permittivity) and pseudo-inverse has to be kept in mind. Again, this is related to the historical artefact of the naming of $\mi$, as mentioned in footnote~\ref{fn:historylesson}.
The effective metric itself now takes on any of the following forms:
\begin{subequations}
	\begin{align}
		\kl{g_\text{eff}}_{ab} &= \frac{\pdet\kl{\gamma^{\bullet\bullet}}}{-\det \kl{g^{\bullet\bullet}}}t_{ab} + \kle{\gamma^{\bullet\bullet}}^\#_{ab},\\
		&= -\sqrt{\frac{-\det(g^{\bullet\bullet})}{\pdet(\epsilon^{\bullet\bullet})}}\, V_a V_b + \sqrt{\frac{\pdet(\epsilon^{\bullet\bullet})}{-\det(g^{\bullet\bullet})}} [\epsilon^{\bullet\bullet}]^\#_{ab},\label{eq:geffeps0}\\
		&= -\sqrt{\frac{-\det(g^{\bullet\bullet})}{\pdet(\mu^{\bullet\bullet})}}\, V_a V_b + \sqrt{\frac{\pdet(\mu^{\bullet\bullet})}{-\det(g^{\bullet\bullet})}} \mu^{-1}_{ab}.\label{eq:geffmu0}
	\end{align}
\end{subequations}

\subsubsection{Step 6: Four-Dimensional Consistency Conditions, Non-Vanishing Magneto-Electric Effects}
The final step in calculating the actual consistency conditions is now to generalise those of non-vanishing magneto-electric effect, equations~\eqref{eq:conscond3+1zeta}, to the fully covariant formalism. This goes hand in hand with generalising the $0i$ components of the metric~\eqref{eq:beta3+1}, and the results for the Kaluza--Klein decomposition~\eqref{eq:NonZeroKaluzaKlein}. Again, this is a simple (though slightly tedious) application of the rules listed in tabel~\ref{tab:translation}.

First, take a look at what happens to the three-vector $\beta^i$:
\begin{equation}
	\beta^i \qquad \longrightarrow \qquad \beta^e  =\sqrt{ \frac{\pdet(\mu^{\bullet\bullet})}{-\det(g^{\bullet\bullet})  }} \; \eps^{ecad} \; \mu^{-1}_{\,bc}\; \zeta_a{}^{b}  V_d.
\end{equation}
We can immediately see that the four-vector $\beta^e$ satisfies
\begin{equation}
	\beta^e\, V_e = 0,
\end{equation}
a transversality result we can immediately put to use to infer that
\begin{equation}
	\mu^{-1}_{\,ij}\beta^i \beta^j \qquad \longrightarrow \qquad \mu^{-1}_{\,ab}\beta^a\beta^b.
\end{equation}
From this we can derive the inverse effective metric:\footnote{Had we chosen to turn $\beta^i$ into the equivalent tensorial form \smash{$\tilde{\beta}^{ab} \defi t^{a}{}_c [g_\text{eff}^{-1}]^{cd} h_d{}^b$}, the transversality would have been $\tilde{\beta}^{ab}V_b = 0$, and the combination $\beta^a V^b$ would be equal to $\tilde{\beta}^{ba}$.}
\begin{align}
	\gi^{ab} =& - \sqrt{\frac{\pdet(\mu^{\bullet\bullet})}{-\det(g^{\bullet\bullet})}}  \left( 1-  \mu^{-1}_{cd} \beta^c \beta^d\right)\; V^a V^b \nonumber\\&+  
	V^a \beta^b + \beta^a V^b + 
	\sqrt{\frac{-\det(g^{\bullet\bullet})}{\pdet(\mu^{\bullet\bullet})}}\; \mu^{ab}.
\end{align}

The consistency condition is simply turned into the fully Lorentz-invariant, covariant equation
\begin{equation}
	\epsilon^{ab} = \mu^{ab} (1 - \mu^{-1}_{\,cd} \beta^c \beta^d )  + \beta^a \beta^b. 
\end{equation}

For the corresponding effective metric itself, we can use the fact that $\gamma$ and $\gamma^\#$ will again be orthogonal to $V$. The somewhat long expression we get is
\begin{subequations}
	\begin{align}
		\left[g_\text{eff}\right]_{ab} = \frac{-\det \left(g_{\,\bullet\bullet}\right)}{\mathrm{pdet}\left(\left[\gamma^{\bullet\bullet}\right]^\#_{\bullet\bullet}\right)} &\left( t_{ab} - V_{a}\left[\gamma^{\bullet\bullet}\right]^\#_{bc}\beta^c - V_{b}\left[\gamma^{\bullet\bullet}\right]^\#_{ac}\beta^c \vphantom{\frac{\mathrm{pdet}\left(\left[\gamma^{\bullet\bullet}\right]^\#_{\bullet\bullet}\right)}{-\det \left(g_{\,\bullet\bullet}\right)}}\right.\nonumber\\
		&\left. + \frac{\mathrm{pdet}\left(\left[\gamma^{\bullet\bullet}\right]^\#_{\bullet\bullet}\right)}{-\det \left(g_{\,\bullet\bullet}\right)} \left[\gamma^{\bullet\bullet}\right]^\#_{ab} - \left[\gamma^{\bullet\bullet}\right]^\#_{ac}\beta^c \left[\gamma^{\bullet\bullet}\right]^\#_{bd}\beta^d\right).
	\end{align}
	More specifically, in terms of $\mu$,
	\begin{align}
		\left[g_\text{eff}\right]_{ab} =& \sqrt{\frac{\mathrm{pdet}\left[\left[\mu^{-1}\right]^{\bullet\bullet}\right]}{-\det \left(g^{\bullet\bullet}\right)}} t_{ab} - \left( V_{a} \,\mu^{-1}_{\,bc}\,\beta^c + V_{b}\,\mu^{-1}_{\,ac}\,\beta^c\right)\nonumber\\
		&  + \sqrt{\frac{-\det \left(g^{\bullet\bullet}\right)}{\mathrm{pdet}\left(\left[\mu^{-1}\right]^{\bullet\bullet}\right)}}\left(\mu^{-1}_{\,ab} - \,\mu^{-1}_{\,ac}\,\beta^c \,\mu^{-1}_{\,bd}\beta^d\right).
	\end{align}
	Alternatively, we can also write this in terms of $\epsilon$ as
	\begin{align}
		\left[g_\text{eff}\right]_{ab} = - \sqrt{\frac{-\det g^{\bullet\bullet}}{\mathrm{pdet}\epsilon^{\bullet\bullet}}}\left(1-\epsilon_{cd}^\#\beta^c\beta^d\right) V_aV_b - V_a \epsilon_{bd}^\# \beta^d - V_b \epsilon_{ad}^\# \beta^d + \sqrt{\frac{\mathrm{pdet}\epsilon^{\bullet\bullet}}{-\det g^{\bullet\bullet}}} \epsilon^\#_{ab},
	\end{align}
	where now
	\begin{equation}
		\beta^e  =\sqrt{ \frac{\mathrm{pdet}(\epsilon^{\bullet\bullet})}{-\det(g^{\bullet\bullet})  }} \; \varepsilon^{ecad} \; \epsilon^{\#}_{\,bc}\; \zeta_a{}^{b}  V_d.
	\end{equation}
\end{subequations}

\subsubsection{Step 7: Physical Reduction To Case of Vanishing Magneto-Electric Effects}
In this last step, we want to partly justify the presentation of a covariant derivation of consistency relations under the assumption of vanishing magneto-electric effects. What we will derive, concretely, is the fact that there has to be a reference frame (pointwise), for which any consistency condition reduces to that of a vanishing magneto-electric matrix. 

To see this, note that on physical grounds, the \enquote{light-cones} of $g_\text{eff}$ will have to lie inside the light-cones of the physical metric $g$. Therefore, for any \emph{physical} four-velocity $U^a$, the quantity
\begin{equation}
	Q= [g_\text{eff}]_{ab} U^a U^b
\end{equation}
will be negative. Now look for the minimum of $Q$ by solving the Lagrange multiplier problem
\begin{equation}
	L = [g_\text{eff}]_{ab} \,U^a U^b - \lambda (g_{ab}\, U^a U^b +1),
\end{equation}
and call this minimum $V$. Adopting Riemann normal coordinates ($g_{ab}\to \eta_{ab}$) and going to the rest-frame of $V$ (so $V^a\to(1,0,0,0)$) we can block-diagonalize the effective metric
\begin{equation}
	\kle{g_\text{eff}}_{ab} = \begin{pmatrix}
	-\lambda & 0 \\ 0 & \kle{g_\text{eff}}_{ij}
	\end{pmatrix},
\end{equation}
with inverse
\begin{equation}
	\kle{g_\text{eff}}^{ab} = \begin{pmatrix}
	-\ed{\lambda} & 0 \\ 0 & \kle{g^{-1}_\text{eff}}^{ij}
	\end{pmatrix}.
\end{equation}

In particular, this means that, for this effective metric, there exists a rest-frame for an observer with four-velocity $V$ such that \emph{in this rest-frame the magneto-electric effects vanish}. This is exactly the result we were looking for: In the reference frame of this specific observer(!), we are allowed to use the much simpler analysis of step 5! Let us therefore call this the \emph{natural rest-frame} of the given medium.\footnote{Indeed, Perlick in his monograph \cite{PerlickRay} makes use of the existence of this frame from the very beginning, his definition~2.1.1. Magneto-electric effects are excluded from his analysis from the start. While our approach certainly is not yet as sophisticated, it should be possible to repeat his WKB-styled analysis in a similar manner including magneto-electric effects.}

Thus, another possible approach to the problem is this: Assume we have found the four-velocity $V$ of this natural rest-frame for our given effective metric. We then define the corresponding permittivity as $\epsilon_V$ and the corresponding permeability as $\mu^{-1}_V$. The natural question to follow this with is: What, then, would be the constitutive matrices $\epsilon$, $\mi$, and $\zeta$ of another observer with four-velocity $W$ in terms of these $\epsilon_V$ and $\mu^{-1}_V$? The answer is basically relegated to the appendix~\ref{sec:frames}: There we give the general result for changing reference frames, which trivially will cover also the current context. It is, however, of little use, to quote the resulting (cumbersome) formulae at this place, as their discussion is of little consequence --- unless we were to look at specific forms of a specific metric to be mimicked. This certainly provides an attractive avenue for future research.

The exclusion of magneto-electric effects from the start can also be further explained through results from classical electrodynamics, see \cite{Post}: It is known that only non-local effects can give rise to non-vanishing magneto-electric tensors $\zeta$. Under the lens of Lorentz transformations, these non-local effects could then be turned into dissipative phenomena. While both non-local effects (think helical molecules), as well as dissipation (most obviously in the form of electrical resistance) play an important role in classical electrodynamics, it is far less easy to reconcile these phenomena with a \emph{local}, relativistic theory. Attempts to incorporate this will easily lead to all kinds of irksome pathological behaviour of the theory, or the need to complexify Lorentzian manifolds (which has a long tradition of working \emph{nearly}, but not fully). Ironically, a closer look at the behaviour of wave equations on curved backgrounds reveals these to actually exhibit dissipation: Without dissipation, our understanding of superradiance in curved space-times would not go far \cite{LNPSuperradiance}. However, this happens on the level of the wave equation --- something that has not been written down at this stage. The space-time analogy was fully algebraic, not analytical. A closer look at this distinction will be made in section~\ref{sec:PDEana}.

It is also worth mentioning that similar lines of reasoning can be called upon in the context of pre-metric electrodynamics to derive the metric instead of the metric being an ingredient to electrodynamics \cite{HehlLaemmerzahl}. We also consider our approach a good link between the two fields of relativity on the one hand, and electromagnetism on the other: The former usually attempt to stay clear of macroscopic electrodynamics (with an uneasy feeling that it cannot be formulated covariantly), while the latter often see little use in the more general formalism of relativity, and thus stick to a 3+1-dimensional approach. Let us now turn to another interconnection of these two fields: Transformation optics.

\section{Distinguishing Covariant Analogue Gravity and Covariant Transformation Optics}\label{sec:trafooptics}
Transformation optics has a slightly different question in mind than our approach to analogue space-times through electrodynamics, as already their names suggest. As described in \cite{CovOptMet1,CovOptMet2,CovOptMet3,TrafoOpticsCartographDistort}, the difference is essentially encoded in two separate transformations considered: For the sake of argument, let us not mix these for the time being. In the case of analogue space-times we want to place on a given space-time $(M,g)$ a medium described by $Z$ such that the triple $(M,g,Z)$ is algebraically equivalent to a different space-time\footnote{We shall, for the sake of brevity, omit the fact that for most practical implementations of an analogue space-time the mimicked metric $\tilde{g}$ would not be defined on all of $M$. Rather, it would be defined on a subset $\tilde{M}$ thereof. Technically, this issue in the details can be evaded by just taking a small enough open subset of both the physical, Lorentzian background $(M,g)$ and the mimicked space-time $(\tilde{M},\tilde{g})$. Of course, the mimicked space-time itself as a physical space-time might have a very different base manifold altogether.} $(M,\tilde{g})$. The point, as described several times now, is precisely to simulate such a different space-time.

In transformation optics, in contrast, one rather tries to mimic with the medium $Z$ a transformation $T: (M,g) \to (M,g)$. These transformations are not constrained to be diffeomorphisms or isometries \cite{CovOptMet3}. For example, the invisibility cloak frequently evoked in transformation optics \enquote{blows up} a point to a fully-fledged hole hiding its content. Phrased drastically: The medium mimicking a black hole would not do the job of a medium bringing to life the afore-mentioned cloaking transformation, as it would (and should!) introduce lensing effects. Nevertheless, overlap exists, as the questions considered in the cited work of Thompson, Cummer, and Frauendiener shows. Especially the aforementioned clear distinction between coordinate transformations on the one hand, and medium induced transformation might be worthwhile to study closer, to see how much of this analysis can be used for our case. Early beginnings of this project can be found in section~\ref{sec:PDEana} further below.

\subsection{Moving Isotropic Media}\label{sec:iso}
As moving media are a frequent theme of transformation optics, see \cite{TrafoOptics2,CovOptMet2,CovOptMet3}, this is a natural touchstone of our formalism. In this spirit, let us have a closer look at some aspects of moving media, again following the discussion in \cite{CovEffMetrics}. An isotropic medium with no magneto-electric effects moving with four-velocity $V^a$ has \emph{in its rest frame} permittivity tensor and permeability tensor given by the following equations:
\begin{subequations}
	\begin{equation}
	\epsilon^{ab} = \epsilon (g^{ab}+V^a V^b ) = \epsilon h^{ab} ,
	\end{equation}
	and
	\begin{equation}
	[\mu^{-1}]^{ab} = \mu^{-1} (g^{ab}+V^a V^b ) = \mu^{-1} h^{ab}.
	\end{equation}
\end{subequations}
Inserting this in the Bel-decomposed constitutive tensor $Z^{abcd}$ yields, according to equation~\eqref{eq:ZBel2},
\begin{align}
	Z^{abcd} = -\frac{\epsilon}{2}& (  V^a V^c h^{bd} + V^b V^d h^{ac} -V^a V^d h^{bc}\nonumber\\& - V^b V^c h^{ad}) + \frac{\mu^{-1}}{2} ( h^{ac} h^{bd} - h^{ad} h^{bc}).
\end{align}
This can be rearranged to get
\begin{align}
Z^{abcd} = {\frac{\mu^{-1}}{2}} &\kle{ \left(h^{ac} - \epsilon\mu V^a V^c\right) \left(h^{bd} - \epsilon \mu V^b V^d\right)\right.\nonumber\\
	&\left.- \left(h^{ad} - \epsilon\mu V^a V^d\right) \left( h^{bc} - \epsilon\mu  V^b V^c\right)},\label{eq:Ziso}
\end{align}
which in turns lends itself to two different applications: The first is to derive again the consistency condition~\eqref{eq:conscond3+1}. The second is to get fully covariant expressions for the magneto-electric effect of moving media. We shall do both consecutively in the following short subsections.

\subsubsection{The Consistency Condition}
Taking from equation~\eqref{eq:actioneff} that an effective metric would mean
\begin{equation}
	Z^{abcd} = \sqrt{\frac{\det (\g)}{\det (g)}}\kl{\gi^{ac}\gi^{bd} - \gi^{ad}\gi^{bc}},
\end{equation}
and comparing this with the just derived equation~\eqref{eq:Ziso}, we see that the existence of an effective metric $\g$ would imply
\begin{equation}
	\sqrt[4]{\frac{\det(\g)}{\det(g)}} \;  \gi^{ab}  = \mu^{-1/2} \left(h^{ab} - \epsilon\mu V^a V^b\right).
\end{equation}
Taking determinants on both sides, we get the following equivalent of the previously derived consistency condition~\eqref{eq:conscond3+1} in the special case of an isotropic medium:
\begin{equation}
	-1 = -\frac{\epsilon}{\mu}.
\end{equation}
If the isotropic medium fulfils this condition we can then immediately write down the inverse effective metric as
\begin{equation}
	\gi^{ab}  \propto \left(h^{ab} - \epsilon\mu V^a V^b\right),
\end{equation}
or more specifically as
\begin{equation}
	\gi^{ab}  =  (\epsilon\mu)^{-1/4} \left(h^{ab} - \epsilon\mu V^a V^b\right).
\end{equation}

\subsubsection{The Magneto-Electric Effect of Moving Media}
Instead of looking for the possibility for an effective metric describing the constitutive tensor, we can also use the results of section~\ref{sec:frames} to see what \enquote{constitutive matrices} an observer, who is not comoving to the natural reference frame of the medium, would measure. To this end, let us look at the equations~\eqref{eq:consmatW}, again, with $W^a$ denoting the four-velocity of the observer. First, we shall calculate the permittivity $\epsilon_W^{ab}$. After some algebra, equation~\eqref{eq:epsW} is evaluated to be
\begin{align}
	\epsilon_W^{bd} =& -2 Z^{abcd} W_a W_c,\\
=& \mu^{-1} (g^{bd}+W^b W^d) + (\epsilon-\mu^{-1}) \left( g^{bd} (V\cdot W)^2 \right.\nonumber\\&\left.- (W^b V^d+V^b W^d)(V\cdot W) - V^b V^d\right).
\end{align}
Defining the spatial projection operator
\begin{equation}
	h_W^{bd} \defi g^{bd} + W^b W^d,
\end{equation}
and realizing that
\begin{align}
	h_W^{be} h_{ef} h_W^{fd} =& g^{bd} + [1+(V\cdot W)^2] W^b W^d \nonumber\\&+ (V\cdot W) [W^b V^d + V^d W^b] + V^b V^d,
\end{align}
we can even simplify $\epsilon_W^{ab}$ further to 
\begin{subequations}
	\begin{align}
		\epsilon_W^{bd} 
		=&  \mu^{-1} (h_W^{bd}) \nonumber\\
		&-(\epsilon-\mu^{-1})\left[ h_W^{be} h_{ef} h_W^{fd}\right. \left.-  [1+(V\cdot W)^2] h_W^{bd}\right],\\
		=&\epsilon \; h_W^{bd}  +(\epsilon-\mu^{-1})  \kle{(V\cdot W)^2 h_W^{bd}    - h_W^{be} h_{ef} h_W^{fd}}.\label{eq:epsiso}
	\end{align}
\end{subequations}

For $[\mu_W^{-1}]^{bd}$ it is helpful to realize that $h^{ab}-\epsilon\mu V^a V^b$ is for the following calculational needs the inverse of a (Lorentzian) metric $\mathcal{G}_{ab}$.\footnote{On a purely formal level it is of the form of the inverse Gordon metric\cite{lrrAnalogue}, even though at this stage we have not yet imposed the consistency condition which may or may not hold. And given most materials' properties it most likely will not! On the other hand, the Gordon metric \emph{does} have general validity in the ray optics limit, as opposed to wave optics. This does not mean, that refractive indices have no meaning in wave optics --- see section~\ref{sec:refractive}.\label{fn:GordonMetric}} Therefore, it will have an associated Levi-Civita tensor (density) $\eps^\mathcal{G}$. This then means that we can \enquote{pictorially} --- meaning we forget numerical factors and physical coefficients like $\mu^{-1}$ --- rewrite the defining equation~\eqref{eq:muW} to showcase the tensorial dependencies:
\begin{subequations}
	\begin{align}
		\kle{\mu_W^{-1}}^{\bullet\bullet} &\simeq \kl{\ast \kl{\mathcal{G}^{-1}\mathcal{G}^{-1} - \mathcal{G}^{-1}\mathcal{G}^{-1}}\ast }^{\bullet\bullet\bullet\bullet}W_\bullet W_\bullet,\\ 
		&\simeq \eps^{\bullet\bullet}{}_{\bullet\bullet} \kl{\mathcal{G}^{-1}\mathcal{G}^{-1} - \mathcal{G}^{-1}\mathcal{G}^{-1}}^{\bullet\bullet\bullet\bullet}\eps^{\bullet\bullet}{}_{\bullet\bullet}W_\bullet W_\bullet,\\
		&\simeq [g^{-1}]^{\bullet\bullet}[g^{-1}]^{\bullet\bullet}[g^{-1}]^{\bullet\bullet}[g^{-1}]^{\bullet\bullet} \hspace{-1.2em}\underbrace{\eps_{\bullet\bullet\bullet\bullet}\eps_{\bullet\bullet\bullet\bullet}}_{\simeq \sqrt{\frac{\det g}{\det \mathcal{G}}}^2\; \eps^\mathcal{G}_{\bullet\bullet\bullet\bullet}\eps^\mathcal{G}_{\bullet\bullet\bullet\bullet}}\hspace{-1.2em}\kl{\mathcal{G}^{-1}\mathcal{G}^{-1} - \mathcal{G}^{-1}\mathcal{G}^{-1}}^{\bullet\bullet\bullet\bullet} W_\bullet W_\bullet,\\
		&\simeq \frac{\det g}{\det \mathcal{G}} [g^{-1}]^{\bullet\bullet}[g^{-1}]^{\bullet\bullet}[g^{-1}]^{\bullet\bullet}[g^{-1}]^{\bullet\bullet} \underbrace{\eps^\mathcal{G}_{\bullet\bullet\bullet\bullet}\eps^\mathcal{G}_{\bullet\bullet\bullet\bullet}\kl{\mathcal{G}^{-1}\mathcal{G}^{-1} - \mathcal{G}^{-1}\mathcal{G}^{-1}}^{\bullet\bullet\bullet\bullet}}_{\simeq \kl{\mathcal{G}\mathcal{G} - \mathcal{G} \mathcal{G}}_{\bullet\bullet\bullet\bullet}} W_\bullet W_\bullet,\\
		&\simeq \frac{\det g}{\det \mathcal{G}} \kl{\kle{g^{-1}\mathcal{G}g^{-1}}\kle{g^{-1}\mathcal{G}g^{-1}}-\kle{g^{-1}\mathcal{G}g^{-1}}\kle{g^{-1}\mathcal{G}g^{-1}}}^{\bullet\bullet\bullet\bullet} W_\bullet W_\bullet.
	\end{align}
\end{subequations}
Now $\frac{\det g}{\det \mathcal{G}}$ evaluates to $\epsilon\mu$ and
\begin{equation}
	\kle{g^{-1}\mathcal{G}g^{-1}}^{\bullet\bullet} = g^{\bullet\bullet} + \kl{1- \ed{\epsilon\mu}} V^\bullet V^\bullet.
\end{equation}
With this we can then perform a similar analysis to the one for $\epsilon^{ab}$ and arrive at
\begin{equation}
	[\mu^{-1}_W]^{bd}  = \frac{h_W^{bd}}{\mu}  +(\mu^{-1}-\epsilon)  \kl{ (V\cdot W)^2 h_W^{bd}    - h_W^{be} h_{ef} h_W^{fd}}.\label{eq:muiso}
\end{equation}
Finally, starting from equation~\eqref{eq:zetaW} we arrive, again after some algebra, at the equation
\begin{equation}
	\zeta_W^{ac} = (\epsilon-\mu^{-1}) (V\cdot W) \kl{\epsilon^{acef}W_e V_f} \label{eq:zetaiso}
\end{equation}
for the magneto-electric matrix $\zeta_W^{ac}$.

Note that this calculation reproduces several important physical insights:
\begin{enumerate}
	\item If we pull out a factor $\epsilon$ in front of the right-hand side of equation~\eqref{eq:zetaiso}, the remainder of the right-hand sides will contain a factor of $1-\nicefrac{1}{\epsilon\mu} = 1-\nicefrac{1}{n^2}$ --- which nicely reproduces the Fresnel--Fizeau effect in flat space.
	\item Similarly, in flat space and if both the observer and the natural reference frame of the medium are inertial frames, note that $(V\cdot W)^2 = \gamma^2$ is just the Lorentz factor we expect second-rank tensors like the \enquote{constitutive matrices} to have.
	\item Finally, equation~\eqref{eq:zetaiso} gives the well-known result that a moving medium will have magneto-electric effects, even if it would not be at rest. Again, this is tightly related to the Fresnel--Fizeau effect, but is a more general result.
	\item Inverted, this last point also illustrates how a medium of the right optical properties simulate the optics of moving media. This is where transformation optics would often place the emphasis.
	\item Also, isotropy is lost under a change of observer. This happens even for inertial observers in Minkowski space and is intimately connected to the appearance of magneto-electric effects.
\end{enumerate}

\section{Bespoke Meta-Materials}\label{sec:bespoke}
With the theoretical discussion of the above sections in place, it is time to turn to more mundane matters: Actual examples of constitutive matrices for space-times. This step is the first relevant step for preparing actual experiments. While we shall not be as ambitious to actually propose concrete experiments on concrete analogue space-times based on the methods described so far, we shall nevertheless perform this first step of highlighting some possibilities of analogue space-times and the constitutive matrices they would require. For this we will employ different methods requiring either brute force or some case-by-case sophistication.

The brute force methods presented at first are most general, most straightforward, but also least illuminating as the produced output will rarely lend itself to simple interpretation in geometrical or physical terms. Achieving the latter --- simple interpretations --- is the goal of two subsections. As the analysis becomes often more transparent in (quasi-)Cartesian coordinates, this will be our entry point into this second approach to bespoke meta-materials for analogue space-times. The restriction to (quasi-)Cartesian coordinates in particular means no requirement to keep track of metric components when raising and lowering indices with the background metric.

In either case, quasi-Cartesian or fully curvilinear, we will make a particularly simple choice of mapping the laboratory coordinates to those of the effective space-time. We will simply and fully identify them on an interval of their coordinate range (unspecified, for the time being). We will also limit ourselves in both cases to flat Minkowski background metrics.

As with the consistency conditions earlier, the general possibility of performing this procedure is nothing new, see many of the other earlier references, and particularly \cite{Delphenich}. However, it seems to us that so far little emphasis has been put on how general this procedure actually is. A cynical point of view would be that everything is solved once the constitutive law~\eqref{eq:Zeff} has been written down. It seems that in many cases the attention was quickly drawn towards the general impossibility of representing every material as a space-time --- the question that leads naturally to the consistency conditions as described above. To us, and from an experimental point of view, it seems much more natural to also emphasize the other, trivial direction: That every space-time can be mimicked by a (meta-)material. We firmly believe that the cynicist's point of view would be misguided. 

Before starting this, there is, however, one technical point that bears keeping in mind: The simplest phrase to capture this issue stems from a paper by Fathi and Thompson \cite{TrafoOpticsCartographDistort} where it is called a \emph{cartographic distortion}. That these distortions play indeed a physical role will then be shown in section~\ref{sec:HawkingLab}.

\subsection{Cartographic Distortions}
The coordinates usually used to write down a given metric are coordinates on a given manifold. For the argument's sake, let us say we are considering the ordinary curvature coordinates of a Schwarzschild space-time. They are used to write the Schwarzschild metric in the form described in equation~\eqref{eq:Schwarzschild}. But if one were to realise this space-time as an analogue space-time, these coordinates would have a slightly different meaning: They now have to be \enquote{labelled} using the background manifold's coordinates --- the simplest case being that of approximating the background metric of a hypothetical laboratory as a Minkowski space-time. This view is now a bit closer in nature to that encountered in transformation optics: While there it is the medium that is used to mimic a given transformation, here the labelling through the background manifold's coordinates will have to mimic the coordinates of the space-time one wants to simulate. These coordinates of the analogue space-time are, at the end of the day, defined through a patch of $\mathbb{R}^4$ and a transformation (not necessarily a diffeomorphism!). This allows us to put the differences of the two fields more succinctly: In transformation optics the medium mimics a transformation, in an analogue space-time laboratory the background space-time has to mimic a transformation. While the last sentences might seem slightly repetitive, we hope that the slight variations might help with the picture in mind. A sketch of this idea can be found in figure~\ref{fig:analoguecoords}. In that figure, we tried to emphasize in particular the possibly complicated relation between the \emph{same} spatial laboratory coordinate (for all experimental purposes usually considered time-translation invariant) and the varying time-slices in the effective space-time these coordinates label. Obviously, the simplest such transformation imaginable is an identity transformation. But even then one has to be careful: For example, what has been Cartesian in the laboratory may not be Cartesian in the effective space-time. These issues will be mentioned several times. 

In the attempt to find helpful mappings between the two space-times $(M_\text{eff},g_\text{eff})$ and $(M_\text{lab},g_\text{lab})$, it is tempting and easy to oversimplify the transformation behaviour of both differential and algebraic (tensorial) quantities. The most obvious relation to fail is that the laboratory connection will not be metric-compatible with the effective metric and \emph{vice versa}. Neither will it be helpful (in most cases) to simply think of these transformations in terms of diffeomorphisms: For a diffeomorphism both pull-back and push-forward can be related to the Jacobian matrix (and its inverse) --- a relation that does not necessarily have to hold under general transformations, especially for connections. Again, the link to classic problems of cartography is apparent. 

\begin{figure}
	\includegraphics[width=\textwidth]{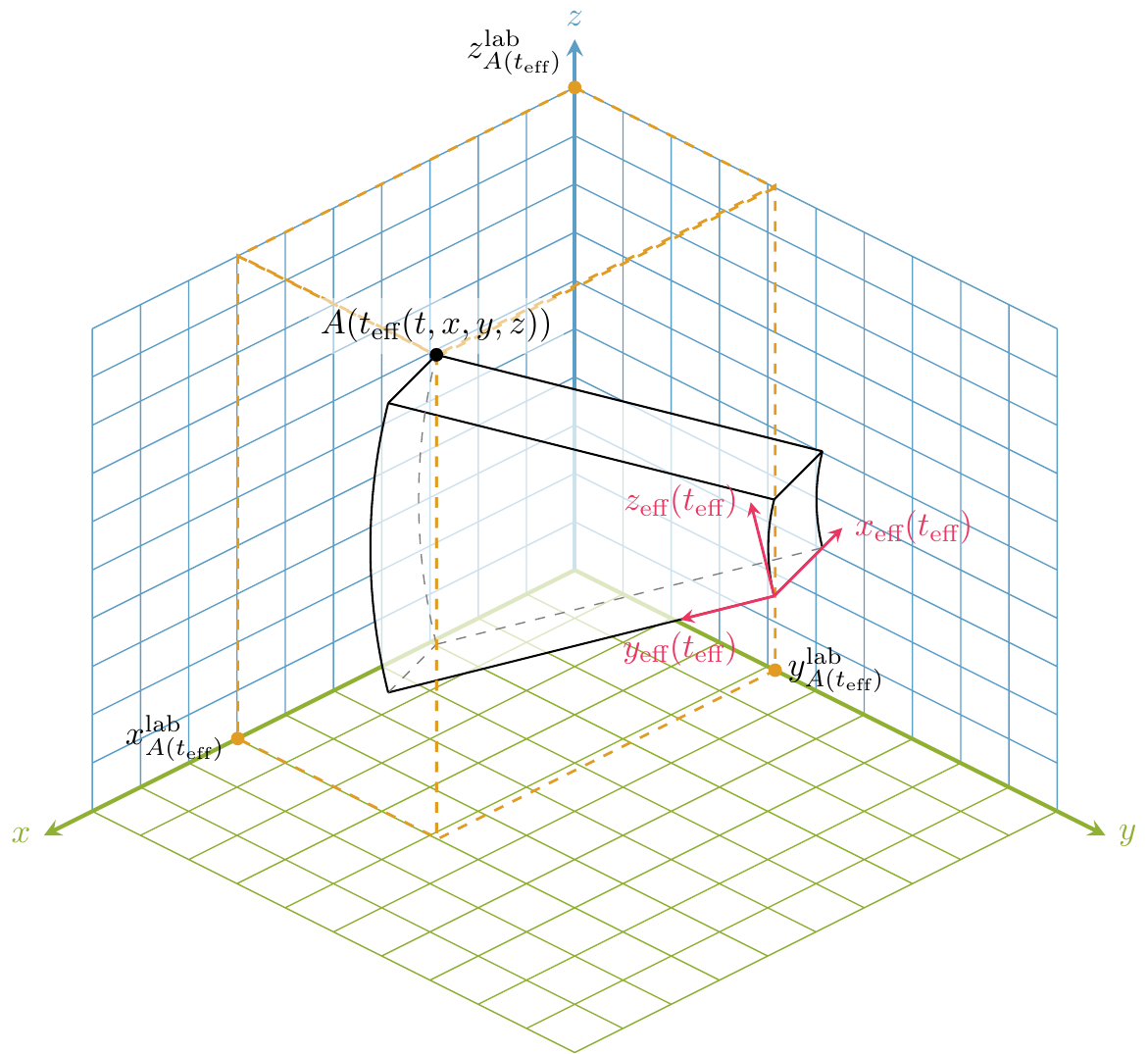}
	\caption[Sketch of Difference between Analogue and Laboratory Coordinates]{A sketch of the difference between analogue and laboratory coordinates. Note that every instance of $t_\text{eff}$ will depend on the laboratory coordinates $t,x,y,z$. The meaning of a given laboratory coordinate can change with respect to the effective space-time's time coordinate.}
	\label{fig:analoguecoords}
\end{figure}

This distinction is more than vacuous sophistry: As we will discuss in section~\ref{sec:HawkingLab} (from the algebraic analogy viewpoint of the last sections), any attempt at actually measuring Hawking radiation (or similar processes) in an experiment based on these electromagnetic analogue space-times will have to face this issue before any measurement can be conclusively linked to what is sought. Section~\ref{sec:PDEana} will then attempt to look more closely at the dynamical situation involving the different notions of Maxwell's equations in the effective space-time on the one hand, and on the other hand of the laboratory space-time. It should not come as a surprise that the transformation between the space-times will not be enough to rid the macroscopic laboratory Maxwell equation of material properties, to straightforwardly yield the vacuum Maxwell equations.

Before heading into the examples, it is worth repeating that the whole analysis of the previous section~\ref{sec:CovEMana} was completely and fully on the wave optics level. At no stage was a ray optics approximation employed. This statement remains true in the following. However and again, this by no means implies that the measured electromagnetic fields can immediately be identified with the fields as they would have been seen from within the analogue space-time.
\FloatBarrier
\subsection{Brute Force Methods}\label{sec:brute}
Given that \emph{any} metric can be used to form a constitutive tensor --- either argued for through trivial mathematics involving the insertion in equation~\eqref{eq:Zvacuum} or the equivalent physical argument that any space-time metric should admit Maxwellian electrodynamics ---, the corresponding constitutive tensors can easily and immediately turned into constitutive matrices readily available for laboratory (or theory) use. The only additional step involves making use of either the covariant or 3+1-dimensional, defining equations for these constitutive matrices, equations~\eqref{eq:defconstmatrices} or~\eqref{eq:defconst4matrices}. However, given the amount of index handling, this is best delegated to a computer algebra system. The advantage of this is obviously easily available expressions for the constitutive matrices of any given metric. On the downside, this comes along with the disadvantage of usually very unintuitive results which might have been easier to understand (and maybe even implement) with a different technique of derivation. (This alternative approach will be discussed in the next two subsections, subsection~\ref{sec:quasicart} and subsection~\ref{sec:curvilinear}.) In the present section, we will have a quick look at the brute force method and some of its results. The Maple2016 code used for achieving this was also tested by cross-checking the results of the next subsections; since, as mentioned, the alternative results in expressions easier to understand, we omit the cumbersome, CAS-gained version, and simply report their agreement. We still make a trivial identification of the coordinates on both effective and background space-time. Nevertheless, the brute-force methods will employ both rectilinear and curvilinear background coordinates. (In the parlance of the coming subsection on analytic methods for deriving meta-material mimics: We use \enquote{quasi-Cartesian} and \enquote{quasi-spherical} coordinates.)

\subsubsection{Gödel Space-Time}
The Gödel space-time is one of the most famous pathological space-times: It is a cosmological model showing locally causal behaviour, but globally is infested, similar to the Taub--NUT family, by closed time-like curves. The base manifold is simple $\mathbb{R}^4$. Slightly modifying the form given in \cite{GriffithsPodolsky}, we define its line element as
\begin{equation}
	\dif s^2 = -\kl{\dif t + e^{\omega x}\dif y}^2 + \dif x^2 + \ed{2}e^{2\omega x}\dif y^2 + \dif z^2.\label{eq:goedel}
\end{equation}
The Lorentzian manifold is geodesically complete, free of curvature singularities, and stationary. The matter which would source this metric, would have to fulfil
\begin{equation}
	4\pi (\rho-p) = -\Lambda.
\end{equation}
As the metric~\eqref{eq:goedel} already suggests, we will be using a flat, Cartesian background metric $\eta$ (to raise and lower indices). This profoundly simplifies possible index changes.

First, insert the metric in equation~\eqref{eq:Zvacuum} defining the constitutive tensor. Second, use this in equations~\eqref{eq:CovConstMatrices}. Third, one finds that the constitutive matrices needed to mimic this pathological metric would be
\begin{subequations}
	\begin{align}
		\epsilon^{ij} &= \begin{pmatrix}
							\ed{\sqrt{6}} e^{\omega x} & 0 &0\\
							0 &\frac{\sqrt{2}}{\sqrt{3}} e^{-\omega x}& 0\\
							0 & 0& \ed{\sqrt{6}} e^{\omega x}
						 \end{pmatrix},\\
		\mu^{ij} &= \begin{pmatrix}
				\frac{\sqrt{3}}{\sqrt{2}}e^{\omega x} & 0 & 0\\
				0 & \frac{\sqrt{2}}{\sqrt{3}}e^{-\omega x} & 0\\
				0 & 0 & \frac{\sqrt{3}}{\sqrt{2}}e^{\omega x}
		\end{pmatrix},\\
		\zeta^{i}{}_j &= \begin{pmatrix}
			0 &0 & -\frac{\sqrt{2}}{\sqrt{3}}\\
			0&0&0\\
			\frac{\sqrt{2}}{\sqrt{3}}&0&0
		\end{pmatrix}.
	\end{align}
\end{subequations}
Despite the rather complicated causal behaviour, see for example \cite{HawEll,VisserWormholes,GriffithsPodolsky}, the structure of the optical properties of a mimicking medium turn out to be of a rather simple kind. It is worthwhile to compare this result with later, not brute-forced results regarding general metrics --- one cannot help but notice (despite absence of staticity) the similarity to our results for static, spherically symmetric space-times in section~\ref{sec:curvilinear}. In fact, while not spherically symmetric, the Gödel space does have axisymmetry and is homogeneous.

Note that the causal issues of the \emph{effective} Gödel space-time are unproblematic: In the present context, the background space-time enforces causality, and anything unusual happening in the framework of the effective space-time will be comparable to the method of active noise control: The speed of sound is causally irrelevant, as it is the electrical signal speed that allows that particular technology.

Also, just as the general discussion of the Kaluza--Klein decomposition in section~\ref{sec:curvilinear} shows, we can observe (and anticipate) the appearance of a zero eigenvalue for $\zeta$ which will lie in the direction of $\beta$ (as it was defined in the Kaluza--Klein decomposition).

\subsubsection{The Taub--NUT Space-Time}
Let us shortly return to the example of another unphysical space-time, expounded in section~\ref{sec:unphysical}. There we considered the line element
\begin{align}
	\dif s^2 =& - \left( \frac{r^2-2mr-a^2}{r^2+a^2}\right) (\dif t - 2 a \cos\theta\; \dif \phi)^2 
	+ \left( \frac{r^2+a^2}{r^2-2mr-a^2}\right) \dif r^2 
	\nonumber\\
	&\qquad+ (r^2+a^2) (\dif\theta^2 +\sin^2\theta\; \dif \phi^2).\label{eq:TaubNUTre}
\end{align}
As this medium is captured in an easier way using a spherical background metric, this is a first case where a computer-aided approach to equation~\eqref{eq:Zeff} and its subsequent use for defining constitutive matrices becomes a great boon. We take insert metric~~\eqref{eq:TaubNUTre} in equation~\eqref{eq:Zeff}, and then use (in a computerised way) again equations~\eqref{eq:CovConstMatrices}. Now, we can give the following constitutive tensors as meta-material mimics for this space-time:
\begin{subequations}
	\begin{align}
		\epsilon^{ij}=& \mathrm{diag}\left(
			-\frac{\kl{r^4+6a^2 r^2-8ma^2r - 3a^4}\cos^2\theta - (r^2+a^2)^2}{(r^2+a^2)r^2\sin^2\theta},\right.\nonumber\\
			&\left.\frac{\kl{r^4+6a^2 r^2-8ma^2r - 3a^4}\cos^2\theta - (r^2+a^2)^2}{(r^2+a^2)(a^2+2mr-r^2)r^2\sin^2\theta},
			\frac{r^2+a^2}{r^2(r^2-2mr-a^2)\sin^2\theta}\right)
		,\\
		\mu^{ij} =& \begin{pmatrix}
			\frac{r^2+a^2}{r^2} &0 &0\\
			0& \frac{r^2+a^2}{r^2\kl{r^2-2mr-a^2}} & 0\\
			0 & 0& \frac{r^2+a^2}{r^2\kl{r^2-2mr-a^2}\sin^2\theta}
		\end{pmatrix},\\
		\zeta^{i}{}_{j}=& \begin{pmatrix}
			0 & -\frac{2a\cos\theta(r^2-2mr-a^2)}{(r^2+a^2)\sin\theta}&0\\
			\frac{2a\cos\theta}{(r^2+a^2)\sin\theta} & 0 &0 \\
			0&0&0
		\end{pmatrix}.
	\end{align}
\end{subequations}
As our observations would be similar to those in the previous example, let us just reiterate the absence of causal problems. Anything happening in the effective space-time's metric is still constrained causally by the background metric of the laboratory. We just refer to the general results of the coming section for observations beyond this.

\subsubsection{Rindler Space-Time}
To compare with the literature, let us have a try at the line element analysed by Reznik in \cite{ReznikRefIndexHorizons}, see also (6.154) in \cite{BirrellDavies}. This is (one version of) the Rindler space-time, describing accelerated motion in Minkowski space-time. The acceleration results in an observational horizon limiting the accelerated observer to observations in one Rindler wedge. It also gives rise to the celebrated Unruh effect, analogous to the Hawking effect, and mentioned in the introduction. Note that the present example is extremely simple: The methods of section~\ref{sec:quasicart} easily arrive at the same results, compare equations~\eqref{eq:RindlerAnalytic} to \eqref{eq:RindlerConstitutive}.

The line element is
\begin{equation}
	\dif s^2 = - \alpha^2 z^2 \dif t^2 + \dif x^2 + \dif y^2 + \dif z^2.\label{eq:Rindler}
\end{equation} 
The parameter $\alpha$ is the proper acceleration in $z$-direction. Note that this form can also be achieved for black hole space-times in the near-horizon limit, see \cite{FabbriNS05}. It should come as no surprise that it is frequently occurring both in analyses of the genuine Unruh effect as well as rephrasings of the Hawking effect. For the background space-time, we again choose Minkowski space-time in Cartesian form.

We get for permittivity, permeability and magneto-electric tensor
\begin{subequations}\label{eq:RindlerBrute}
	\begin{align}
		\epsilon^{ij} &= \begin{pmatrix}
		\ed{\alpha z} & 0 &0\\ 0&\ed{\alpha z} & 0\\0&0&\ed{\alpha z}
		\end{pmatrix},\\
		\mu^{ij} &= \begin{pmatrix}
		\ed{\alpha z} & 0 &0\\ 0&\ed{\alpha z} & 0\\0&0&\ed{\alpha z}
		\end{pmatrix},\\
		\zeta^i{}_{j} &= 0.
	\end{align}	
\end{subequations}
This is now again a simple calculation both by hand or by computer, following the previous examples fully.

As no off-diagonal terms appear in the metric~\eqref{eq:Rindler}, it should come as no surprise that the magneto-electric tensor vanishes. Adding that we are in Cartesian coordinates, we immediately regain the standard consistency condition, as in equation~\eqref{eq:conscond3+1}. This is also the form used in \cite{ReznikRefIndexHorizons}.

\subsubsection{Dirty Black Holes}
Following the metric as discussed in section~\ref{sec:dirty}, we now give the material properties for mimicking a metric of the quite general form
\begin{equation}
	\dif s^2 = -e^{-2\Phi(r)}\kl{1-\frac{b(r)}{r}}\dif t^2 + \frac{\dif r^2}{1-\frac{b(r)}{r}} + r^2(\dif \theta^2 + \sin^2\theta\dif\phi^2).
\end{equation}
While we focus on its use as a \enquote{dirty black hole}, this family of metrics also easily encompasses wormhole space-times \cite{MorrisThorne,VisserWormholes}. Whereas these are causally worrisome in astrophysical contexts, this is again of no concern in an analogue space-time. As before, the example of active noise control serves as a comparison.

As all manipulations necessary for the derivation of the electromagnetic properties of the mimic are purely algebraic in nature, the $r$-dependence of both $b(r)$, as well as of $\Phi(r)$ are only a notational complication. For this reason, we will from now on omit them in the expressions. The results following the usual brute-force procedure are:
\begin{subequations}
	\begin{align}
		\epsilon^{ij} &= \begin{pmatrix}
			e^{\Phi} & 0 &0\\
			0 &\frac{e^\Phi}{(r-b)r}& 0\\
			0 & 0& \frac{e^\Phi}{(r-b)r\sin^2\theta}
		\end{pmatrix},\\
		=\mu^{ij} &= \begin{pmatrix}
			e^{\Phi} & 0 &0\\
			0 &\frac{e^\Phi}{(r-b)r}& 0\\
			0 & 0& \frac{e^\Phi}{(r-b)r\sin^2\theta}
		\end{pmatrix},\\
		\zeta^{ij} &= 0.
	\end{align}
\end{subequations}
As an example of static, spherically symmetric space-times, it is useful to compare this with the discussions of both static and spherical space-times in the following applications using calculations \enquote{by hand}. (As the Schwarzschild solution in curvature coordinates is contained in the general form of a dirty black hole, its corresponding material properties are easily recognised.) As we are encountering here a metric with vanishing $\beta$, the equality of permittivity and permeability is manifest.

\subsection{Quasi-Cartesian Coordinates} \label{sec:quasicart}
\begin{sloppypar}As mentioned, the simplest approach to deriving bespoke meta-material properties straight from the consistency conditions --- allowing one to work with matrices alone without invoking the fourth-rank tensor $Z$ involved in their definition --- not only identifies laboratory and effective space-time coordinates, it further only considers quasi-Cartesian ones. The reason for calling these coordinates only quasi-Cartesian becomes apparent once one looks at their meaning in the two space-times in use: While the coordinates are, given our assumption of a flat background (laboratory) metric, orthonormal in the laboratory, the corresponding quasi-Cartesian coordinates these are mapped to in the effective space-time need not be orthonormal. What we will look at in this chapter are the following effective space-times as starting point:\end{sloppypar}
\begin{itemize}
	\item Another look at the Rindler space-time
	\item Several different coordinate systems for the Schwarzschild geometry:
		\begin{itemize}
			\item Cartesian Curvature coordinates
			\item Cartesian Painlevé--Gullstrand coordinates
			\item Cartesian Kerr--Schild form
			\item Cartesian Gordon form
			\item Cartesian isotropic form
		\end{itemize}
	\item General remarks on static, spherically symmetric effective space-times
	\item Two examples for the geometries which include the Kerr geometry
		\begin{itemize}
			\item Cartesian Kerr--Schild form
			\item Cartesian Doran form
		\end{itemize}
\end{itemize}
With these in place, it would be easy to adapt according to needs, \emph{e.g.}, to Reissner--Nordström or Kerr--Newman space-times.

With the outline in place, let us begin by making some (very short) general comments on quasi-Cartesian coordinates for the most straightforward case of the coordinates in the effective space-time: An effective space-time with spherical symmetry. These are given by
\begin{equation}
	x^a = (t,x,y,z) = (t,r\sin\theta\cos\phi,r\sin\theta\sin\phi,r\cos\theta).\label{eq:spherical}
\end{equation}
Then we can define the spatial (3-)vector of unit length in $r$-direction by setting
\begin{equation}
	\hat r_i \defi \left(\frac{x}{r}, \frac{y}{r}, \frac{z}{r} \right),
\end{equation}
with the familiar $r=r(x,y,z)=\sqrt{x^2+y^2+z^2}$. This allows us to define a projection operator in the spatial slices of the full (effective) space-time unto the space orthogonal to the $r$-direction (\emph{i.e.}, two-spheres),
\begin{equation}
	P^i{}_{j} \defi \delta^i{}_{j} - \hat r^i \hat r_j.\label{eq:projectionP}
\end{equation}

This example, however, has an obvious deficit: It cannot be horizon-penetrating. In order for a coordinate system to be horizon-penetrating, $g_{tt}$ becoming 0 somewhere means that the off-diagonal parts of the metric, $\beta^i$, must not vanish there, for the coordinate patch to extend $g$ smoothly beyond this. We shall see that this can still be achieved even in quasi-Cartesian coordinates, but it is not as simple as the above practice example of coordinates. In particular, horizon-penetrating coordinates immediately imply non-vanishing magneto-electric effects. As these are even harder to manufacture with prescribed spatial dependency than permittivity and permeability are, this will definitely lower their practical relevance. But as the production of the corresponding materials is not the point of this thesis, and even theoretical inquiries have merit, we will show several examples involving horizon-penetrating coordinates in the following.

There are two reasons for introducing this ahead of the more concrete examples: One, most of them \emph{are} spherically symmetric. Two, noticing the presence of projection operators is a recurring theme which is worth highlighting as early as possible.

The following work is based on \cite{Bespoke1}; as we fix the background metric to be $\eta$, we refrain from keeping track of the index \enquote{${}_\text{eff}$} on the effective space-time metric $g_\text{eff}$ and just refer to it as $g$. Later, in the curvilinear examples in section~\ref{sec:curvilinear}, more notational care has to be taken.

\subsubsection{Rindler Space-Time}
As a first example, let us take again a look at the line element
\begin{equation}
	\dif s^2 = - \alpha^2 z^2 \dif t^2 + \dif x^2 + \dif y^2 + \dif z^2.\label{eq:RindlerAnalytic}
\end{equation}
The reason for this choice is that it is quick and simple, and a good way to illustrate some of the steps occurring frequently in the following example.

Instead of just relying on the results of a CAS, our goal is to arrive at the same results by means of analytic techniques. For this we want to make use of the decompositions as shown in equation~~\eqref{eq:g3+1mu} and equation~\eqref{eq:g3+1eps}, but before we can do this we need to remind ourselves of a key ingredient in deriving these equations: In setting the determinant of background metric and effective metric equal to each other --- see equation~\eqref{eq:ConfChoice} --- we made use of the conformal freedom of the underlying electromagnetic theory. A quick look at equation~\eqref{eq:RindlerAnalytic}, however, reveals this to have determinant $\det g = \alpha^2 z^2$. Thus, let us choose a conformal factor of $\sqrt{|\alpha z|}^{\,-1}$, resulting in the metric
\begin{equation}
	\dif s^2 = - (|\alpha z|)^{3/2} \dif t^2 + \sqrt{|\alpha z|}^{\,-1}\kl{\dif x^2 + \dif y^2 + \dif z^2}.\label{eq:RindlerConf}
\end{equation}
conformally equivalent to our original question. Now we can simply read off the determinant of $\mu^{-1}$ in equation~\eqref{eq:g3+1mu} from the $tt$-component as
\begin{equation}
	\sqrt{\det\mu^{\circ\circ}}^{\,-1} = (|\alpha z|)^{3/2}.
\end{equation}
Then the $ij$-components translate to the condition
\begin{equation}
	(|\alpha z|)^{-3/2} \mu^{-1}_{ij} = \sqrt{|\alpha z|}^{-1} \delta_{ij},
\end{equation}
whence
\begin{equation}
	\mu^{-1}_{ij} = |\alpha z| \delta_{ij}.
\end{equation}
The vanishing of $\beta$ then immediately allows the identification of
\begin{subequations}\label{eq:RindlerConstitutive}
	\begin{align}
		\epsilon^{ij} = \mu^{ij} =& \ed{|\alpha z|} \delta^{ij},\\
		\zeta^i{}_j =& 0.
	\end{align}
\end{subequations}
The appearance of the absolute value compared to the previous, brute-forced calculation is irrelevant as $z=0$ corresponds to the horizon where the optical properties diverge in the first place and hence force us to limit attention to one side of the horizon. This is reminiscent of the sfreeituation of the Schwarzschild geometry in isotropic coordinates, see the example below.

\subsubsection{Schwarzschild I: Cartesian Curvature Coordinates}
Rephrasing the original Schwarzschild metric in curvature coordinates as given in equation~\eqref{eq:Schwarzschild} in terms of the coordinates defined in equation~\eqref{eq:spherical}, and making use of the projection operator $P_{ij}$ as defined in equation~\eqref{eq:projectionP}, we arrive at the following, quasi-Cartesian form of the Schwarzschild metric:
\begin{equation}
	\dif s^2 = g_{ab} \dif x^a \dif x^b=  - (1-2m/r) \dif t^2 + \frac{(\hat r_i \; \dif x^i)^2}{1-2m/r} + P_{ij} \dif x^i \dif x^j. \label{eq:pseudo-cartesian}
\end{equation}
Note that this metric has determinant $-1$.

Comparing this with the form~\eqref{eq:g3+1mu} for the effective metric derived previously, it is easy to see that the following three identities have to hold:
\begin{subequations}
	\begin{align}
		\sqrt{\det(\mu^{\circ\circ})}^{-1} &= 1-2m/r,\label{eq:BespokeDet1}\\ 
		\beta^i  &= 0,\\
		\sqrt{\det(\mu^{\circ\circ})} \; \mu^{-1}_{\,ij} &=  \frac{\hat r_i \; \hat r_j}{1-2m/r} + P_{ij}.
	\end{align}
\end{subequations}
After some rearranging, we can solve this for $\mu$ and $\mu^{-1}$, remembering that in our scenario indices are raised and lowered using Kronecker deltas:
\begin{subequations}
	\begin{align}
		\mu^{-1}_{\,ij}  &= (1-2m/r)P_{ij} + \hat r_i \hat r_j,\\
		\mu_{\,ij}  &= (1-2m/r)^{-1} P_{ij} + \hat r_i \hat r_j.
	\end{align}
\end{subequations}
It is easy to see that this fulfils the condition on the determinant, equation~\eqref{eq:BespokeDet1}; we also have already fully specified the properties of our bespoke meta-material mimic! As the off-diagonal parts, the $\beta^i$ in the KK decomposition, of the metric~\eqref{eq:pseudo-cartesian} are zero, the magneto-electric tensor trivially vanishes. Summarising, we have:
\begin{subequations}
	\begin{align}
		\epsilon_{\,ij} = \mu_{\,ij}  &= (1-2m/r)^{-1} P_{ij} + \hat r_i \hat r_j,\\
		\zeta_{ij} &= 0.
	\end{align}
\end{subequations}

Before continuing on to more bespoke meta-material mimics, it is worthwhile to point out two features of this first result:
\begin{itemize}
	\item Were we to return to a triad based on spherical coordinates we would get that
	\begin{subequations}
		\begin{equation}
			\epsilon_{rr} = \mu_{rr} = 1
		\end{equation}
		in the radial direction, and
		\begin{equation}
			\epsilon_{\hat\theta\hat\theta}=\epsilon_{\hat\phi\hat\phi} = \mu_{\hat\theta\hat\theta}=\mu_{\hat\phi\hat\phi} = (1-2m/r)^{-1}  > 1
		\end{equation}
		along the angular directions. More concretely, any part of $\epsilon$ or $\mu$ will be equal to or larger than 1. Note that this is precisely the form Reznik reports in \cite{ReznikRefIndexHorizons} in his reference~[18] (reference~[11] in his preprint), apart from the naming of coordinates and their identification with triad components.
		\end{subequations}
	\item The product of $\epsilon$ and $\mu$ being the square of the refractive index, the previous point translates into a statement on the refractive index in any direction:
	\begin{equation}
		n \geq 1.
	\end{equation}
	\item This bound on the refractive index extends and agrees with a previous result by Reznik \cite{ReznikRefIndexHorizons}: When approaching the horizon, any electromagnetic analogue must have a diverging refractive index.
\end{itemize}

\subsubsection{Schwarzschild II: Cartesian Painlevé--Gullstrand Coordinates}
A second example shall demonstrate the effects of horizon-penetrating coordinates, and the resulting non-vanishing magneto electric effects. For this we employ the Painlevé--Gullstrand form of the Schwarzschild space-time:
\begin{equation}
	\dif s^2 = g_{ab} \dif x^a \dif x^b = - \dif t^2 + \delta_{ij} \left(\dif x^i - \sqrt{2m/r} \; \hat r_i \,\dif t\right)  \left(\dif x^j - \sqrt{2m/r} \; \hat r_j \,\dif t\right).
\end{equation}
We omitted the steps taken to derive this quasi-Cartesian form from the traditional one in spherical coordinates --- the discussion preceding the present examples covers that sufficiently. Its determinant once more is $-1$.

Again, taking equation~\eqref{eq:g3+1mu} as the starting point allows one to read off the following three identifications:
\begin{subequations}
	\begin{align}
		\sqrt{\det(\mu^{\circ\circ})}^{-1} &= 1-2m/r,\\
		\mu^{-1}_{ik}  \beta^k  &= \sqrt{2m/r} \; \hat r_i,\\
		\delta_{ij} &= \sqrt{\det(\mu^{\circ\circ})}\left(\mu^{-1}_{\,ij} - \left[\sqrt{2m/r} \; \hat r_i\right]\left[\sqrt{2m/r} \; \hat r_j\right]\right).
	\end{align}
	The last equation can be rephrased as
	\begin{equation}
		\mu^{-1}_{\,ij} = (1-2m/r)\delta_{ij} + (2m/r) \hat r_i \hat r_j,\\
	\end{equation}
\end{subequations}
and using the previously introduced projection operator $P_{ij}$ anew, we can rewrite this in a similar vein as before:
\begin{subequations}
	\begin{align}
		\mu^{-1}_{\,ij}  &= (1-2m/r)P_{ij} +  \hat r_i \hat r_j,\\
		\mu_{\,ij}  &= (1-2m/r)^{-1} P_{ij} +  \hat r_i \hat r_j.
	\end{align}
\end{subequations}
From this we can derive the corresponding magneto-electric tensor as
\begin{equation}
	\zeta_{ij} = -\ed{2} \sqrt{\frac{2m}{r}} \eps_{ijk} \hat r^k.
\end{equation}

The simplest way to find the permittivity is certainly by looking at equation~\eqref{eq:g3+1eps}, the decomposition of the metric in terms of the permittivity itself. This immediately results in $\epsilon_{ij} = \delta_{ij}$. Other, less direct paths can be taken, for example:
\begin{equation}
	\epsilon^{ij} =  
	\mu^{ij}\, (1 - \mu^{-1}_{\,kl} \beta^k \beta^l )  + \beta^i \beta^j = 
	\mu^{ij} (1-2m/r) + (2m/r) \hat r^i \hat r^j = \delta^{ij}.
\end{equation}
Either way, we now can summarise the results:
\begin{subequations}
	\begin{align}
		\epsilon_{ij} &= \delta_{ij},\\
		\mu_{\,ij}  &= (1-2m/r)^{-1} P_{ij} +  \hat r_i \hat r_j,\\
		\zeta_{ij} &= -\ed{2} \sqrt{\frac{2m}{r}} \eps_{ijk} \hat r^k.
	\end{align}
\end{subequations}

Let us list conclusions from these results, some recurring from the first example:
\begin{itemize}
	\item The permeability is the same as in Cartesian curvature coordinates. The explanation for this can be found when looking at the origin of the Painlevé--Gullstrand coordinates: They are gained from a coordinate transformation only changing the time-coordinate according to $t \to t+f(r)$. As the permeability is the purely spatial part of the constitutive tensor $Z$, such a transformation cannot change $\mu$, compare the various decompositions in appendix~\ref{sec:ortho}.
	\item The permittivity, on the other hand, is the purely time-projected part of the constitutive tensor $Z$, and concomitantly has to change.
	\item While $\epsilon=1$ now is constant, $\mu$ remains $\geq 1$.
	\item Correspondingly, $n\geq 1$ still holds.
	\item As $\mu$ still diverges on the horizon, so will the refractive index $n$.
\end{itemize}

\subsubsection{Schwarzschild III: Cartesian Kerr--Schild Form}
The Kerr--Schild form of the Schwarzschild space-time is given by
\begin{equation}
	g_{ab} = \eta_{ab} + {\frac{2m}{r}} \ell_a \ell_b,  \qquad \text{where~}   \ell_a = ( -1, \hat r_i),
	\qquad \det(g_{ab})=-1.
\end{equation}
Further below, when discussing the corresponding case of the Kerr metric in (quasi-Cartesian) Kerr--Schild form, we will also make a few general comments on the Kerr--Schild form which are valid here, too. Continuing as in the previous two examples, we look at equation~\eqref{eq:g3+1mu} to find
\begin{subequations}
	\begin{align}
		\sqrt{\det(\mu^{\circ\circ})}^{-1} &= 1-2m/r,\label{eq:BespokeSchwarzschildKerrSchild1}\\
		\mu^{-1}_{ik}  \beta^k  &= (2m/r) \; \hat r_i,\\
		\delta_{ij} + (2m/r)\; \hat r_i\; \hat r_j &= 
		\sqrt{\det(\mu^{\circ\circ})}\left(\mu^{-1}_{\,ij} - \left[(2m/r) \; \hat r_i\right]\left[(2m/r) \; \hat r_j\right]\right).\label{eq:BespokeSchwarzschildKerrSchild2}
	\end{align}
\end{subequations}
Inserting the determinant~\eqref{eq:BespokeSchwarzschildKerrSchild1} in the third equation~\eqref{eq:BespokeSchwarzschildKerrSchild2}, we can solve for $\mu^{-1}$:
\begin{equation}
	\mu^{-1}_{\,ij} = (1-2m/r)[\delta_{ij} +  (2m/r)\; \hat r_i\; \hat r_j ] + [(2m/r) \; \hat r_i][(2m/r) \; \hat r_j].
\end{equation}
This can be quickly simplified to the (by now) usual pair of equations
\begin{subequations}
	\begin{align}
		\mu^{-1}_{\,ij}  &= (1-2m/r)P_{ij} +  \hat r_i \hat r_j,\\
		\mu_{\,ij}  &= (1-2m/r)^{-1} P_{ij} +  \hat r_i \hat r_j. 
	\end{align}
\end{subequations}
By heeding that the only possible off-diagonal components can come from $\ell_0 \ell_i$, the magneto-electric tensor $\zeta$ is quickly evaluated to
\begin{equation}
	\zeta_{ij} = -\frac{m}{r} \,\varepsilon_{ijk} \,\hat r^k.
\end{equation}
For $\epsilon$, we shall only mention that the quickest method of calculation involves again equation~\eqref{eq:g3+1eps}, but that the same result can be glanced by noticing from the consistency condition~\eqref{eq:conscond3+1zeta} that
\begin{subequations}
	\begin{align}
		\epsilon^{ij} &= \mu^{ij}\, (1 - \mu^{-1}_{\,kl} \beta^k \beta^l )  + \beta^i \beta^j, \\
		&=
		\mu^{ij} (1-[2m/r]^2) + (2m/r)^2 \hat r^i \hat r^j,\\
		&= (1+2m/r) P^{ij} +\hat r^i \hat r^j.
	\end{align}
\end{subequations}

Summarising the results for the three constitutive matrices of the Cartesian Kerr--Schild form of the Schwarzschild metric, we have
\begin{subequations}
	\begin{align}
		\epsilon_{\,ij} &= (1+2m/r)  P_{ij}+ \hat r_i\; \hat r_j,\\
		\mu_{\,ij}  &= (1-2m/r)^{-1} P_{ij} + \hat r_i \hat r_j,\\
		\zeta_{ij} &= -\frac{m}{r} \varepsilon_{ijk} \hat r^k.
	\end{align}
\end{subequations}

We end the discussion of this example, as before, with a list of observations and conclusions.
\begin{itemize}
	\item As the Kerr--Schild form can be found from the curvature coordinates by a transformation of the type $t \to t+f(r)$, $\mu$ does not change (like in Painlevé--Gullstrand coordinates).
	\item $\zeta$, does again change, as it is the \enquote{mixed} part of the orthogonal decomposition of $Z$.
	\item Again, we have that $\epsilon, \mu\geq 1$.
	\item The same conclusions as before hold for the corresponding refractive index.
\end{itemize}

\subsubsection{Schwarzschild IV: Cartesian Gordon Form}
The Gordon form is inspired, as the name hopefully suggests, from the Gordon metric whose form we explicitly saw in the discussion of the moving media, in section~\ref{sec:iso}. However, there it was more a happy accident simplifying our analysis, as mentioned in footnote~\ref{fn:GordonMetric}. In the present variation on the theme of the Schwarzschild space-time, just as in its historical origin, we shall concretely link the notion of a refractive index and the resulting light propagation to the propagation of light in a curved space-time. However, unlike large parts of Gordon's original analysis in \cite{GordonMetric}, we will not need simplifying assumptions like the geometric optics approximation to complete our analysis; at least for the purposes of this example. More on this form of the Schwarzschild metric can be found in references~\cite{RosquistGordonMetric,GiacomelliLiberatiRotBHRelAnalogues}.\footnote{\cite{RosquistGordonMetric} also presents coordinate transformations allowing to find the limiting cases of the metric as given here for $n\to 1$ and $n\to \infty$. This can be used to interpret the Gordon form of the Schwarzschild metric as an interpolation between Kerr--Schild on the one hand, and Painlevé--Gullstrand on the other. This certainly gives the current placement of this example a very natural context.}

The Schwarzschild metric in Gordon form reads
\begin{align}
	g_{ab} &=\sqrt{n} \left( \eta_{ab} + [1-n^{-2}]V_a V_b\right), \quad  \text{where}\label{eq:GordonSchwarzschild}\\
	V_a &= \left(- \sqrt{1+2m/r}, \sqrt{2m/r} \; \hat r_i\right).\nonumber
\end{align}
Here, $n$ acts as a conformal factor forcing the metric determinant to be $-1$, which can also be interpreted as a \emph{constant} refractive index. Together with the four-velocity $V_a$ this then gives rise to the interpretation of a fluid with constant refractive index and four-velocity $V$, which is championed in \cite{RosquistGordonMetric}. While the naming already suggests it, it requires a bit of work to relate the last remaining, unexplained parameter $m$ to the physical mass. We start this demonstration by looking at
\begin{equation}
	g_{tt} = - \sqrt{n}(1-[1-n^{-2}] (1+2m/r)) =  - \sqrt{n} \left(n^{-2} - [1-n^{-2}] (2m/r)\right).
\end{equation}
It follows that
\begin{equation}
	g_{tt} =  - n^{-3/2} \left(1 - [n^{2}-1] (2m/r)\right) \propto - \left(1 - [n^{2}-1] (2m/r)\right) ,
\end{equation}
which then leads us to the actual, physical mass in terms of $n$ and $m$:
\begin{equation}
	\frac{G m_\text{physical}}{ c^2} = (n^2-1) m.
\end{equation}
Note that (due to the demand of unimodularity and conformal freedom) the single physical parameter $m_\text{physical}$ is now encapsulated in \emph{two} parameters. The information content stays the same. The idea behind this identification is that by looking at the fall-off of the various components of the metric, we can simply read off the ADM mass in the present case. The same result can be gained by performing the coordinate transformation
\begin{equation}
	\dif t = n \dif t_\text{physical} + \kl{1-\frac{2G m_\text{physical}}{r}}^{-1}\sqrt{\frac{2G m_\text{physical}}{r}\kl{n^2-1+\frac{2G m_\text{physical}}{r}}}\dif r.
\end{equation}

The underlying assumption of the fluid model is isotropy of the refractive index $n$ in the rest frame of the fluid. This means that in that particular frame, also $\epsilon_V$, $\mu_V$, and $\zeta_V$ are isotropic. We, however, are not interested in the natural rest-frame of the fluid, we want to know the corresponding constitutive matrices of the laboratory frame. Certainly, since we are given both four-velocities (the fluid's is specified together with the metric in equation~\eqref{eq:GordonSchwarzschild}, while the laboratory's frame\footnote{One of the velocities should be renamed, for example to $W$ (in agreement with the notation of our appendix~\ref{sec:frames}), but this would be rather confusing notation at this point.} is defined through $V_a = (-1,0,0,0)$) we could follow through with the change of frame analysis of appendix~\ref{sec:frames}, but this will neither be helping physical understanding nor be simple. We thus again repeat the analysis along the lines of the previous examples.

Accordingly, again comparing terms with equation~\eqref{eq:g3+1mu}, we are able to glean the following identities:
\begin{subequations}
	\begin{align}
		\sqrt{\det(\mu^{\circ\circ})}^{-1} =& \sqrt{n}\left\{1-[1-n^{-2}] (1+2m/r)\right\},\label{eq:GordonDet}\\
		\mu^{-1}_{ik}  \beta^k  =& \sqrt{n} [1-n^{-2}] \sqrt{(1+2m/r)(2m/r)}\; \hat r_i,\label{eq:GordonMuBeta}\\
		\sqrt{n}\left\{ \delta_{ij} + [1-n^{-2}](2m/r) \hat r_i \hat r_j \right\} =& \nonumber\\\sqrt{\det(\mu^{\circ\circ})}&\left(\mu^{-1}_{\,ij} - \left[n(1-n^{-2})^2(1+2m/r)(2m/r)\hat r_i \hat r_j\right]\right).
	\end{align}
\end{subequations}
Multiplying the first and the third equation, 
\begin{gather}
	\left(\mu^{-1}_{\,ij} - \left[n(1-n^{-2})^2(1+2m/r)(2m/r)\hat r_i \hat r_j\right]\right) = \nonumber\\
	n \left\{1-[1-n^{-2}] (1+2m/r)\right\} \left\{ \delta_{ij} + [1-n^{-2}](2m/r) \hat r_i \hat r_j \right\},
\end{gather}
we can gain an equation for $\mu^{-1}$:
\begin{subequations}
	\begin{align}
		\mu^{-1}_{\,ij} =
		n &\left\{1-[1-n^{-2}] (1+2m/r)\right\} \left\{ \delta_{ij} + [1-n^{-2}](2m/r) \hat r_i \hat r_j \right\}\nonumber\\
		&+\left[n(1-n^{-2})^2(1+2m/r)(2m/r)\hat r_i \hat r_j\right],\\
		= \hphantom{n}& n \left\{1-[1-n^{-2}] (1+2m/r)\right\}  P_{ij} + n^{-1}\; \hat r_i \hat r_j.\label{eq:GordonMuInv}
	\end{align}
\end{subequations}
This then can be inverted to
\begin{equation}
	\mu_{\,ij} = \frac{P_{ij} }{ n \left\{1-[1-n^{-2}] (1+2m/r)\right\} } + n\; \hat r_i \hat r_j.\label{eq:GordonMu}
\end{equation}

For the remaining two matrices, $\zeta$ and $\epsilon$, we will begin with the former. In order to calculate the magneto-electric tensor, we calculate $\beta$ by first reminding us of the determinant~\eqref{eq:GordonDet}; second, as $\beta$ has to be radial (directly seen from spherical symmetry, see also the first non-Schwarzschild, and very general example below), the contraction of $\mu^{-1}$, as in equation~\eqref{eq:GordonMuInv}, with $\beta$ will be again just proportional to $\beta$:
\begin{equation}
	\mu^{-1}_{ik}  \beta^k  = n^{-1}  \beta_i.
\end{equation}
Together with equation~\eqref{eq:GordonMuBeta} we are thus led to
\begin{equation}
	\beta_i =  n^{3/2} [1-n^{-2}] \sqrt{(1+2m/r)(2m/r)}\; \hat r_i,
\end{equation}
from which we are able to calculate
\begin{subequations}
	\begin{align}
		\zeta^{ij} &= -\ed{2}  \sqrt{\det[\mu^{-1}]}\epsilon^i{}_{kl} \beta^l \mu^{kj}, \\
		           &= -\ed{2}  \left( [n^{2}-1] \sqrt{(1+2m/r)(2m/r)} \right)\epsilon^{ij}{}_k \; \hat r^k .\label{eq:GordonZeta}
	\end{align}
\end{subequations}
To calculate the final matrix, the permittivity, we start by calculating
\begin{subequations}
	\begin{align}
		\beta_i \beta_j=&  n^{3} [1-n^{-2}]^2 (1+2m/r)(2m/r)\; \hat r_i \hat r_j, \\
		=& n^{-1} [n^{2}-1]^2 (1+2m/r)(2m/r)\; \hat r_i \hat r_j,
	\end{align}
\end{subequations}
implying
\begin{equation}
	\mu^{-1}_{ik}  \beta_i \beta^k =n^{-2} [n^2-1]^2(1+2m/r)(2m/r).
\end{equation}
These last two equations are the remaining ingredients to calculate $\epsilon$ straight from the consistency condition~\eqref{eq:conscond3+1zeta}:
\begin{subequations}
	\begin{align}
		\epsilon^{ij} =& \mu^{ij}\, \left(1 - \mu^{-1}_{\,kl} \beta^k \beta^l \right)  + \beta^i \beta^j, \\
		=& \mu^{ij} \left(1-n^{-2} [n^2-1]^2(1+2m/r)(2m/r)\right) \nonumber\\
		&+  n^{-1}[n^2-1]^2(1+2m/r)(2m/r) \hat r^i \hat r^j, \\
		=& \frac{(1-n^{-2}[n^2-1]^2(1+2m/r)(2m/r)) }{ n \left\{1-[1-n^{-2}] (1+2m/r)\right\}} P^{ij}\nonumber \\
		&\quad+ \Big( n  (1-n^{-2}[n^2-1]^2(1+2m/r)(2m/r))  \nonumber\\
		&\qquad + (n^{-1}[n^2-1]^2(1+2m/r)(2m/r)) \Big) \hat r^i \hat r^j,\\
		=& n \{ 1+[1-n^{-2}] (2m/r) \} P^{ij} + n \hat r^i \hat r^j.
	\end{align}
\end{subequations}

This allows us to collect the results for the electromagnetic properties of the medium:
\begin{subequations}
	\begin{align}
		\epsilon_{ij} &=  n \{ 1+[1-n^{-2}] (2m/r) \} P_{ij}  + n \; \hat r_i \hat r_j,\\
		\mu_{\,ij}  &= \frac{P_{ij} }{ n \left\{1-[1-n^{-2}] (1+2m/r)\right\} } + n\; \hat r_i \hat r_j,\\
		\zeta_{ij}  &= -\ed{2}  \left( [n^{2}-1] \sqrt{(1+2m/r)(2m/r)} \right)\epsilon_{ijk} \; \hat r^k .
	\end{align}
\end{subequations}

We close again by making a list of these results' features:
\begin{itemize}
	\item First, note that, looking at equations~(\ref{eq:GordonMuInv},\ref{eq:GordonMu}) we can tell that
	\begin{equation}
		\mu^{-1}_{\,ij}  \stackrel{r\to\infty}{\longrightarrow} n^{-1} \delta_{ij}, \qquad\text{while}\qquad \mu_{ij} \stackrel{r\to\infty}{\longrightarrow} n \delta_{ij}. 
	\end{equation}
	\item Likewise, for $\zeta$ we have
	\begin{equation}
		\zeta^{ij} \stackrel{r \to \infty}{\longrightarrow} 0
	\end{equation}
	from equation~\eqref{eq:GordonZeta}.
	\item For the permittivity's asymptotic behaviour we get that
	\begin{equation}
		\epsilon^{-1}_{\,ij}  \stackrel{r\to\infty}{\longrightarrow} n^{-1} \delta_{ij}, \qquad \text{and}\qquad \epsilon_{ij} \stackrel{r\to\infty}{\longrightarrow} n \; \delta_{ij}. 
	\end{equation}
	\item As before, $\epsilon\geq 1$, as well as $\mu\geq1$.
	\item This time $\mu$ diverges on the horizon while $\epsilon$ does not.
	\item The location of the horizon is calculated by solving
	\begin{equation}
		1-[1-n^{-2}] (1+2m/r)=0,
	\end{equation}
	giving that in terms of the \enquote{parametric mass} $m$
	\begin{equation}
		r_\text{H}=2(n^2-1)m.
	\end{equation}
	\item Also note again, that we here assume a real-valued refractive index. While any other choice would break our formalism, it is noteworthy that astrophysical situations in general relativity do involve dissipation (in the form of backscattering), and thus will at least in some prescriptions necessarily lead to complex-valued refractive indices. More on this will be mentioned briefly in section~\ref{sec:refractive}.
\end{itemize}

\subsubsection{Schwarzschild V: Cartesian Isotropic Form}
Our last example specifically dealing with the Schwarzschild metric concerns itself with the Cartesian isotropic form:
\begin{equation}
	g_{ab} = - \left(\frac{1-{\frac{m}{2r}}}{1+{\frac{m}{2r}}}\right)^2 \dif t^2 + \left(1+{\frac{m}{2r}}\right)^4 |\dif\vec x|^2, \qquad \text{where~} r = |\vec x|. 
\end{equation}
While all of the previous examples had (in Cartesian form!) a determinant of $-1$, this by no means holds true now for the isotropic coordinate form. Nonetheless, this is of little concern as the context we are examining is macroscopic electrodynamics in 3+1 dimensions, which is conformally invariant. We can therefore pull out a conformal factor by rewriting the metric as
\begin{equation}
	g_{ab} =  \sqrt[4]{  \left(1-{\frac{m}{2r}}\right)^2\,\left(1+{\frac{m}{2r}}\right)^{10}} \; 
	\left(
	-\sqrt[4]{\frac{\left(1-{\frac{m}{2r}}\right)^6}{\left(1+{\frac{m}{2r}}\right)^{18}}} \; \dif t^2 
	+ \sqrt[4]{\frac{\left(1+{\frac{m}{2r}}\right)^6}{\left(1-{\frac{m}{2r}}\right)^2}} \; |\dif\vec x|^2
	\right),
\end{equation}
and then making use of the conformal invariance to discard this overall factor:
\begin{subequations}
	\begin{align}
		g_{ab} &= 
		\left(
		-\sqrt[4]{\frac{\left(1-{\frac{m}{2r}}\right)^6}{\left(1+{\frac{m}{2r}}\right)^{18}}} \; \dif t^2 
		+ \sqrt[4]{\frac{\left(1+{\frac{m}{2r}}\right)^6}{\left(1-{\frac{m}{2r}}\right)^2}}\;  |\dif \vec x|^2
		\right), \\
		&=
		- B^{-6} \; \dif t^2 + B^2 \; |\dif\vec x|^2.
	\end{align}
\end{subequations}
This is the form already known and employed for similar purposes to ours since de~Felice's paper \cite{deFeliceTrafoOptics}.

\begin{sloppypar}As these coordinates are, again, not horizon-penetrating, deductions from equation~\eqref{eq:g3+1mu} are fairly simple:\end{sloppypar}
\begin{subequations}
	\begin{align}
		\sqrt{\det(\mu^{\circ\circ})}^{-1} &= B^{-6},\label{eq:IsotropicDet}\\
		\beta^k  &= 0,\\
		B^2 \delta_{ij} &= \sqrt{\det(\mu^{\circ\circ})}\;\mu^{-1}_{\,ij}.
	\end{align}
\end{subequations}
We can immediately calculate and collect all three constitutive matrices in one go:
\begin{subequations}\label{eq:IsotropicResults}
	\begin{align}
		\mu^{-1}_{\,ij} &= B^{-4} \delta_{ij}, &&\text{or} &\mu_{\,ij}  &= B^{+4} \delta_{ij},\label{eq:IsotropicMu}\\
		\epsilon_{\,ij} &= B^4 \delta_{ij},  &&\text{or} & \epsilon^{-1}_{\,ij} &= B^{-4} \delta_{ij},\\
		\zeta_{ij} &= 0.&&&&
	\end{align}
\end{subequations}
Note that equation~\eqref{eq:IsotropicMu} for the permeability trivially fulfils the condition~\eqref{eq:IsotropicDet} on its determinant. More explicitly than in equations~\eqref{eq:IsotropicResults}, the permeability and permittivity evaluate to
\begin{equation}
	\epsilon_{\,ij} = \mu_{\,ij}  =  {\frac{\left(1+{\frac{m}{2r}}\right)^3}{\left|1-{\frac{m}{2r}}\right|}}\;\delta_{ij}.
\end{equation}

After a comparatively short analysis we are already in the position to make comments regarding this example:
\begin{itemize}
	\item As \begin{equation}
		B^4 =  \sqrt{\frac{\left(1+{\frac{m}{2r}}\right)^6}{\left(1-{\frac{m}{2r}}\right)^2}} = {\frac{\left(1+{\frac{m}{2r}}\right)^3}{\left|1-{\frac{m}{2r}}\right|}}> 1,
	\end{equation}
	we have the new result that $\epsilon=\mu>1$, compared to our earlier examples where equality was also possible.
	\item Compared to our earlier results, there is another, more basic and qualitative (mathematical) difference: We actually made use of the conformal freedom of the electromagnetic context within which we were operating. \emph{Internally}, that is, within the effective space-time itself, this is also of little consequence for the effect most analogue space-time experiments are looking for: The Hawking effect. As shown by Jacobson and Kang in \cite{JacobsonKang93}, the Hawking temperature is conformally invariant.\footnote{Slightly related --- especially if the above example of the Gordon--Schwarzschild metric is taken seriously as a mix of electromagnetic and fluid analogue --- is the result of Hossenfelder and Zingg, see \cite{HossenfelderZinggConformal}, that conformal rescalings can in certain fluid analogues be absorbed in field redefinitions.} Thus the Hawking temperature of an analogue black hole will be conformally invariant in the coordinates of the effective space-time. Sadly, any non-conformal relation to the laboratory coordinates will turn this into a veritable mess. This will be discussed in depth below in section~\ref{sec:sparseEM}.
	\item Again, it is useful to take note of the equivalence between (1) coordinates that are \emph{not} horizon-penetrating, (2) Vanishing magneto-electric effects, and (3) a metric with no $0i$ components.\footnote{For appropriately chosen index ranges distinguishing clearly spatial from temporal indices.}
\end{itemize}

\subsubsection{Staticity and Spherical Symmetry}
After a rather extensive list of examples modelled on the Schwarzschild geometry, it is worthwhile to put some thought into the more general case of an arbitrary, static and spherically symmetric (effective) space-time. This is also the easiest way to affirm the argument used in the example of the Cartesian Gordon form above that $\beta$ has to be in the radial direction, as the Schwarzschild space-time is the prime example of this class of space-times.

Just by demanding spherical symmetry and adopting our (quasi-)Cartesian framework, we can deduce that the tensors involved fulfil
\begin{subequations}
	\begin{align}
		\epsilon_{ij} &= \epsilon_\perp P_{ij} + \epsilon_\parallel \; \hat r_i \hat r_j,\\
		\mu_{ij} &= \mu_\perp P_{ij} + \mu_\parallel \; \hat r_i \hat r_j,
	\end{align}
\end{subequations}
while the corresponding determinants are
\begin{subequations}
	\begin{align}
		\det(\epsilon_{ij}) &=  \epsilon_\perp^2\; \epsilon_\parallel,\\
		\det(\mu_{ij}) &=  \mu_\perp^2\; \mu_\parallel.
	\end{align}
\end{subequations}
Spherical symmetry applied to the three-vector $\beta$ then demands that
\begin{equation}
	\beta^i =  \beta \; \hat r^i.
\end{equation}

The consistency conditions~\eqref{eq:conscond3+1zeta}, that is, $\epsilon^{ij} = \mu^{ij}\, (1 - \mu^{-1}_{\,kl} \beta^k \beta^l )  + \beta^i \beta^j$, then tightly constrain the possible constitutive tensors:
For the permittivity and the permeability
\begin{subequations}
	\begin{equation}
		\epsilon_\perp = \mu_\perp \left(1- {\frac{\beta^2}{\mu_\parallel}}\right),
	\end{equation}
	and
	\begin{equation}
		\epsilon_\parallel = \mu_\parallel,
	\end{equation}
\end{subequations}
on the one hand, while on the other hand, for the magneto-electric tensor we have
\begin{subequations}
	\begin{align}
		\zeta^{ij} &= -\ed{2}  \sqrt{\det[\mu^{-1}]}\epsilon^i{}_{kl} \beta^l \mu^{kj},\\
		&= -\ed{2}  {\frac{\beta}{\sqrt{\mu_\parallel}}} \; \epsilon^{ij}{}_k \; \hat r^k .
	\end{align}
\end{subequations}

This can be summarised in terms of the projection operator $P_{ij}$ as
\begin{subequations}
	\begin{align}
		\epsilon_{ij} &=  \mu_\perp \left(1- {\frac{\beta^2}{\mu_\parallel}}\right) P_{ij} + \mu_\parallel \; \hat r_i \hat r_j,\\
		\mu_{ij} &= \mu_\perp P_{ij} + \mu_\parallel \; \hat r_i \hat r_j,\\
		\zeta^{ij} &= -\ed{2}  {\frac{\beta}{\sqrt{\mu_\parallel}}} \; \epsilon^{ij}{}_k \; \hat r^k .
	\end{align}
\end{subequations}
It is useful to note that this is exactly the form taken by our results for the Painlevé--Gullstrand form, the Kerr--Schild form, the Gordon form, and (in a trivial way) even the isotropic form. Similarly, it would be possible to rephrase the previous brute-force result for dirty black hole space-times in these quasi-Cartesian terms.

\subsubsection{Kerr I: Cartesian Kerr--Schild Form}
A metric in Kerr--Schild form is a fairly general class of metric, see \cite{ExactSolsEinstein}, such that
\begin{equation}
	g_{ab}  = \eta_{ab} + 2\Phi \ell_a \ell_b,   \qquad \ell_a = (-1,\ell_i),  \qquad ||\ell_i||=1,
	\qquad \det(g_{ab})=-1.
\end{equation}
Here, $\Phi$ is a scalar function and $\ell_i$ a three-vector both to be specified depending on the concrete space-time under examination. We adapted the more general formalism found in the literature to fit into our (quasi-)Cartesian framework. The general formalism would only demand $\ell$ to be null, while we have to add more restrictions. It is important to note that the four-form $\ell_a$ in this class of metrics is null w.r.t. both the flat Minkowski metric $\eta_{ab}$ appearing here, as well as the full metric $g_{ab}$ under consideration. The inverse to this metric is just
\begin{equation}
	g^{ab}  = \eta^{ab} - 2\Phi \ell^a \ell^b.
\end{equation}
To show that (a) this indeed is the inverse, and (b) $\ell$ is null for both metrics, one proceeds in the following way: First, one shows that indeed, $\ell$ is null for $\eta$. Then, one raises the indices of $g$ with $\eta$ and shows that this produces indeed an inverse to $g$. It then remains to be shown, that the same result could have been obtained by raising the indices with $g$. For this it is enough to check that the inverse of $g$ as a formal power series around $\eta$ in terms of $\ell$ will terminate after the first term due to the former argument; the uniqueness of the inverse then provides the rest of the argument.

The Kerr metric is one metric capable of taking the Kerr--Schild form, see \cite{ExactSolsEinstein,KerrFest}. In this case, the one-form $\ell_a$ will be the Kerr null congruence, $\Phi$ is simply the gravitational potential. Explicitly, the parameters of the Kerr--Schild form for the Kerr metric itself are given by
\begin{equation}
	\Phi = \frac{mr^3}{ r^4+a^2 z^2} = \frac{m}{ r(1+a^2 z^2/r^4)},
\end{equation}
and
\begin{equation}
	\ell_a = \left(1 ,  \frac{rx  + ay  }{ a^2 + r^2} +\frac{ry-ax }{ a^2 + r^2} , {\frac{z}{r}} \right),
\end{equation}
while the function $r(x,y,z)$ is only implicitly\footnote{Strictly speaking, a quartic \emph{could} still be solved exactly. Whether the result is helpful in any meaningful way is a very different matter.} defined through the equation
\begin{equation}
	x^2+y^2+z^2 = r^2 + a^2\left[1-{\frac{z^2}{r^2}}\right].\label{eq:r-KS}
\end{equation}
It is important to remember that $r$ plays the role of a function depending on the quasi-Cartesian coordinates $x,y,z$ --- it is not to be understood as a coordinate in its own right, as in equation~\eqref{eq:Kerr}.

To proceed, we only take two points of data immediately from the decomposition~\eqref{eq:g3+1mu} of $g$ in terms of $\mu$:
\begin{subequations}
	\begin{align}
		\sqrt{\det(\mu^{\circ\circ})}^{-1} &= 1-2\Phi,\\
		\mu^{-1}_{\,jk}\beta^k  = 2\Phi \ell_j.
	\end{align}
\end{subequations}
Writing the spatial part of the metric as
\begin{equation}
	\left(\delta_{ij} +2\Phi \ell_i \ell_j \right) 
	=
	\sqrt{\det(\mu^{\circ\circ})} (\mu^{-1}_{\,ij} - 4\Phi^2 \ell_i \ell_j),
\end{equation}
we can then solve for $\mu^{-1}$ and get that
\begin{subequations}
	\begin{align}
		\mu^{-1}_{\,ij}  &= (1-2\Phi)(\delta_{ij} +2\Phi \ell_i \ell_j )  +4\Phi^2 \ell_i \ell_j,\\
		&=(1-2\Phi)(\delta_{ij}) - 2\Phi \ell_i \ell_j.
	\end{align}
\end{subequations}

Unlike the spherically symmetric case, we have now to be much more careful in choosing our projection operator in order to perform an analysis similar to the ones seen before. For this purpose, let us define
\begin{equation}
	P_{ij} \defi \delta_{ij} - \ell_i \ell_j.
\end{equation}
We can now express both $\mu$ and $\mu^{-1}$ in terms of this new projection operator, 
\begin{subequations}
	\begin{align}
		\mu^{-1}_{\,ij}  &= (1-2\Phi)P_{ij} +\ell_i \ell_j,\\
		\mu_{\,ij}  &= (1-2\Phi)^{-1} P_{ij} + \ell_i \ell_j = \frac{\delta_{ij} - 2\Phi \ell_i\ell_j}{ 1-2\Phi},
	\end{align}
\end{subequations}
which also allows us to read off $\beta$ as
\begin{equation}
	\beta^i =2 \Phi \ell^i.
\end{equation}
This gives the determinant of $\mu^{-1}$ as $(1-2\Phi)^2$, in accordance with the condition previously derived from equation~\eqref{eq:g3+1mu}. Thus, we are in a position to calculate the permittivity:
\begin{subequations}
	\begin{align}
		\epsilon^{ij} =& \mu^{ij}\, (1 - \mu^{-1}_{\,kl} \beta^k \beta^l )  + \beta^i \beta^j,\\
		=&( (1-2\Phi)^{-1} P_{ij} + \ell_i \ell_j) (1-4\Phi^2)+ 4\Phi^2 \ell_i \ell_j,\\
		=& P_{ij} (1+2\Phi) +  \ell_i \ell_j,
	\end{align}
\end{subequations}
while its determinant
\begin{equation}
	\det(\epsilon^{ij}) = (1+2\Phi)^2,
\end{equation}
and its inverse fulfils
\begin{subequations}
	\begin{align}
		[\epsilon^{-1}]_{ij} &= \frac{P_{ij} }{ (1+2\Phi)} +  \ell_i \ell_j,\\
		& = \frac{\delta_{ij}+ 2\Phi \ell_i \ell_j}{ 1+2\Phi}.
	\end{align}
\end{subequations}

For the magneto-electric tensor we now only have to do a reasonably straightforward calculation:
\begin{subequations}
	\begin{align}
		\zeta^{ij} &= -\ed{2}\;
		\left( \frac{ \epsilon^{i}{}_{kl}\beta^l  \mu^{kj} }{ \sqrt{\det(\mu^{\circ\circ})}} \right),\\
		&= -\ed{2}\;\left(  [1-2\Phi] \epsilon^{i}{}_{kl} [2\Phi\ell^l] \left[\frac{\delta^{kj} - 2\Phi \ell^k\ell^j}{ 1-2\Phi}\right]\right),\\
		&= - \Phi \epsilon^{ij}{}_{l} \ell^l.
	\end{align}
\end{subequations}
As before, even though not explicitly mentioned, all analyses equally well could have started from the decomposition~\eqref{eq:g3+1eps} of $g$ in terms of $\epsilon$, and then making a similar derivation of $\mu$. Either case will result in, basically, proving the consistency conditions on a case-by-case basis.

Let us summarise our results:
\begin{subequations}
	\begin{align}
		\epsilon^{ij} &=  (1+2\Phi) P^{ij} +  \ell^i\ell^j,\\
		\mu^{ij} &= \frac{P^{ij} }{ 1-2\Phi} + \ell^i\; \ell^j,\\
		\zeta^{ij} &= - \Phi \epsilon^{ij}{}_{l} \ell^l.
	\end{align}
\end{subequations}

Observations for this example are:
\begin{itemize}
	\item It bears repeating: This analysis is more general than the name (and the application of our immediate concern) suggests.
	\item The inverse permittivity is related to the three-metric $g_{ij}$ in the simple way
	\begin{equation}
		[\epsilon^{-1}]_{ij} = \frac{g_{ij}}{ \det(g_{ij})}.
	\end{equation}
	\item As under usual conditions the \enquote{gravitational potential} $\Phi$ will be $\geq 0$, the eigenvalues of permittivity and permeability will be $\geq1$. 
	\item When $g_{00}=0$, or equivalently, $2\Phi = 1$, two of the eigenvalues of $\mu$ will diverge. As $g_{00}$ defines the ergo-surface, this means that compared to the non-rotating black hole described by the Schwarzschild solution, in the Kerr metric the divergences of the optical properties move from the horizon to the ergo-surface. As ergo-surface and horizon agree in the non-rotating case, this difference could not be observed there.
	\item As the ergo-region is closely related to super-radiance \cite{SuperradianceReview,LNPSuperradiance}, the divergence of the optical properties should not come as a surprise.
	\item A similar, intuitive reason for these divergences is that a laboratory observer will be \enquote{superluminal} (only w.r.t the effective metric) behind the ergo-surface.
\end{itemize}

\subsubsection{Kerr II: Cartesian Doran Form}
The last example of our quasi-Cartesian method of calculating the material properties of media mimicking a given effective space-time shall be concerned with metrics which can be described by the Doran form. More on this form than presented here can be found, for example, in \cite{Doran,RiverModelBH,KerrFest}. The Doran form reads
\begin{subequations}
	\begin{align}
		g_{ab} &= \eta_{ab} + F^2 V_a V_b + F(V_a S_b + S_a V_b),\\
		&= \eta_{cd} (\delta^c{}_a+ FS^c V_a) (\delta^d{}_b + FS^d V_b).
	\end{align}
\end{subequations}
Here, $V$ and $S$ are with respect to the background Minkowski metric $\eta$ four-orthogonal, time-like and space-like unit vectors, respectively. $S^0=0$, and hence the three-vector $S^i$ is of unit norm with respect to our quasi-Cartesian three-metric. Likewise, this results in the spatial parts of $S$ and $V$ to be three-orthogonal w.r.t. this quasi-Cartesian three-metric. From the four-orthogonality we can furthermore deduce that
\begin{equation}
	\det(\delta^a{}_{b} + FS^a V_b) = 1,
\end{equation}
and therefore also
\begin{equation}
	\det(g_{ab})=-1.
\end{equation}
We can additionally see that the requirement of $S^0=0$ also implies a vanishing lapse (as specified in an ADM decomposition). A vanishing lapse is a general feature of the Doran form. 

Again, the Kerr metric is one particular and especially important representative of this class of metrics. For the Kerr metric the following three identifications are needed:
\begin{subequations}
	\begin{align}
		F &= \sqrt{\frac{2mr}{ r^2+ a^2}},\\
		V_a &= \sqrt{\frac{r^2(r^2+a^2)}{ r^4+a^2 z^2}} \left( 1, \frac{ay}{ r^2+a^2}, \frac{-ax}{ r^2+a^2}, 0\right),\\
		S_a &= \sqrt{\frac{r^2(r^2+a^2)}{ r^4+a^2 z^2}} \left( 0, \frac{r x}{ r^2+a^2}, \frac{r y}{ r^2+a^2}, {\frac{z}{r}}\right).
	\end{align}
\end{subequations}
The quantity $r(x,y,z)$ is defined just as in the previous example of the quasi-Cartesian Kerr--Schild form, equation~\eqref{eq:r-KS}. For ease of reading, we again give this equation determining $r$ implicitly:
\begin{equation}
	x^2+y^2+z^2 = r^2 + a^2\left[1-{\frac{z^2}{r^2}}\right].
\end{equation}
It is worth mentioning that the Doran form does have another link to the Kerr--Schild form, as
\begin{equation}
	F \; (V_a + S_a ) = \sqrt{\Phi} \; \ell_a,
\end{equation}
that is, the two one-forms as defined above add together to something proportional to the Kerr null congruence.

Let us return back to the general case: Since we can see that
\begin{equation}
	(\delta^a{}_d - F  S^a V_d ) (\delta^d{}_b + FS^d V_b) = \delta^a{}_b,
\end{equation}
we can also invert the metric to:
\begin{subequations}
	\begin{align}
		[g^{-1}]^{ab} &= \eta^{cd} (\delta^a{}_c - F  S^a V_c )(\delta^b{}_d - F  S^b V_d ),\\
		&= \eta^{ab} - F^2 S^a S^b - F( S^a V^b + V^a S^b).
	\end{align}
\end{subequations}

This then can be turned into statements about the three-projections of both metric and inverse metric. However, as $\beta$ is non-vanishing, it is important to note that these projections will not be inverses of each other. They are
\begin{subequations}
	\begin{align}
		g_{ij} &= \delta_{ij} + F^2 V_i V_j + F[V_i S_j+S_iV_j],\\
		[g^{-1}]^{ij} &= \delta^{ij} - F^2 S^i S^j - F[V^i S^j+S^iV^j].
	\end{align}
\end{subequations}
By defining $V\defi \abs{V^i}$, we can write $V^i$ as $\abs{V} \hat{V}^i$, and this in turn allows us to write $\beta$ as
\begin{equation}
	\beta^i = [g^{-1}]^{0i} = - F V^0 S^i = - F \sqrt{1+V^2} S^i.
\end{equation}

For the resulting $3\times3$ matrices the following similarity relations hold (under an appropriate orthogonal transformation):
\begin{subequations}
	\begin{align}
		g_{ij} &\sim \left(\begin{array}{ccc}1+F^2 V^2 & FV & 0\\ FV & 1 & 0\\0&0&1\end{array}\right),\\
		[g^{-1}]^{ij} &\sim \left(\begin{array}{ccc}1 & -FV & 0\\ -FV & 1-F^2 & 0\\0&0&1\end{array}\right).
	\end{align}
\end{subequations}
As determinants are unchanged under similarity transformations, this gives us that
\begin{subequations}
	\begin{align}
		\det(g_{ij}) &= 1,\\
		\det([g^{-1}]^{ij}) &= 1-F^2(1+V^2).
	\end{align}
\end{subequations}

The $3\times3$ susceptibility tensors can then be calculated by looking at
\begin{subequations}
	\begin{align}
		\mu^{ij} &=  \frac{ [g^{-1}]^{ij} }{ \det([g^{-1}]^{\circ\circ})},\\
		[\epsilon^{-1}]_{ij}  &=\frac{g_{ij}}{ \det(g_{\circ\circ})},\\
		\zeta^{ij} &= -\ed{2}\; \left(  \epsilon^{i}{}_{kl}[g^{-1}]^{0l}  [g^{-1}]^{kj}  \right).
	\end{align}
\end{subequations}
This implies for the inverse of the permittivity
\begin{equation}
	[\epsilon^{-1}]_{ij}  =  \delta_{ij} + F^2 V_i V_j + F(V_i S_j+S_iV_j).
\end{equation}
But since we are interested particularly in the permittivity itself, we need to invert this identity. For this we again make use of (orthogonal) similarity transformations to note that
\begin{equation}
	[\epsilon^{-1}]_{ij}  \sim \left(\begin{array}{ccc}1+F^2 V^2 & FV & 0\\ FV & 1 & 0\\0&0&1\end{array}\right),
\end{equation}
and hence
\begin{equation}
	\epsilon^{ij}  \sim \left(\begin{array}{ccc}1 & -FV & 0\\ -FV & 1+F^2V^2 & 0\\0&0&1\end{array}\right).\label{eq:CartDoranepsOrth}
\end{equation}
We see that $\det(\epsilon_{ij})=1$. Keeping careful track of the orthogonal transformation, we can then deduce that
\begin{equation}
	\epsilon^{ij}  = \delta^{ij} - F(V^iS^j+S^iV^j) + F^2 V^2 S^iS^j.
\end{equation}
The permeability $\mu$ can be found by simply inserting the results for $\epsilon$ and $\epsilon^{-1}$ into the consistency condition solved for $\mu$ as in equation~\eqref{eq:conscond3+1mu}:
\begin{equation}
	\mu^{ij} = \frac{ \delta^{ij} - F^2 S^i S^j - F(V^i S^j+S^iV^j)}{1-F^2(1+V^2)}.
\end{equation}
It remains to calculate the magneto-electric tensor:
\begin{subequations}
	\begin{align}
		\zeta^{ij} &= -\ed{2}\; \left(  \epsilon^{i}{}_{kl}[g^{-1}]^{0l}  [g^{-1}]^{kj}  \right),\\
		&=+ \ed{2}\; \left(  \epsilon^{i}{}_{kl} F V^0 S^l  \left[\delta^{kj} + F^2 S^k S^j - F(V^k S^j+S^k V^j)\right]  \right).
	\end{align}
	As $ \epsilon^{i}{}_{kl} S^k S^l = 0$, this can be further simplified to
	\begin{align}
		\zeta^{ij} &= \frac{FV^0}{2}\; \left(  \epsilon^{i}{}_{kl} S^l  \left[\delta^{kj}  - FV^k S^j\right]  \right),\\
		&= {\frac{F\sqrt{1+V^2}}{2}}\; \left(  \epsilon^{ij}{}_{l} S^l  - FV [\epsilon^{i}{}_{kl} \hat V^k S^l] S^j \right).
	\end{align}
\end{subequations}

We thus can summarise the situation for an arbitrary metric in Doran form with the following three constitutive matrices:
\begin{subequations}
	\begin{align}
		\epsilon^{ij}  &= \delta^{ij} - F(V^iS^j+S^iV^j) + F^2 V^2 S^iS^j,\\
		\mu^{ij} &= \frac{ \delta^{ij} + F^2 S^i S^j - F[V^i S^j+S^iV^j]}{ 1+F^2(1-V^2)},\\
		\zeta^{ij} &= {\frac{F\sqrt{1+V^2}}{2}}\; \left(  \epsilon^{ij}{}_{l} S^l  - FV [\epsilon^{i}{}_{kl} \hat V^k S^l] S^j \right).
	\end{align}
\end{subequations}

To conclude, we note the following observations:
\begin{itemize}
	\item Looking at equation~\eqref{eq:CartDoranepsOrth}, we see that the eigenvalues of $\epsilon_{\,ij}$ will be purely real and positive.
	\item From this it also follows that the permittivity will have one eigenvalue $1$, one eigenvalue $\geq 1$, but one eigenvalue between $0$ and $1$. This will make the production of an appropriate (meta-)material slightly more complicated than those of the previous examples.
	\item Making use of yet another orthogonal transformation, we can rephrase $\mu$ in a way that allows us to get a closer look at its eigenvalues, as
	\begin{equation}
		\mu^{ij} \sim \ed{1-F^2(1+V^2)} \left(\begin{array}{ccc}1 & -FV & 0\\ -FV & 1-F^2 & 0\\0&0&1\end{array}\right).
	\end{equation}
	We observe that the nature of the eigenvalues now depends on the sign of $1-F^2(1+V^2)$: $\mu$ either has three positive, or two positive and one negative eigenvalues.
	\item A closer look at $g_{00}$ reveals that
	\begin{subequations}
		\begin{align}
			g_{00} &= \eta_{00} + F^2 V_0 V_0, \\&= -1 + F^2(1+V^2),\\ &= -\{1-F^2(1+V^2)\}.
		\end{align}
	\end{subequations}
	This means that $1-F^2(1+V^2)$ flips sign at the ergo-surface as it goes through zero. This corresponds to the divergence of two of the eigenvalues of $\mu$. Note that this situation is similar to what we observed in the previous example of the Kerr--Schild form and its ergo-surface.
\end{itemize}

\subsection{Curvilinear Coordinates} \label{sec:curvilinear}
The next step in adding complications is to consider curvilinear coordinates, both on the background as well as on the effective space-time. We shall, however, not change our assumption of a flat background. This time, we will restrict ourselves to three large, but still special cases regarding the analogue space-time:
\begin{itemize}
	\item Space-times which can be fit in Boyer--Lindquist coordinates.
	\item A second look at the Kaluza--Klein decomposition: This might seem no loss in generality, until one remembers our flat background. The background will present only a mild restriction on possible effective space-times, but still effecting helpful simplifications compared to the analysis of section~\ref{sec:conscond}.
	\item Static, spherically symmetric space-times in spherical polar coordinates, which is an extension of the previous analysis in quasi-Cartesian coordinates.
\end{itemize}

While in principle only of immediate relevance to the first and third example, we shall use this introductory part as an opportunity to set up our coordinates, and distinguish between laboratory and effective space-time coordinates. In a sense, especially within the context of the third example of static, spherically symmetric (effective) space-times, much of the explicit analysis to be presented could be labelled \enquote{quasi-spherical}, in analogy to the \enquote{quasi-Cartesian} coordinates of the previous subsection.

The laboratory coordinates are, again, assumed to correspond identically to the coordinates in the effective space-time, and will be named $(t,r,\theta,\phi) \ifed (t,\xi^i)$, where they are made explicit. While the naming is reminiscent of spherical polar coordinates (and intended to be), there would be plenty of coordinate systems more one might use like this: Oblate spheroidal, prolate spheroidal, cylindrical coordinates (where it would be customary to relabel $\theta\to z$), parabolic cylindrical, paraboloidal, elliptic cylindrical, ellipsoidal, bipolar, toroidal, conical, or general orthogonal coordinates. Their traditional labelling usually is different from that of spherical polar coordinates, but it is not uncommon, either, to copy the naming of spherical polar coordinates to other coordinate systems, as seen, for example, in many of standard coordinate systems for the Kerr geometry (and its closely related cousins).

Likewise, as we will always assume a flat background, we shall again label the metric as $\eta$. However, as the coordinates are now allowed to be curvilinear, the associated metric will not be unimodular anymore. Instead, we have:
\begin{equation}
	\eta_{ab} = \left(\begin{array}{c|c} 
	-1 & 0\\
	\hline
	0 & \eta_{ij}(\xi)\\
	\end{array}\right),  \qquad \det(\eta_{\bullet\bullet}) = - \det(\eta_{\circ\circ}).
\end{equation}
Note that we still restrict the time-component of the coordinates. We will keep raising and lowering indices with \emph{this} flat metric. The complication compared to our previous quasi-Cartesian examples lies in the fact that now the spatial indices are not simply raised and lowered with a Kronecker-$\delta$ any more, instead additional metric components come in.

Looking back at the definition of the constitutive tensor, equation~\eqref{eq:Zeff}, and adjusting to the current circumstances, we have
\begin{equation}
	Z^{abcd} = \ed{2}\frac{\sqrt{\det \g}}{\sqrt{\det  \eta}} \kl{\gi^{ac}\gi^{bd} - \gi^{ad}\gi^{bc}}.\label{eq:defZcurvi}
\end{equation}
This will be the starting point of the construction of our analogues in this subsection.

\subsubsection{Boyer--Lindquist Space-Times}
Physically interesting, \emph{general} stationary axisymmetric space-times can be written in Boyer--Lindquist form, if the space-time is circular.\footnote{For the technical definition of circularity, we refer to \cite{Heusler}, as this would lead to far outside of our line of inquiry. Put somewhat handwavingly, it is a statement about the integrability of the angular Killing vector field (present due to axisymmetry) and the temporal Killing vector field (present due to stationarity). This links commutation properties of the two Killing vector fields, the Frobenius theorem on their integrability, and properties of the Lorentzian manifold itself.}\textsuperscript{,}\footnote{From a more physical point of reasoning, one wants the replacement of both $\phi\to -\phi$ and $t\to -t$ to leave the metric invariant. This corresponds to the demand that the combination of both does not change the direction of rotation, while only one would.} In quasi-spherical coordinates the Boyer--Lindquist form is equivalent to stating that the metric can be given as
\begin{equation}
	g^\text{eff}_{ab} = \left(\begin{array}{c|cc|c} 
	g^\text{eff}_{tt}  & 0 & 0& g^\text{eff}_{t\phi}\\
	\hline
	0 & g^\text{eff}_{rr} &0 &0\\
	0&0&g^\text{eff}_{\theta\theta}&0\\
	\hline
	g^\text{eff}_{t\phi}&0&0&g^\text{eff}_{\phi\phi}
	\end{array}\right).\label{eq:BoyerLindquist}
\end{equation}
Note that we already adapted to a notation highlighting our application of the Boyer--Lindquist form: We are not interested in changing the background metric $\eta$ to it, rather we are only interested in the effective space-time to be in this form. However, we will omit the adjective \enquote{effective} most of the time in the text to reduce clutter, when a confusion with the background space-time's metric $\eta$ is ruled out by context.

The inverse metric for this form is then
\begin{subequations}
\begin{align}
	\gi^{ab} &= 
	\left(\begin{array}{c|cc|c} 
	g^\text{eff}_{\phi\phi}/ g_2  & 0 & 0& - g^\text{eff}_{t\phi} /g_2 \\
	\hline
	0 & 1/g^\text{eff}_{rr} &0 &0\\
	0&0&1/g^\text{eff}_{\theta\theta}&0\\
	\hline
	-g^\text{eff}_{t\phi} /g_2 &0&0&g^\text{eff}_{tt} /g_2
	\end{array}\right),\\
	&=
	\left(\begin{array}{c|c} 
	\gi^{00} & \gi^{0j}  \\
	\hline
	\gi^{0i}  & \gi^{ij}\\
	\end{array}\right),
\end{align}
where we defined the following shorthand for the subdeterminant involving temporal parts of the metric,
\begin{equation}
	g_2 \defi g^\text{eff}_{tt} \,g^\text{eff}_{\phi\phi}-\kl{g^\text{eff}_{t\phi}}^2.	
\end{equation}
\end{subequations}
This last definition also allows us to write the determinant of the metric as $\det(g^\text{eff}_{ab})= g_2\,g^\text{eff}_{rr}\,g^\text{eff}_{\theta\theta}$. From this expression for the determinant, and by looking at the original metric, as in equation~\eqref{eq:BoyerLindquist}, we can infer that the horizon will be at $g_2 = 0$, while the ergo-region is at $g^\text{eff}_{tt}=0$. $g_2=0$ coincides and therefore is equivalent to $g^\text{eff}_{rr} = \infty$.

This allows us to calculate the permittivity:
\begin{subequations}
	\begin{align}
		\epsilon^{ij} &=  -\sqrt{\frac{\det(g^\text{eff})}{\det(\eta)}} \; 
		\left( \gi^{ij}  \gi^{00}- \gi^{0i}  \gi^{0j}  \right),
		\\
		&= -\sqrt{\frac{\det(g^\text{eff})}{\det(\eta)}} \left(\begin{array}{cc|c} 
		g^\text{eff}_{\phi\phi}/ (g_2 g^\text{eff}_{rr}) &0 &0\\
		0&g_{\phi\phi}/ (g_2 g^\text{eff}_{\theta\theta})&0\\
		\hline
		0&0&(g^\text{eff}_{\phi\phi}/ g_2) (g^\text{eff}_{tt} /g_2) - (g^\text{eff}_{t\phi} /g_2)^2
		\end{array}\right)^{ij},\\
		&= -\sqrt{\frac{\det(g^\text{eff})}{\det(\eta)}} \; {\frac{1}{g_2}} \; \left(\begin{array}{cc|c} 
		g^\text{eff}_{\phi\phi}/ g^\text{eff}_{rr} &0 &0\\
		0&g^\text{eff}_{\phi\phi}/ g^\text{eff}_{\theta\theta}&0\\
		\hline
		0&0&1
		\end{array}\right)^{ij}.
	\end{align}
\end{subequations}
This far and compared to the previous (quasi-)Cartesian analysis, there have been only minor changes by inserting the Boyer--Lindquist form in the definition of the permittivity matrix in terms of the constitutive tensor, equation~\eqref{eq:defconstmatrices}. Physical insight now allows us to go further: As we are only interested in the domain of outer communication\footnote{Mathematically speaking, one could look for the analogue for the space-time behind an event horizon. However, the nature of the event horizon would make any analogy to an astrophysical black whole flimsy at best.}, we now make use of the fact that there (that is, outside the horizon) both $\det\g$ and $g_2$ are negative. $\det\eta$, on the other hand, is everywhere negative. We then have the final result:
\begin{equation}
	\epsilon^{ij} = \sqrt{\frac{g^\text{eff}_{rr}g^\text{eff}_{\theta\theta}}{g_2 \det(\eta)}} 
	\left(\begin{array}{cc|c} 
		g^\text{eff}_{\phi\phi}/ g^\text{eff}_{rr} &0 &0\\
		0&g^\text{eff}_{\phi\phi}/ g^\text{eff}_{\theta\theta}&0\\
		\hline
		0&0&1
	\end{array}\right)^{ij}.
\end{equation}

For the permeability, a similar start using the constitutive tensor~\eqref{eq:defZcurvi} requires more thought:
\begin{subequations}
	\begin{align}
		[\mu^{-1}]^{ij} &=  \ed{2}\;\eps^i{}_{kl}\,\eps^j{}_{mn}\,Z^{klmn},
		\\
		&=\ed{4}\;\eps^i{}_{kl}\,\eps^j{}_{mn} \sqrt{\frac{\det(g^\text{eff})}{\det(\eta)}} \; 
		\left(\gi^{km}  \gi^{ln} - \gi^{kn} \gi^{lm} \right),
		\\
		&=\ed{2}\;\eps^i{}_{kl}\,\eps^j{}_{mn} \sqrt{\frac{\det(g^\text{eff})}{\det(\eta)}} \; 
		\left(\gi^{km}  \gi^{ln} \right).
	\end{align}
\end{subequations}
At this point, it becomes helpful to rephrase results in terms of the original Levi-Civita (pseudo-)tensor \emph{density},
\begin{equation}
	\tilde{\eps}_{ijk} = \mathrm{signum}\kl{ijk},
\end{equation}
which relates to the Levi-Civita (pseudo-)tensors appearing in the above expressions in the following way
\begin{subequations}
	\begin{align}
		\eps^i{}_{kl}  &= \eta^{ip} \sqrt{\det(\eta_{ij})} \; \tilde{\eps}_{ijk}, \\
		&=  \eta^{ip} \sqrt{-\det(\eta_{ab})} \; \tilde{\eps}_{ijk}, \\
		&=  \eta^{ip} \sqrt{-\det(\eta)} \; \tilde{\eps}_{ijk}.
	\end{align}
\end{subequations}
This allows us to rewrite the permeability in terms of the Levi-Civita (pseudo-)tensor density:
\begin{equation}
	[\mu^{-1}]^{ij} =\ed{2}\; \eta^{ip} \eta^{jq}  [-\det(\eta)] \sqrt{\frac{\det(g^\text{eff})}{\det(\eta)}} \; \tilde{\eps}_p{}_{kl}\,\tilde{\eps}_q{}_{mn} 
	\left(\gi^{km}  \gi^{ln} \right).
\end{equation}
Furthermore, since
\begin{align}
	\tilde{\eps}_r{}_{kl}\,\tilde{\eps}_r{}_{mn} 
	\left(\gi^{km}  \gi^{ln} \right) &= 2 \gi^{\theta\theta}  \gi^{\phi\phi} &&= 2 g^\text{eff}_{tt}/(g^\text{eff}_{\theta\theta} g_2),\\
	\tilde{\eps}_\theta{}_{kl}\,\tilde{\eps}_\theta{}_{mn} 
	\left(\gi^{km}  \gi^{ln} \right) &= 2 \gi^{rr}  \gi^{\phi\phi} &&= 2 g^\text{eff}_{tt}/(g^\text{eff}_{rr} g_2),\\
	\tilde{\eps}_\phi{}_{kl}\,\tilde{\eps}_\phi{}_{mn} 
	\left(\gi^{km}  \gi^{ln} \right) &= 2 \gi^{rr}  \gi^{\theta\theta} &&= 2 /(g^\text{eff}_{rr} g^\text{eff}_{\theta\theta}),
\end{align}
we know from $\gi^{ij}$ being diagonal that also $\tilde{\eps}_p{}_{kl}\,\tilde{\eps}_q{}_{mn} \left(\gi^{km}  \gi^{ln} \right)$ will be diagonal (as 3-matrices!). For the permeability we thus have
\begin{equation}
	[\mu^{-1}]^{ij} = \eta^{ip} \eta^{jq}  \sqrt{\det(g^\text{eff}) \det(\eta)} 
	\left(\begin{array}{cc|c} 
	g^\text{eff}_{tt}/(g^\text{eff}_{\theta\theta} g_2)&  0 &0\\
	0&g^\text{eff}_{tt}/(g^\text{eff}_{rr} g_2) & 0\\
	\hline
	0&0&1/(g^\text{eff}_{rr} g^\text{eff}_{\theta\theta})
	\end{array}\right)_{pq}.
\end{equation}
The matrix inversion of this is
\begin{equation}
	\mu_{ij} = \eta_{ip} \eta_{jq}  \ed{\sqrt{\det(g^\text{eff}) \det(\eta)}}
	\left(\begin{array}{cc|c} 
	(g_{\theta\theta} g_2)/g^\text{eff}_{tt}&  0 &0\\
	0&(g^\text{eff}_{rr} g_2)/g^\text{eff}_{tt} & 0\\
	\hline
	0&0&g^\text{eff}_{rr} g^\text{eff}_{\theta\theta}
	\end{array}\right)^{pq},
\end{equation}
which reads with raised indices
\begin{subequations}
	\begin{align}
		\mu^{ij} &=   \ed{\sqrt{\det(g^\text{eff}) \det(\eta)}}
		\left(\begin{array}{cc|c} 
		(g^\text{eff}_{\theta\theta} g_2)/g^\text{eff}_{tt}&  0 &0\\
		0&(g^\text{eff}_{rr} g_2)/g^\text{eff}_{tt} & 0\\
		\hline
		0&0&g^\text{eff}_{rr} g^\text{eff}_{\theta\theta}
		\end{array}\right)^{ij},\\
		&=   \frac{g^\text{eff}_{rr} g^\text{eff}_{\theta\theta}}{\sqrt{\det(g^\text{eff}) \det(\eta)}}
		\left(\begin{array}{cc|c} 
		g_2/(g^\text{eff}_{tt} g^\text{eff}_{rr})&  0 &0\\
		0&g_2/(g^\text{eff}_{tt} g^\text{eff}_{\theta\theta}) & 0\\
		\hline
		0&0&1
		\end{array}\right)^{ij},\\
		&=   \sqrt{\frac{g^\text{eff}_{rr} g^\text{eff}_{\theta\theta}}{g_2 \det(\eta)}}
		\left(\begin{array}{cc|c} 
		g_2/(g^\text{eff}_{tt} g^\text{eff}_{rr})&  0 &0\\
		0&g_2/(g^\text{eff}_{tt} g^\text{eff}_{\theta\theta}) & 0\\
		\hline
		0&0&1
		\end{array}\right)^{ij}.
	\end{align}
\end{subequations}

Finally, for the magneto-electric matrix, 
\begin{equation}
	\zeta^{ij} = -\ed{2}\; \sqrt{\frac{\det(g^\text{eff})}{\det(\eta)}} \;  \left(  \eps^{i}{}_{kl}\gi^{0l}  \gi^{kj}  \right),
\end{equation}
it proves useful to lower an index, and then to make use of the explicit expressions for $\gi^{0l}$ and $\eps_{ik\phi}$ (in that order) to have
\begin{subequations}
	\begin{align}
		\zeta_{\,i}{}^j &= -\ed{2}\; \sqrt{\frac{\det(g^\text{eff})}{\det(\eta)}} \;  \left(  \eps_{i}{}_{kl}\gi^{0l} \gi^{kj}  \right),\\
		&= \ed{2}\; (g^\text{eff}_{t\phi} /g_2) \sqrt{\frac{\det(g^\text{eff})}{\det(\eta)}} \;  \left(  \eps_{i}{}_{k\phi} \gi^{kj}  \right),\\
		&= \ed{2}\; (g^\text{eff}_{t\phi} /g_2) \sqrt{\frac{\det(g^\text{eff})}{\det(\eta)}} \; \sqrt{-\det(\eta)}
		\left(\begin{array}{cc|c} 
		0 & 1/g^\text{eff}_{\theta\theta} &0\\
		-1/g^\text{eff}_{rr}&0&0\\
		\hline
		0&0&0
		\end{array}\right)_i{}^{\;j},\\
		&= \ed{2}\; g^\text{eff}_{t\phi} \sqrt{\frac{- g^\text{eff}_{rr} g^\text{eff}_{\theta\theta}}{g_2}} 
		\left(\begin{array}{cc|c} 
		0 & 1/g^\text{eff}_{\theta\theta} &0\\
		-1/g^\text{eff}_{rr}&0&0\\
		\hline
		0&0&0
		\end{array}\right)_i{}^{\,j}.
	\end{align}
\end{subequations}

To summarise, we have for the general Boyer--Lindquist form the following constitutive tensors:
\begin{subequations}\label{eq:BLsummary}
	\begin{align}
		\epsilon^{ij} &= \sqrt{\frac{g^\text{eff}_{rr}g^\text{eff}_{\theta\theta}}{g_2 \det(\eta)}} 
		\left(\begin{array}{cc|c} 
		g^\text{eff}_{\phi\phi}/ g^\text{eff}_{rr} &0 &0\\
		0&g^\text{eff}_{\phi\phi}/ g^\text{eff}_{\theta\theta}&0\\
		\hline
		0&0&1
		\end{array}^{ij}\right),\\
		\mu^{ij} &=   \sqrt{\frac{g^\text{eff}_{rr} g^\text{eff}_{\theta\theta}}{g_2 \det(\eta)}}
		\left(\begin{array}{cc|c} 
		g_2/(g^\text{eff}_{tt} g^\text{eff}_{rr})&  0 &0\\
		0&g_2/(g^\text{eff}_{tt} g^\text{eff}_{\theta\theta}) & 0\\
		\hline
		0&0&1
		\end{array}\right)^{ij},\\
		\zeta_{\,i}{}^j&= \ed{2}\; g^\text{eff}_{t\phi} \sqrt{\frac{- g^\text{eff}_{rr} g^\text{eff}_{\theta\theta}}{g_2}} 
		\left(\begin{array}{cc|c} 
		0 & 1/g^\text{eff}_{\theta\theta} &0\\
		-1/g^\text{eff}_{rr}&0&0\\
		\hline
		0&0&0
		\end{array}\right)_i{}^{\,j}.
	\end{align}
\end{subequations}

Let us finish with a few general remarks on the results at this point:
\begin{itemize}
	\item To repeat what was already used in the derivation of the results, the horizon is at $g_2 = 0$ (or equivalently, $g^\text{eff}_{rr} = \infty$). 
	\item Similarly, the ergo-region is at $g^\text{eff}_{tt}=0$.
	\item All constitutive matrices retrieved are conformally invariant under conformal transformations of the effective metric $\g$. This is easily seen from inserting the appropriate factors of $\Omega^2$ in equations~\eqref{eq:BLsummary}. 
	\item This naturally also corresponds to the conformal invariance of the inverses of these matrices, if well-defined (or seen as a Moore--Penrose pseudo-inverse).
	\item We see that $\epsilon^{\phi\phi}=\mu^{\phi\phi}$. This is reminiscent of the similar feature in spherically symmetric space-times, see section~\ref{sec:quasicart}.
	\item \emph{All} components of the permittivity are well-defined down to the horizon (remembering that there $g_2=0$, and $g^\text{eff}_{rr}=\infty$).
	\item While the above remark shows that similarly $\mu^{\phi\phi}$ is equally well-defined even at the horizon, the other non-zero components of the permeability, $\mu^{rr}$ and $\mu^{\theta\theta}$, are not --- they diverge at the ergo-surface where $g^\text{eff}_{tt}=0$. This is again a sign of the general property that (parts of) the optical properties will have to diverge on the horizon or ergo-surface.
	\item The magneto-electric tensor is, like the permittivity, well-defined down to the horizon.
	\item The difference of permittivity and permeability,
	\begin{equation}
		\epsilon^{ij} - \mu^{ij} = 
		\sqrt{\frac{g^\text{eff}_{rr} g^\text{eff}_{\theta\theta}}{g_2 \det(\eta)}}  \; \frac{ \kl{g^\text{eff}_{t\phi}}^2}{ g^\text{eff}_{tt}}
		\left(\begin{array}{cc|c} 
		1/g^\text{eff}_{rr} &0 &0\\
		0&1/g^\text{eff}_{\theta\theta}&0\\
		\hline
		0&0&0
		\end{array}\right)^{ij},
	\end{equation}
	depends on the strength of rotation, as encoded in the off-diagonal piece $g^\text{eff}_{t\phi}$. If this is taken to zero, we regain the consistency condition~\eqref{eq:conscond3+1} and $\zeta = 0$. This is, again, the old result for static space-times, as to be expected.
	\item As $\det\zeta = 0$ (independent of position of indices) each possible index arrangement of this tensor has an eigenvector to eigenvalue zero. In the present case, this eigenvector is along the $\phi$-direction and parallel to $\beta$.
	\item Squaring the magneto-electric tensor,
	\begin{equation}
		(\zeta^2)_i{}^k = \zeta_i{}^j \zeta_j{}^k= \ed{4}\; {\frac{\kl{g^\text{eff}_{t\phi}}^2 }{ g_2}} 
		\left(\begin{array}{cc|c} 
		1 & 0&0\\
		0 &1&0\\
		\hline
		0&0&0
		\end{array}\right)_i{}^{\;k},
	\end{equation}
	we see that this is proportional to a projection operator onto the directions perpendicular to $\beta$, \emph{i.e.}, to $\gi^{0i}$. This relation to projection operators is another reminiscence of the quasi-Cartesian analysis of section~\ref{sec:quasicart}.
	\item With the trace of this square
	\begin{equation}
		\mathrm{tr}(\zeta^2) = \zeta_i{}^j \zeta_j{}^i= \ed{2}\; {\frac{\kl{g^\text{eff}_{t\phi}}^2 }{ g_2}},
	\end{equation}
	we have a scalar invariant (at least under spatial coordinate transformations) for the strength of the magneto-electric effect. As the original magneto-electric tensor, so is this invariant well-defined down to the horizon.
\end{itemize}

\subsubsection{A Second Look at Arbitrary, Kaluza--Klein-Decomposed Space-Times}
As the \enquote{threading}\footnote{For this phrasing, see, for example, \cite{TaleOfTwoVelocities}, or the references for this given in \cite{Bespoke1,Bespoke2}.} used for a Kaluza--Klein decomposition is at least locally always possible (unlike the choice of Boyer--Lindquist coordinates above), it is useful to have a close and careful look at how much can be told from just this decomposition alone.

Hence, let us start with the Kaluza--Klein decomposition of the inverse metric, though this time with a slightly different notation compared to section~\ref{sec:conscond}. The present notation will slightly improve the appearance of the formulae to come:
\begin{equation}
	\gi^{ab} =  \left(\begin{array}{cc}
	-\alpha^{-2} + \gamma^{-1}_{kl} \beta^k \beta^l & \beta^j\\ 
	\beta^i & \gamma^{ij} 
	\end{array}\right),
\end{equation}
such that the determinant of the metric (not the inverse metric as above!) now reads
\begin{equation}
	\det(g_\text{eff})=-\alpha^2\det(\gamma^{-1}).
\end{equation}

Quickly moving through the evaluation of the permittivity, we have:
\begin{subequations}
	\begin{align}
		\epsilon^{ij} &=  -\sqrt{\frac{\det(g_\text{eff})}{\det(\eta)}} \; 
		\left( \gi^{ij}  \gi^{00}- \gi^{0i}  \gi^{0j}  \right),
		\\
		&= + \sqrt{\frac{\det(g_\text{eff})}{\det(\eta)}}  
		\left\{(\alpha^{-2} - \gamma^{-1}_{kl} \beta^k \beta^l )\, \gamma^{ij} + \beta^i \beta^j\right\},\\
		&= {\ed{\alpha \sqrt{|\det(\eta)|\det(\gamma^{pq})} }} 
		\left\{(1 - \alpha^2 \gamma^{-1}_{kl} \beta^k \beta^l )\, \gamma^{ij} + \alpha^2 \beta^i \beta^j\right\}.
	\end{align}
\end{subequations}

As before, the permeability requires more effort to arrive at easily manageable identities. Again, we shall make use of the Levi-Civita tensor \emph{density} $\tilde{\eps}$ to simplify appearances. Inserting this in the definitions, we have
\begin{subequations}
	\begin{align}
		[\mu^{-1}]^{ij} &=  \ed{2}\;\eps^i{}_{kl}\,\eps^j{}_{mn}\,Z^{klmn},
		\\
		&=\ed{4}\;\eps^i{}_{kl}\,\eps^j{}_{mn} \sqrt{\frac{\det(g_\text{eff})}{\det(\eta)}} \; 
		\left(\gi^{km}  \gi^{ln} - \gi^{kn} \gi^{lm} \right),
		\\
		&=\ed{2}\;\eps^i{}_{kl}\,\eps^j{}_{mn} \sqrt{\frac{\det(g_\text{eff})}{\det(\eta)}} \; 
		\left(\gamma^{km}  \gamma^{ln} \right),\\
		&=\ed{2}\; \eta^{ip} \eta^{jq}  [-\det(\eta)] \sqrt{\frac{\det(g_\text{eff})}{\det(\eta)}} \; \tilde{\eps}_p{}_{kl}\,\tilde{\eps}_q{}_{mn} 
		\left(\gamma^{km}  \gamma^{ln} \right).
	\end{align}
\end{subequations}
Lowering indices twice, and making use of the matrix identity~\eqref{eq:Cramer}, we can change this to
\begin{subequations}
	\begin{align}
		[\mu^{-1}]_{ij} &=\ed{2}\; [-\det(\eta)] \sqrt{\frac{\det(g_\text{eff})}{\det(\eta)}} \; \tilde{\eps}_i{}_{kl}\,\tilde{\eps}_j{}_{mn} 
		\left(\gamma^{km}  \gamma^{ln} \right),\\
		&=\; [-\det(\eta)] \sqrt{\frac{\det(g_\text{eff})}{\det(\eta)}} \; \det(\gamma^{pq}) \;  [\gamma^{-1}]_{ij},\\
		&=\; \alpha \sqrt{|\det(\eta)|\det(\gamma^{pq})} \;  [\gamma^{-1}]_{ij}.
	\end{align}
\end{subequations}
This can now be inverted to yield
\begin{equation}
	\mu^{ij} =\; {\ed{\alpha \sqrt{|\det(\eta)|\det(\gamma^{pq})} }} \;  \gamma^{ij}.
\end{equation}

The single appearance of a Levi-Civita pseudo-tensor in the definition of the magneto-electric tensor greatly diminishes our ability to simplify matters at the current level of generality. We have:
\begin{subequations}
	\begin{align}
		\zeta^{ij} &= -\ed{2}\; \sqrt{\frac{\det(g_\text{eff})}{\det(\eta)}} \;  \left(  \eps^{i}{}_{kl}\gi^{0l}  \gi^{kj}  \right),\\
		&=
		-\ed{2}\; \sqrt{\frac{\det(g_\text{eff})}{\det(\eta)}} \;  \left(  \eps^{i}{}_{kl} \beta^l \gamma^{kj}  \right).
	\end{align}
\end{subequations}
To reduce the amount of implicit use of background metrics to lower or raise indices, it proves convenient to lower the first index in the above result:
\begin{equation}
	\zeta_{\,i}{}^j = -\ed{2}\; \sqrt{\frac{\det(g_\text{eff})}{\det(\eta)}} \;  \left(  \eps_{ikl}\beta^l \gamma^{kj}  \right).
\end{equation}

We can summarise the state of affairs as
\begin{subequations}\label{eq:KKfinal}
	\begin{align}
		\epsilon^{ij}&= {\ed{\alpha \sqrt{|\det(\eta)|\det(\gamma^{pq})} }} 
		\left\{(1 - \alpha^2 \gamma^{-1}_{kl} \beta^k \beta^l )\, \gamma^{ij} + \alpha^2 \beta^i \beta^j\right\},\\
		\mu^{ij} &=\; {\ed{\alpha \sqrt{|\det(\eta)|\det(\gamma^{pq})} }} \;  \gamma^{ij},\label{eq:KKmu}\\
		\zeta_{\,i}{}^j &= -\ed{2}\; \sqrt{\frac{\det(g_\text{eff})}{\det(\eta)}} \;  \left(  \eps_{ikl}\beta^l \gamma^{kj}  \right).\label{eq:KKzeta}
	\end{align}
\end{subequations}

Despite the generality, we can notice a few conclusions:
\begin{itemize}
	\item Despite the appearance of determinants and occasional use of the Levi-Civita tensor \emph{densities}, it is important to keep in mind that we carefully set up the formalism in such a way to only yield true $T^2_0$ tensors. As the present example is the most general of our discussion of bespoke meta-materials, we repeat this here alone --- the comment applies equally well to any of the other curvilinear examples.
	\item Likewise, the electromagnetic properties listed in equations~\eqref{eq:KKfinal} are all invariant under conformal transformations of the effective metric.
	\item Let us remind ourselves of equation~\eqref{eq:muepsdegen}:
	\begin{equation}
		\mi_{jl} \beta^l = [\epsilon^{-1}]_{jl} \beta^l.
	\end{equation}
	As this was a general statement, it will hold true in the present case, too. It tells us that the electromagnetic properties along $\beta = \gi^{0i}$ are degenerate. This corresponds to the observation that $\epsilon^{\phi\phi} = \mu^{\phi\phi}$ in the Boyer--Lindquist case, and similar observations in section~\ref{sec:quasicart}.
	\item Again, we can observe that the difference of permittivity and permeability,
	\begin{equation}
		\epsilon^{ij}  -\mu^{ij} = - \frac{\alpha}{ \sqrt{|\det(\eta)|\det(\gamma^{pq})} } 
		\left\{( \gamma^{-1}_{kl} \beta^k \beta^l )\, \gamma^{ij} -  \beta^i \beta^j\right\},
	\end{equation}
	will vanish if $\beta^i \to 0$, and thus we again regain $\epsilon^{ij} = \mu^{ij}, \zeta^{ij} =0$ as consistency conditions, just as we did in the original analysis of section~\ref{sec:conscond}.
	\item Contracting equation~\eqref{eq:KKzeta} with $\beta^i$, we see that $\beta^i$ is an eigenvector of eigenvalue zero for $\zeta_i{}^j$. From this it also follows that $\det\zeta_i{}^j = 0$.
	\item We can also investigate in this general setting the previously introduced scalar invariant (under spatial coordinate transformations)\footnote{This certainly can be promoted to a scalar under full four-dimensional coordinate transformations, following step~4 of section~\ref{sec:conscond}.} $\mathrm{tr}(\zeta^2)$ for the strength of magneto-electric effects. We get:
	\begin{subequations}
		\begin{align}
			\mathrm{tr}(\zeta^2) &= \zeta_i{}^j \zeta_j{}^i= \ed{4}\; {\frac{\det(g_\text{eff})}{\det(\eta)}}   
			\left(  \eps_{i}{}_{kl}\beta^l \gamma^{kj}  \right)  
			\left(  \eps_{j}{}_{mn}\beta^n \gamma^{mi}  \right),\\
			&=  \ed{2}\; {\frac{\det(g_\text{eff})}{\det(\eta)}}   \;  [-\det(\eta)] \; \det(\gamma^{pq}) 
			[\gamma^{-1}]_{ij} \beta^i \beta^j,\\
			&=  \ed{2}\;\alpha^2 \;
			[\gamma^{-1}]_{ij} \beta^i \beta^j.
		\end{align}
	\end{subequations}
\end{itemize}

\subsubsection{Static, Spherically Symmetric Space-Times in Spherical Polar Coordinates}
Let us now revisit the example of static, spherically symmetric space-times discussed previously in the quasi-Cartesian approach, however now in the more befitting spherical polar coordinates. While more appropriate given the assumed symmetry, these coordinates will complicate the discussion by the associated appearance of various metric components in both determinants and the raising and lowering of indices.

Let us write our flat background metric as
\begin{equation}
	(\dif s_\text{lab})^2 = - \dif t^2 +  [R'(r) \dif r]^2 + R(r)^2\{\dif \theta^2+\sin^2\theta \, \dif\phi^2\}.
\end{equation}
This is now, unlike the (quasi-)Cartesian version, not unimodular:
\begin{equation}
	\det(\eta_{ab}) = -R'(r)^2\,R(r)^4\,\sin^2\theta \neq -1.
\end{equation}
While we do identify the coordinates of laboratory and effective space-time identically, we also allow the most general form of spherical symmetry for the effective metric:
\begin{equation}
	(\dif s)^2 = g_{tt} \,\dif t^2 + 2 g_{tr} \,\dif t \dif r + g_{rr} \,\dif r^2 + R(r)^2\{\dif\theta^2+\sin^2\theta \, \dif\phi^2\}.
\end{equation}
The benefit of this very general form of the metric(s) is that it allows for several different standard forms of metrics in spherical symmetry to be captured simultaneously: While the standard choice is $R(r)=r$, which corresponds in the background just to standard spherical polar coordinates, and in the analogue space-time of, for example, the Schwarzschild geometry to curvature coordinates, different choices can be encoded in this, too. For example, the Schwarzschild back hole in isotropic coordinates would correspond to $g^\text{eff}_{rr} = R(r)^2$.

We now write this effective metric again in terms of a Kaluza--Klein form, according to the notation introduced in the previous example:
\begin{equation}
	\gi^{ab} =  \left(\begin{array}{c|c|cc}
	-\alpha^{-2} + \beta^2/\gamma & \beta & 0 & 0\\
	\hline
	\beta & \gamma &0&0\\
	\hline
	0&0& R^{-2} & 0 \\
	0&0&0&R^{-2} (\sin^2\theta)^{-1}
	\end{array}\right). 
	\label{eq:g-KKsph}
\end{equation}
For the determinants, we now have
\begin{subequations}
	\begin{align}
		\det(g^\text{eff}_{ab}) &= - \alpha^2 \gamma^{-1} R(r)^4\,\sin^2\theta,\\
		\det(g^\text{eff}_{ab}) /\det(\eta_{ab}) &= \alpha^2 \gamma^{-1} R'(r)^{-2}.
	\end{align}
\end{subequations}

It is fairly straightforward to gain the permittivity from this:
\begin{subequations}
	\begin{align}
		\epsilon^{ij} &=  -\sqrt{\frac{\det(g^\text{eff})}{\det(\eta)}} \; 
		\left( \gi^{ij}  \gi^{00}- \gi^{0i}  \gi^{0j}  \right),
		\\
		&= + \frac{\alpha}{\sqrt{\gamma}|R'(r)|}
		\left(\begin{array}{c|cc}
		\alpha^{-2} \gamma &0 &0\\
		\hline
		0&(\alpha^{-2} - \beta^2/\gamma)  R^{-2} & 0 \\
		0&0&(\alpha^{-2} - \beta^2/\gamma)  R^{-2} (\sin^2\theta)^{-1}
		\end{array}\right),\\
		&= + {\frac{\sqrt{\gamma}}{\alpha|R'(r)|}}
		\left(\begin{array}{c|cc}
		1 &0 &0\\
		\hline
		0& (1- \alpha^{2} \beta^2/\gamma)  \gamma^{-1} R^{-2} & 0 \\
		0&0&(1- \alpha^{2} \beta^2/\gamma) \gamma^{-1} R^{-2} (\sin^2\theta)^{-1}
		\end{array}\right).
	\end{align}
\end{subequations}
As the factors of $\sin^{-2}(\theta)$ are rather cumbersome, it is helpful to employ the help of an orthonormal dyad (or orthonormal \enquote{zweibein}):
\begin{equation}
	\epsilon^{\hat i \hat j}  = + {\frac{\sqrt{\gamma}}{\alpha|R'(r)|}}
	\left(\begin{array}{c|cc}
	1 &0 &0\\
	\hline
	0& (1- \alpha^{2} \beta^2/\gamma)  \gamma^{-1} R^{-2}& 0 \\
	0&0&(1- \alpha^{2} \beta^2/\gamma) \gamma^{-1} R^{-2}
	\end{array}\right). 
\end{equation}

For the permeability, it is easiest to start with the result calculated in the previous example, equation~\eqref{eq:KKmu}. Unwrapping the definitions of and choices made in the quantities involved, one arrives at two equivalent results:
\begin{subequations}
	\begin{align}
		\mu^{ij} &=\; {\ed{\alpha \sqrt{|\det(\eta)|\det(\gamma^{pq})} }} \;  \gamma^{ij},\\
		&= + \frac{1}{\alpha\sqrt{\gamma}|R'(r)|}
		\left(\begin{array}{c|cc}
		\gamma &0 &0\\
		\hline
		0& R^{-2} & 0 \\
		0&0&R^{-2} (\sin^2\theta)^{-1}
		\end{array}\right), \qquad \text{or}\\
		&= + {\frac{\sqrt{\gamma}}{\alpha|R'(r)|}}
		\left(\begin{array}{c|cc}
		1 &0 &0\\
		\hline
		0& \gamma^{-1} R^{-2} & 0 \\
		0&0&\gamma^{-1}  R^{-2}   (\sin^2\theta)^{-1}
		\end{array}\right).
	\end{align}
\end{subequations}
Repeating the introduction of an orthonormal zweibein, this can be rewritten as
\begin{equation}
	\mu^{\hat i \hat j} = + {\frac{\sqrt{\gamma}}{\alpha|R'(r)|}}
	\left(\begin{array}{c|cc}
	1 &0 &0\\
	\hline
	0& \gamma^{-1} R^{-2} & 0 \\
	0&0&\gamma^{-1} R^{-2} 
	\end{array}\right).
\end{equation}

Lastly, we calculate the magneto-electric tensor. As with the permeability, we start from the result for the Kaluza--Klein decomposed form given in the previous example, equation~\eqref{eq:KKzeta}.
\begin{subequations}
	\begin{align}
		\zeta_{\,i}{}^j &= -\ed{2}\; \sqrt{\frac{\det(g_\text{eff})}{\det(\eta)}} \;  \left(  \eps_{ikl}\gi^{0l}  \gi^{kj}  \right),\\
		&= -{\frac{\beta}{2}}\; \sqrt{\frac{\det(g_\text{eff})}{\det(\eta)}} \;  \left(  \eps_{i}{}_{kr} \gamma^{kj}  \right),\\
		&=-{\frac{\beta}{2}}\; \sqrt{\det(g_\text{eff})} \;  \left(  \tilde{\eps}_{ikr} \gamma^{kj}  \right),\\
		&= -{\frac{\beta}{2}}\;  \frac{\alpha |R'| R^2 \sin\theta }{\sqrt{\gamma}} \;  R^{-2} \;
		\left(\begin{array}{c|cc}
		0 &0 &0\\
		\hline
		0 & 0 & (\sin^2\theta)^{-1} \\
		0&-1& 0
		\end{array}\right),\\
		&= -{\frac{\beta}{2}}\;  {\frac{\alpha |R'|}{\sqrt{\gamma}}} \;  \;
		\left(\begin{array}{c|cc}
		0 &0 &0\\
		\hline
		0 & 0 & (\sin\theta)^{-1} \\
		0&-\sin\theta& 0
		\end{array}\right).
	\end{align}
\end{subequations}
Finally, let us rephrase this by adopting an orthonormal zweibein again:
\begin{equation}
	\zeta_{\,\hat i}{}^{\hat j} = -{\frac{\beta}{2}}\;  {\frac{\alpha |R'|}{\sqrt{\gamma}}} \;  \;
	\left(\begin{array}{c|cc}
	0 &0 &0\\
	\hline
	0 & 0 & 1 \\
	0&-1& 0
	\end{array}\right).
\end{equation}

We summarise by listing the results after adoption of an orthonormal zweibein, given that the corresponding quantities are slightly more insightful and less cluttered:
\begin{subequations}\label{eq:sphcurvi}
	\begin{align}
		\epsilon^{\hat i \hat j}  &= + {\frac{\sqrt{\gamma}}{\alpha|R'(r)|}}
		\left(\begin{array}{c|cc}
		1 &0 &0\\
		\hline
		0& (1- \alpha^{2} \beta^2/\gamma)  \gamma^{-1} R^{-2}& 0 \\
		0&0&(1- \alpha^{2} \beta^2/\gamma) \gamma^{-1} R^{-2}
		\end{array}\right),\\
		\mu^{\hat i \hat j} &= + {\frac{\sqrt{\gamma}}{\alpha|R'(r)|}}
		\left(\begin{array}{c|cc}
		1 &0 &0\\
		\hline
		0& \gamma^{-1} R^{-2} & 0 \\
		0&0&\gamma^{-1} R^{-2} 
		\end{array}\right),\\
		\zeta_{\,\hat i}{}^{\hat j} &= -{\frac{\beta}{2}}\;  {\frac{\alpha |R'|}{\sqrt{\gamma}}} \;  \;
		\left(\begin{array}{c|cc}
		0 &0 &0\\
		\hline
		0 & 0 & 1 \\
		0&-1& 0
		\end{array}\right).
	\end{align}
\end{subequations}

We finish also this last example with some general remarks:
\begin{itemize}
	\item As in the previous curvilinear examples (and based on \emph{general} results), all our results listed in equations~\eqref{eq:sphcurvi} are invariant under conformal transformations of the effective metric $g_\text{eff}$.
	\item The fact that permittivity and permeability are degenerate along $\beta^i$ is this time exemplified by the equation $\epsilon^{rr} = \mu^{rr}$.
	\item This time, let us show the transition to the static consistency conditions of equation~\eqref{eq:conscond3+1} by taking the dyadic expression for the difference of permittivity and permeability:
	\begin{equation}
		\epsilon^{\hat i \hat j}  -\mu^{\hat i \hat j} = 
		+ {\frac{\sqrt{\gamma}}{\alpha|R'(r)|}} \frac{\alpha^2\beta^2}{\gamma^2 R^2} 
		\left(\begin{array}{c|cc}
		0 &0 &0\\
		\hline
		0& 1& 0 \\
		0&0&1
		\end{array}\right). 
	\end{equation}
	\item The zero eigenvalue direction of $\zeta_{i}{}^j$ is the radial direction, the determinant remains, as to be expected, zero.
	\item For the three-scalar previously introduced to characterise magneto-electric effect strength, we have this time the equations
	\begin{subequations}
		\begin{align}
			\mathrm{tr}(\zeta^2) &= \zeta_i{}^j \zeta_j{}^i= \ed{4}\; {\frac{\det(g_\text{eff})}{\det(\eta)}}   
			\left(  \eps_{ikl}\beta^l \gamma^{kj}  \right)  
			\left(  \eps_{jmn}\beta^n \gamma^{mi}  \right),\\
			&=  \ed{2}\; {\frac{\det(g_\text{eff})}{\det(\eta)}}   \;  [-\det(\eta)] \; \det(\gamma^{pq}) 
			[\gamma^{-1}]_{ij} \beta^i \beta^j,\\
			&=  \ed{2}\; \frac{\alpha^2 \beta^2}{\gamma} (R')^2.
		\end{align}
	\end{subequations}
	\item One can now also identify components of the above equations with the results for dirty black holes, given at the end of the discussion of brute force methods. Put differently, the present analysis of static, spherically symmetric metrics encompasses also that as a special case. Regaining those previous results is now a mere exercise in inserting terms.
\end{itemize}

\subsection{Hawking Temperature and Laboratory Coordinates} \label{sec:HawkingLab}  
The most general way to give the Hawking temperature, as described in section~\ref{sec:Hawking}, is in terms of the surface gravity $\kappa$ at the exterior horizon:
\begin{equation}
	T = \frac{\hbar}{2\pi} \kappa(r_+).
\end{equation}
Note that for our convenience we set $\kB=c=1$ in this discussion, unlike in section~\ref{sec:Hawking}. As one of the goals of the analogue space-time program is to make the Hawking effect experimentally accessible, the natural question in the present context is how to relate the Hawking temperature to the electromagnetic properties of the black hole space-time mimic. Put differently, one is looking for the functional relation
\begin{equation}
	\kappa(r_+) = \kappa(\epsilon,\mu,\zeta).
\end{equation}
While we do have equation~\eqref{eq:g3+1mu} and equation~\eqref{eq:g3+1eps}, identifying the surface gravity for an arbitrary effective metric on this level is equivalent to the question what the most general form of a time-like Killing vector field on \emph{some} metric is. It is doubtful that this can be done in closed form. Therefore, we will focus our attention on one particular type of metric, namely a general spherically symmetric and static one. Afterwards, we specialise to the Schwarzschild case contained in this type of metric. We assume the form of the metric to be
\begin{equation}
	\dif s^2 = -f \dif t^2 + f^{-1} \dif r^2 + r^2\dif \Omega^2.\label{eq:metricwithf}
\end{equation}
For this metric, the time-like Killing vector field is easily found and given --- it is identical to the vector field $\partial_t$ associated to the coordinate time $t$. Thus we know that
\begin{subequations}
	\begin{align}
		\kappa(r_+) &= \sqrt{-\frac{t^{a;b} t_{a;b}}{2}},\\
		&=\ed{2} f'(r_+).\label{eq:kappaf}
	\end{align}
\end{subequations}

Combining this metric with the results of equation~\eqref{eq:geffeps0} or equation~\eqref{eq:geffmu0} then gives that\footnote{We could have also started from either equation~\eqref{eq:g3+1mu} or equation~\eqref{eq:g3+1eps}. But we have to be very careful: In deriving either, we made heavy use of our conformal freedom. This would utterly obfuscate the inclusion of factors of determinants of the background metric $g$.}
\begin{subequations}
	\begin{align}
		f(r) & = \sqrt{\det g \det \mu^{-1}},\\
		&=\sqrt{\det g \det \epsilon^{-1}},
	\end{align}
\end{subequations}
Then inserting this in equation~\eqref{eq:kappaf} allows us to identify the Hawking temperature in terms of either permeability or permittivity:
\begin{subequations}\label{eq:HawkingEff}
	\begin{align}
		T_\text{H} &= \left.\kl{\frac{\hbar}{4\pi \sqrt{\det g\det\epsilon}}}_{,r_\text{eff}}\right\lvert_{r_\text{eff}=r_+},\\
		&= \left.\kl{\frac{\hbar}{4\pi \sqrt{\det g \det\mu}}}_{,r_\text{eff}}\right\lvert_{r_\text{eff}=r_+}.
	\end{align}
\end{subequations}
It is worth pointing out (again) that the constitutive tensors are invariant under conformal transformations of the effective metric $g_\text{eff}$.\footnote{Note that this is not true for conformal transformations of the laboratory metric $g$!} Naturally, this means that also the determinant of $\epsilon$ and $\mu$ are conformally invariant, and hence the Hawking temperature as given in equations~\eqref{eq:HawkingEff}. This is in full agreement with the result of Jacobson and Kang \cite{JacobsonKang93}, that the Hawking temperature in general should be conformally invariant. In preparation for the next step, we already started highlighting the importance of differentiating between laboratory and effective coordinates in previous sections.

Let us be more concrete about the effective metric under consideration, and choose $f(r)=1-2M/r$ in equation~\eqref{eq:metricwithf}, as appropriate for an effective Schwarzschild black hole, see section~\ref{sec:Schwarzschild} (but setting $G=c=1$). The constitutive tensors of the corresponding meta-material mimic are then easily calculated using the brute-force dirty black hole results (see section~\ref{sec:brute}), via the general results in curvilinear background coordinates for static, spherically symmetric space-times (see section~\ref{sec:curvilinear}), or entirely (and quickly) from scratch. They are:
\begin{subequations}
	\begin{align}
	\epsilon^{ab} &= \begin{pmatrix}
	1 & 0 & 0\\ 0 & \ed{r(r-2M)} & 0 \\ 0 & 0 & \ed{r\sin^2\theta(r-2M)}
	\end{pmatrix},\\
	\mu^{ab} &= \begin{pmatrix}
	1 & 0 & 0\\ 0 & \ed{r(r-2M)} & 0 \\ 0 & 0 & \ed{r\sin^2\theta(r-2M)}
	\end{pmatrix},\\
	\zeta^a{}_b &= 0.
	\end{align}
\end{subequations}
However, the temperature as given in equations~\eqref{eq:HawkingEff} is unlikely to be immediately measured. In a laboratory, as described at the beginning of the present section, the coordinates would be different to those of the effective space-time. In particular, the relation is unlikely to be a four-dimensional conformal transformation of the analogue space-time metric. To illustrate this, let us suppose we stretch or shrink the radial coordinate when going from laboratory coordinates to those of the effective space-time,
\begin{subequations}
	\begin{align}
	(t,r,\theta,\phi)_\text{lab} \qquad&\longrightarrow \qquad (t,ar_\text{lab},\theta,\phi)_\text{eff},\\
	r_\text{eff} &= a r_\text{lab}.
	\end{align}
\end{subequations}
The example of the Schwarzschild metric would now read in the stretched coordinates
\begin{equation}
	\dif s^2_\text{eff} = - \kl{1-\frac{2M}{ar_\text{lab}}}\dif t^2 + a^2 \kl{1-\frac{2M}{ar_\text{lab}}}^{-1} \dif r_\text{lab}^2 + a^2 r_\text{lab}^2\dif \Omega^2.
\end{equation}
There are several important points to be made here: First note, that while it \emph{could} be viewed as a simple coordinate transformation, this would ignore the physical significance of $r_\text{lab}$. This is, after all, the distance measured in a laboratory, thus lending particular significance to it. Second, note that this does \emph{not} correspond to a conformal transformation of an effective Schwarzschild metric. Third, note that when performing this calculation, it is absolutely vital to not fix conformal factors to simplify calculations, as the scale $a$ will break this choice. Since only the $r$-coordinate differs in the laboratory coordinate system from the corresponding one in the effective space-time, we only place appropriate index labels (\enquote{eff}, \enquote{lab}) on $r$. This then corresponds to electromagnetic tensors
\begin{subequations}
	\begin{align}
		\epsilon^{ab} &= \begin{pmatrix}
			a & 0 & 0 \\ 0 & \frac{a^2}{(ar_\text{lab}-2M)r_\text{lab}}& 0 \\ 0&0&\frac{a^2}{(ar_\text{lab}-2M)r_\text{lab}\sin^2\theta}
		\end{pmatrix},\\
		\mu^{ab} &= \begin{pmatrix}
			a & 0 & 0 \\ 0 & \frac{a^2}{(ar_\text{lab}-2M)r_\text{lab}}& 0 \\ 0&0&\frac{a^2}{(ar_\text{lab}-2M)r_\text{lab}\sin^2\theta}
		\end{pmatrix},\\
		\zeta^a{}_{b} &= 0.
	\end{align}
\end{subequations}

If we now follow through with the calculation of the Hawking temperature, we see that for $a=1$, which is the \enquote{standard} Schwarzschild metric, equations~\eqref{eq:HawkingEff} correctly evaluate to
\begin{equation}
	T_\text{H} = \frac{\hbar}{8\pi M}.
\end{equation}
However, if we keep $a$ arbitrary, we now get
\begin{equation}
	T_\text{H} = \frac{\hbar}{8\pi a^{9/2} M}.
\end{equation}
This is the temperature measured in the laboratory. As we see, if we attempt to miniaturise a microscopic black hole geometry to fit inside a tabletop laboratory setting, the scale factor will actually increase the Hawking temperature (as $a\ll 1$) to ludicrous values. On the other hand, if we attempted to mimic an actual astrophysical black hole (at mass regimes at which we know or assume their existence), we would have to blow the radial coordinate up with $a\gg 1$, greatly diminishing the Hawking temperature. We give some values for the temperature as they would be measured in the laboratory in table~\ref{tab:TH}. Already the temperature of a solar mass black hole becomes laughably small, not to mention that of a black hole of the mass of Sagitarius A$^*$, here taken to be $M_{\text{Sgr A}^*}=\num{4e6} M_\odot$: A temperature in the range of $\SI{100}{\pico\kelvin}$ can be achieved in the context of Bose--Einstein condensates,\footnote{For this, we refer to the thesis \cite{LowestTemp}. The upcoming results of the Cold Atom Laboratory aboard the International Space Station will likely further lower this particular record, see \url{https://coldatomlab.jpl.nasa.gov/}.} but only much more modest $\SI{6}{\milli\kelvin}$ are achievable for macroscopic objects,\footnote{Experimentally achieved by the CUORE collaboration at the INFN Gran Sasso National Laboratory, see \url{https://www.interactions.org/node/12905}, accessed at 10:41am, June $12^\text{th}$.} while the lowest naturally occurring temperatures observed to date are about $\SI{1}{\kelvin}$ in the Boomerang nebula\footnote{Source: \url{https://apod.nasa.gov/apod/ap071228.html}, accessed at 10:07am, June $12^\text{th}$.} due to gas expansion. However, notice that one still wants to keep to astronomical objects as effective masses --- already Earth's effective temperature is reasonable, and even that of Venus might be achievable. Depending on the experimental set-up (cooling, accuracy of temperature measurements) even the comparatively small values for Uranus might still be realisable. We included the proton mass and $\SI{1}{\kilogram}$ only for illustrative purposes. Modelling an elementary particle by black hole space-times would be riddled with \emph{many} more issues in the first place --- including naked singularities, and that it is \emph{charged} and \emph{has spin} while we are calculating values for uncharged, non-rotating Schwarzschild black holes.\footnote{Other things include, but are not limited to: Much less than a bit of information in the Bekenstein entropy, energy peak frequencies higher than the rest mass, all the problems galore in league with naked singularities, not to mention that its (Kerr--Newman) ring singularity would have a radius already excluded by experiments on electrons\dots}

Let us reiterate: While masses slightly above the mass of the Earth give experimentally achievable temperatures, \emph{these are the temperatures of the analogues}. A real, \emph{astrophysical} object of comparable mass would have the well-known, low Hawking temperatures given in the third column of table~\ref{tab:TH}. Likewise, the experimental situation would not be concerned with any masses as given in the first column. These masses only hold in the analogue space-time (as we work with an effective Schwarzschild geometry, it does not matter which mass concept precisely we invoke, so \enquote{effective} ADM masses are sufficient). The only physical mass involved will be that of the optical medium used as an effective space-time.

This strongly implies that coordinate artefacts will be of an even more pernicious nature than in classical general relativity: Already general relativity itself had almost from its infancy to grapple with the question which effects are real and which are just due to a choice of coordinates --- examples being both the event horizon and the existence of gravitational waves (which was only settled once the sticky-bead-argument became accepted). The additional layer of coordinate transformations (from the laboratory space-time to the effective space-time or \emph{vice versa}) will only serve, sadly, to exacerbate this further. Underneath this tangle, however, will lie valuable lessons for effects of curved space-time physics, both classical and quantum mechanical in nature.

\begin{table}
	\centering
	\begin{tabular}{||c|c|c|c||}\hline\hline
		$M[\si{\kilo\gram}]$ & $r_\text{H}[\si{\meter}]$ & $T_\text{H}[\si{\kelvin}]$  & $T_\text{lab}[\si{\kelvin}]$\\\hline
		$M_p = \num{1.6726e-27}$ & $\num{2.4841e-54}$ & $\num{7.3355e49}$ & $\num{3.8651e286}$\\
		$\num{1}$ & $\num{1.4852e-27}$ & $\num{1.2269e23}$ & $\num{2.0693e139}$\\
		$M_{\leftmoon} = \num{7.342e22}$ & $\num{1.0904e-4}$ & $\num{1.6711}$ & $\num{3.5796e13}$\\
		$M_{\mars} = \num{6.4171e23}$& $\num{9.5306e-4}$ & $\num{0.1912}$ & $\num{2.3738e8}$\\
		$M_{\venus} = \num{4.8685e24}$ & $\num{7.2306e-3}$ & $\num{2.5202e-2}$ & $\num{3428.8}$\\
		$M_\oplus = \num{5.9736e24}$ & $\num{8.8719e-3}$ & $\num{2.0539e-2}$ & $\num{1113.1}$\\
		$M_{\uranus} = \num{8.6832e25}$ & $\num{0.1290}$ & $\num{1.4130e-3}$ & $\num{4.4986e-4}$\\
		$M_\odot = \num{1.9886e30}$ & $\num{2953.4}$ & $\num{6.1700e-8}$ & $\num{4.7191e-28}$\\
		$M_{\text{Sgr A}^*} = \num{7.9542e36}$ & $\num{1.1813e10}$ & $\num{1.5425e-14}$ & $\num{2.3043e-64}$\\\hline\hline
	\end{tabular}
	\caption[Table of Comparison of Analogue and Astrophysical Hawking Temperatures]{Table comparing the Hawking temperature $T_\text{H}$ as \enquote{observed} within the effective (Schwarzschild) space-time itself with the actually observed temperature $T_\text{lab}$ for different values of the black hole mass $M$. The value of $a$ is chosen such that the radius of the event horizon in the laboratory is $\SI{10}{\centi\meter}$. Thus, $a$ is simply $\num{10}$ times the value of $r_\text{H}$ in meters.}
	\label{tab:TH}
\end{table}

\FloatBarrier
\section{A Tale of Two Wave Equations} \label{sec:PDEana} 
It was already frequently alluded to that the understanding of the partial differential equation for macroscopic electrodynamics, as it appears in our electromagnetic analogue, will have two very distinct interpretations. Similarly, we added in section~\ref{sec:bespoke} that the relation between laboratory coordinates and effective space-time coordinates will introduce a new layer of complication. Let us, for the sake of readability, in this section again \enquote{ignore} the latter issue, so we can focus on the former. Remember, that \enquote{ignoring} the coordinate complication corresponds to linking the coordinates by the identity map. The identity map giving the transformation $(M_\text{eff}, \g)\to (M_\text{lab},g_\text{lab})$ will certainly simplify things below, but even if we identify the coordinates $1:1$, this may be only possible on a subset of the coordinate patch the laboratory is working in: The simplest example for such a subset is a doughnut-shaped optical material as our mimic. The laboratory most certainly is described by a simply connected coordinate patch, the effective metric's coordinate patch cannot be simply connected. These coordinate identifications hence need not even be diffeomorphisms. These considerations will play a role even with the given $1:1$ correspondence, before we even introduce the equally important cartographic distortions.

A second simplification we shall occasionally assume in the present section is the availability of a four-potential $A_a$. While this does mean that we have to fix topological constraints, we find it unlikely that experiments will immediately start operating with situations where a four-potential cannot be defined consistently on the whole analogue space-time. This holds true even more since the experimental realisation will most likely involve only a small part (if one wants, a \enquote{single chart}) of the (maximal analytic extension of the) analogue space-time where we can appeal to locality arguments to guarantee the existence of $A_a$.

Before we confuse more than we clear up, let us reiterate points of previous sections: The reason behind this complication is that our analogy of the present chapter is of an algebraic, not an analytic nature. Many analogies (like those linking space-time physics with fluid flows, see \cite{lrrAnalogue}) start from the get-go on the level of the underlying wave equations as partial differential equations. Equation~\eqref{eq:Zeff}, the foundation of most of the electromagnetic analogue space-times considered in this thesis, on the other hand is algebraic --- we cannot expect the analogue to extend quite as straightforwardly on the level of the partial differential equation governing the analogue. It is also useful to contrast this with the analogy considered in the next section~\ref{sec:refractive}: There we actually work out the analogy based on differential equations in the first place, but lack a geometric picture for the analogy. The presently demonstrated complications thus will not arise there.

This far, we have entirely confined our attention to strictly Lorentzian geometry, a triple $(M,g,\Gamma)$: A base manifold $M$ of dimension $4$ together with a symmetric, Lorentzian (signature 3, non-degenerate) covariant tensor of valence two, and lastly a symmetric affine connection on $TM$ that is metrically compatible with $g$. The last ingredient is so ingrained in physics, that it is often not even stated, and people occasionally just talk of $(M,g)$. This will not suffice in what follows. Specifically, we will quickly see that we cannot expect metric compatibility any more. This introduces a new tensor
\begin{equation}
	q^\text{lab/eff}_{bca} \defi \nabla_a^\text{lab/eff} g^\text{eff/lab}_{bc},\label{eq:defnonmetricity}
\end{equation}
the \emph{non-metricity tensor.} Note that already this definition could only be written out as it was done, \emph{if the laboratory and effective coordinates are identified}. Real laboratory situations are likely to complicate even the definition of the non-metricity involved! Since this complicates our Ricci calculus (our index manipulation) already considerably, it is worthwhile to stop and think for a second about the second possible complication: Torsion,
\begin{equation}
	T^{m}{}_{ab} \defi \ed{2}\left[\Gamma_\text{eff}\right]_{[ab]}^m.
\end{equation}
This might seem overly complicated, but we shall see that it only adds one more term to nearly all resulting equations. Since this is the simpler part (being purely algebraically defined), let us start with its contribution.

As the inhomogeneous Maxwell equation behind our analogue space-time is (using that analogue space-time's connection) most easily represented as
\begin{equation}
	\nabla_{a}^{\text{eff}} F^{ab}_\text{eff} = J^b_\text{eff},
\end{equation}
we will be working in the effective space-time from now on, and see how this equation changes if looked at using the laboratory space-time's Levi-Civita connection\footnote{For the Levi-Civita connection we shall use the notation of a Christoffel symbol of the second kind.} $\begin{Bmatrix}a\\bc\end{Bmatrix}_{\text{lab}}$. Both connections' symmetry will not be touched by the cartographic distortions between laboratory labelling and effective space-time coordinates. Their metricity will be seen to be an entirely different matter. We will see clearly how the unavoidable short-coming of using a wrong connection --- the possibility of the appearance of the non-metricity tensor --- enters our analysis.

Remember that we said, when discussing equations~(\ref{eq:defA1},\ref{eq:defA2}), that all forms for the field strength tensor appearing in equation~\eqref{eq:defA2} are totally equivalent. There is one caveat to this: This holds only for \emph{torsion-free} connections. This will be true if we take as connection in the effective space-time the unique effective Levi-Civita connection $\begin{Bmatrix}a\\bc\end{Bmatrix}_{\text{eff}}$ to define $F$ in terms of $\nabla^\text{eff}_a A^\text{eff}_b$ as appearing in the analogue space-time. Likewise it is true if we define it as living in the laboratory space-time and use the laboratory space-time's Levi-Civita connection $\begin{Bmatrix}a\\bc\end{Bmatrix}_{\text{lab}}$. Now, one might want to find analogue space-times not just for standard Lorentzian geometry, but also theories with torsion, as in \cite{SpinAndTorsion,HehlObukTorsionHowToEdyn}.  But if we then define $F$ with a non-symmetric connection in the covariant derivative of the four-potential a torsion term will appear:
\begin{subequations}
	\begin{align}
		F^{\text{eff}}_{ab} &\defi \nabla^{\text{eff}}_a A_b^{\text{eff}} - \nabla^{\text{eff}}_b A_a^{\text{eff}},\\
		&= \partial_a A_b^{\text{eff}} - \partial_b A_a^{\text{eff}} + 2 T^m{}_{ab} A_m^{\text{eff}}.\label{eq:Ftorsion}
	\end{align}
\end{subequations}
We can see that this would break the gauge invariance of $F$ without additional changes to the effective space-time's electromagnetism, see again \cite{SpinAndTorsion,HehlObukTorsionHowToEdyn}. This far, this is still on the algebraic level: The anti-symmetrisation of the connection is just plain subtraction. Torsion is actually \emph{not} what we will be most concerned with, but rather the non-metricity.

Non-metricity, in contrast, is not algebraic, but differential in nature --- as its definition~\eqref{eq:defnonmetricity} clearly proves. Already here we can see how it can come about in vacuum electrodynamics: The indices of $F_{ab}$ are twice raised with a metric. Only if the metric commutes with a covariant derivative can we ignore this raising of indices in the differential equations, like we did above in~\eqref{eq:Ftorsion}.

It is useful at this stage to recall then how to decompose a general connection into non-metricity, torsion, and a reference connection (in our case the effective Levi-Civita connection) \cite{Schouten,VisserDG}:\footnote{We explicitly put the label \enquote{${}^\text{eff}$} on the three-times covariant torsion tensor, as the indices have to be lowered with \emph{the effective metric}. The index lowered appears now as the first covariant index. We omitted the explicit appearance to make the expression more familiar, despite our retained convention that we raise and lower indices with the laboratory background metric, unless $\g$ is explicitly written out.}
\begin{equation}
	\left[\Gamma^\text{lab}\right]^c_{ab} = \begin{Bmatrix}c\\ab\end{Bmatrix}_\text{eff} + \underbrace{\gi^{cm}\kl{\ed{2}\kle{q_{abm}^\text{lab}-2q^\text{lab}_{m(ab)}} + \kle{2T^\text{eff}_{(ab)m}-T^\text{eff}_{mba}}}}_{\ifed \Delta^c{}_{ab}}.\label{eq:GammaLV}
\end{equation}
Furthermore, let us remind ourselves of the general Ricci identity for a tensor\footnote{We will not be as careful as normally with the index placement on $X$, as this is not relevant to the identity. It can be arbitrary, as long as it is done consistently in the identity.}
\begin{align}
	[\nabla^\text{lab}_a,\nabla^\text{lab}_b]X^{c_1\dots c_t}_{d_1\dots d_f} =& \sum_{j=1}^{t} X^{c_1\dots c_{j-1}mc_{j+1}\dots c_t}_{d_1\dots d_f} R^{c_j}{}_{mba} + \sum_{k=1}^{f} X^{c_1\dots c_t}_{d_1\dots d_{k-1}md_{k+1}\dots d_f} R^{m}{}_{d_kba} \nonumber \\ &\qquad + 2T^m{}_{ab}\nabla^\text{lab}_m X^{c_1\dots c_t}_{d_1\dots d_f},\label{eq:Ricci}
\end{align}
as this identity will be alluded to later on.

As said repeatedly, there actually is physically motivated method in the resulting index madness: In the context of analogue space-times, we are naturally interested in $F^\text{eff}_{ab}$ (be the underlying effective manifold torsion-free or with torsion), not $F_{ab}^\text{lab}$. However, experimentally it makes much more sense to use (and possibly even quantise) $A_a^\text{lab}$ or $F^\text{lab}_{ab}$, as this is the experimentally available covector field or field strength tensor. Hence, our goal: We want to write the laboratory, macroscopic, inhomogeneous Maxwell equation in terms of the effective space-time's microscopic inhomogeneous Maxwell equation and its $J^b$. Note that our identification of points on each manifold allows us to set
\begin{equation}
	A_a^\text{lab}(x_\text{lab}) \equiv A_a^\text{eff}(x_\text{eff}) \equiv A_a(x),
\end{equation}
with a bit of abuse of notation, and application of the earlier mentioned tremendous simplification in the transformations considered. (The idea is to maximise generality as much as possible while keeping notational clutter for the moment to a minimum.)

As our look into bespoke meta-materials (see section~\ref{sec:bespoke}) shows, as well as the work done previously by Fathi, Frauendiener, and Thompson in transformation optics (see references~\cite{CovOptMet3,TrafoOpticsCartographDistort}), this is not enough. The analysis done below will \emph{have to be} extended to the more general case of looking into mappings $x_\text{lab}(x_\text{eff})$ and the resulting cartographic distortions. At the time of writing, this remains future work. Again we refer to figure~\ref{fig:analoguecoords} for a depiction of the issue. 

Let us begin tackling the indices:
\begin{subequations}
	\begin{align}
		\nabla_{a}^\text{lab} Z^{abcd} F_{cd}^\text{lab} &= J_\text{lab}^b\\
		&= \kl{\nabla_{a}^\text{lab} Z^{abcd}} F_{cd}^\text{lab} + Z^{abcd} \nabla_{a}^\text{lab} F_{cd}^\text{lab},\\
		&= \cancelto{0}{\kl{\nabla_{a}^\text{eff} Z^{abcd}} F_{cd}^\text{lab}} + \Delta^a{}_{ma} Z^{mbcd} F_{cd}^\text{lab} + \Delta^b{}_{ma} Z^{amcd} F_{cd}^\text{lab}\nonumber\\&\qquad + \Delta{}^c_{ma} Z^{abmd} F_{cd}^\text{lab}  + \Delta^d{}_{ma} Z^{abcm} F_{cd}^\text{lab} + Z^{abcd} \nabla_{a}^\text{lab} F_{cd}^\text{lab},\label{eq:cancel}\\
		&=  \Delta^a{}_{ma} Z^{mbcd} F_{cd}^\text{lab} + \Delta^b{}_{ma} Z^{amcd} F_{cd}^\text{lab} + \Delta{}^c_{ma} Z^{abmd} F_{cd}^\text{lab} + \Delta^d{}_{ma} Z^{abcm} F_{cd}^\text{lab} \nonumber\\&\qquad - \Delta^m{}_{ca} Z^{abcd} F_{md}^\text{lab} -\Delta^m{}_{da} Z^{abcd} F_{cm}^\text{lab} + Z^{abcd} \nabla_a^\text{eff} F_{cm}^\text{lab},\\
		&= J^b_\text{eff} + \text{terms dependent on $\Gamma_\text{eff}$ and its coupling to }F^\text{eff} \nonumber\\&\qquad+\Delta^a{}_{ma} Z^{mbcd} F_{cd}^\text{lab} + \Delta^b{}_{ma} Z^{amcd} F_{cd}^\text{lab} + \Delta{}^c_{ma} Z^{abmd} F_{cd}^\text{lab} \nonumber\\&\qquad + \Delta^d{}_{ma} Z^{abcm} F_{cd}^\text{lab} - \Delta^m{}_{ca} Z^{abcd} F_{md}^\text{lab} -\Delta^m{}_{da} Z^{abcd} F_{cm}^\text{lab}.\label{eq:Ztangle}
	\end{align}
\end{subequations}
Note that with the abbreviation $\Delta^a{}_{bc}$ the torsion only enters in one additional term, when included naively. It thus seems sensible to open up the discussion of analogue space-times mimicking electrodynamics with a coupling of effective gravity (as encoded in torsion, non-metricity, curvature). The additional structural change in the equations involved seems minimal, provided one couples the connection (the \emph{effective} gravitational sector) in simple ways to the \emph{effective} electromagnetic fields. However, as discussed in \cite{HehlObukTorsionHowToEdyn}, most simple ways of coupling the gravitational sector with the electrodynamical sector are fraught with problems. Many of the corresponding proposals would violate either electric charge or magnetic flux conservation. There still are possible ways without violating other well-established cornerstones of classical electrodynamics, but because of the great freedom in parameter space for this, we only had these terms appear as \enquote{terms dependent on $\Gamma_\text{eff}$ and its coupling to $F^\text{eff}$}.

After that choice it would then also be possible to rephrase the result~\eqref{eq:Ztangle} in a variety of ways, depending on which relationship one wishes to bring to the forefront. For example, couplings to the curvature tensors like the Ricci tensor $R_{ab}^\text{eff, LC}$ of the effective metric with respect to its Levi-Civita connection $\begin{Bmatrix}a\\bc\end{Bmatrix}_{\text{eff}}$ could appear. This is not surprising, as it has to be included in curved space-time electrodynamics in the first place (if one can write out the field strength tensor using a four-potential $A_a$ and a covariant derivative) by use of the Weitzenböck identity \cite{GriHar,LawMichSpin} (see also the discussion on page~86 in \cite{BirrellDavies}). Among the many relationships one could pull out of the full expression~\eqref{eq:Ztangle} would be several ones relating it back to other curvature tensors. For example, if one defines the Ricci tensor of the laboratory connection $\Gamma_\text{lab}$ \emph{in the effective Lorentzian manifold} as $R_{ab}^{\text{eff, lab}}$, one can relate the two according to
\begin{equation}
	R_{ab}^\text{eff, LC} = R_{ab}^{\text{eff, lab}} - \nabla_c^\text{eff} \Delta^c{}_{ab} + \nabla_{a}^\text{eff}\Delta^c{}_{cb} - \Delta^c{}_{cd} \Delta^d{}_{ab} + T^c{}_{ad} \Delta^d{}_{cb},
\end{equation}
see equation (4.23) on page~141 in \cite{Schouten}. Similarly, one could ask the question the other way around and look for the Ricci tensor of the Levi-Civita connection belonging to the background metric --- though in all cases considered in this thesis that would result in $0$, as we assume the background laboratory metric to be flat. It might be possible to simplify our result~\eqref{eq:Ztangle} further with this fact. It is here that also the Ricci identity~\eqref{eq:Ricci} can be used to make curvature tensors appear in equation~\eqref{eq:Ztangle}.

While this all seems not particularly helpful, it \emph{does} show that the Maxwell equations of the laboratory and those of the effective space-time are non-trivially connected. This was this section's main point. Before one could claim having measured the analogue space-time electrodynamics of this \emph{algebraic analogue}, one would have to:
\begin{enumerate}
	\item Repeat the above analysis for the correct transformation of laboratory coordinates into effective coordinates, $x_\text{eff}(x_\text{lab})$.
	\item Then disentangle the residual laboratory dynamics encoded in $\Delta^a{}_{bc}$ from the effective space-time dynamics one is interested in.
	\item Finally, compare the measured laboratory dynamics with this and see if the analogue space-time dynamics can be deduced from the data.
\end{enumerate}
Only after this arduous procedure can definitive statements be made about possible detection of analogue space-time electrodynamics, as intended. This becomes even more pronounced if one wanted to look at Hawking radiation in the effective space-time (as this would require additionally a look at quantised fields). And again, this cannot be avoided: Non-metricity arises trivially, unless the effective metric also happens to be flat --- in which case most, if not all reasons for considering them are rendered moot. Lastly, one additional warning: The discussion presented above is \emph{before} one starts including cartographic distortions, too. Note that analytic analogue space-times are less prone to issues of the kind encountered in this section.

\section{Separation Ansatz and Refractive Indices}\label{sec:refractive}
In this section, we shall give an alternative access to models providing analogies building on electromagnetism. Concretely, we will this time work in the analytic analogy of linking the dynamics of PDEs in GR with that of PDEs in electromagnetism. This will lead to refractive index profiles analogous to wave propagation in a curved space-time. It is important to understand that this will not lead to an analogue space-time. Rather it will only constitute an analogy to a part of curved space-time physics, in fact just to a part of physics in a particular metric or class of metrics. It will lack a full geometric picture. We will not use in this thesis the word \enquote{analogue} for this suggested ansatz, and rather refer to it as an \enquote{analogy}.

This approach was inspired by the work of Heading and Westcott \cite{HeadingHyperGeo,WestcottN,HeadingSingFreeIndices}. Their idea is to find exactly solvable (using available special functions) refractive index profiles in horizontally \cite{HeadingHyperGeo} or spherically \cite{WestcottN} stratified media by looking at the wave equation for the part of the electric/magnetic field propagating orthogonally to the stratification layers. Much of their analysis was dedicated to finding further, exactly solvable profiles beyond those known at their time. In our case, the task is framed in a slightly different way: Starting from the wave equation for massless perturbations of the Kerr space-time, we use its separability to reduce the problem to a system of ordinary differential equations. Looking at the \enquote{radial} equation, we then try to recognize the form of a one-dimensional Helmholtz equation for stratified media, which can be written in terms of a refractive index depending on one variable.

There does exist an extension of the methods of Heading and Westcott in \cite{HeunRefIndex} to tackle second order Fuchsian equations with four singular points, that is, to functions of the Heun class, see appendix~\ref{sec:Heun}. We shall follow a slightly different approach which will be less general, and more immediately related to our intended application in the context of analogue space-times.

Following \cite{Bremmer49}, and \cite{MoonSpencer}, such \enquote{target} Helmholtz equations can be found, for example, either in spherical symmetry, or in cylindrical symmetry. We shall focus on these. Regarding the horizontal stratification (\emph{i.e.}, in $z$-direction) in Cartesian coordinates, \cite{BornWolf} gives the resulting Helmholtz equation on pages~55--56. If the stratification of $n$ is pushed solely into the permittivity this is turned into a special case of the discussion to come below. Once the required formalism is in place, we will give the final detail regarding this.

For spherical coordinates the scalar Helmholtz equation can be derived as a scalar equation to be fulfilled by a radially directed Hertzian vector $\vec{\Pi}$. This gives two possible scalars --- one for an electrical dipole in radial direction, $\Pi_{\text{e}}$, and one for the case of a magnetic dipole in radial direction, $\Pi_{\text{m}}$. More concretely, the corresponding electromagnetic fields are determined by
\begin{subequations}
	\begin{align}
		\vec{E} &= \frac{c}{\omega^2 n^2(r)} \nabla \times \nabla \times (\vec{r} \omega/c \Pi_{\text{e}}) e^{-i\omega t},\\
		\vec{H} &= -i \nabla \times (\vec{r} \omega/c \Pi_{\text{e}}) e^{-i\omega t}
	\end{align}
\end{subequations}
for the electric case, and for the magnetic one in the following way:
\begin{subequations}
	\begin{align}
		\vec{E} &= i\frac{c}{\omega n^2(r)} \nabla \times (\vec{r} (\omega/c)^2 \Pi_{\text{m}}) e^{-i\omega t},\\
		\vec{H} &= c^2 \frac{c}{\omega^2 n^2(r)} \nabla \times \frac{\nabla \times (\vec{r} \omega/c \Pi_{\text{e}})}{\omega^2 n^2(r)} e^{-i\omega t}.
	\end{align}
\end{subequations}
The spherical scalar Helmholtz equation with refractive index $n(r)$ then reads
\begin{equation}\label{eq:scHeqnEsph}
	\frac{\dif^2}{\dif r^2}\Pi_{\text{e}} + \kl{\frac{\omega^2n^2(r)}{c^2} - n(r)\frac{\dif^2}{\dif r^2}\kl{\ed{n(r)}} - \frac{l(l+1)}{r^2}}\Pi_{\text{e}}=0
\end{equation}
for the electric case, and
\begin{equation}\label{eq:scHeqnMsph}
	\frac{\dif^2}{\dif r^2}\Pi_{\text{m}} + \kl{\frac{\omega^2n^2(r)}{c^2} - \frac{l(l+1)}{r^2}}\Pi_{\text{m}}=0
\end{equation}
for the magnetic one. Note the occurrence of the separation constant $l(l+1)$ from the separation of variables of the scalar Helmholtz equation. These $l$ can be given the usual interpretation in terms of angular momentum quantum numbers.

In the case of cylindrical coordinates $(r,\theta,z)$, we take a slightly different path and we start by looking at fields independent of $z$ (which is justified as it is the $r$-direction where we are looking for stratification): The $z$-components for $\vec{H}$ and $\vec{E}$ then can be decomposed as
\begin{subequations}
	\begin{align}
		E_z &= \sum_{l=0}^{\infty} a_l r^{-1/2} f_l(r) \cos(l\theta),\\
		H_z &= \sum_{l=0}^{\infty} b_l r^{-1/2} g_l(r) \cos(l\theta).
	\end{align}
\end{subequations}
Here, $a_l, b_l$ are constants, and $f_l(r), g_l(r)$ fulfil the following scalar Helmholtz equations \cite{BurmanCylProp,WestcottCylProp}:
\begin{align}
	\frac{\dif^2 f_l(r)}{\dif r^2} + \kl{\frac{\omega^2}{c^2}n^2(r) - \frac{l^2-\ed{4}}{r^2}}f_l(r) &= 0,\label{eq:scHeqnEcyl}\\
	\frac{\dif^2 g_l(r)}{\dif r^2} + \kl{\frac{\omega^2}{c^2}n^2(r) -\frac{\omega}{cr}\frac{\dif}{\dif r}\kl{r\frac{\dif}{\dif r}\kl{\frac{c}{\omega n(r)}}} - \frac{l^2-\ed{4}}{r^2}}g_l(r) &= 0.\label{eq:scHeqnMcyl}
\end{align}

In principle, one could also try to look at the scalar Helmholtz equation in spheroidal symmetry --- a separation of variables is also possible in this case. However, special care has to be taken when considering this last case: As the vector Helmholtz equation is not separable in spheroidal coordinates, the origin of a spheroidal \emph{scalar} Helmholtz equation in this essentially electromagnetic analogy would have to be carefully justified. It is not unlikely, though, that in a different physical setting an application also for the spheroidal scalar Helmholtz equation can be found along the lines of the following discussion for the first two cases.

All four equations are of the same form, as will be the one granting us our analogy later on. However, looking at equations~(\ref{eq:scHeqnEsph},\ref{eq:scHeqnMcyl}), matching the coefficients of the function $g_l$ or $f_l$ appearing with another ODE of the same type would create a second order ODE for $n(r)$. We will therefore be particularly interested in the cases of equation~\eqref{eq:scHeqnEcyl}, and of equation~\eqref{eq:scHeqnMsph}. Both are of the form
\begin{equation}
	\frac{\dif^2 f(r)}{\dif r^2} + \kl{\frac{\omega^2}{c^2}n^2(r) - \frac{D}{r^2}}f(r) = 0,\label{eq:refindD}
\end{equation}
though for different, but constant values of $D$ and different dependent functions $f(r)$. The horizontal stratification à la Born and Wolf \cite{BornWolf}, with $\mu$ chosen $z$-independent, appears here as a special case of setting $D$ equal to $0$. This case is covered by the spherical Helmholtz equation for $M$, when setting $l=0$.\footnote{From a more technical point of view, one could also change the variable in \cite{BornWolf}, equation~(6), page~56, to get rid of the first-derivative piece with standard methods from the theory of second order ODEs; but we are neither interested in keeping separately track of $\epsilon$ and $\mu$, nor would the resulting ODE have quite the appearance of equation~\eqref{eq:refindD} on which we shall focus.}

To turn any of these scalar Helmholtz equations into a simple analogy to curved space-time physics, we have to take a closer look at the wave equation for massless fields in the Kerr space-time. This wave equation is given by the Teukolsky equation, and we will be following its presentation in \cite{FroNo98},
\begin{align}\label{eq:Teukolsky}
	0=& \kle{\frac{(r^2+a^2)^2}{\Delta} - a^2\sin^2\theta} \frac{\partial^2\Psi}{\partial t^2} + \frac{4Mar}{\Delta}\frac{\partial^2 \Psi}{\partial t \partial \phi} + \kle{\frac{a^2}{\Delta} - \ed{\sin^2\theta}}\frac{\partial^2\Psi}{\partial \phi^2}\nonumber\\
	&-\Delta^{-s} \frac{\partial}{\partial r}\kl{\Delta^{s+1}\frac{\partial\Psi}{\partial r}} - \ed{\sin\theta} \frac{\partial}{\partial\theta} \kl{\sin\theta \frac{\partial \Psi}{\partial \theta}} - 2s\kle{\frac{a(r-M)}{\Delta} + i\frac{\cos\theta}{\sin^2\theta}}\frac{\partial\Psi}{\partial\phi}\nonumber\\
	&-2s \kle{\frac{M(r^2-a^2)}{\Delta} - r -ia\cos\theta}\frac{\partial\Psi}{\partial t} + (s^2\cot^2\theta - s)\Psi.
\end{align}
Here, $s$ is the so-called spin-weight, while the other terms are defined as in section~\ref{sec:Kerr}. The Teukolsky equation describes massless (classical) fields of helicity $s$. Usually, these are interpreted as perturbations of the corresponding metric, and play a role in stability analyses of the metric as a solution of the Einstein equations. We, however, shall in this section consider them as an approximation to the propagation of (massless) radiation emitted or scattered by the black hole --- this is just a rephrasing of the question that generated the Teukolsky equation in the context of black hole stability analysis. Correspondingly, it is just as important for scattering problems, including, but not limited to, the finding of quasi-normal modes.

In order to further distinguish the analogy from the physical laboratory in which the Helmholtz equation occurs, we omit physical constants from the former, while keeping the speed of light $c$ in the latter.

While more general versions of the Teukolsky equation exist --- the separation of variables works in any Petrov type $D$ space-time \cite{GalErt89,HeunTypeD}\footnote{Kodama and Ishibashi further generalised this to higher dimensions in work culminating in \cite{KodamaIshibashiMasterEqn3}.} ---, we shall be concerned as a proof of concept with the special case of the Kerr solution, as described above. These more general differential equations can be distinguished from the version above by referring to them as the \enquote{(Teukolsky) master equation} instead.

The Teukolsky equation~\eqref{eq:Teukolsky} can be separated using the following mode decomposition of $\Psi$:
\begin{subequations}
	\begin{align}
		\Psi &= \sum_{\ell =0}^{\infty} \sum_{m=-\ell}^{\ell} \vphantom{\Psi}_s\Psi_{\ell m},\\
		&= \sum_{\ell =0}^{\infty} \sum_{m=-\ell}^{\ell} \vphantom{R}_sR_{\ell m}(r,\omega) \vphantom{Z}_sZ_{\ell m}(\theta,\phi) e^{-i\omega t},\\
		&= \sum_{\ell =0}^{\infty} \sum_{m=-\ell}^{\ell} \vphantom{R}_sR_{\ell m}(r,\omega) \ed{\sqrt{2\pi}}\vphantom{S}_sS_{\ell m}(\theta) e^{im\phi} e^{-i\omega t}.
	\end{align}
\end{subequations}
The solution for both $\vphantom{R}_sR_{\ell m}(r,\omega)$ and for $\vphantom{S}_sS_{\ell m}(\theta)$ would involve confluent Heun functions, see appendix~\ref{sec:Heun} and \cite{SlavyanovLay00}.

Inserting this ansatz, we get the following equation for $\vphantom{S}_sS_{\ell m}(\theta)$:
\begin{align}
	0=&\ed{\sin\theta} \frac{\dif}{\dif \theta}\kl{\sin\theta\frac{\dif \vphantom{S}_sS_{\ell m}(\theta)}{\dif \theta}} + \kl{E_{\ell m} -s^2}\vphantom{S}_sS_{\ell m}(\theta)\nonumber\\
	& + \kl{a^2\omega^2\cos^2\theta - \frac{m^2+2ms\cos\theta + s^2\sin^2\theta}{\sin^2\theta} - 2a\omega s \cos\theta }\vphantom{S}_sS_{\ell m}(\theta), 
\end{align}
while for $\vphantom{R}_sR_{\ell m}(r)$ we have
\begin{align}\label{eq:radialTeukolsky}
	0=& \Delta^{-s} \frac{\dif}{\dif r} \kl{\Delta^{s+1} \frac{\dif \vphantom{R}_sR_{\ell m}(r)}{\dif r}} + \kl{s(s+1) - E_{\ell m} - a^2\omega^2 + 2am\omega}\vphantom{R}_sR_{\ell m}(r)\nonumber\\
	& + \kl{\frac{[(r^2+a^2)\omega - am]^2 - 2is(r-M)[(r^2+a^2)\omega-am]}{\Delta} + 4is\omega r }\vphantom{R}_sR_{\ell m}(r),
\end{align}
where $E_{\ell m}$ is the separation constant. The values of $E_{\ell m}$ will be taken from \cite{TeukolskyPress74}, where a table of polynomial approximations in $a\omega$ can be found on page~454. Specifically, we will test two modes for photons. The choice fell on photons as they will still retain some fundamental relevance in this particular analogue model --- after all, it is build on macroscopic electromagnetism. The modes we want to look at are $\ell=1, m=0$ and $\ell=6,m=3$. This choice is arbitrary, any other mode would work equally well. The corresponding values for the separation constant $E_{\ell m}$ are
\begin{alignat}{5}
	E_{1,0} &= 2 &&+ 0.00281 a \omega &&- 0.413370 a^2 \omega^2 &&+ 0.021476 a^3 \omega^3 &&- 0.0335098 a^4 \omega^4 \nonumber\\&&&+ 0.0025402 a^5 \omega^5 &&+ 0.00032399 a^6 \omega^6,&&&&\\
	E_{6,3} &= 42&&- 0.14285 a \omega &&- 0.385851 a^2 \omega^2 &&+ 0.003204 a^3 \omega^3 &&+ 0.0002062 a^4 \omega^4 \nonumber\\&&&+ 0.0000197 a^5 \omega^5 &&+ 0.00000229 a^6 \omega^6,&&&&
\end{alignat}
The range of validity of these polynomial fits is $0\leq a\omega \lesssim 3$.

It is worth pointing out at this stage that we are \emph{not} looking at anything related to quasi-normal modes (QNM) \cite{QNMLRR}. The questions resulting in the search for QNM are independent of our aim. As a two-point connection problem in the sense of special function theory underlies QNM analyses (and thus more restrictive boundary conditions), the frequencies of QNM tend to be discrete. What we are looking at is rather the question of arbitrary incoming waves of arbitrary frequency. This corresponds to plane waves in the standard flat-space context. In this analogy, the conditions where QNM arise have the correspondence of the addition of boundary conditions to flat-space examples --- which also there can turn eigenfrequencies discrete.

Having separated the Teukolsky equation, we can now ask a more modest question in the same vein as our previous discussion of an actual (algebraic) space-time analogue: Instead of looking at an analogy for the full wave propagation on a curved background, we restrict ourselves to a mode-by-mode analysis. Then the separation allows us to look for one-dimensional analogies to a particular radial mode; each mode characterised by several separation constants. The analogue would then be simply given as a one-dimensional refractive index profile. 

However, and to our chagrin, the radial Teukolsky equation contains a \emph{complex} potential. While it is certainly not impossible to follow down the lines of complex refractive indices (and at least both Heading and Westcott later did, see \cite{HeadingComplex,WestcottComplex}) --- this would be a rather jarring contrast to our earlier statement of wanting to focus on real-valued optical properties to simplify the (algebraic) analogue space-time picture. It bears repeating that allowing optical properties to be complex-valued, and the resulting dispersion would quickly destroy the Lorentzian properties of the algebraic analogy in the sense of the earlier sections of this chapter. We ignore for a moment that wave equations in black hole space-times indeed exhibit absorption (due to backscattering), and thus would make complex refractive indices in the current context not too surprising. (Again, this is different from the purely algebraic analogy of the main part of this chapter!) To our knowledge, however, refractive indices are easier to manufacture to order if only the real part has to be matched. With the above complex potential, it therefore seems that we reached an impasse, or at least an unpleasant complication. Strictly speaking, the present analogy by no means requires real-valued optical properties, but in order to compare with the earlier results for the algebraic analogue having them would certainly help.

Luckily, however, at least for electromagnetic fields --- that is, $s=\pm1$ --- the Teukolsky equation allows for transformations turning the potential real, and thus available to our desired, one-dimensional space-time analogue. Let us quickly repeat here this process by Detweiler \cite{DetweilerPotential} (while skipping some intermediate steps\footnote{For example, we shall only implicitly use the intermediate step of relating $\vphantom{R}_sR_{\ell m}(r)$ and $\vphantom{R}_{-s}R_{\ell m}(r)$ through the Teukolsky--Starobinsky identities to explain the origin of the following transformations.}):

Instead of $\vphantom{R}_sR_{\ell m}(r)$, consider the following function:
\begin{subequations}
	\begin{align}
		\vphantom{\chi}_1\chi_{\ell m}(r) \defi& p(r) \vphantom{R}_1R_{\ell m}(r) + \frac{p(r)\Delta}{\sqrt{4\lambda_{-1}^2 - 16a^2\omega^2 + 16a\omega m}} \vphantom{R}_{-1}R_{\ell m}(r),\\ =&p(r)\kl{1 + \frac{\Delta}{\sqrt{4\lambda_{-1}^2 - 16a^2\omega^2 + 16a\omega m}}A(r)} \vphantom{R}_1R_{\ell m}(r)\nonumber\\
		&+\frac{p(r)\Delta}{\sqrt{4\lambda_{-1}^2 - 16a^2\omega^2 + 16a\omega m}}B(r)\frac{\dif\vphantom{R}_1R_{\ell m}(r)}{\dif r},
	\end{align}
	where
	\begin{align}
		p(r) \defi& \sqrt{\frac{\lambda_{-1}^2 - 4a^2\omega^2 + 4a\omega m}{\frac{2[am-\omega(r^2+a^2)]^2}{\Delta} -\lambda_{-1} + \sqrt{\lambda_{-1}^2 - 4a^2\omega^2 + 4a\omega m}}},\\
		\lambda_{-1} \defi& E + a^2\omega^2 -2a\omega m,\\
		A(r) \defi& \frac{2}{\Delta^2}\kle{2(am-\omega(r^2+a^2))^2 - \Delta(i\omega r + \lambda_{-1})},\\
		B(r) \defi& -4\frac{am-\omega(r^2+a^2)}{\Delta}i.
	\end{align}
\end{subequations}
With this redefined dependent variable it is now possible to rewrite the ODE for $r$ as
\begin{equation}\label{eq:chiODE1}
	\frac{\dif^2 \vphantom{\chi}_1\chi_{\ell m}}{\dif r^2} + V(r,\omega) \vphantom{\chi}_1\chi_{\ell m} = 0,
\end{equation}
with the potential defined as
\begin{equation}
	V(r,\omega) \defi \frac{[am-\omega(r^2+a^2)]^2}{\Delta^2} - \lambda_{-1} - \frac{[am-\omega(r^2+a^2)]p'' - 2r\omega p'}{(am-\omega[r^2+a^2])p}.
\end{equation}

At this stage it is already possible to make several comments: First, we notice that the complex variables have been pushed from the original ODE into the dependent variable $\vphantom{\chi}_1\chi_{\ell m}$. After this has been done, the complex nature of $\vphantom{\chi}_1\chi_{\ell m}$ can be ignored, as we are now looking only for solutions of equation~\eqref{eq:chiODE1}. If these solutions are complex is irrelevant: As the linear(!) differential operator has only real coefficients any non-trivially complex solution (meaning: both real and imaginary part are unequal 0) will necessarily lead to a second, linear independent solution formed from its complex conjugate --- and thus two linearly independent, \emph{real} solutions become available. Second, unlike the prototype equations~ (\ref{eq:scHeqnEsph}, \ref{eq:scHeqnMsph}, \ref{eq:scHeqnEcyl}, \ref{eq:scHeqnMcyl}) the refractive index is now frequency-dependent. While the resulting dispersion is likely to complicate the experimental realisation, the absence of mode mixing (thanks to the separation) will alleviate this deficit somewhat.

As an aside it is prudent to mention that this foundation on a single mode analysis will make it difficult to check the results reported in \cite{SuperradianceOrNotBig,SuperradianceOrNot} which claim reflection of wave packets instead of superradiance.

The next step is to separate in equation~\eqref{eq:chiODE1} the terms corresponding to the refractive index in our analogue target equations~(\ref{eq:scHeqnEsph}, \ref{eq:scHeqnMsph}, \ref{eq:scHeqnEcyl}, \ref{eq:scHeqnMcyl}) from the terms arising from the choice of coordinates or separation constants. As said before, since the spherical case for the E-field, equation~\eqref{eq:scHeqnEsph}, or the magnetic parts in cylindrical coordinates, equation~\eqref{eq:scHeqnMcyl}, will result in a prohibitively convoluted ODE determining the refractive index, we shall restrict ourselves to the $H$-field in the spherical case, equation~\eqref{eq:scHeqnMsph}, and the electric field in the cylindrical case, equation~\eqref{eq:scHeqnEcyl}. In these two cases, both equations are of the same form, as described above in equation~\eqref{eq:refindD}, with just minor changes in the precise form in which the corresponding separation constants appear: The relevant constant being $D=l(l+1)$ in the former, and $D=l^2-1/4$ in the latter case. Then the corresponding refractive index for creating an analogue for the radial part of the Teukolsky equation is given by
\begin{equation}
	n(r,\omega) = \frac{c}{\omega}\sqrt{V(r,\omega) + \frac{D}{r^2}}.
\end{equation}
Written in such a way, it seems straightforward to give exact expressions for $n(r,\omega)$. This is true, however straightforward does not mean \enquote{manageable}: The (by far and wide) easiest way to achieve an exact expression involves a computer algebra system. We employed both Maple~2017 and Mathematica~11.2 for this purpose.

For the limit $a\to 0$ (corresponding to the Schwarzschild case) the resulting refractive index reads
\begin{equation}
	n_\text{Schwarzschild}(r,\omega) = c\sqrt{\frac{(4D-27)M^2 + 2Mr(E-2D+11) + r^2\kl{r^2\omega^2 - E -4 +D}}{r^2 \omega^2 \kl{r-2M}^2}}
\end{equation}

In figures~\ref{fig:refinSchwarzschild1}--\ref{fig:refinKerr3} we can see the results for different values of modes --- both of the separated Teukolsky equation (encapsulated in the separation constant $E$) and the underlying, separated spherical Helmholtz equation (encapsulated in the constant $D$) ---, and angular momentum, as described in their corresponding captions. The mass was set to be $M=1$. With the exception of figure~\ref{fig:refinReIm} we chose to plot only the real part of the refractive index for ease of viewing. While the refractive index does turn purely imaginary for low frequencies, for any value of $\omega$ there is a neighbourhood of the horizon in which it becomes real and positive again. However, any approach of the horizon leads to a divergence of the refractive index. Changing the separation constant $D$ of the Helmholtz equation can change the low-frequency behaviour, as noticeable in figure~\ref{fig:refinKerr2}, if it \enquote{outperforms} the contributions of $E_{\ell m}$. The $\omega$ range of the fits is limited by the polynomial approximation used for $E_{\ell m}$. The exception was the Schwarzschild case, where the range of $\omega$ could be chosen freely, since $a=0$.

However, for any mass, any black hole rotation, any (non-zero) frequency, any mode, and any separation constant $D$, it holds that
\begin{equation}
	\lim\limits_{r\to\infty} n(r,\omega) = 1. 
\end{equation}
The refractive indices hence pass the necessary consistency check: At spatial infinity, the refractive index has to reproduce the asymptotic flatness corresponding to the flat space Minkowski value of $n=1$.

On the other hand, we saw that the refractive index diverges (to $+\infty$) when approaching the horizon. This is difficult to see in some of the figures, as the divergence close to $\omega\sim 0$ is very rapid and difficult to capture numerically. The numerical issues aside, this observation is in full agreement with the previously mentioned results of \cite{ReznikRefIndexHorizons} and section~\ref{sec:bespoke}: The optical properties have to diverge on the horizon/ergo-surface. In the case of the horizon, this mimics the vanishing coordinate speed of light there. For experimental purposes, this divergence constitutes the biggest bane of the proposed analogy --- the refractive index of available materials rarely exceeds 4, usually only achieved by using meta-materials and then only in small bandwidths. This may, however, be enough for exploring interesting physics in this analogy: As shown in \cite{BHQuantumAtmosphere}, interesting parts of the renormalised stress-energy tensor for the Hawking effect can be found safely away from the location of the horizon.

\begin{figure}
	\centering
	\includegraphics[width=.48\textwidth]{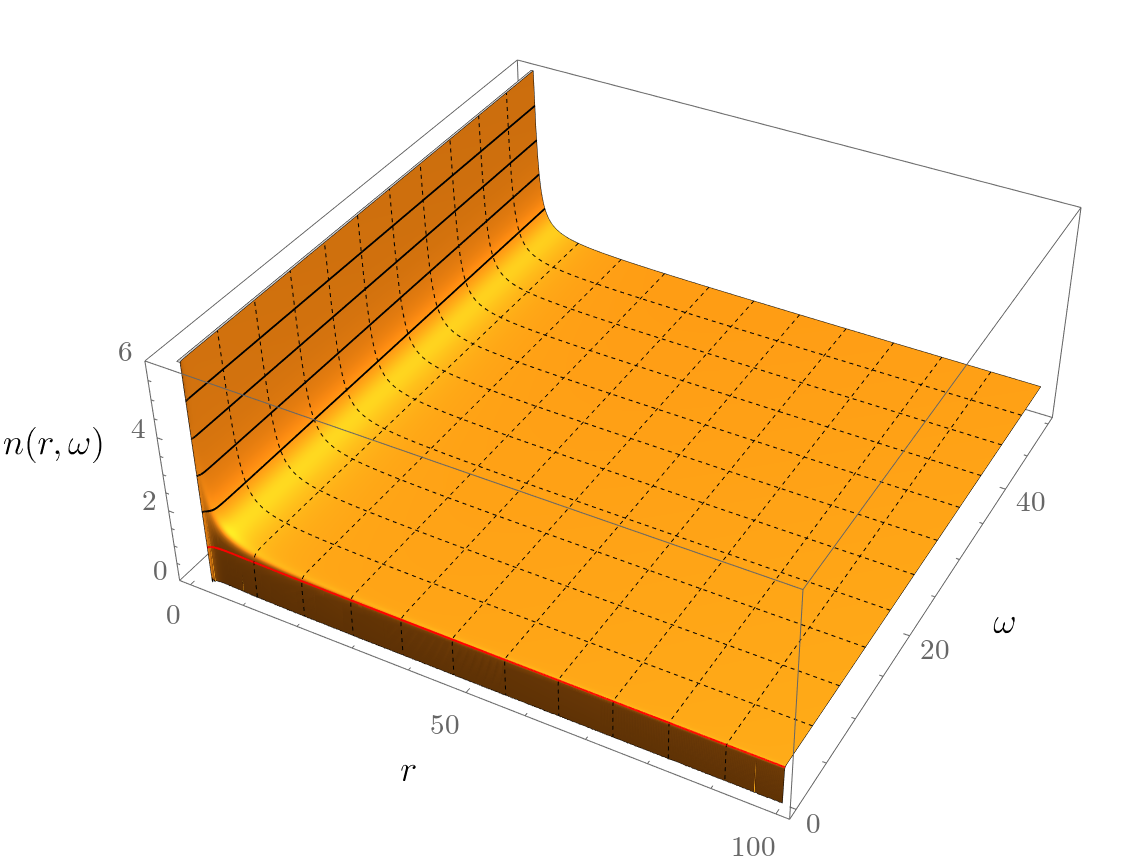}~\includegraphics[width=.48\textwidth]{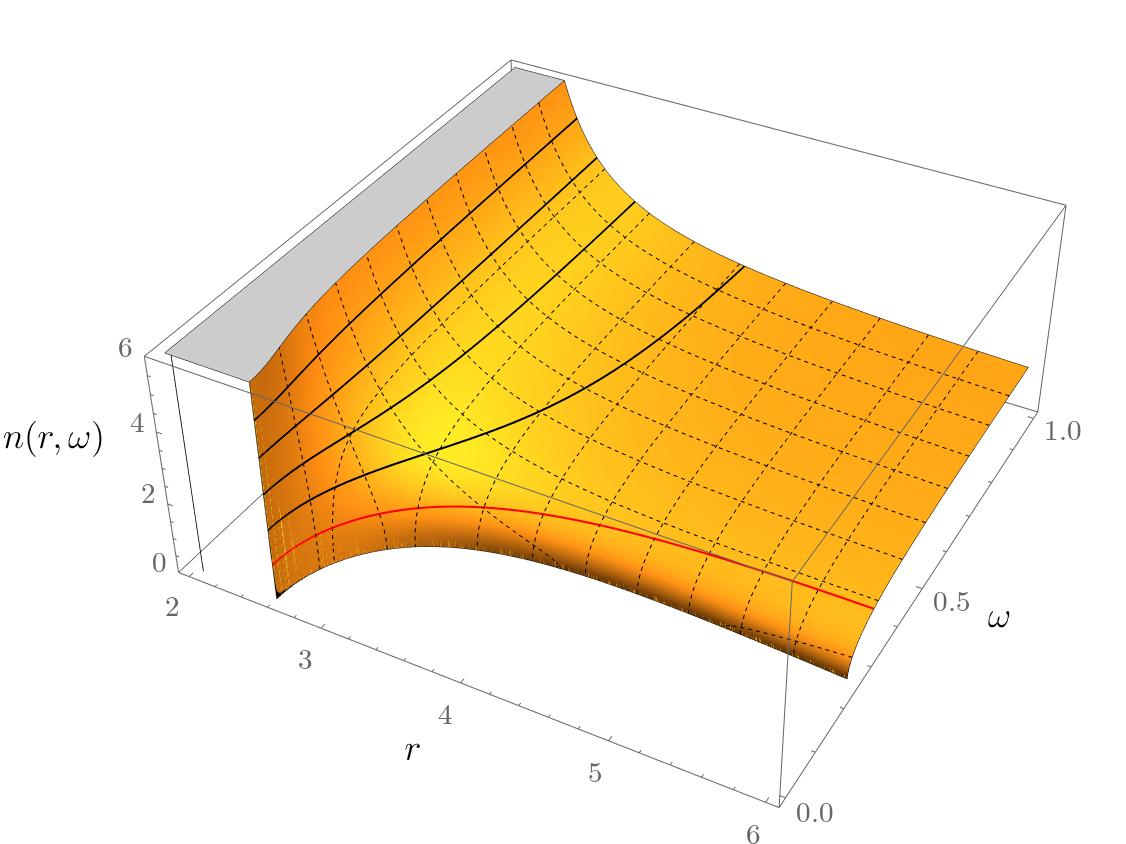}\\
	\includegraphics[width=.48\textwidth]{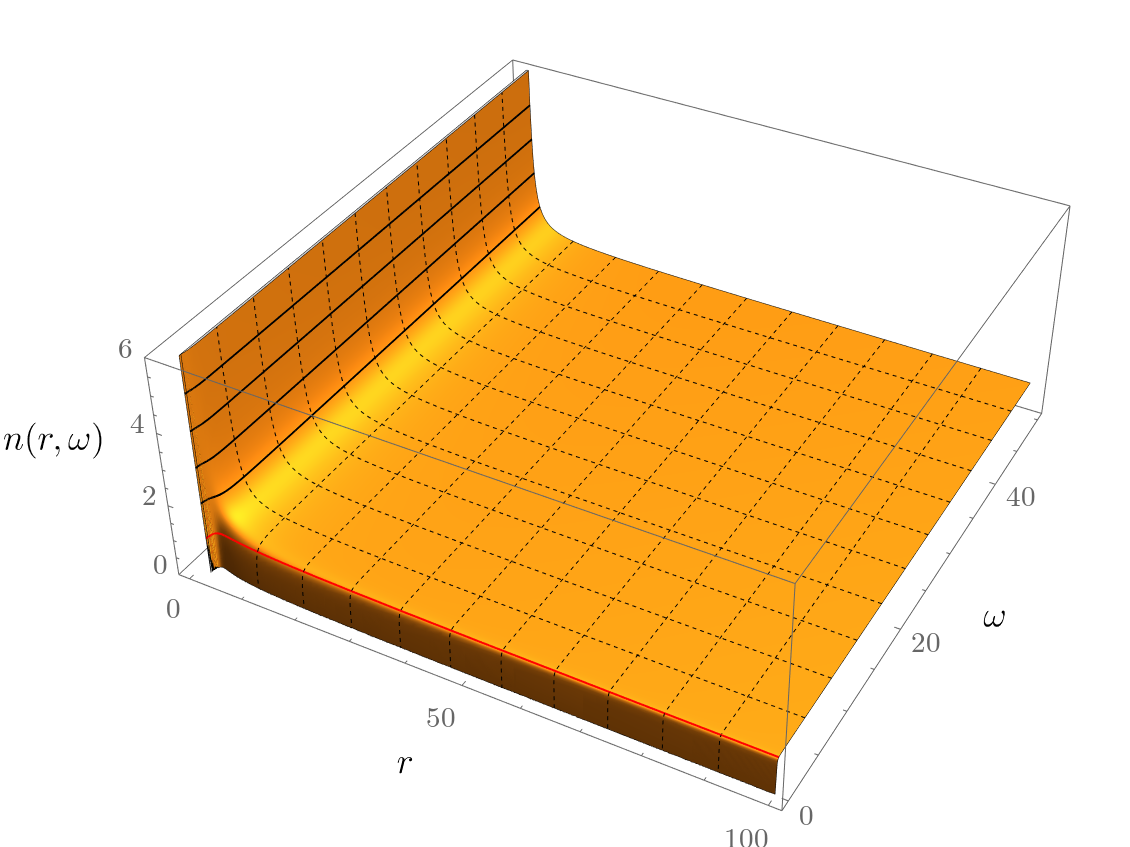}~\includegraphics[width=.48\textwidth]{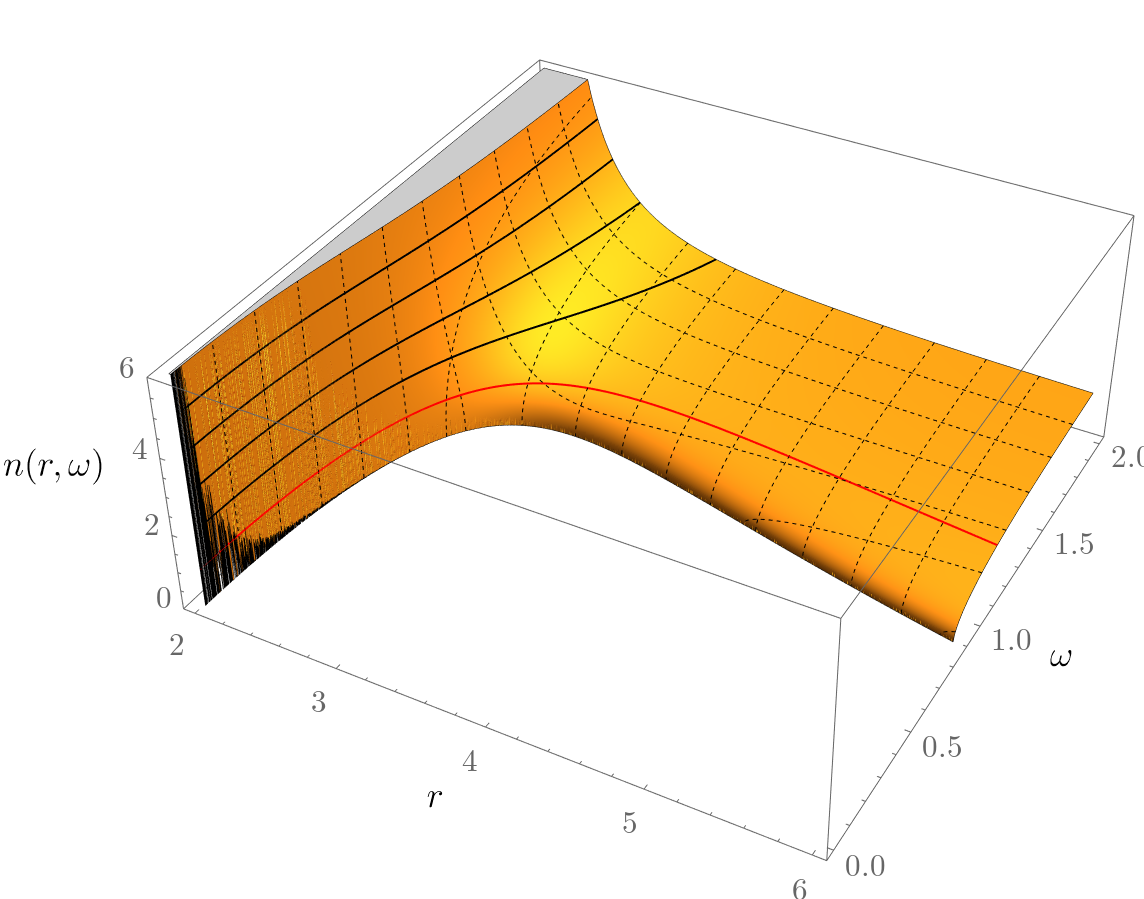}
	\caption[Plots of Indices of Refraction for Schwarzschild Analogy]{Indices of refraction for the Schwarzschild case, mass $M=1$, angular momentum $a=0$, and separation constants $D=0$. Dashed lines correspond to lines of constant $\omega$ or $r$. The red line is the contour of index of refraction equal to 1, that is, the vacuum value. Black lines are lines of constant $n\in\{2,3,4,5\}$, the cut-off is taken at $n=6$. \emph{In the left column:} A look at the larger picture. \emph{In the right column:} A zoom on smaller values of $r$ and $\omega$. \emph{Top row:} $l=1$ and $m=0$. \emph{In the bottom row:} $l=6$, $m=3$. The rugged structures visible in the bottom right plot for very small $r$ and $\omega$ are numerical artefacts.}
	\label{fig:refinSchwarzschild1}
\end{figure}

\begin{figure}
	\centering
	\includegraphics[width=.48\textwidth]{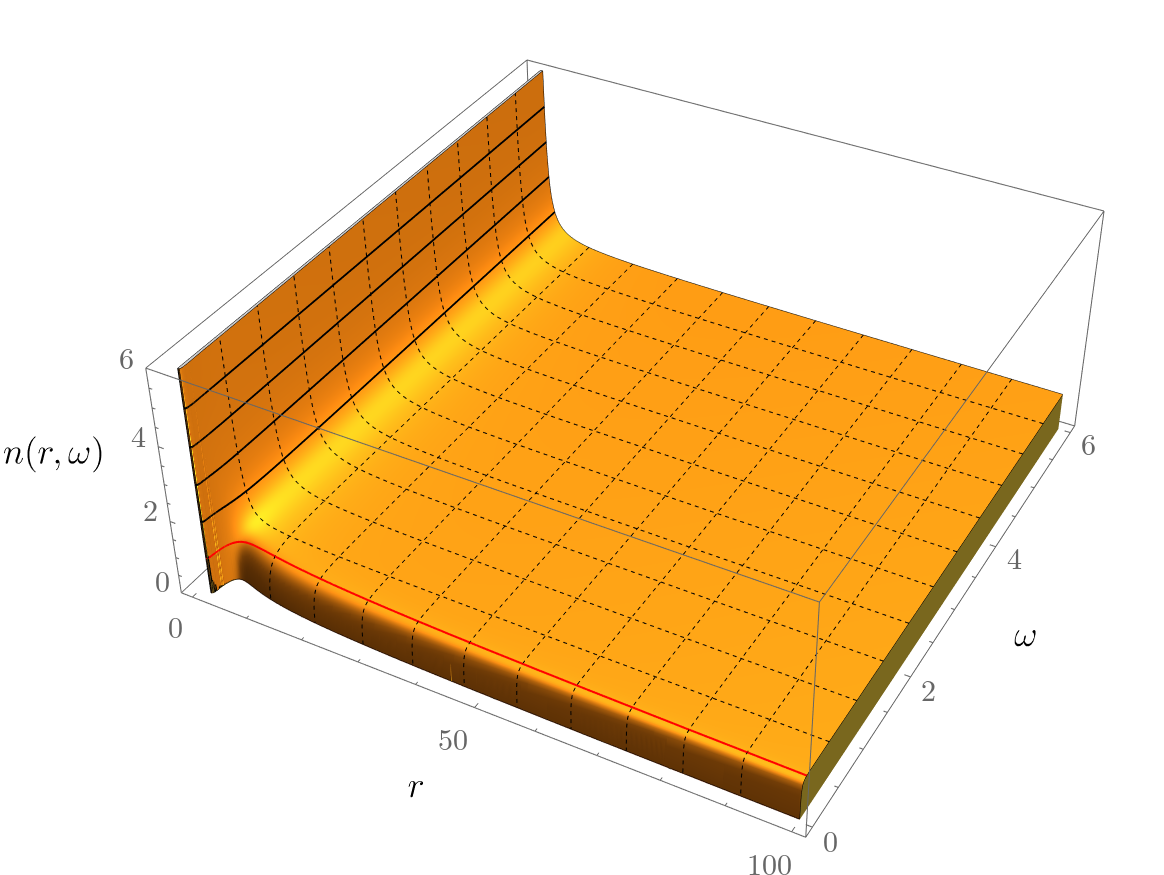}~\includegraphics[width=.48\textwidth]{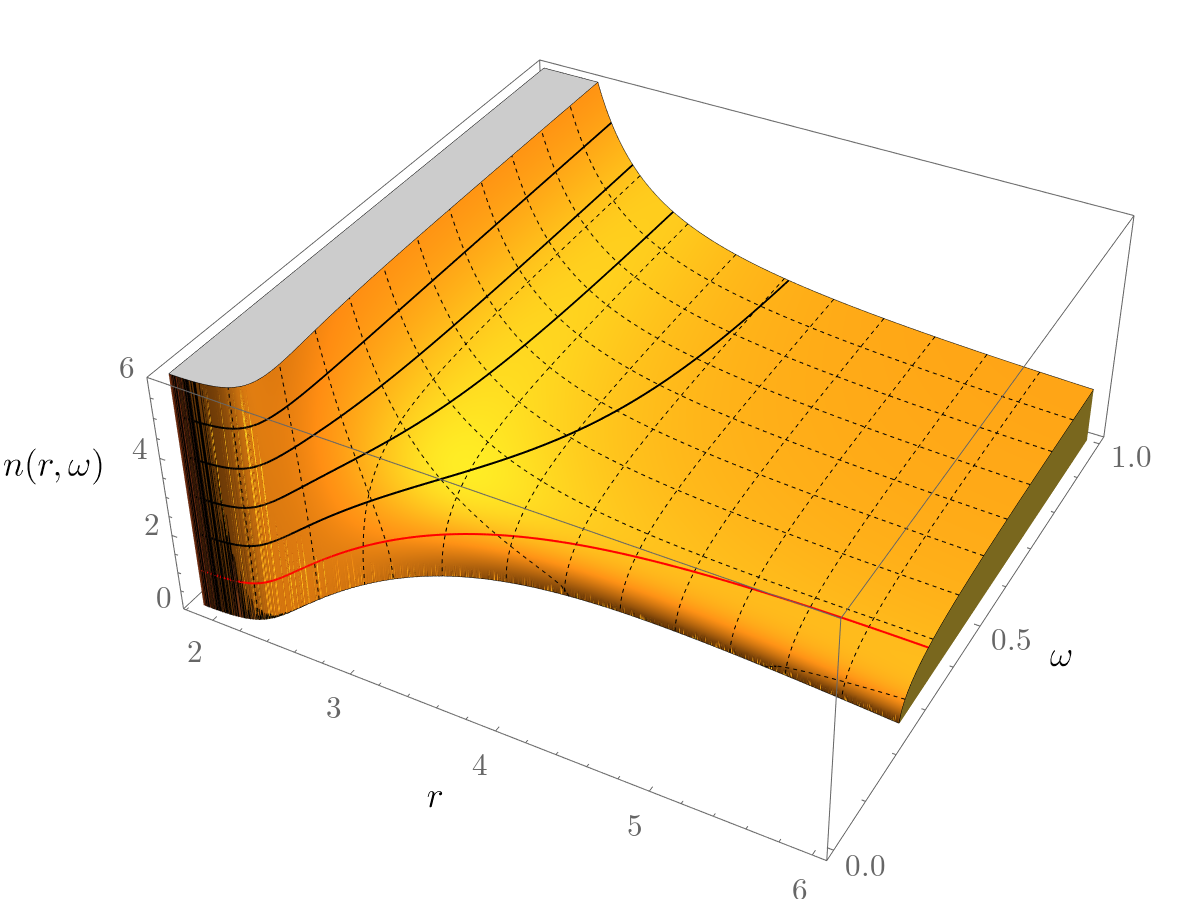}\\
	\includegraphics[width=.48\textwidth]{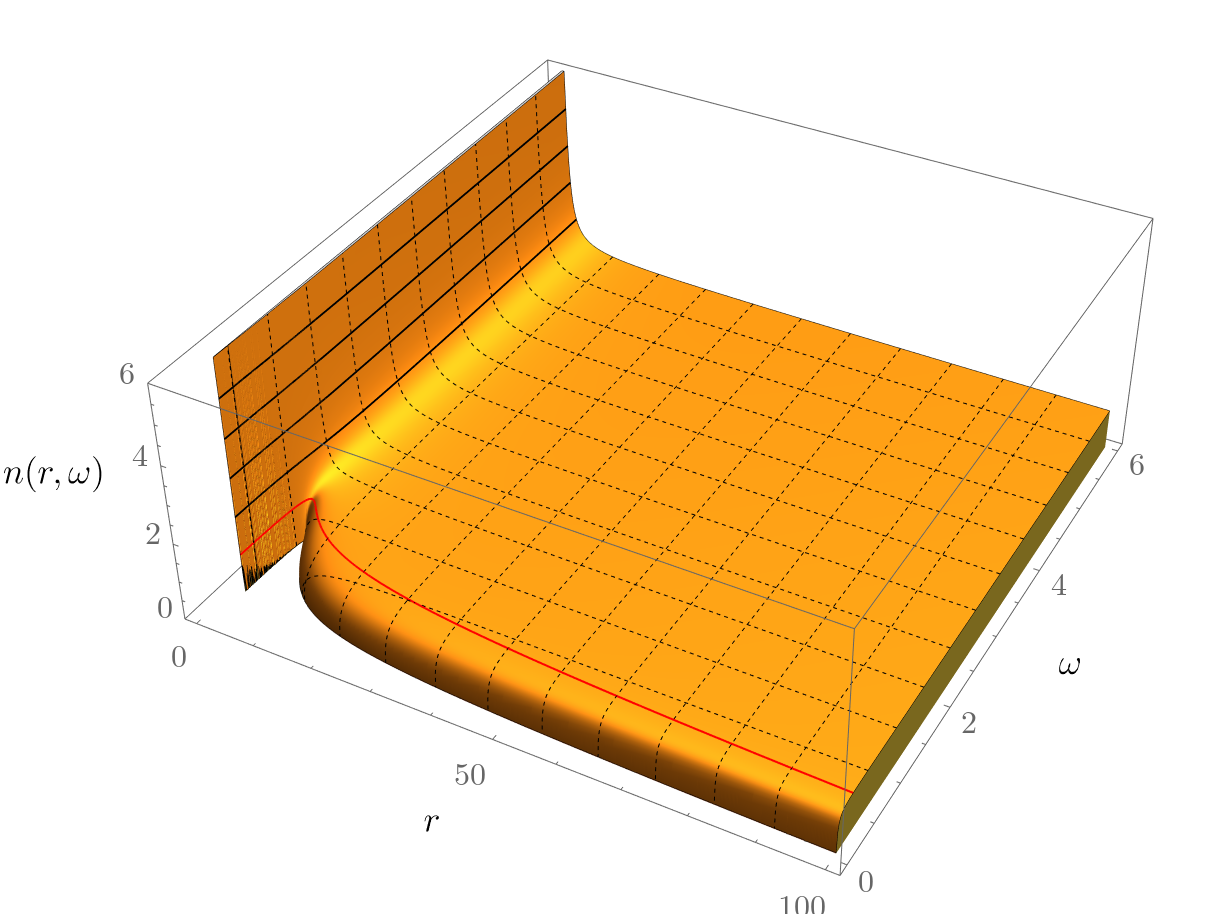}~\includegraphics[width=.48\textwidth]{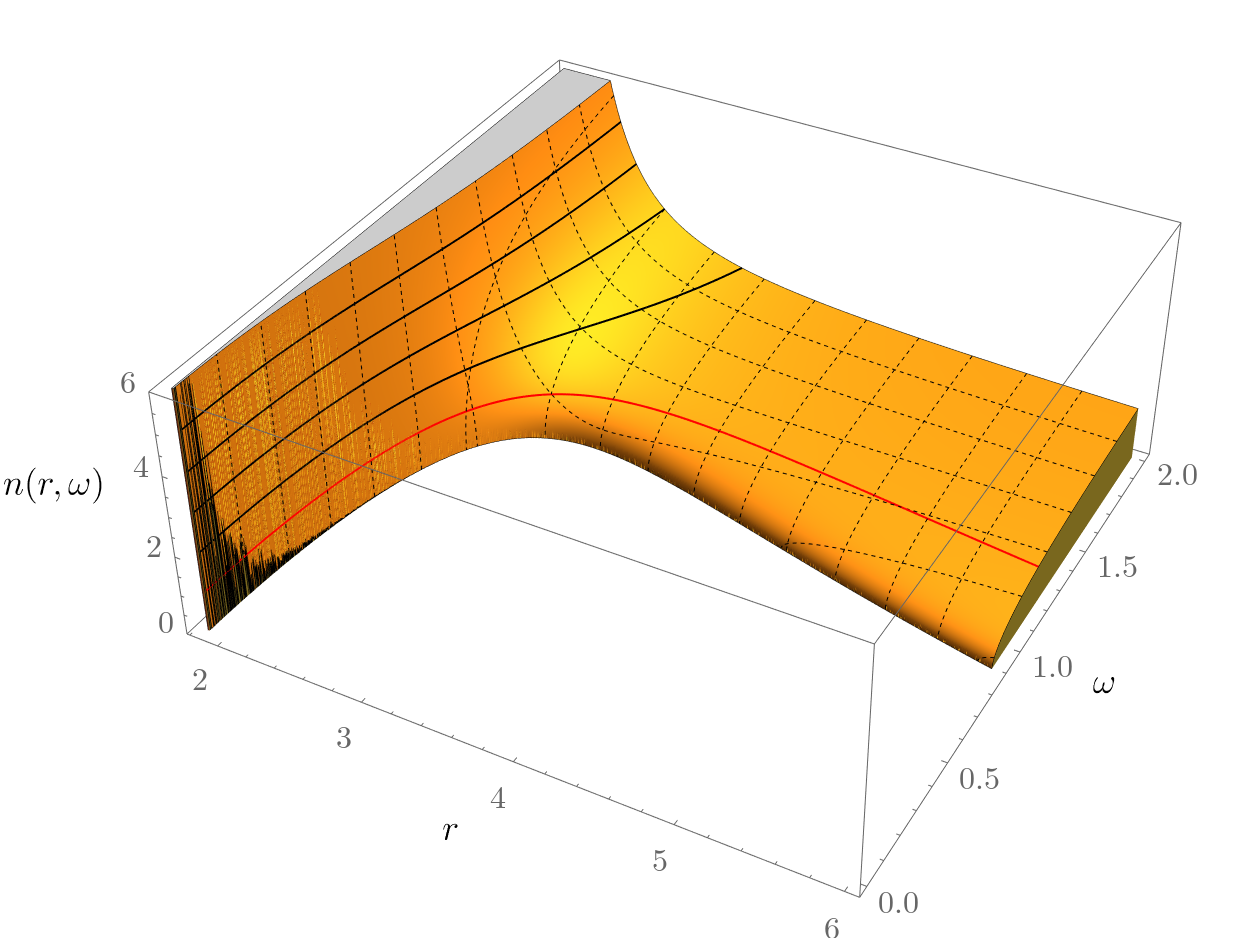}
	\caption[Plots of Indices of Refraction for Kerr Analogy, Part I]{Indices of refraction for the Kerr case, mass $M=1$, angular momentum $a=0.5$, and separation constants $D=0$. Dashed lines correspond to lines of constant $\omega$ or $r$. The red line is the contour of index of refraction equal to 1, that is, the vacuum value. Black lines are lines of constant $n\in\{2,3,4,5\}$, the cut-off is taken at $n=6$. \emph{In the left column:} A look at the larger picture. \emph{In the right column:} A zoom on smaller values of $r$ and $\omega$. \emph{Top row:} $l=1$ and $m=0$. \emph{In the bottom row:} $l=6$, $m=3$. As we used the data from \cite{TeukolskyPress74} for the separation constants of the Teukolsky equation, this time $\omega$ cannot exceed $6$. As the bottom right picture indicates, the fact that the diverging, real refractive index does not reach $\omega=0$ is a numerical artifact.}
	\label{fig:refinKerr1}
\end{figure}

\begin{figure}
	\centering
	\includegraphics[width=.48\textwidth]{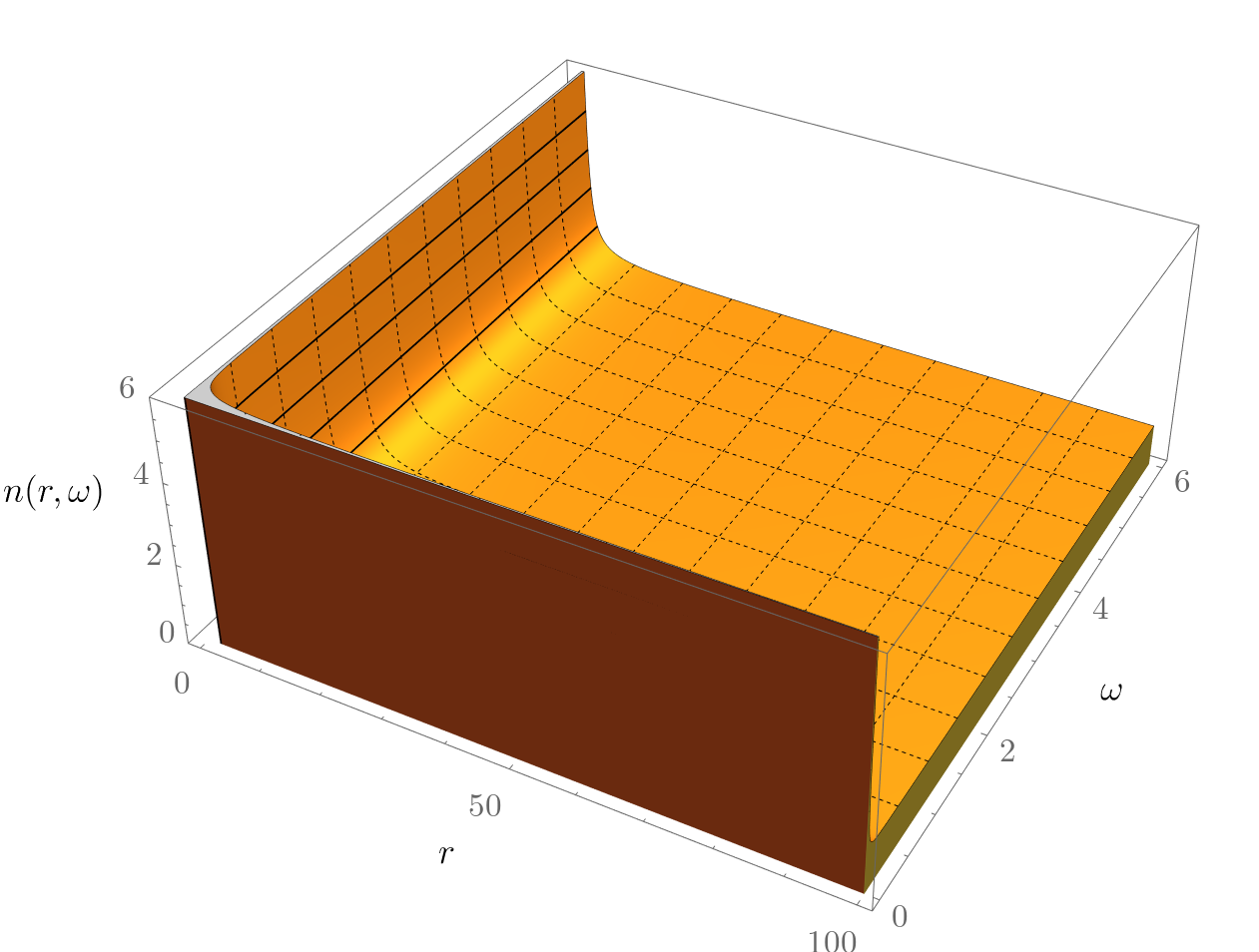}~\includegraphics[width=.48\textwidth]{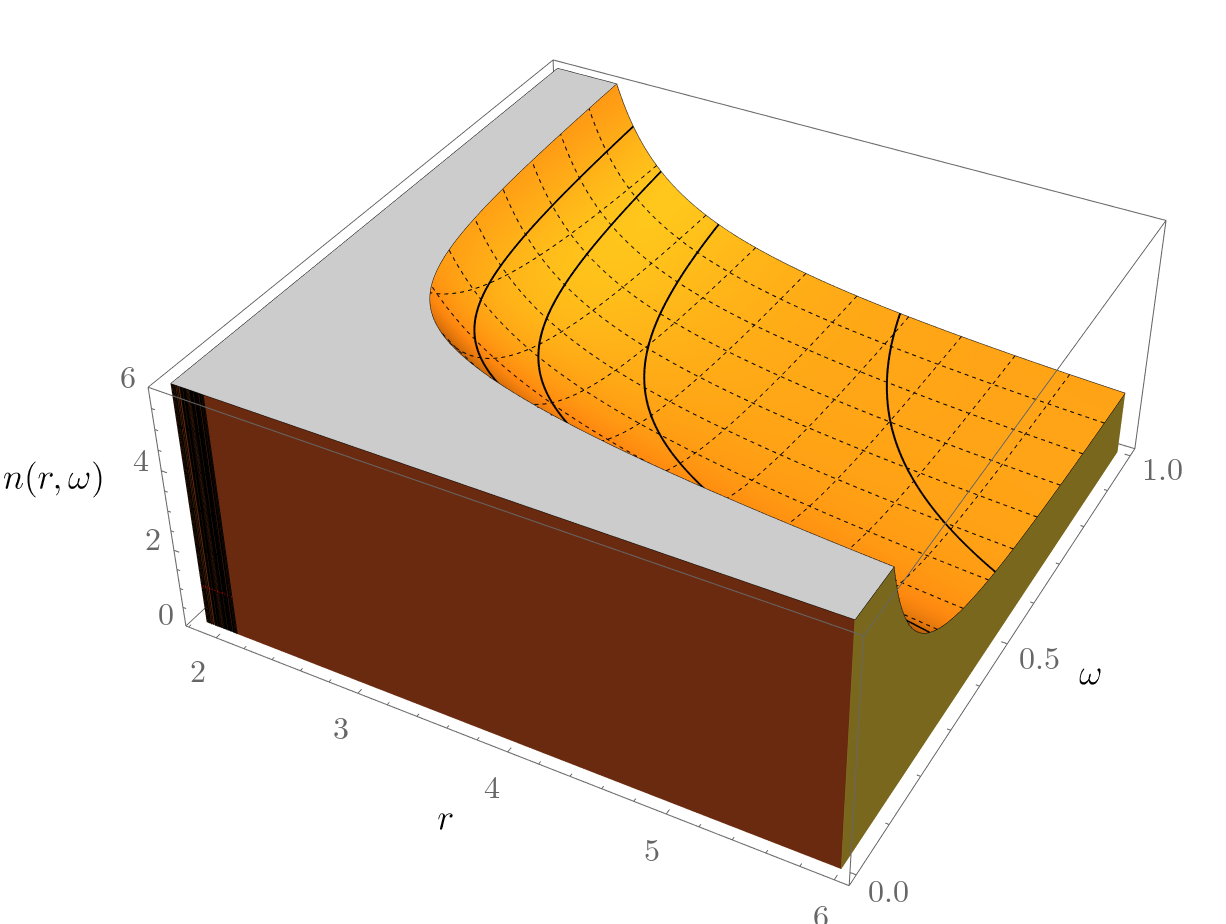}\\
	\includegraphics[width=.48\textwidth]{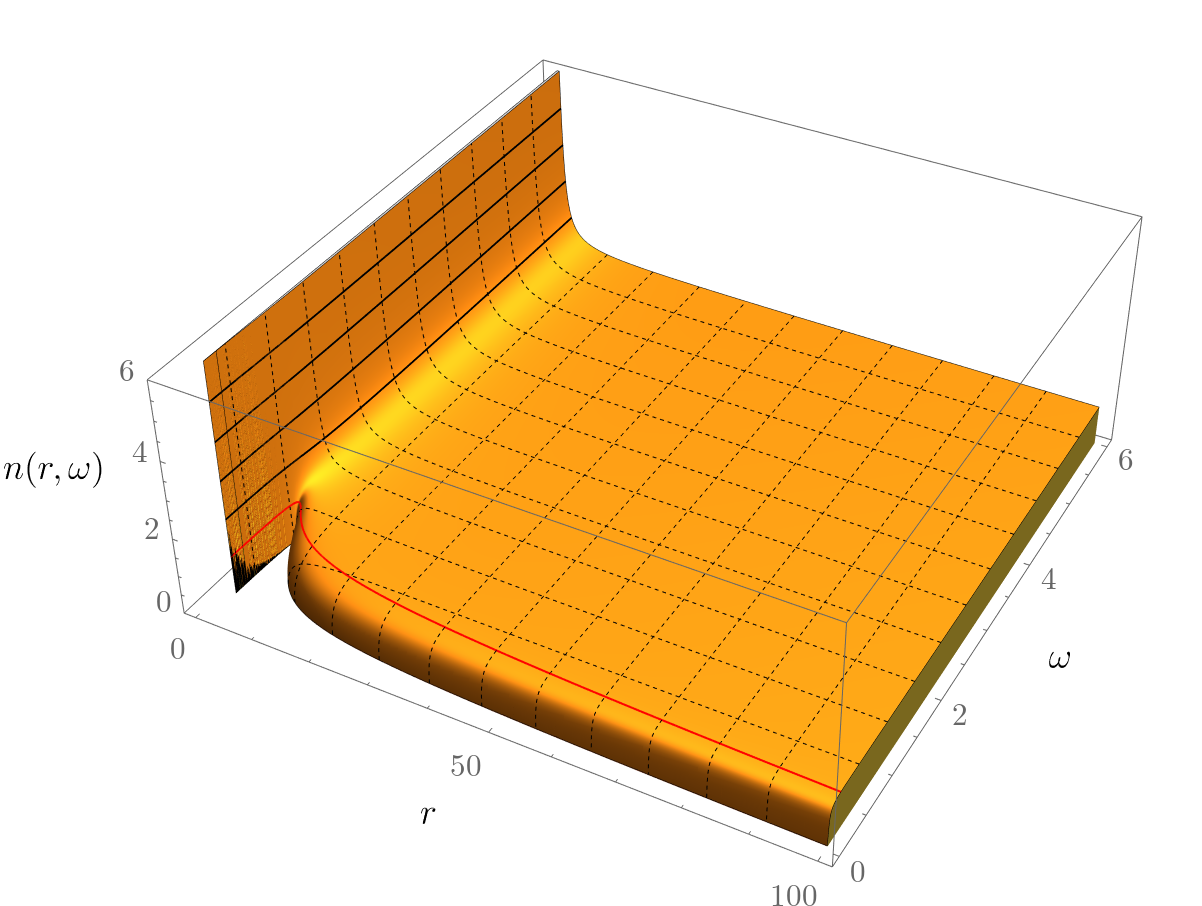}~\includegraphics[width=.48\textwidth]{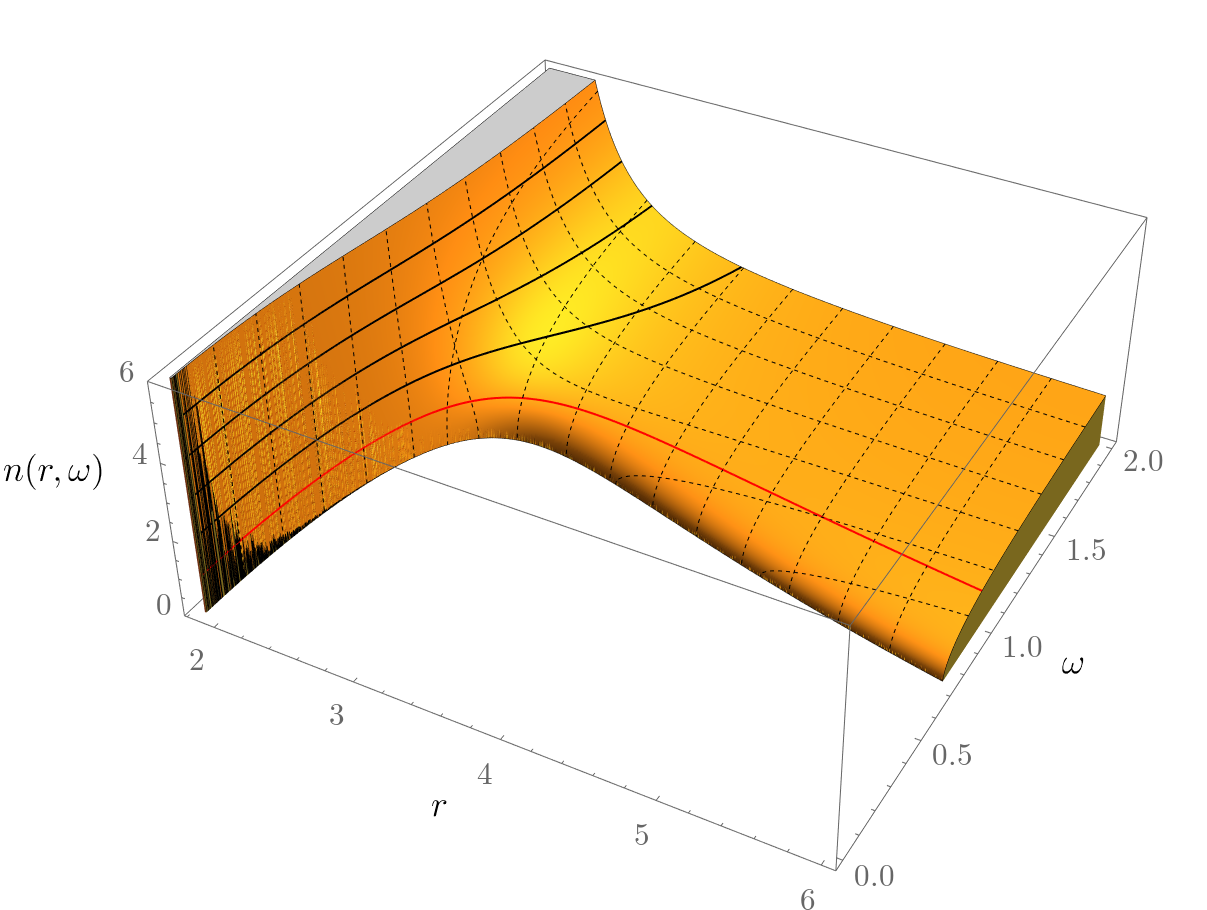}
	\caption[Plots of Indices of Refraction for Kerr Analogy, Part II]{Indices of refraction for the Kerr case, mass $M=1$, angular momentum $a=0.5$, and separation constants $D=4\cdot(4+1)=20$. Dashed lines correspond to lines of constant $\omega$ or $r$. The red line is the contour of index of refraction equal to 1, that is, the vacuum value. Black lines are lines of constant $n\in\{2,3,4,5\}$, the cut-off is taken at $n=6$. \emph{In the left column:} A look at the larger picture. \emph{In the right column:} A zoom on smaller values of $r$ and $\omega$. \emph{Top row:} $l=1$ and $m=0$. \emph{In the bottom row:} $l=6$, $m=3$. As we used the data from \cite{TeukolskyPress74} for the separation constants of the Teukolsky equation, this time $\omega$ cannot exceed $6$. As the bottom right picture indicates, the fact that the diverging, real refractive index does not reach $\omega=0$ is a numerical artifact.}
	\label{fig:refinKerr2}
\end{figure}

\begin{figure}
	\centering
	\includegraphics[width=.48\textwidth]{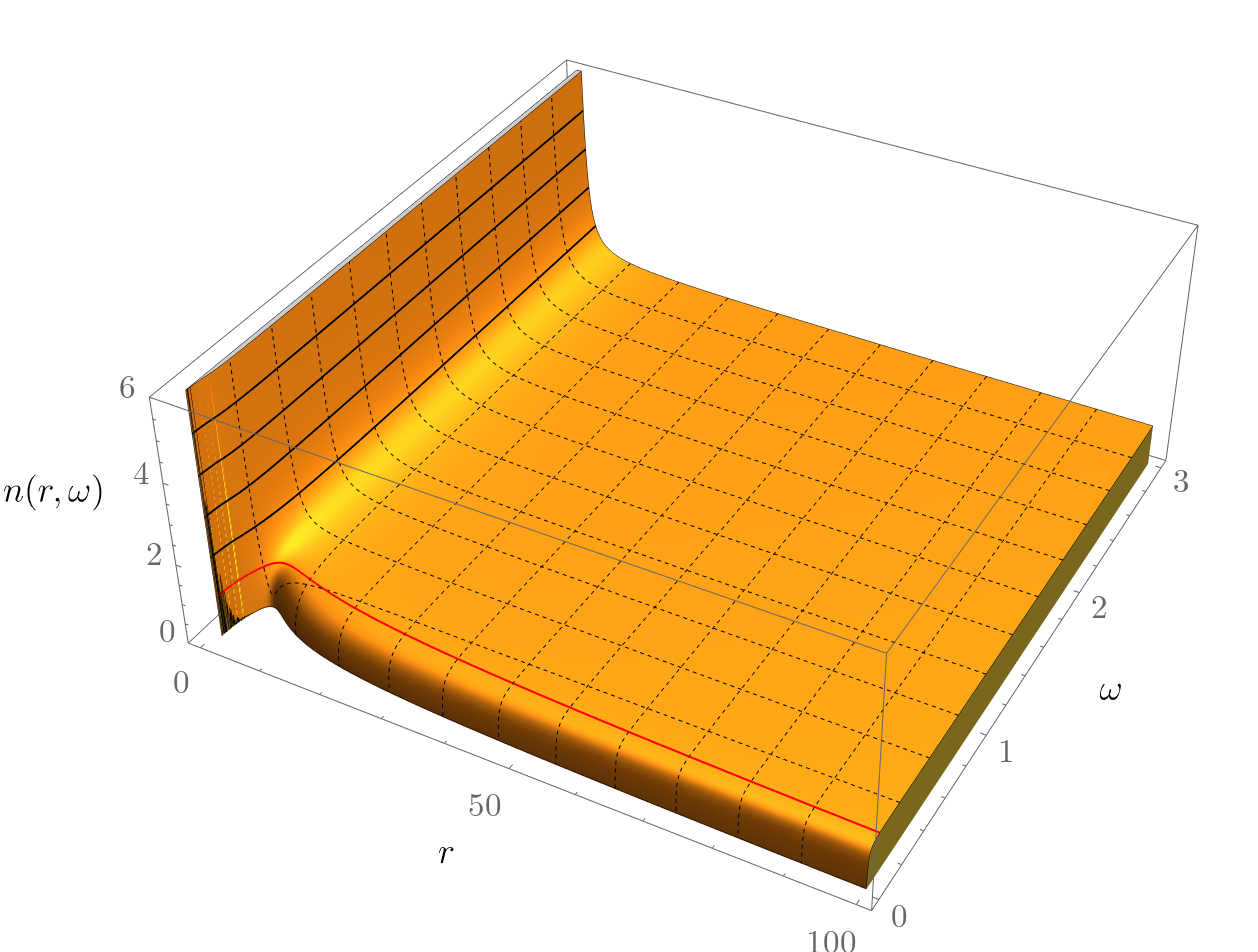}~\includegraphics[width=.48\textwidth]{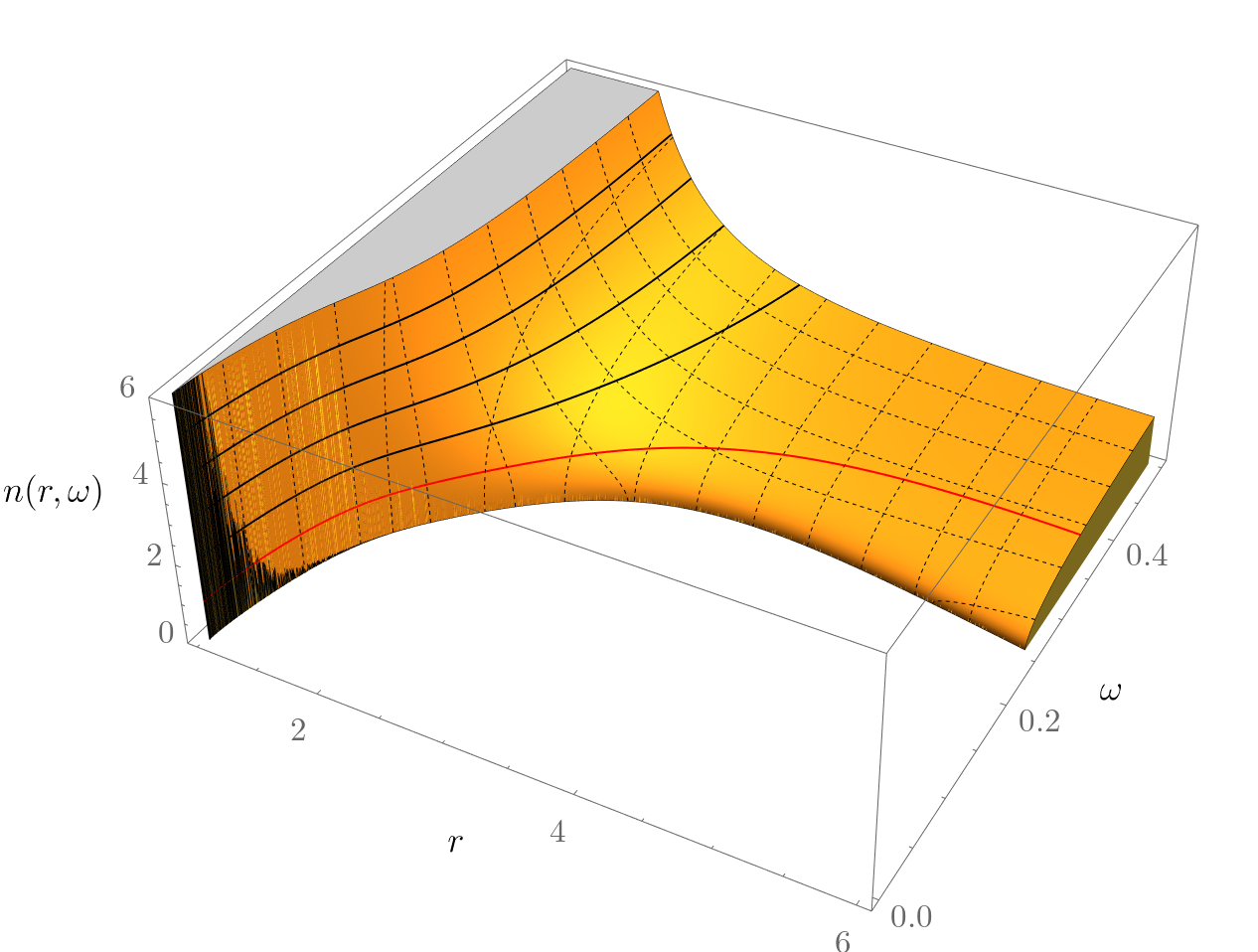}\\
	\includegraphics[width=.48\textwidth]{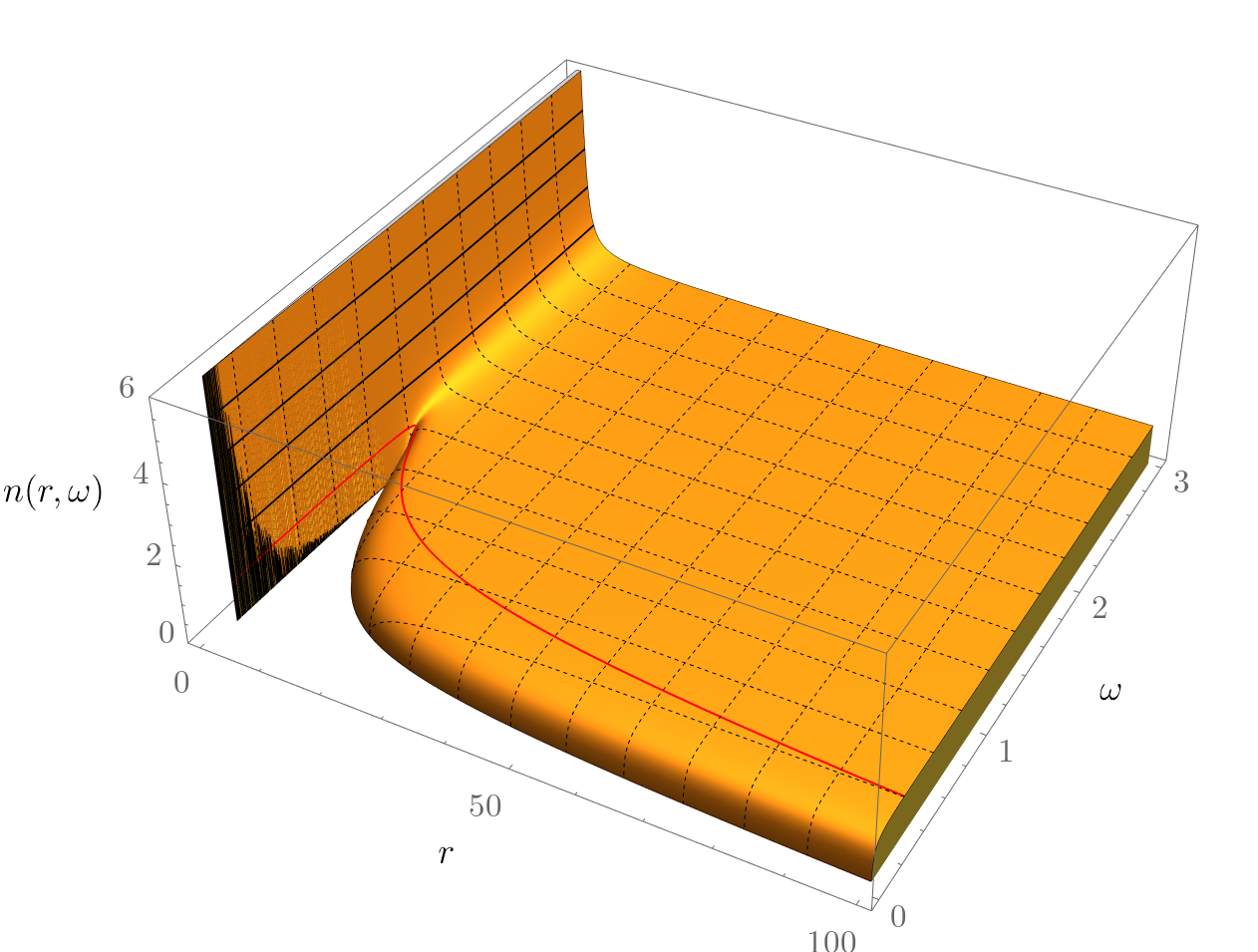}~\includegraphics[width=.48\textwidth]{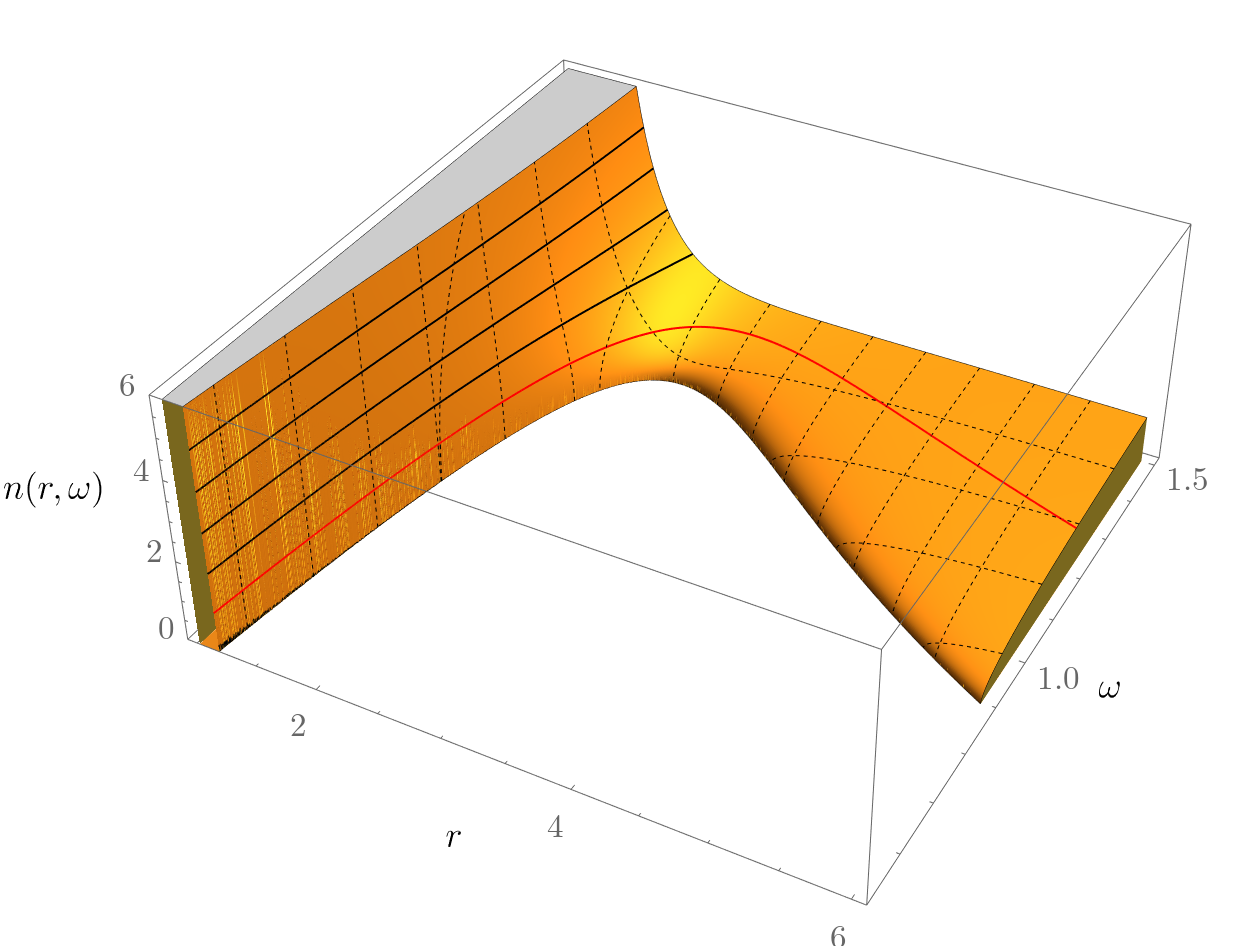}
	\caption[Plots of Indices of Refraction for Kerr Analogy, Part III]{Indices of refraction for the Kerr case, mass $M=1$, angular momentum $a=0.99975$, and separation constants $D=0$. Dashed lines correspond to lines of constant $\omega$ or $r$. The red line is the contour of index of refraction equal to 1, that is, the vacuum value. Black lines are lines of constant $n\in\{2,3,4,5\}$, the cut-off is taken at $n=6$. \emph{In the left column:} A look at the larger picture. \emph{In the right column:} A zoom on smaller values of $r$ and $\omega$. \emph{Top row:} $l=1$ and $m=0$. \emph{In the bottom row:} $l=6$, $m=3$. As we used the data from \cite{TeukolskyPress74} for the separation constants of the Teukolsky equation, this time $\omega$ cannot exceed $\approx 3$. As the bottom right picture indicates, the fact that the diverging, real refractive index does not reach $\omega=0$ is a numerical artifact.}
	\label{fig:refinKerr3}
\end{figure}

\begin{figure}
	\includegraphics[width=.48\textwidth]{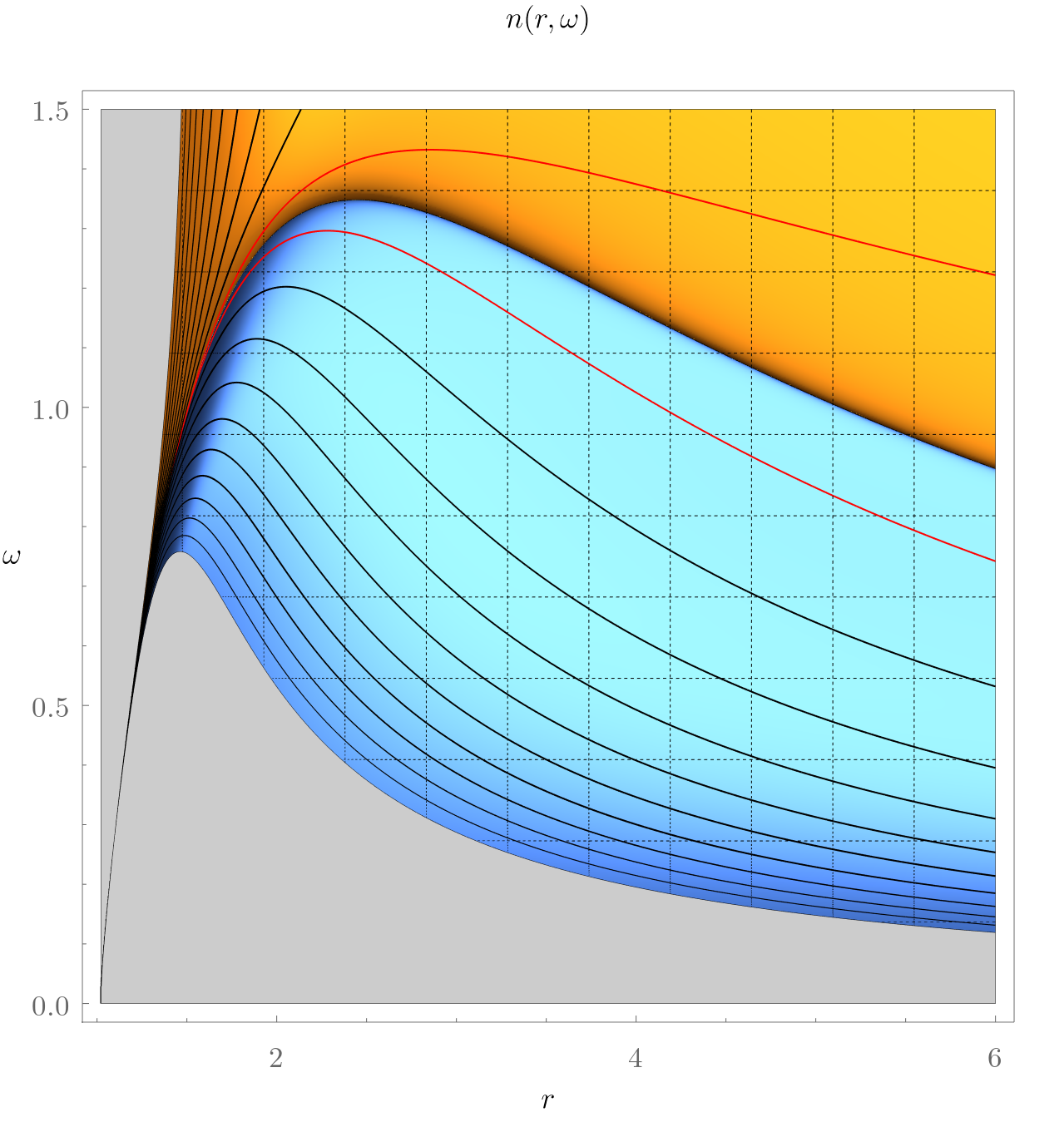}~\includegraphics[width=.48\textwidth]{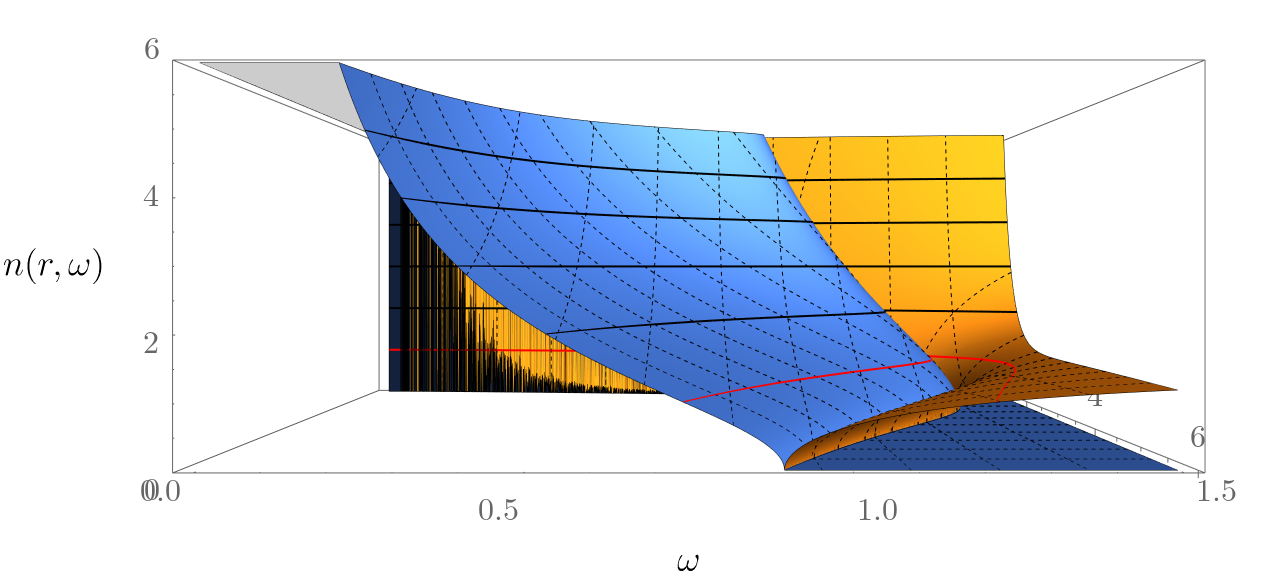}
	\caption[Plots of Real and Imaginary Part of Index of Refraction for Kerr Analogy]{Real part (orange) and imaginary part (blue) of the Index of refraction for the Kerr case, mass $M=1$, angular momentum $a=0.99975$, and separation constants $D=0$. Dashed lines correspond to lines of constant $\omega$ or $r$. The red line is the contour of index of refraction equal to 1, that is, the vacuum value. Black lines are lines of constant $n\in\{2,3,4,5\}$, the cut-off is taken at $n=6$. \emph{Left:} View from above, showing contour lines. \emph{Right:} View from the front, demonstrating how the imaginary part vanishes in the ranges shown in the remaining figures.}
	\label{fig:refinReIm}
\end{figure}
 	
\chapter{Sparsity}\label{ch:sparsity}
\epigraph{\enquote{That's really disconcerting. Especially when you know the root cause of the heat is radiation.}}{Andy Weir, \emph{The Martian}}
The idea of sparsity is to provide a measure to distinguish radiation processes of different nature. In our case, this means distinguishing, on the one hand, a classical black body as considered in traditional thermodynamics (and used to derive the Planck spectrum) from black hole radiation on the other hand. The latter is usually compared to the former, but as mentioned in the introductory chapter, by doing so, one peculiar feature of it is lost. This feature is most easily described in terms of the new notion of \enquote{sparsity}. More colloquially, sparsity is exactly what it sounds like: It measures how sparse radiation is, that is, how well-separated the radiated particles are. In order to measure this, two notions of time scales or length scales are necessary: One is given by the typical time/length scale of the radiation itself, while the other is derived from how often the radiation process happens.
 	
In order to develop this idea, it is advisable to start with the assumption of flat space. This we will do in section~\ref{sec:sparseflat}. Already in this theatre we can highlight the difference between black body radiation and black hole radiation, and we shall demonstrate this on bosons, fermions, and classical particles following Boltzmann statistics. With the notion of sparsity in place, we can then have a closer look at the necessary changes for a fully general relativistic implementation in sections~\ref{sec:sparsity} and \ref{sec:ndim} --- though full covariance is outright impossible, as will be mentioned, too. The general relativistic application will provide generous hunting grounds: Not only will the different particle species from their flat-space-time introduction see continued use, also different space-times, chemical potentials and their interrelation to the phenomenon of superradiance will be encountered. This discussion being focussed on the 3+1-dimensional case, we shall then extend the discussion to the spherically symmetric solution for $D$-dimensional relativity, the Tangherlini space-time. This chapter's closing section~\ref{sec:sparseEM} concludes with a discussion of the further complications arising from applying the notion of sparsity to the analogue space-times introduced in the previous chapter~\ref{ch:analogues}. At the time of writing the calculation of sparsity in analogue space-times remains an open problem.

While the consequences of Hawking radiation being sparse were known already in the early works immediately following Hawkings derivation of black hole evaporation \cite{PageThesis,Page1,Page2,Page3}, this understanding was not capitalised on. Partly, a subsequent focus on high temperature regimes can be considered responsible for this \cite{OliensisHill84,BHEvapJets,NatureBHgamma,BHPhotospheresPage,BremsstrahlungBH,MG12PageMacGibbonsCarr}.

In order to further stress this idiosyncratic property of the otherwise thermal-as-expected radiation in black hole evaporation, we introduced the term \enquote{sparsity} in \cite{HawkFlux2,HawkFlux1} to label it. Gray and Visser then further refined this first work to also capture the influence of grey body factors through numerical methods, as described in \cite{SparsityNumerical} and further below. We will see that the results are of the form
\begin{equation}
	\text{sparsity} = \frac{\text{dimensionless number}}{\text{degeneracy factor of radiation}} \times \frac{(\text{thermal wavelength})^2}{\text{characteristic area of radiator}}.
\end{equation}
The physical content of the measure \enquote{sparsity} is entirely contained in the last factor, while the \enquote{dimensionless number} varies only mildly with differing options for defining sparsity.

A similar analysis of time scales as used in our definition of \enquote{sparsity} appears in \cite{BekensteinMukhanovQuantumBH}, however, there the input is more speculative: For one, the horizon is quantised to give rise to a notion of black hole spectra. If that is the right quantity to quantise, and how precisely a quantum theory of black holes (derived from some theory of quantum gravity) would influence the existence or nature of a horizon, is all far from clear. For the other, the argument makes heavy use of the notion of particle creation close to the horizon. This argument indeed has a long tradition going back all the way to Hawking's own work, and is related to the trans-Planckian problem: If one follows outgoing wave packets back to the horizon, they are all vastly blue-shifted, mostly to trans-Planckian energy regimes and thus way outside the range of quantum field theory on curved space-times. However, since studies of renormalised energy-momentum tensors usually show maxima in flux and particle creation (more or less) safely away from the horizon \cite{BHQuantumAtmosphere,RSETKerr}, it is dubious how much one can rely on arguments founded on the trans-Planckian issues. Bekenstein's and Mukhanov's result --- in our parlance: \enquote{the sparsity of the Hawking flux} --- however agrees with our analysis.

\section{Defining Sparsity in Flat Space}\label{sec:sparseflat}
While it is tempting to immediately jump to full generality when defining \enquote{sparsity}, including all effects actually will rather obfuscate the exact origin of sparsity, namely the Hawking effect itself. Any complication arising from curved space-time influences, or inclusion of chemical potentials and particle masses, however, will actually not change the results significantly. In order to stress this, we will try to develop the concept from a very coarse approximation, then step-by-step reintroducing technicalities and generalisations. This will show that the effect of black hole radiation being much sparser than the radiation of a black body of similar temperature already exists in the radical simplification that this evaporation process is happening in flat space, utterly ignoring the geometric origin of black holes. The reintroduction of general relativity will not change these results. In many cases, the opposite will actually hold, and reintroducing curved space-time physics increases the sparsity beyond what one would expect from the flat space-time starting point.

\subsection{Bosonic Particles as Entry Point}

In order to formalise the colloquialism of the introduction to this chapter, let us have a look at the most readily available example of a radiation process, and how to define sparsity of it: Black body radiation. Remember that the differential number flux $\dif \Upgamma_n$ for massless, bosonic particles within a given wave number range $\dif^3\,\vec{k}$ of a black body of temperature $T$ into a surface element $\dif A$ with surface normal $\hat{n}$ is given by
\begin{equation}\label{eq:dGamma}
	\dif \Upgamma_{n} = \frac{g}{(2\pi)^3} \frac{c (\hat{k}\cdot \hat{n})}{\exp(\hbar c k / \kB T) - 1} \dif^3 \vec{k}\, \dif A.
\end{equation}
Here, $k$ is the absolute value of $\vec{k}$. As scalar and non-scalar bosons have a differing degeneracy, we included a spin degeneracy factor $g$ --- being 1 for scalar and 2 for non-scalar massless bosons.\footnote{This argument has to be refined both for massive particles, as well as when transitioning to higher-dimensional general relativity.} In order to easily distinguish this differential number flux (and later also energy flux) from the Gamma function, the former are denoted with an upright $\Upgamma$, while the latter with an italic $\Gamma$.

Assuming a finite surface, we can integrate equation~\ref{eq:dGamma}, yielding
\begin{subequations}
	\begin{align}
		\Upgamma_{n} =& \int_{0}^{A} \int_{-\infty}^{\infty} \int_{-\infty}^{\infty} \int_{-\infty}^{\infty} \frac{g}{(2\pi)^3} \frac{c \cos\theta}{\exp(\hbar c k / \kB T) - 1} \dif^3 k \dif \vec{A}',\label{eq:Upgamma}\\
		=& \int_{0}^{2\pi} \int_{0}^{\frac{\pi}{2}} \int_{0}^{\infty} \frac{g}{(2\pi)^3} \frac{c k^2\,\cos\theta\, A}{\exp(\hbar c k / \kB T) - 1} \dif k \sin \theta \,\dif \theta \dif\phi,\\
		=& \frac{g\, \zeta(3)}{4\pi^2} \frac{\kB^3 T^3}{\hbar^3\,c^2} A.\label{eq:3.4}
	\end{align}
\end{subequations}
Note that in the last step we made use of the integral definition~\ref{eq:zetaint} of the Riemann zeta function. We should also explain the integration a bit more: For each infinitesimal part $\dif \vec{A}$ of the horizon's surface area $A$ the integral over $\dif^3k$ gives the radiation \emph{away} from the radiating surface --- thus only a integration of $\theta$ from $0$ to $\pi/2$. This integration then happens for each infinitesimal area piece $\dif A$ separately, and each infinitesimal, directed area element gives a new axis with respect to which we take $\theta$. As the integration does not change, the integration over the scalar $\dif A$ can be pulled out and just gives a multiplicative factor of $A$. Phrased in terms of the more standard language of thermodynamics, we calculate the number of particles radiated away per unit time and unit area, and then integrate over the full area. More traditionally, this area would be flat and corresponding to a small hole in a cavity, see for example \cite{SchwablSM}.

The quantity $\Upgamma_n$ thus gives the number of emitted particles per unit time. Hence its inverse gives the average time $\tau_\text{gap}$ between emission of two subsequent particles:
\begin{equation}\label{eq:taugap}
	\tau_\text{gap} \defi \ed{\Upgamma_{n}}.
\end{equation}
This $\tau_\text{gap}$ will in all practical situations be much larger than calculated as above from $\Upgamma_n$. In fact, unless one builds (\enquote{could build}) a Dyson sphere surrounding the black hole, not all particles will be registered and one will have to wait until a particle is actually emitted towards the observer. For \emph{our} purposes, this is perfectly fine. We want to underestimate this time scale, as it will decrease the soon-to-be introduced sparsity. If a radiation process turns out to be sparse in our sense, it will turn out to be even sparser in any experimentally viable scenario.

The time scale to compare this $\tau_\text{gap}$ with should now be linked with the localisation time scale of the particle, $\tau_\text{loc}$. For a single quantum, this is reasonably simple to estimate from half its wavelength \cite{deBroglie}. Similar reasoning can also be found in statistical mechanics in the study of the thermal (deBroglie) wavelength $\lambda_{\text{thermal}}$, see appendix~B.2 of \cite{SchwablSM}. While more general arguments based on uncertainty relations\footnote{Which, as a statement about canonically conjugate variables, can be stated for any system, not just quantum mechanical systems! Thus, this general argument regarding the localisation will go through for any kind of wave phenomena. However, if one would want to aim for stronger definitions of localisation nothing prevents one from going above the lower bound given in the uncertainty relation.} can be made, in the following, we shall take a more heuristic approach.

For this, we define our localisation time scales as
\begin{equation}
	\tau_\text{loc} \defi \ed{\nu_{\text{peak, average, }q,d}} = \frac{2\pi}{\omega_{\text{peak, average, }q,d}},
\end{equation}
meaning, we have an ensemble of different time scales to choose from: $\tau_{\text{loc, peak, }q,d}$, $\tau_{\text{loc, average, }q,d}$, and $\tau_{\text{loc, average, }E}$. Here, \enquote{peak} refers to peaks of a spectrum, \enquote{average} to the average of that spectrum, while $d$ denotes \emph{which} spectrum concretely we are referring to --- number flux density or energy flux density will be most frequently used. The label $q$ refers to which quantity's peak or average with respect to the spectrum (or flux) $d$ is taken. The energy flux density is simply given as
\begin{equation}\label{eq:dGammaE}
	\dif \Upgamma_{E} \propto g \frac{c k^3 }{\exp(\hbar c k / \kB T) - 1} \dif k,
\end{equation}
where we need not worry about the overall proportionality factors, as they will drop out in \emph{our} calculations.\footnote{Note that this differs from $\Upgamma_{n}$ whose proportionality constants will determine its size in a given system of units --- which is precisely what we are interested in. $\Upgamma_{E}$, on the other hand will only serve to calculate peak frequencies.} Obviously, these are not the only choices possible. The definition is given in terms of frequency $\nu$ rather than angular frequency $\omega$, as this results in a further reduction of sparsity, turning our estimates to be even more conservative.\footnote{This additional step was suggested by an anonymous referee for \cite{HawkFlux1}.}

Before continuing with the definition of sparsity itself, let us dwell for a moment on the localisation time scales and calculate the ones we choose, focussing on bosons for the time being. Let us start with two localisation time scales based on the peak frequency (equivalently: energy) of a spectrum. As for finding local extrema (in this case maxima), pre-factors are of no relevance to their location in the number flux density or the energy flux density. In our case, we need to find the local maxima of $k^2 / (e^{\hbar c k/ \kB T}- 1)$ and $k^3 / (e^{\hbar c k/ \kB T}- 1)$, respectively for number flux and energy flux. A not too difficult and short calculation reveals that this can be solved using the Lambert W-function, see appendix~\ref{sec:Lambert}. The results are
\begin{align}
	\omega_{\text{peak, }E, n} &= \frac{\kB T}{\hbar} \kl{2+W(-2e^{-2})},\\
	\omega_{\text{peak, }E, E} &= \frac{\kB T}{\hbar} \kl{3+W(-3e^{-3})}.
\end{align}

Another time scale we shall use is related to the average frequency of the number flux. Reminding ourselves of the definition of the average of a spectrum $\Upgamma_{d}(k)$ as
\begin{equation}
	\omega_{\text{average},E,d} \defi \frac{\int \overbrace{c~ k}^\omega \nicefrac{\dif \Upgamma_{d}}{\dif k} \dif k}{\int \nicefrac{\dif \Upgamma_{d}}{\dif k} \dif k},
\end{equation}
we can give the average frequency (energy) as
\begin{equation}
	\omega_{\text{average, }E, n} = \frac{\pi^4}{30\zeta(3)} \frac{\kB T}{\hbar}. \label{eq:omEav4D1}
\end{equation}
Let us then introduce yet another time scale; this time, we will base it on the \enquote{average wave number} of the \enquote{wave number flux}:
\begin{equation}
	\omega_{\text{average, }k,k} = \frac{12\zeta(3)}{\pi^2} \frac{\kB T}{\hbar}.
\end{equation}
The reason for its introduction will become apparent in our short foray into including particle masses in the emission process.

The general idea for the comparison was to define some figure of merit 
\begin{equation}
	\eta = \tau_\text{gap}/\tau_\text{loc}.
\end{equation} 
At this point, we are thus able to define the following four differing notions of sparsity,
\begin{align}
	\eta_{\text{peak, }E,n} &\defi \frac{\omega_{\text{peak, }E, n}}{2\pi \Upgamma_{n}},\label{eq:etapeakn}\\
	\eta_{\text{peak, }E,E} &\defi \frac{\omega_{\text{peak, }E, E}}{2\pi \Upgamma_{n}},\label{eq:etapeakE}\\
	\eta_{\text{average, }k,k} &\defi \frac{\omega_{\text{average, }k,k}}{2\pi \Upgamma_{n}},\label{eq:etaaverk}\\
	\eta_{\text{average, }E, n} &\defi \frac{\omega_{\text{average, }E,n}}{2\pi \Upgamma_{n}},\label{eq:etaaverE}
\end{align}
and can immediately collect our results for bosons up to this point:
\begin{align}
	\eta_{\text{peak, }E,n} &= 2 \pi \frac{(2+W(-2 e^{-2}))}{\zeta(3)} \ed{g A} \kl{\frac{\hbar c}{\kB T}}^{2} &=& \frac{(2+W(-2e^{-2}))}{2\pi\zeta(3)} \frac{\lambda_\text{thermal}^{2}}{g A},\label{eq:etapn4b} \\
	\eta_{\text{peak, }E,E} &= 2 \pi \frac{(3+W(-3 e^{-3}))}{\zeta(3)} \ed{g A} \kl{\frac{\hbar c}{\kB T}}^{2} &=& \frac{(3+W(-3 e^{-3}))}{2\pi\zeta(3)} \frac{\lambda_\text{thermal}^{2}}{g A}, \label{eq:etapE4b}\\
	\eta_{\text{average, }k,k} &= \frac{24}{\pi gA}\kl{\frac{\hbar c}{\kB T}}^2 &=& \frac{6}{\pi^3}\frac{\lambda_\text{thermal}^2}{gA}\label{eq:etaavn4b},\\
	\eta_{\text{average, }E,n} &= \frac{\pi^5}{15\, \zeta(3)^2}\frac{1}{Ag} \kl{\frac{\hbar c}{\kB T}}^2 &=& \frac{\pi^3}{60\, \zeta(3)^2}\frac{\lambda_\text{thermal}^2}{Ag} \label{eq:etaavE4b},
\end{align}
where we have introduced the thermal wavelength $\lambda_{\text{thermal}} \defi 2\pi\hbar c/(\kB T)$ mentioned earlier.

Before turning our attention to fermions, there is one last variant of sparsity to be introduced: Bolometric, or \enquote{binned} sparsity. Its definition is
\begin{equation}\label{eq:defbinned}
	\eta_{\text{binned}} \defi \ed{\int \frac{2\pi}{ck}\frac{\dif \Upgamma_n}{\dif k}\dif k}.
\end{equation}
The physical idea behind this definition --- and for the alternative description as \enquote{bolometric}\footnote{If it would not lead to confusion when comparing the present text to the published version in \cite{HawkFlux1}, the present thesis would be using the index \enquote{bolometric} instead of \enquote{binned}.} --- is easiest seen when the integral in the denominator is written as a Riemann sum using the midpoint rule:
\begin{equation}
	\int \frac{2\pi}{ck}\frac{\dif \Upgamma_n}{\dif k}\dif k = \lim_{k_0 \to \infty} \lim_{N \to \infty}\sum_{i=1}^{N} \underbrace{\frac{2\pi}{ck_i}}_{\nu_i^{-1}} \left.\frac{\dif \Upgamma_n}{\dif k}\right\vert_{k=k_i} \underbrace{\frac{k_0}{N}}_{\Delta k},  \label{eq:binnedRiemann}
\end{equation}
where we set $k_i = \frac{2i-1}{2N}k_0$. Thus, the number flux spectrum is first separated into (infinitesimal) wavenumber bins and the number flux into each bin is compared to the frequency corresponding to that wave number, giving a measure for \enquote{number (of emitted particles) per wavenumber}. As high values in each wavenumber bin will result in small chances of localising single particles of that wavenumber, low values would correspond to a high sparsity in that bin. By summing, \emph{i.e.}, integrating, and inverting we therefore gain a new measure for sparsity.

This additional measure has a very useful property which we shall exploit in section~\ref{sec:superradiance}: If two processes add to the number of emitted particles per wavenumber bin, they can in the context of this particular sparsity measure be seen as \enquote{in parallel} (as in the sense of electric circuit analysis) --- if $N$ processes with corresponding number flux density $\dif \Upgamma_i$ occur at the same time, the total binned sparsity $\eta_{\text{binned}}$ can be calculated as
\begin{equation}
	\eta_{\text{binned}} = \sum_{i=1}^{N} \eta_{\text{binned, }i}.\label{eq:binnedsum}
\end{equation}

In the case of the bosonic spectrum~\eqref{eq:dGamma}, the binned sparsity~\eqref{eq:defbinned} is quickly evaluated to
\begin{equation}
	\eta_{\text{binned}} = \frac{24}{\pi gA}\kl{\frac{\hbar c }{\kB T}}^{2} = \frac{6}{\pi^3}\frac{\lambda_\text{thermal}^{2}}{gA}.\label{eq:etabin4b}
\end{equation}
At this point it is worthwhile to notice that this binned sparsity has the same value as the one defined via the average frequency $\omega_{\text{average, }k,k}$. This turns out to be a general link between $\eta_{\text{average, }k,k}$ and the binned sparsity, independent of particle type, or the possible presence of a chemical potential. Let us prove this quickly. The wavenumber density used for calculating $\omega_{\text{average, }k,k}$ is
\begin{equation}
	\Upgamma_{k} \propto \int_{0}^{\infty} \frac{x}{e^x - 1}\dif x,
\end{equation}
where $x$ is a dimensionless version of the wavenumber $k$. We quickly see that the average wavenumber with respect to $\Upgamma_{k}$ results in just $\Upgamma_{n}$. Thus 
\begin{equation}
	\omega_{\text{average, }k,k} = \frac{\Upgamma_{n}}{\Upgamma_{k}}.
\end{equation}
Thus, if we compare the time scale $\tau_{\text{loc, average, }k,k}$ given by $\omega_{\text{average, }k,k}$ with $\tau_{\text{gap}}$, $\Upgamma_{n}$ cancels and we are left with the same integral as for $\eta_{\text{binned}}$. This would carry over to higher dimensions, so long as the dispersion relation remains that of a free particle. We will, for the time being, keep both sparsities $\eta_{\text{binned}}$ and $\eta_{\text{average, }k,k}$, as they are a good pair to look at when incorporating particle rest mass in our analysis. We will see that the physical construction will lead to very different results in that case. Overall, $\eta_{\text{binned}}$ is the more useful one, though: It retains the property~\eqref{eq:binnedsum} even when changing the dispersion relation (\emph{e.g.} by including a rest mass), while $\eta_{\text{average, }k,k}$ does, by its very definition, not. For this reason, we will discuss only $\eta_{\text{binned}}$ in the higher dimensional case in section~\ref{sec:ndim}.

It seems tempting to consider a sixth sparsity measure, $\eta_{\text{binned, }E}$: Take the idea of the binned sparsity $\eta_{\text{binned}}$ as introduced above, modify it for the energy flux density to appear instead of the number flux density. However, this will not gain us new information: Instead of $\dif\Upgamma_n/\nu$ the relevant integrand would also have to compare energies of a given wavenumber bin: $\dif\Upgamma_{E}/(\hbar \omega)= \dif\Upgamma_{E}/(h \nu)= \dif\Upgamma_{E}/(\hbar c k)$. This obviously has not the right physical dimensions yet, as it would still have dimensions of a flux. Were we to normalise also with respect to each frequency of each wavenumber bin, we would end up with just the $\eta_{\text{binned}}$ we tried to deviate from in the first place. While it might be possible to alter the definition by comparing to different notions of frequency or notions of energy, we fail to see equally obvious choices without introducing additional physical assumptions on either the radiation or the context in which it appears.

Note that all these considerations are semi-classical (they only take a quantum mechanical result and run with it in a classical way) and semi-analytic (we approximate the spectra both by limiting ourselves to black body radiation, and looking at the geometric optics approximation). That these quantities give useful results can thus only be decided once we compare them with previous measures of different sources. Either way, with the different definitions of \enquote{sparsity} in place, it is time to look at their values for different types of particles. Once that has been achieved, one can try to unify, generalise, and compare. Anticipating section~\ref{sec:firstcomparison}, let us mention that the properties of black holes enter at two points in the calculation of sparsity: Through the Hawking temperature and through the area of the black hole's horizon.

\subsection{Sparsity of Fermionic Particles}
The difference between fermions and bosons in the expressions of our different definitions for sparsity only comes in through the different densities $\dif\Upgamma_{d}$. We shall encounter this difference and its importance later again, once chemical potentials are in place. For fermions, the number flux density changes to
\begin{equation}\label{eq:dGammaF}
	\dif \Upgamma_n = \frac{g}{(2\pi)^3} \frac{c (\hat{k}\cdot \hat{n})}{\exp(\hbar c k / \kB T) + 1} \dif^3 \vec{k}\, \dif A,
\end{equation}
where the difference in the denominator is basically an incarnation of the spin-statistics theorem/\allowbreak Pauli's exclusion principle \cite{StreaterWightman,SchwablSM}.

Repeating the calculations (and adjusting the corresponding definitions of sparsity by using the right, now fermionic flux densities) previously done for bosons then results in the following five sparsities for fermions in a flat space-time:
\begin{align}
	\eta_{\text{peak, }E,n} &= 8 \pi \frac{(2+W(-2 e^{-2}))}{3\zeta(3)} \ed{g A} \kl{\frac{\hbar c}{\kB T}}^{2} &=& \frac{2(2+W(-2e^{-2}))}{3\pi\zeta(3)} \frac{\lambda_\text{thermal}^{2}}{g A},\label{eq:etapn4f} \\
	\eta_{\text{peak, }E,E} &= 8 \pi \frac{(3+W(-3 e^{-3}))}{3\zeta(3)} \ed{g A} \kl{\frac{\hbar c}{\kB T}}^{2} &=& \frac{2(3+W(-3 e^{-3}))}{3\pi\zeta(3)} \frac{\lambda_\text{thermal}^{2}}{g A}, \label{eq:etapE4f}\\
	\eta_{\text{average, }k,k} &= \frac{48}{\pi gA}\kl{\frac{\hbar c}{\kB T}}^2 &=& \frac{12}{\pi^3}\frac{\lambda_\text{thermal}^2}{gA}\label{eq:etaavn4f},\\
	\eta_{\text{average, }E,n} &= \frac{14\pi^5}{135\, \zeta(3)^2}\frac{1}{Ag} \kl{\frac{\hbar c}{\kB T}}^2 &=& \frac{7\pi^3}{270\, \zeta(3)^2}\frac{\lambda_\text{thermal}^2}{Ag} \label{eq:etaavE4f},\\
	\eta_{\text{binned}} &= \frac{48}{\pi gA}\kl{\frac{\hbar c }{\kB T}}^{2} &=& \frac{12}{\pi^3}\frac{\lambda_\text{thermal}^{2}}{gA}. \label{eq:etabin4f}
\end{align}
Again, note the agreement of $\eta_{\text{binned}}$ and $\eta_{\text{average, }k,k}$. It is also worth noticing that the sparsity increased compared to the corresponding values for bosons (if the degeneracy $g$ is the same). This, again, is a general feature and can be traced back to the Pauli exclusion principle: For most of its lifetime, a black hole will radiate particles of low energy. While for bosons an arbitrary amount of particles can inhabit the same energy level, and thus, low energy levels, fermions are prevented from this by the exclusion principle.

\subsection{Sparsity and \enquote{Boltzmannian} Particles}
A last \enquote{type} of particles constitutes particles who \enquote{forgot} their inherent type: For high energy particles, both the Bose--Einstein and the Fermi--Dirac statistics can be approximated by Boltzmann statistics. However, these can also be considered in their own right (though, physically speaking, they can only be encountered in the context of an approximation scheme), which we shall do here. The benefit of doing this is that we can find another stringent bound on both of the other, physically more plausible cases --- once from above, once from below. To see this, define a new \enquote{Boltzmannian} number flux density $\dif \Upgamma_n$:
\begin{equation}\label{eq:dGammaBoltz}
	\dif \Upgamma_n = \frac{g}{(2\pi)^3} \frac{c (\hat{k}\cdot \hat{n})}{\exp(\hbar c k / \kB T)} \dif^3 \vec{k}\, \dif A.
\end{equation}
All integrals involving this new $\dif \Upgamma$ encountered in the calculation of our five different sparsities now turn out to be of a much simpler form, and averages and peaks agree with each other.
\begin{align}
	\eta_{\text{peak, }E,E} &= 6 \pi \ed{g A} \kl{\frac{\hbar c}{\kB T}}^{2} &&= \frac{3}{2\pi} \frac{\lambda_\text{thermal}^{2}}{g A},\label{eq:etapn4bolt}\\
	&= \eta_{\text{average, }E,n}, &&\\
	\eta_{\text{peak, }E,n} &= 4 \pi \ed{g A} \kl{\frac{\hbar c}{\kB T}}^{2} &&= \frac{1}{ \pi} \frac{\lambda_\text{thermal}^{2}}{g A}, \label{eq:etapE4bolt}\\
	&= \eta_{\text{average, }k,k} &&= \eta_{\text{binned}}.
\end{align}
In this case, only two different values for the different definitions of sparsity appear.

\subsection{First Generalisations}\label{sec:firstgen}
Already at this point and in flat space-times we can make a few generalisations helping in our physical understanding of the notion of sparsity.

\subsubsection{Unifying Different Number Flux Densities}
First of all, using the results from the appendix~\ref{sec:polylogs} it is possible to unify our results so far to

\begin{subequations}\label{eq:4dpolylogs}
\begin{align}
	\eta_{\text{average, }E,n} &= 6 \pi \frac{\Li{4}(-s)/(-s)}{(\Li{3}(-s)/(-s))^2} \ed{g A} \kl{\frac{\hbar c}{\kB T}}^2 &=& \frac{3}{2 \pi} \frac{\Li{4}(-s)/(-s)}{(\Li{3}(-s)/(-s))^2} \frac{\lambda_{\text{thermal}}^2}{g A},\\
	\eta_{\text{average, }k,k} &= 4 \pi \frac{(-s)}{\Li{2}(-s)} \frac{1}{g A} \kl{\frac{\hbar c}{\kB T}}^2 &=& \ed{\pi} \frac{(-s)}{\Li{2}(-s)} \frac{\lambda_{\text{thermal}}^2}{g A},\\
	\eta_{\text{peak, }E,E} &= 2 \pi \frac{3 + W(s 3 e ^{-3})}{\Li{3}(-s)/(-s)} \frac{1}{g A} \kl{\frac{\hbar c}{\kB T}}^2 &=& \ed{2 \pi} \frac{3 + W(s 3 e ^{-3})}{\Li{3}(-s)/(-s)} \frac{\lambda_{\text{thermal}}}{g A},\\
	\eta_{\text{peak, }E,n} &= 2 \pi \frac{2 + W(s 2 e ^{-2})}{\Li{3}(-s)/(-s)} \frac{1}{g A} \kl{\frac{\hbar c}{\kB T}}^2 &=& \ed{2 \pi} \frac{2 + W(s 2 e ^{-2})}{\Li{3}(-s)/(-s)} \frac{\lambda_{\text{thermal}}}{g A},\\
	\eta_{\text{binned}} &= 4 \pi \frac{(-s)}{\Li{2}(-s)} \frac{1}{g A} \kl{\frac{\hbar c}{\kB T}}^2 &=& \ed{\pi} \frac{(-s)}{\Li{2}(-s)} \frac{\lambda_{\text{thermal}}}{g A},
\end{align}
\end{subequations}
where $s\in \{-1,0,+1\}$ corresponds to the sparsity results of bosons, Maxwell--Boltzmann particles, and fermions, respectively. Modifying these results to include chemical potentials (see the next section below) or curved space-time effects will show these preliminary results from flat space-time to appear as ingredients of the more general (and thus more reliable) sparsities. Two exceptions to this are worth mentioning: First, if we want to include mass for the emitted particle, the arising integrals still can be attacked analytically to \emph{some} degree --- however the results prove to be neither enlightening nor appear our previous results as simple parts of the whole. Nevertheless, we shall still be able to provide estimates where the latter property of flat-space sparsities will come up. Second, to include superradiant scenarios we will have to confine our analysis to the binned sparsity definition. As we want to be able to distinguish Hawking radiation from radiation arising from superradiance, the property~\eqref{eq:binnedsum} will prove to be an integral part of that analysis.\footnote{As the averaged wavenumber sparsity $\eta_{\text{average, }k,k}$ by its very definition can not be expected to retain this property in general, as mentioned above, it is important to explicitly refer to the binned sparsity here!}

This completes a unified picture of the particle  densities in the new guise of \enquote{sparsity} --- at first glance, and in flat, four-dimensional space. This notwithstanding, the next subsections shall provide two further, already alluded-to flat-space extensions.

\subsubsection{Introducing Chemical Potentials --- Superradiance}
As a quick glance at appendix~\ref{sec:polylogs} shows, additive terms in the exponents of the densities $\dif \Upgamma_{d}$ considered so far can be covered with ease with the equations~\eqref{eq:4dpolylogs}. And this inclusion's raison d'etre is not derived out of a misguided sense of reaching the highest mathematical generality possible: Rather, these additive terms are precisely of the form of chemical potentials, thus lending physical import to their export out of tables of special functions.

Chemical potentials in the context of sparsity are relevant as only their presence ensures that the phenomenon of superradiance can be captured. Let us set the stage by a very brief, one-paragraph summary of superradiance: While the concrete parlance of \enquote{superradiance} seems to be most deeply ingrained in the fields linked to relativity, superradiance as such does neither require special nor general relativity, though both highlight famous examples, see, for example \cite{SuperradianceReview,GenSuperradiance,DispersiveSuperradiance,SuperradianceFlux,LNPSuperradiance}; \cite{SuperradianceReview} also mentions the term's genesis outside of relativity. Two of the most important examples are given by Zel'dovich's rotating cylinder which exemplifies superradiance in the context of (special relativistic) electromagnetism on the one hand, and, on the other hand, the Kerr geometry (obviously of general relativistic nature). The former can be seen as the conceptual precursor of the latter. We refer to the just-given references for the Zel'dovich cylinder. The latter we will discuss in section~\ref{sec:superradiance}, as we are concerned with superradiance's incarnation in general relativity. Here, the inclusion of angular momenta, charge, or both leads to the Kerr--Newman family of black hole solutions to Einstein's field equations, see section~\ref{sec:genBH}. Already in the derivations of black hole radiation of Hawking himself \cite{HawkingEffect2} can we find the corresponding inclusion of this effect on these space-times. And it is here that they re-emerge as chemical potentials being additional terms in the Planck-like part of the spectrum:\footnote{They likewise appear as additional terms in the generalised first and second laws of black hole thermodynamics.}
\begin{equation}
	\frac{1}{e^{\frac{\hbar \omega}{\kB T}}\pm 1} \quad\longrightarrow \quad\frac{1}{e^{\frac{\hbar\omega  - q V_{\text{H}} - \hbar m \Omega_{\text{H}}}{\kB T}}\pm 1},
\end{equation}
where $q$ is the charge of particle emitted, $V_\text{H}$ is the electric potential at the horizon, $m$ is the angular momentum quantum number of the emitted particle,\footnote{As we consider only \emph{either} massive particles \emph{or} angular momentum quantum numbers, we deem it unproblematic that they share the same symbol.} and $\Omega_{\text{H}}$ is the angular frequency of the horizon. This explains our lumping together of chemical potentials and superradiance. To simplify our notation, let us write this previous substitution as
\begin{equation}
	\frac{1}{e^{\frac{\hbar \omega}{\kB T}}\pm 1} \quad\longrightarrow \quad\frac{1}{e^{\frac{\hbar \omega  - \mu}{\kB T}}\pm 1},
\end{equation}
for a general chemical potential $\mu$. It is important to note that the inclusion of other chemical potentials than the two\footnote{Three, if one were to allow for magnetic charges. Their inclusion follows the same principles as the two already discussed.} already mentioned (electric charges and angular momenta) may result (from a formal and technical viewpoint) in the necessity to re-examine the Hawking effect, as already our two examples are accompanied by a change of the metric, if not the whole dynamics: While the Kerr geometry, necessary for the sensible inclusion of angular momentum, means changing the solution of the vacuum Einstein equations, the inclusion of charges even means the change from the Einstein field equations to the Einstein--Maxwell system of equations.

As we noted earlier, with the help of our sparsity expressions in terms of polylogarithms the introduction of chemical potentials is just a minor variation of the results previously achieved:
\begin{subequations}\label{eq:4dpolylogsmu}
\begin{align}
	\eta_{\text{average, }E,n} &= 6 \pi \frac{\Li{4}(-s\exp(\mu/\kB T))}{\Li{3}(-s\exp(\mu/\kB T))} \ed{g A} \kl{\frac{\hbar c}{\kB T}}^2, \\&= \frac{3}{2 \pi} \frac{\Li{4}(-s\exp(\mu/\kB T))}{\Li{3}(-s\exp(\mu/\kB T))} \frac{\lambda_{\text{thermal}}^2}{g A},\\
	\eta_{\text{average, }k,k} &= 4 \pi \frac{(-s)}{\Li{2}(-s\exp(\mu/\kB T))} \frac{1}{g A} \kl{\frac{\hbar c}{\kB T}}^2, \\&= \ed{\pi} \frac{(-s)}{\Li{2}(-s\exp(\mu/\kB T))} \frac{\lambda_{\text{thermal}}^2}{g A},\\
	\eta_{\text{peak, }E,E} &= 2 \pi \frac{3 + W(s 3 \exp(-3+\mu/\kB T))}{\Li{3}(-s\exp(\mu/\kB T))/(-s)} \frac{1}{g A} \kl{\frac{\hbar c}{\kB T}}^2, \\&= \ed{2 \pi} \frac{3 + W(s 3 \exp(-3+\mu/\kB T))}{\Li{3}(-s\exp(\mu/\kB T))/(-s)} \frac{\lambda_{\text{thermal}}}{g A},\\
	\eta_{\text{peak, }E,n} &= 2 \pi \frac{2 + W(s 2 \exp(-2+\mu/\kB T))}{\Li{3}(-s\exp(\mu/\kB T))/(-s)} \frac{1}{g A} \kl{\frac{\hbar c}{\kB T}}^2, \\&= \ed{2 \pi} \frac{2 + W(s 2 \exp(-2 + \mu/\kB T))}{\Li{3}(-s\exp(\mu/\kB T))/(-s)} \frac{\lambda_{\text{thermal}}}{g A},\\
	\eta_{\text{binned}} &= 4 \pi \frac{(-s)}{\Li{2}(-s\exp(\mu/\kB T))} \frac{1}{g A} \kl{\frac{\hbar c}{\kB T}}^2, \\&= \ed{\pi} \frac{(-s)}{\Li{2}(-s\exp(\mu/\kB T))} \frac{\lambda_{\text{thermal}}}{g A},
\end{align}
\end{subequations}
where we have, for improved legibility, omitted the division by $-s$ in the expressions for $\eta_{\text{average, }E,n}$, as for this sparsity the division of two polylogarithms already takes care of the removal of the singularity at $s=0$.

\subsubsection{Including Particle Mass}\label{sec:masses}
The inclusion of rest masses for the emitted particles significantly complicates matters. The relation between particle wavevector $k$ and angular frequency $\omega$ now changes to
\begin{equation}
	\hbar \omega = \sqrt{\hbar^2 c^2 k^2 + m^2c^4},
\end{equation}
such that our integrals appearing in averaged $\eta$'s become some version of
\begin{equation}
	\int_{0}^{\infty} \frac{k^q\sqrt{m^2c^4+\hbar^2c^2k^2}^{\,r}}{\exp\kl{\sqrt{m^2c^4+\hbar^2c^2k^2}/\kB T_\text{H}}+s}\dif k,\label{eq:massiveintegrand}
\end{equation}
where $q,r\in\mathbb{Z}$.\footnote{At least in the case of $s=0$, the Boltzmann case, the results below are known, or require very little steps involving a CAS or integral table.} In total, we require an evaluation of four separate integrals, and it is here that the difference between $\eta_{\text{average, }k,k}$ and $\eta_{\text{binned}}$ appears: The former is explicitly in terms of the wavenumber $k$, the latter involves actually comparison of each wavenumber bin to its average \emph{frequency}. The four integrals we need to evaluate are, written in a dimensionless version:
\begin{subequations}\label{eq:dimlessI}
\begin{align}
	I_1 &= \int_{0}^{\infty} \frac{x^2}{\exp\kl{\sqrt{x^2+z^2}} \pm 1}\dif x,\\
	I_2 &= \int_{0}^{\infty} \frac{\sqrt{x^2+z^2}x^2}{\exp\kl{\sqrt{x^2+z^2}} \pm 1}\dif x,\\
	I_3 &= \int_{0}^{\infty} \frac{x}{\exp\kl{\sqrt{x^2+z^2}} \pm 1}\dif x,\\
	I_4 &= \int_{0}^{\infty} \frac{x^2/\sqrt{x^2+z^2}}{\exp\kl{\sqrt{x^2+z^2}} \pm 1}\dif x.
\end{align}
\end{subequations}

We already have to give up on the peaked $\eta$'s as we are not aware of any closed form solution of setting the derivative of the integrand appearing in equation~\eqref{eq:massiveintegrand} to zero. The exception to this is for Boltzmann particles, where we can find
\begin{align}
	\omega_{\text{peak, }E,n}^\text{Boltzmann} = \frac{\kB T_\text{H}}{c\hbar}\sqrt{2+\frac{c^4m^2}{\kB^2 T_\text{H}^2} + 2 \sqrt{1+\frac{c^4m^2}{\kB^2 T_\text{H}^2}}}.
\end{align}
It \emph{is} possible to write a closed form solution to $\omega_{\text{peak, }E,E}^\text{Boltzmann}$, too --- however, the resulting root of a cubic has a tendency to confuse rather than to enlighten, as complex parameters appear in them. A quick look at the corresponding spectra, however, shows that there is a real-valued maximum, as was expected. While certainly not exciting, we show the general behaviour for dimensionless mass $z$ and dimensionless wavevector $k$ in figure~\ref{fig:spectra} for this purpose. For most practical purposes one would have to rely on numerical methods for evaluating peaked sparsities.

Let us now have a look at the integral
\begin{equation}
	I_1 = \int_{0}^{\infty} \frac{x^2}{\exp\kl{\sqrt{x^2+z^2}} \pm 1}\dif x.
\end{equation}
This is the integral appearing in the number flux. The first step is to realise that this can be written as a geometric series:
\begin{equation}
	\int_{0}^{\infty} \frac{x^2}{\exp\kl{\sqrt{x^2+z^2}} \pm 1}\dif x = \sum_{n=0}^{\infty} (\mp 1)^n \int_{0}^{\infty} x^2 \exp\kl{-(n+1)\sqrt{x^2+z^2}} \dif x.
\end{equation}
Now integrate by parts (the boundary terms vanish),\footnote{For this we take $u=x$ and $v' = x \exp\kl{-(n+1)\sqrt{x^2+z^2}}$.} and change the variable by setting $x=z\sinh(t)$:
\begin{align}
	&\sum_{n=0}^{\infty} (\mp 1)^n \int_{0}^{\infty} x^2 \exp\kl{-(n+1)\sqrt{x^2+z^2}} \dif x = \nonumber\\
	&\qquad \sum_{n=0}^{\infty} (\mp 1)^n \int_{0}^{\infty}\kl{\frac{1+(n+1)\sqrt{x^2+z^2}}{(n+1)^2}} \exp\kl{-(n+1)\sqrt{x^2+z^2}} \dif x,\\
	&=  \sum_{n=0}^{\infty} (\mp 1)^n \int_{0}^{\infty} z\kl{\frac{1+(n+1)z\cosh(t)}{(n+1)^2}}\exp\kl{-(n+1)z\cosh(t)}\cosh(t)\dif t,\\
	&= \sum_{n=0}^{\infty} (\mp 1)^n \left[ \frac{z}{(n+1)^2} \int_{0}^{\infty} \exp\kl{-(n+1)z\cosh(t)}\cosh(t)\dif t\right.\nonumber\\
	&\qquad + \left.\frac{z^2}{2(n+1)} \int_{0}^{\infty}\klg{ \exp\kl{-(n+1)z\cosh(t)} + \exp\kl{-(n+1)z\cosh(t)} \cosh(2t)}\dif t\right].\label{eq:massiveintmiddle}
\end{align}
In the last step we made use of
\begin{equation}
	\cosh^2 (t) = \ed{2}\kl{1+\cosh(2t)}.
\end{equation}
At this point we can then use facts about modified Bessel functions of the second kind, concretely, equation~\eqref{eq:Kint}:
\begin{equation}
	\eqref{eq:massiveintmiddle} = \sum_{n=0}^{\infty} (\mp 1)^n \frac{z}{n+1}\kl{\frac{K_1((n+1)z)}{(n+1)z}} + \frac{z}{2} \kle{K_0((n+1)z)+K_2((n+1)z)}.
\end{equation}
Making use of the recursion relation~\eqref{eq:Krecursion}, this gives as a final result that
\begin{equation}
	\int_{0}^{\infty} \frac{x^2}{\exp\kl{\sqrt{x^2+z^2}} \pm 1}\dif x = \sum_{n=0}^{\infty} (\mp 1)^n \frac{z^2}{(n+1)^2} K_2((n+1)z).
\end{equation}

A similar calculation, this time without the need to introduce a geometric series, will reveal that the similar Boltzmann case (\emph{i.e.}, $s=0$) reads simply
\begin{equation}
	\int_{0}^{\infty} \frac{x^2}{\exp\kl{\sqrt{x^2+z^2}}}\dif x = z^2 K_2(z).
\end{equation}
If we took $0^0$ to represent $\lim_{x\to 0^+} x^x$, we could abuse this notation to rewrite the above equations for the number flux in a nutshell as
\begin{equation}
	\int_{0}^{\infty} \frac{x^2}{\exp\kl{\sqrt{x^2+z^2}} + s}\dif x = \sum_{n=0}^{\infty} (-s)^n \frac{z^2}{(n+1)^2} K_2((n+1)z).\label{eq:numberKfinal}
\end{equation}

The next integral is for the case of the energy flux,\footnote{Note that the energy is proportional to the frequency, which is proportional to $\sqrt{x^2+z^2}$! This integral also reappears when calculating the average frequency for the \emph{number} flux.}
\begin{equation}
	I_2 = \int_{0}^{\infty} \frac{\sqrt{x^2+z^2}x^2}{\exp\kl{\sqrt{x^2+z^2}} \pm 1}\dif x.
\end{equation}
We start again with a geometric series ansatz, giving
\begin{equation}
	\int_{0}^{\infty} \frac{\sqrt{x^2+z^2}x^2}{\exp\kl{\sqrt{x^2+z^2}} \pm 1}\dif x = \sum_{n=0}^{\infty} (\mp 1)^n \int_{0}^{\infty} \sqrt{x^2+z^2}x^2 e^{-(n+1)\sqrt{x^2+z^2}} \dif x.\label{eq:massiveEflux}
\end{equation}
Then we again substitute $x=z\sinh(t)$ and make more use of hyperbolic identities:
\begin{align}
	\eqref{eq:massiveEflux} =& \sum_{n=0}^{\infty} (\mp 1)^n z^4 \int_{0}^{\infty} \cosh^2(t) \sinh^2(t)\exp\kl{-(n+1)z\cosh(t)}\dif t,\\
	=& \sum_{n=0}^{\infty} (\mp 1)^n z^4 \int_{0}^{\infty} \kle{\frac{4\cosh(2t) + \cosh(4t) + 3}{8} - \frac{1+\cosh(2t)}{2}}e^{-(n+1)z\cosh(t)}\dif t,\\
	=& \sum_{n=0}^{\infty} (\mp 1)^n \frac{z^4}{8} \kl{K_4((n+1)z) - K_0((n+1)z)}.
\end{align}
As the Boltzmann case is again simply the first sum term ($n=0$),
\begin{equation}
	\int_{0}^{\infty} \frac{\sqrt{x^2+z^2}x^2}{\exp\kl{\sqrt{x^2+z^2}}}\dif x = \dots = \frac{z^4}{8} \kl{K_4(z) - K_0(z)},
\end{equation}
we can, like in the massless case, summarise the situation for the energy flux (and average number flux frequency!) as
\begin{equation}
	\int_{0}^{\infty} \frac{\sqrt{x^2+z^2}x^2}{\exp\kl{\sqrt{x^2+z^2}} + s}\dif x = \sum_{n=0}^{\infty} (- s)^n \frac{z^4}{8} \kl{K_4((n+1)z) - K_0((n+1)z)},\label{eq:energyKfinal}
\end{equation}
where, again as in the massless case, $s\in\{-1,0,1\}$ corresponds to bosons, Boltzmann particles, and fermions, respectively. It also includes the same abusive notation, $0^0=1$ --- its occurrence will henceforth not be commented on any more.

Now we come to the sole integral required for $\eta_{\text{average, }k,k}$,
\begin{equation}
	I_3 = \int_{0}^{\infty} \frac{x}{\exp\kl{\sqrt{x^2+z^2}} +s}\dif x.
\end{equation}
The geometric series in this case actually results in a (more or less) elementary integral:
\begin{subequations}
	\begin{align}
		I_3 &= \sum_{n=0}^{\infty} (-s)^n \int_{0}^{\infty} x \exp\kl{-(n+1)\sqrt{x^2+z^2}} \dif x,\\
		&= \sum_{n=0}^{\infty} (-s)^n \frac{1+(n+1)z}{(n+1)^2} e^{-(n+1)z}.
	\end{align}
\end{subequations}
This turns out to be the only remaining series which can be evaluated:
\begin{equation}
	I_3 = z\frac{\Li{1}\kl{-s e^{-z}}}{(-s)} + \frac{\Li{2}\kl{-s e^{-z}}}{(-s)}.
\end{equation}
\begin{figure}
	\centering
	\includegraphics[width=.9\textwidth]{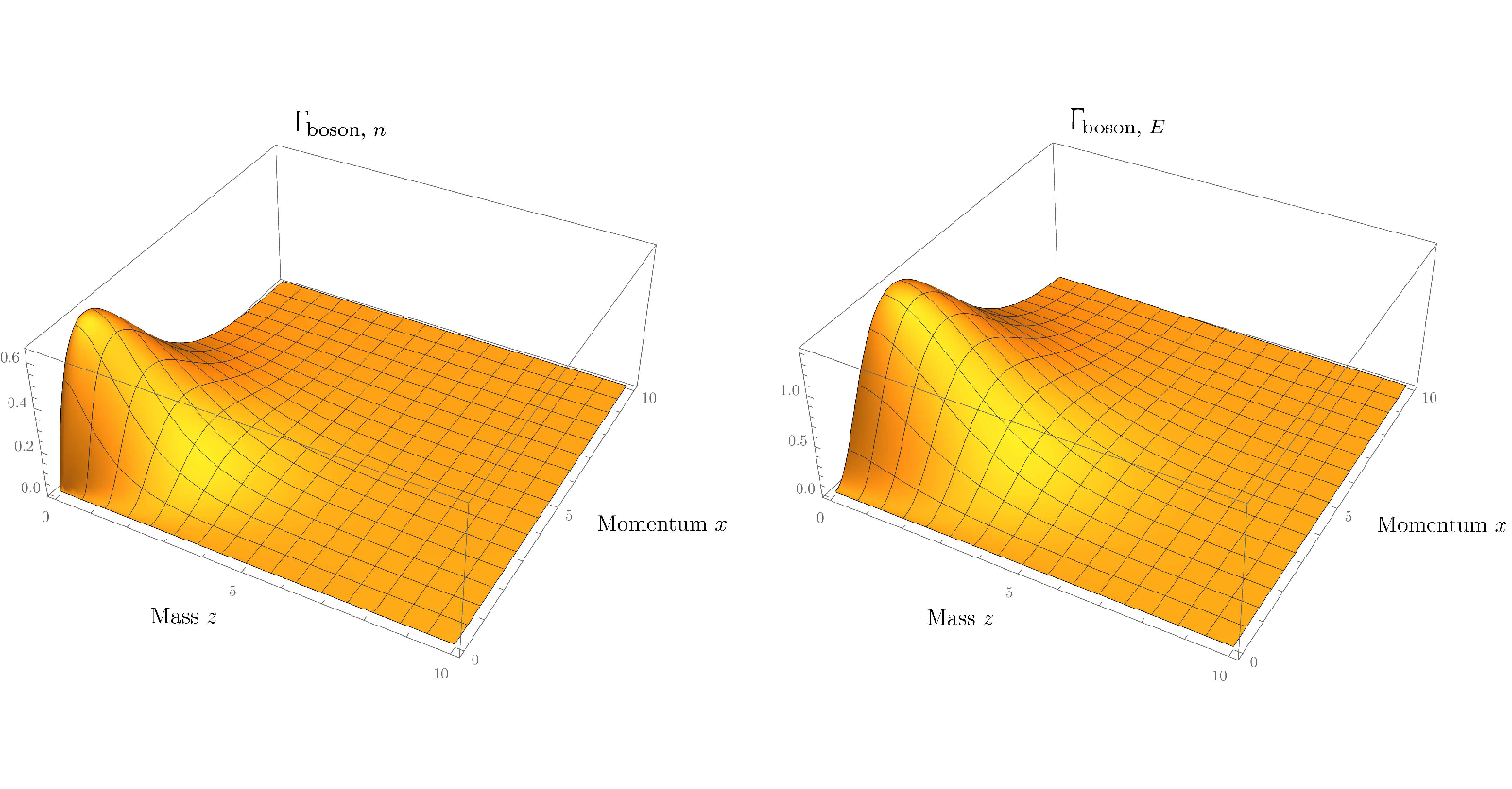}
	\includegraphics[width=.9\textwidth]{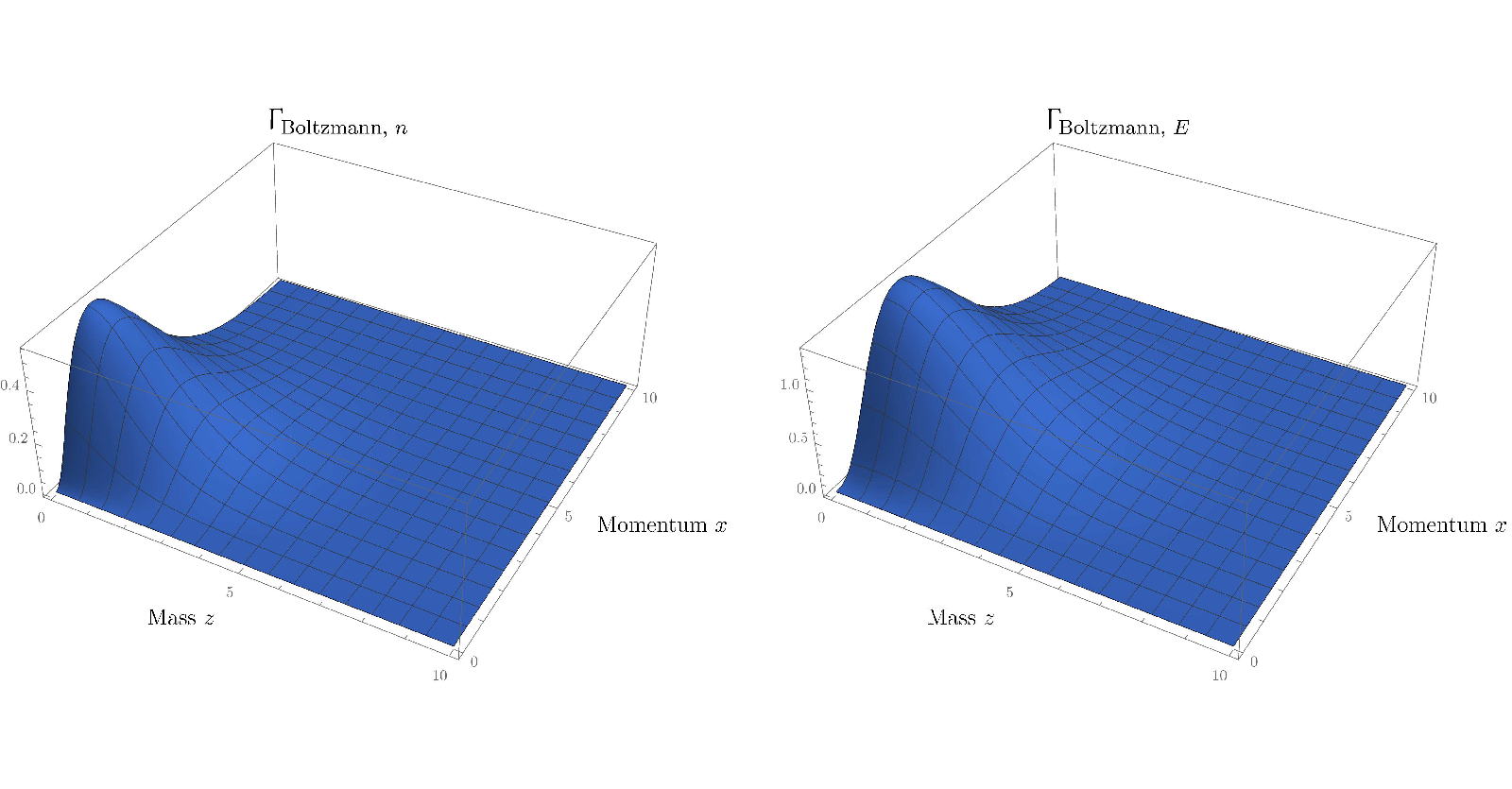}
	\includegraphics[width=.9\textwidth]{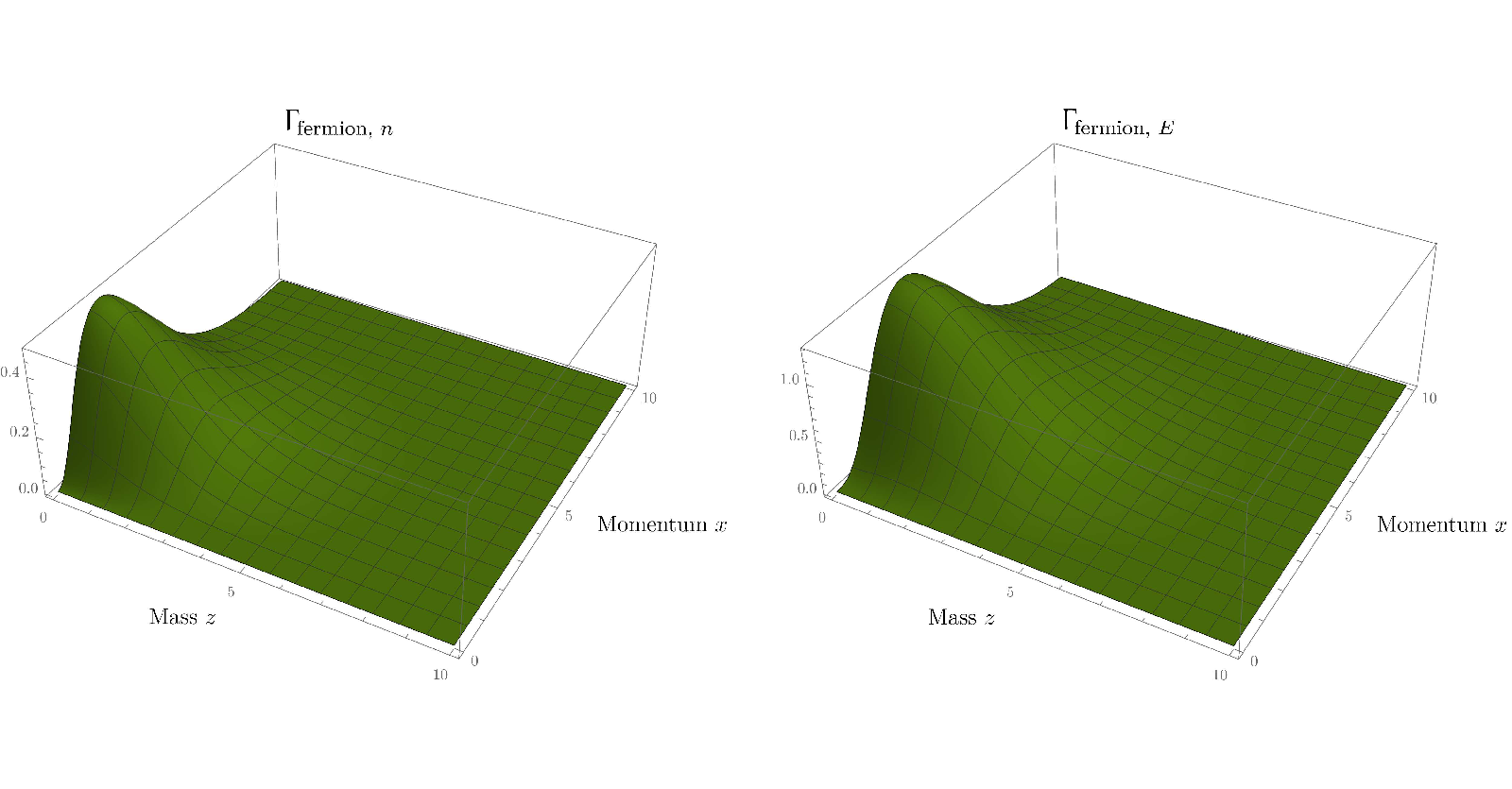}
	\caption[Plots of Number and Energy Flux Spectra for Massive Particles]{The dimensionless, massive spectra for the various particle types.}
	\label{fig:spectra}
\end{figure}

Lastly, we need to evaluate yet another integral in the case of the binned sparsity, $\eta_{\text{binned}}$. It is vital at this point to remember that the change of dispersion relation ($k\to\sqrt{k^2+m^2c^2}$) changes the frequency of the wave number bins we use to weigh each bin in equation~\eqref{eq:binnedRiemann}. As before, we will include the Boltzmann case right from the start. This time, the relevant integral (in a dimensionless form) is
\begin{equation}
	I_4 = \int_{0}^{\infty} \ed{\sqrt{x^2+z^2}}\frac{x^2}{\exp\kl{\sqrt{x^2+z^2}} + s}\dif x.
\end{equation}
When rewriting this as a geometric series, we then perform a change of variable to make use of the relation~\eqref{eq:Kool}:
\begin{subequations}
	\begin{align}
		I_4 &= \sum_{n=0}^{\infty} (-s)^n \int_{0}^{\infty} \frac{x^2}{\sqrt{x^2+z^2}} \exp\kl{-(n+1)\sqrt{x^2+z^2}} \dif x,\\
		&= \sum_{n=0}^{\infty} (-s)^n z^2 \int_{0}^{\infty} \sinh^2(t) e^{-(n+1)z\cosh(t)}\dif t,\\
		&= \sum_{n=0}^{\infty} (-s)^n \frac{z}{n+1} K_1((n+1)z).
	\end{align}
\end{subequations}

In view of our previous inclusion of a chemical potential, we could check how far we can extend this into the present, massive particle analysis. The answer is: All the previous results can be generalised to include a chemical potential. For practical reasons, let us consider in the following a dimensionless $\tilde{\mu} \defi \mu/(\kB T_\text{H})$. First of all, the maxima of the Boltzmann distribution will not change, as $\tilde{\mu}$ will only constitute a multiplicative factor to the fluxes. The result~\eqref{eq:numberKfinal} will change to
\begin{equation}
	\int_{0}^{\infty} \frac{x^2}{\exp\kl{\sqrt{x^2+z^2}+\tilde{\mu}} + s}\dif x = -\sum_{n=0}^{\infty} s^n e^{-(n+1)\tilde{\mu}} \frac{z^2}{(n+1)^2} K_2((n+1)z).
\end{equation}
All other appearances of series of modified Bessel functions would be changed in just the same way, by an overall factor of $\exp(-(n+1)\tilde{\mu})$ per sum term. The chemical potential would influence the polylogarithmic expressions appearing in the integral for $\eta_{\text{average, }k,k}$ just as it did in the massless case. However, given that the expressions for the sparsities to be derived will be long and arduous enough --- as far as they can be written down exactly at all ---, we will refrain from including the chemical potential in the following. If the need should arise to include them, the above expressions should be ample information to fill in the details.

Making use of the expressions above in the definitions~(\ref{eq:etapeakn}--\ref{eq:etaaverE}) and \eqref{eq:defbinned}, we arrive at
\begin{subequations}
	\begin{flalign}
		\eta_{\text{peak, }E,n} &= \frac{4\pi\omega_{\text{peak, }n}}{\sum_{n=0}^{\infty} (-s)^n \frac{z^2}{(n+1)^2} K_2((n+1)z)}\ed{g A} \kl{\frac{\hbar c}{\kB T}}^{2},&\\
	 	&= \frac{\omega_{\text{peak, }n}}{\pi\sum_{n=0}^{\infty} (-s)^n \frac{z^2}{(n+1)^2} K_2((n+1)z)}\frac{\lambda_\text{thermal}^{2}}{g A},&\\
	 	\addtocounter{parentequation}{1}\setcounter{equation}{0}
		\eta_{\text{peak, }E,E} &= \frac{4\pi\omega_{\text{peak, }E}}{\sum_{n=0}^{\infty} (-s)^n \frac{z^2}{(n+1)^2} K_2((n+1)z)}\ed{g A} \kl{\frac{\hbar c}{\kB T}}^{2},&\\
			  &= \frac{\omega_{\text{peak, }n}}{\pi\sum_{n=0}^{\infty} (-s)^n \frac{z^2}{(n+1)^2} K_2((n+1)z)}\frac{\lambda_\text{thermal}^{2}}{g A},&\\
			  \addtocounter{parentequation}{1}\setcounter{equation}{0}
		\eta_{\text{average, }E,n} &= 4\pi\frac{\sum_{n=0}^{\infty} (- s)^n \frac{z^4}{8} \kl{K_4((n+1)z) - K_0((n+1)z)}}{\kle{\sum_{n=0}^{\infty} (-s)^n \frac{z^2}{(n+1)^2} K_2((n+1)z)}^2}\ed{g A} \kl{\frac{\hbar c}{\kB T}}^{2},&\\
			  &= \frac{\sum_{n=0}^{\infty} (- s)^n \frac{z^4}{8} \kl{K_4((n+1)z) - K_0((n+1)z)}}{\pi \kle{\sum_{n=0}^{\infty} s^n \frac{z^2}{n+1} K_2((n+1)z)}^2}\frac{\lambda_\text{thermal}^{2}}{g A},&\\
			  \addtocounter{parentequation}{1}\setcounter{equation}{0}
		\eta_{\text{average, }k,k} &= \frac{4\pi}{z\frac{\Li{1}\kl{-s e^{-z}}}{(-s)} + \frac{\Li{2}\kl{-s e^{-z}}}{(-s)}}\ed{g A} \kl{\frac{\hbar c}{\kB T}}^{2},&\\
		&= \frac{1}{\pi\kl{z\frac{\Li{1}\kl{-s e^{-z}}}{(-s)} + \frac{\Li{2}\kl{-s e^{-z}}}{(-s)}}}\ed{g A} \kl{\frac{\hbar c}{\kB T}}^{2},&\\
		\addtocounter{parentequation}{1}\setcounter{equation}{0}
		\eta_{\text{binned}} &= \frac{4\pi}{\sum_{n=0}^{\infty} (-s)^n \frac{z}{n+1} K_1((n+1)z)}\ed{g A} \kl{\frac{\hbar c}{\kB T}}^{2},&\\
							&= \frac{1}{\pi\kle{\sum_{n=0}^{\infty} (-s)^n \frac{z}{n+1} K_1((n+1)z)}}\frac{\lambda_\text{thermal}^{2}}{g A}.&
	\end{flalign}
\end{subequations}
Here, $z=mc^2/\kB T_\text{H}$. When one writes the definitions of sparsities in terms of the dimensionless integrals~\eqref{eq:dimlessI} and keeps the $2\pi$ appearing every time in the denominator of the definitions, one needs a factor of $8\pi^2$ in each expression to get back the right numerical values. This is most easily verified by looking at the equivalent kind of calculation in the massless case. 

If one takes astrophysical black holes (say, of $M=1\, M_\odot$ and above), the value of $z$ will be large, even for masses of elementary particles. Thus for the early stage of the evaporation process, one can look at the limit $z\to \infty$. A quick look at the asymptotic behaviour of modified Bessel functions of the second kind, equation~\eqref{eq:asymptoticK}, results for all sparsities involving these special functions in an asymptotic behaviour of the kind $z^{n/2}e^z$ for some $n\in \mathbb{N}$, thus indeed diverging, as expected on physical grounds --- low temperatures make any massive emission less likely. The only remaining sparsity to check is $\eta_{\text{average, }k,k}$, whose denominator likewise converges to $0$ for $z\to\infty$. We see that mass, like other changes, increases sparsity.

\subsection{First Comparisons} \label{sec:firstcomparison}
At this stage we are able to do a first comparison of our semi-classical, semi-analytic flat space-time results. The different particles (if we gloss over the fact that it is a bit far-fetched to consider particles obeying Maxwell--Boltzmann statistics fundamental) fulfil
\begin{equation}
	\eta_{\text{fermion}} > \eta_{\text{Boltzmann}} > \eta_{\text{boson}},
\end{equation}
with each different definition of sparsity $\eta$ (binned, averaged frequencies, peak number, \dots) following these inequalities separately. And even independent of the actual numerical values, by appealing to the spin-statistics theorem (and the role of the Boltzmann distribution) these inequalities have merit, being just inverse statements about the number of emitted particles in thermal radiation.

While we provided some expressions for the case of including particle rest masses in the spectra, the resulting equations are hardly edifying. It therefore seems sensible to look for more helpful bounds that allow making statements about the physics changes when introducing these rest masses. For all sparsities besides the binned version, one has that
\begin{equation}
	f(z) = {\frac{\int_0^\infty  \frac{x^2 }{ \exp\left(\sqrt{z^2+x^2}\right) \mp 1 }  \; \dif x}{ \int_0^\infty  \frac{x^2 }{ \exp\left(x\right) \mp 1 } \; \dif x}}  \leq 1.
\end{equation}
This is the factor changing $\Upgamma_{\text{massless, }n}$ to $\Upgamma_{\text{massive, }n}$, thus will appear in all four sparsities, and will always increase the corresponding sparsity. Also, while not explicitly evaluable, both the flux peaks and the flux averages (for energy and number flux) will increase. Both results taken together imply that the sparsity will increase.

As for the binned sparsity, we have that similarly the integrand in $1/\eta_{\text{binned}}$ will be dampened, as
\begin{equation}
	\ed{k} \to \ed{\sqrt{k^2 c^2 + m^2 c^4/\hbar^2}}.
\end{equation}
Thus for all sparsities we gain the inequality
\begin{equation}
	\eta_{\text{massive}} > \eta_{\text{massless}}.\label{eq:massivemassless}
\end{equation}

The important introduction of grey body factors will be deferred until we discuss black hole (curved) space-times: As the grey body factors arise from the wave equation on the curved background, the grey body factors we are interested in make much more sense in this fuller setting. Also, we will be able to say little about how the introduction of general grey body factors will change peak or average frequencies. In the absence of superradiance, we at least can say that the grey body factors will have to be between $0$ and $1$, as they give the correction to the emissivity,\footnote{Often, frequency independence is demanded for a grey body factor. In view of our coming application, we shall refrain from doing so.} (with a black body having an emissivity of $1$) --- so they will at least decrease $\Upgamma_{n}$. The inclusion of them in our context will require numerical procedures for which we will rely on the work performed by Gray and Visser, and presented in \cite{SparsityNumerical} where this was done for non-rotating Schwarzschild black holes.

At the same time, the general appearance of the area of the emitting body and its comparison with the thermal wavelength allows the back-of-the-envelope argument that normally for a black \emph{body} $\lambda_{\text{thermal}}/A$ will prove to be reasonably small. In fact, terming it \enquote{unreasonably small} might be even more appropriate if one were to insert the values for the sun, assuming it to be a black body, as
\begin{equation}
	\lambda_{\text{thermal}}^2 \ll A.\label{eq:herecomesthesun}
\end{equation}
Such a small sparsity is equivalent to a tremendous amount of particles present in the radiation. Inserting expressions for Hawking radiation, the converse holds. Very little particles are present, thus the term \enquote{sparsity}. We will provide concrete values to this in section~\ref{sec:sparsity} once we made the analysis more watertight by including general relativistic effects not present up till this point of the discussion. This qualitative difference is one of several ways of recognizing the same fundamental difference between the two radiation types. And this is not just a numerical accident of little consequence, as it is closely related to the difference between classical and quantum statistical systems: Compare this, for example, with the corresponding discussion in terms of concentrations in \cite{KittelKroemer}. There, classicality is defined through a statement equivalent to
\begin{equation}
	\lambda_{\text{thermal}}^3 \ll V,
\end{equation}
where $V$ is the volume of the system considered. Thanks to the monotonicity of the length scales involved, this is entirely equivalent to our observation in equation~\eqref{eq:herecomesthesun}.

While the arguments above try to look specifically at the question of how classical the radiation is, it is possible to look at the issue from a perspective more amenable to quantum theory. For example, in \cite{BHradNonclass}, the average occupation number is calculated in a fully quantum mechanical Fock space treatment. The resulting occupation numbers end up being so low, that a comparison to classical radiation seems untenable. This is also a fairly straightforward result from deducing Hawking radiation from decoherence arguments --- here the low average occupation number is actually an immediate by-product of the derivation, see \cite{KieferDecoherenceHawkingRad}.

At this point, we therefore expect the radiation of the Hawking effect to be non-classical in counterposition to a black body's (\enquote{like} the sun) classical radiation. Let us now delve further into the discussion of this counterposition's quantification. Along the way we will see that our daring approximations (semi-classical, geometric optics/\allowbreak DeWitt approximation, black body) indeed reproduce the cited results, thus providing an intuitive approach to them.

\section{3+1-Dimensional General Relativity}\label{sec:sparsity}

Apart from differences in particle type, apart from allusions to the general relativistic role chemical potentials will play, the relevance of the introduced notion of sparsity has remained vague: While we made frequent references to previous arguments of scales involved that are able to distinguish Hawking radiation from \enquote{classical} black body radiation (comparing system size with thermal wavelength, arguments related to the density of state, \dots), it still remains imperative to actually fully implement and evaluate the introduced measures of sparsity in the relativistic setting. This section shall settle this score, and we will provide the exact values for these measures. We shall do this both in the context of $3+1$-dimensional relativity, as well as provide an example in higher dimensional space-times. Strictly speaking, the Einstein equations enter our discussion only in the context of one particular application in $3+1$ dimensions --- nevertheless we shall be talking about \enquote{general relativity}, as this is the most frequently encountered and basic example for the occurrence of curved space-times. Please bear with this slightly overly restrictive terminology for the sake of brevity. Also in this section, the analysis will stay semi-classical and semi-analytic.

As the frequent mentioning of \enquote{$3+1$} already alludes to, the general relativistic setting still requires the choice of an observer. Neither the Hawking effect itself, nor the definition of sparsity (through the spectra) can be phrased in a fully covariant manner. This is part of the reason why a direct application to analogue space-times will prove troublesome.

Unsurprisingly, the biggest change to the sparsity results gained in the previous section stems from the exchange of a flat space-time for a curved one. This precedes even considering the Hawking process itself: As the previously derived expressions for the sparsity of radiation can be rephrased in terms involving the area of the emitting body, it is important to have a good handle on quantifying this area. As the quantification of an area involves immediately the metric, and the radiation involves geodesic or wave equations on the corresponding geometry, this first step is mostly geometric in nature. Luckily for us, the first mentioned complication due to curved space-times --- possible changes when calculating an area --- do not occur in the situations we concern ourselves with: As our approach to calculating the integral over the area in, for example, equation~\eqref{eq:Upgamma} does not depend on how the resulting surface area $A$ was calculated, we neatly sidestep this issue. However, as each geometry will come with its own geodesic equations, ray optics will change. As we are concerned with black hole space-times, we will be including this effect by changing the effective area of the radiating object: Instead of using the area of the outer-most horizon, we will be using this area changed by a factor turning the emitting area into the cross-sectional area of the given geometry.

The next step, then, can also be understood as stemming from the introduction of geometry --- or more precisely: Geometr\emph{ies}. This not only will change the cross-sections of black holes (and thus the area one would want to consider), it also changes the shape and size of the outer-most horizon, the surface gravity at the horizon, and hence its Hawking temperature. Now, 3+1-dimensional general relativity comes with extra information pertinent to the sparsity as encountered in the Hawking effect: Both surface gravity (and hence the Hawking temperature, as described in section~\ref{sec:Hawking}) and the area of vacuum black holes encountered can be completely characterised by mass, angular momentum, and charge. As this is courtesy of the black hole uniqueness theorems, this situation changes more or less slightly when considering black holes surrounded by/in connection with matter (so-called \enquote{dirty black holes} \cite{DirtyBHs}), or when going to other space-time dimensions. Both variations will be discussed further below. The Schwarzschild case will prove to be particularly nice: All mass dependence cancels completely in the final sparsity results.

After discussing our inclusion of cross-sectional areas some more, we will then give the results for different black hole space-times. In doing so, we will encounter a slight complication in the form of the previously mentioned chemical potentials: The possibility of superradiance. To adequately incorporate this effect, we will need to carefully discriminate between superradiant and non-superradiant geometries. Thanks to the property~\eqref{eq:binnedsum} of the binned sparsity, we can neatly distinguish different radiation channels contributing to (or destroying) sparsity. We will therefore only consider this particular measure of sparsity in the superradiant case as the other measures will not allow easy identification of superradiance and its separation from the actual Hawking effect.

\subsection{Cross-Sections} \label{sec:3+1cross}
We still want to use, as a first approximation, the geometric optics approximation, as described in section~\ref{sec:geoapp}, for the radiation emitted in the Hawking process. While in flat space-time all radiation would leave a radiator of surface area $A$, in a curved space-time, specifically in a black hole space-time, part of the emitted radiation will not be able to reach observers at $r\to\infty$. This characteristic is captured by the capture cross-section already introduced in section~\ref{sec:cross}. Before proceeding, we need to know how this cross-section enters the notion of sparsity, precisely.

To do this, it is useful to consider our goal: We want to show the sparsity of Hawking radiation. If we now (grossly) overestimate the radiation, and the result is still sparse, we will have proved sparsity for the actual process even more. This is where our analysis of section~\ref{sec:cross} comes in: As we have seen there in the case of the Schwarzschild space-time, the cross-sectional area for radiation capture is larger than the area of the horizon $A$. This manifests itself in the statement that
\begin{equation}
	c_\text{eff} > 1.
\end{equation}
This holds true for massive particles, see figure~\ref{fig:ceff}, and also in the limit $\beta\to 0$, see equation~\eqref{eq:capturemassive}, where $c_\text{eff} = 1/\beta^2$ --- which diverges for $\beta\to0$. Likewise, we have for massless particles that $c_\text{eff} = 27/16>1$. This does not change (much) if we include the in section~\ref{sec:cross} mentioned effects of rotating black holes. For these, for Reissner--Nordström black holes, and for dirty black holes we will keep using the factor $27/16$ as a good safety net to bound our sparsity from below.

The other issue mentioned in section~\ref{sec:cross}, however, should be approached more cautiously: If we indeed want to include wave optical effects --- as we will once we discuss superradiance more carefully below --- we need to include grey body factors at some point. The simple way out, to which our geometric optics inspired approach effectively corresponds, is to invoke the DeWitt approximation, as mentioned in section~\ref{sec:geocomplications}. This is also an alternative way of introducing the factor $c_\text{eff}$ nearly from first principles, starting straight on the level of the spectra $\Upgamma_{n/E}$. In hindsight, however, this derivation would face the same fundamental issue as the geometric optics approximation: The low occupation numbers preclude both an appeal to the geometric optics approximation or the DeWitt approximation. It is comforting that the results still work out nicely --- despite being too much classical physics for the problem at hand.

This subsection's main point is this: As we want to be as conservative as possible in our estimate of sparsity of Hawking radiation, we use the capture (and later absorption) cross-section instead of the horizon area to be the area appearing in the area integral in the defining equations of sparsity. Our choice of $c_\text{eff}$ corresponds effectively to a sphere of radius $3\sqrt{3}/2 r_\text{H}$ on which the Hawking radiation originates. This certainly leads to mild tension with some calculations related to the fully quantum field theoretic Hawking effect. There, a long and ongoing dispute relates to where the maximum particle production happens: Near the horizon, or far from it, see \cite{BHQuantumAtmosphere} and its references. If the particle production really \emph{does} happen close to the horizon, our estimates will definitely underestimate sparsity. This would be our intention. On the other hand, \cite{BHQuantumAtmosphere} claims that particle production happens close to $r=4GM$. Then our choice of $c_\text{eff}$ would still not provide a lower bound. We will stick to it: The ubiquity of traditional calculations, both in the wave regime, and in the ray optics regime, of cross-sections leading to $c_\text{eff} = 27/16$ are the most common. If Hawking radiation indeed comes into existence outside of this, we will see that our estimates are still safe. Already at the current point the discussion \emph{where} the radiation of the Hawking effect originates is settled on the order of magnitude. Thus minor differences will not be able to undermine sparsity results, nor is this to be expected given the quantum theoretic knowledge of its sparsity \cite{BHradNonclass,KieferDecoherenceHawkingRad}.

If we include particle mass, the situation becomes a bit murkier: As a quick look at either equations~(\ref{eq:sigmamass},\ref{eq:sigmamassslow}) or figure~\ref{fig:ceff} reveals, for massive particles the effective capture cross-section can be made arbitrarily large. Thus it will make little sense to take this to be the area radiating in the Hawking effect. Rather, one should keep the choice from before. The reason for this can again be found from arguments of renormalised stress-energy tensors as in the already cited \cite{BHQuantumAtmosphere}: We expect the maximum of the renormalised stress-energy tensor to be close to the black hole (in the sense that $2GM/c^2$ to $5GM/c^2$ is a good estimate), and independent of one's stance on the particular radial coordinate where this maximum happens, the calculations so far agree on this. But even if one still used the effective capture cross-section to lower the sparsity further, other physical reasons would invalidate such an argument. The only stable, massive particles known to us are fermions --- which cannot have a classical limit in the sense of section~\ref{sec:geoquantum} as the Pauli exclusion principle prohibits high occupation numbers in the first place. Any massive boson is either not fundamental, but a composite particle, or decays too rapidly to have a long range meaning, classical or otherwise. The only remaining loop hole would be the decay products --- but we consider this possibility too contrived. 

In summary, given the inequalities and ruminations relating massless and massive cases in section~\ref{sec:firstcomparison}, particularly inequality~\eqref{eq:massivemassless}, there seems little purpose to include values for massive particles in the ensuing presentation of results. This is especially true given that we would introduce another free parameter (the emitted particle's mass) which we would need to decide upon in order to generate numerical values of sparsity.

One further coincidence is worth mentioning: The original definition of sparsity involved an integral over the full area which necessarily meant including radiation that will not reach a given observer. The cross-section, on the other hand, is immediately observer-dependent. This is most clearly seen by reminding ourselves of the observer-dependent shadows appearing in the Kerr geometry. Thus despite resulting in a more conservative estimate of sparsity (as we increased the emitting area) it is also physically the more plausible calculation --- it actually is concerned with an area associated with the black hole that an observer can measure.

\subsection{Non-Superradiant Approximation}\label{sec:nonsuper}
In the absence of superradiance, the principal analysis is simple: The only quantities we need to know are
\begin{itemize}
	\item the black hole's area,
	\item and the black hole's thermal wavelength, which can easily be calculated from knowledge of the Hawking temperature $T_\text{H}$ (which in turn is calculated from the surface gravity at the horizon).
\end{itemize}
These steps are reasonably simple, and we will exemplify this by considering
\begin{itemize}
	\item the Schwarzschild space-time,
	\item the Reissner--Nordström space-time,
	\item and general dirty black hole space-times.
\end{itemize}
From a purely technical point of view, though, we already capture the situation in the Reissner--Nordström geometry by the inclusion of dirty black holes, but the additional ingredients in the latter case (energy conditions) tend to obfuscate the physics. This makes the Reissner--Nordström geometry a good first step towards a broader understanding. All calculations will include the factor $c_\text{eff}$ derived from capture cross-section considerations.

\subsubsection{Schwarzschild Space-Time}
The simplest case, and thus a good cornerstone of the discussion of sparsity is obviously a simple, uncharged, and non-rotating black hole; the Schwarzschild solution. In view of the flat space-time analysis of section~\ref{sec:sparseflat}, and the subsequent introduction to changes for curved space-times of the beginning of this section, we only need to evaluate
\begin{equation}
	\eta_\text{Schwarzschild} = \frac{27}{16} \eta_\text{flat}(g,s),
\end{equation}
where $g$ is the degeneracy factor of the particle, and $s$ is the spin parameter introduced in section~\ref{sec:firstgen} when unifying the different particle types. $\eta_\text{flat}$ is a general stand-in for either of the five introduced sparsities $\eta_{\text{average, }E,n}, \eta_{\text{average, }k,k}, \eta_{\text{peak, }E,E}, \eta_{\text{peak, }E,n}$, or $\eta_{\text{binned}}$.

Let us first compute the common feature of all sparsities:
\begin{subequations}
	\begin{align}
	 	\frac{\lambda_{\text{thermal}}^2}{A_\text{eff}} &= \frac{16}{27} \frac{\kl{\frac{2\pi\hbar c}{\kB T_\text{H}}}^2}{A_\text{H}},\\
	 	&= \frac{16}{27} \frac{(8\pi^2 r_\text{H})^2}{4\pi r_\text{H}^2},\\
	 	&= \frac{256}{27}\pi^3.\label{eq:factor4error}
	\end{align}
\end{subequations}
\begin{sloppypar}\noindent This factor now is inserted in the corresponding sparsity values. Note that equation~\eqref{eq:factor4error} corrects an error in \cite{HawkFlux1}: There, a factor of $4$ is missing, thus the values given for the sparsity in \cite{HawkFlux2,HawkFlux1} underestimate the sparsity much more than intended. The values in this thesis can be considered errata to these papers.\end{sloppypar}

It is worth noting that the Schwarzschild black hole's properties conspire to actually lead to a sparsity independent of the black hole's mass, as the appearance of mass in thermal wavelength and area exactly cancel. This will not be true in more general space-times, or when introducing particle mass (as the mass is made dimensionless in the relevant integral with the help of the Hawking temperature).

The results are summarised in tables~\ref{tab:SchwarzschildB}--\ref{tab:Schwarzschildbo}. As $\eta_{\text{average, }n}=\eta_{\text{binned}}$ for massless particles, we refrained from listing both. In these tables we also included the values obtained for various particles in numerical studies, based on \cite{SparsityNumerical} and \cite{Page1}. The numerical values for fermions and Boltzmann particles have been calculated specifically for this thesis, using the code developed for bosons and kindly provided by Finnian Gray. Given the nature of the Regge--Wheeler equation \cite{FroNo98}, one should not trust these numerics when applied to other particles species, as the Regge--Wheeler equation is \emph{a priori} only valid for integer $s$. \emph{A posteriori}, the values at least seem to agree for fermions, and are roughly of the same kind as our semi-analytic results for Boltzmann particles.

\begin{table}
	\begin{center}
		\begin{tabular}{||c||c||c|c|c|c||}
			\hline
			\hline
			Bosons &  g & $\eta_{\text{peak, }E,n}$ & $\eta_{\text{peak, }E,E}$ & $\eta_{\text{average, }E,n}$ &
			$\eta_\text{binned}$ \\
			\hline
			\hline
			Semi-analytic & 1& $\frac{128\pi^2\left( 2 + W(-2e^{-2})\right)}{27\zeta(3)}$ & $\frac{128\pi^2\left( 3 + W(-3e^{-3})\right)}{27\zeta(3)}$ & 
			$\frac{64\pi^6}{405\zeta(3)^2}$& $\frac{512}{9}$
			\\
			\hline
			Value & 1& $\num{62.0307}$ & $\num{109.823}$ & $\num{105.141}$ &  $\num{56.8889}$
			\\
			\hline\hline \hline
			$s=0$ & 1& $\num{82.58}$ & $\num{111.3}$ & $\num{107.1}$ & $\num{65.24}$
			\\
			\hline 
			$s=1$ & 2 & $\num{984.2}$ & $\num{1038}$ & $\num{978.2}$ & $\num{865.2}$
			\\
			\hline 
			$s=2$ & 2 & $\num{20310}$ & $\num{20880}$ & $\num{19860}$ & $\num{18770}$
			\\
			\hline\hline
		\end{tabular}
	\end{center}
	\caption[Table of Bosonic Sparsities in Schwarzschild Geometry]{Black body results of the various figures of merit introduced to measure sparsity for massless bosons in the Schwarzschild case. The numerical results including grey body factors are based on \cite{SparsityNumerical} and \cite{Page1}. Numerical results rounded to four significant digits.}
	\label{tab:SchwarzschildB}
\end{table}

\begin{table}
	\begin{minipage}{\linewidth}
		\begin{center}
			\begin{tabular}{||c||c||c|c|c|c||}
				\hline
				\hline
				Fermions &  g & $\eta_{\text{peak, }E,n}$ & $\eta_{\text{peak, }E,E}$ & $\eta_{\text{average, }E,n}$ &
				$\eta_\text{binned}$ \\
				\hline
				\hline
				Semi-analytic & g& $\frac{512\pi^2\left( 2 + W(2e^{-2})\right)}{81g\zeta(3)}$ & $\frac{512\pi^2\left( 3 + W(3e^{-3})\right)}{81g\zeta(3)}$ & 
				$\frac{896\pi^6}{3645g\zeta(3)^2}$& $\frac{1024}{ 9g}$
				\\
				\hline
				Value & g& $\num{115.097}/g$&$\num{162.497}/g$ & $\num{163.553}/g$ & $\num{113.778}/g$
				\\
				\hline
				Value & 2& $\num{57.5487}$ & $\num{81.2485}$ & $\num{81.7767}$ & $\num{56.8889}$
				\\
				\hline\hline \hline
				{$s=1/2$} & 2& {$\num{116.2}$} & {$\num{128.6}$} & {$\num{123.9}$} & {$\num{104.6}$}
				\\
				\hline\hline
			\end{tabular}
		\caption[Table of Fermionic Sparsities in Schwarzschild Geometry]{Black body results of the various figures of merit introduced to measure sparsity for massless fermions in the Schwarzschild case. The numerical results including grey body factors are based on \cite{SparsityNumerical} and \cite{Page1}.\protect\footnote{Like the earlier results for bosons, the numerical values for fermions have been calculated using the Regge--Wheeler equation, and using a fermionic distribution on the grey body factors. There might be room for improvement here.}}
		\label{tab:SchwarzschildF}
		\end{center}
	\end{minipage}
\end{table}

\begin{table}
	\begin{minipage}{\linewidth}
	\begin{center}
		\begin{tabular}{||c||c|c|c|c|c||}
			\hline
			\hline
			Boltzmann & g & $\eta_{\text{peak, }E,n}$ & $\eta_{\text{peak, }E,E}$ & $\eta_{\text{average, }E,n}$ &
			$\eta_\text{binned}$ \\
			\hline
			\hline
			Semi-analytic &g &$  \frac{256\pi^2}{27 g}  $ & $\frac{128\pi^2}{9 g}$ & $\frac{128\pi^2}{9 g}$ &  $\frac{256\pi^2}{27 g}  $
			\\
			\hline
			Value &g  & $\num{93.5785}/g$ & $\num{140.368}/g$ &$\num{140.368}/g$ & $\num{93.5785}/g$
			\\
			\hline\hline \hline
			{$s=0$} & g & {$\num{107.4}/g$} & {$\num{135.3}$} & {$\num{135.3}/g$} & {$\num{96.78}$}
			\\
			\hline\hline 
		\end{tabular}
		\caption[Table of Boltzmann Sparsities in Schwarzschild Geometry]{Black body results of the various figures of merit introduced to measure sparsity for massless Boltzmann particles in the Schwarzschild case. The numerical results including grey body factors are based on \cite{SparsityNumerical} and \cite{Page1}.\protect\setcounter{mpfootnote}{1}\footnote{Like the earlier results for bosons, the numerical values for Boltzmann particles have been calculated using the Regge--Wheeler equation, and using a Boltzmann distribution on the grey body factors. This is definitely pushing the limits of what Regge--Wheeler is supposed to model, and should not be trusted too far.}}
		\label{tab:Schwarzschildbo}
	\end{center}
\end{minipage}
\end{table}%

\subsubsection{Numerical Calculations for the Schwarzschild Case}

It is worthwhile to explain a bit the background of the numerical results presented in tables~\ref{tab:SchwarzschildB}--\ref{tab:Schwarzschildbo}: Due to the lack of exact, analytical expressions for grey body factors (equivalently, the lack of a full understanding of Heun functions), if one wants to check sparsity on the wave optics level, without resorting to the DeWitt approximation (or going even the step further and calling on the geometric optics approximation), numerics are indispensable. There are as much options available, as there are ways to solve partial differential or ordinary differential equations --- the latter means one chooses to work on the level of the separated wave equation, the former on the not yet separated wave equation. The method employed in \cite{SparsityNumerical} by Gray and Visser is based on the Regge--Wheeler equation, so is working on the separated differential equation. The method used is inspired by a transfer matrix reformulation of the Regge--Wheeler equation, which in turn is built on a reformulation of it as a Shabat--Zakharov system \cite{BVShabatZakharov}. For more details, we refer to the two papers cited in this paragraph.

As can be seen from the tables~\ref{tab:SchwarzschildB}--\ref{tab:Schwarzschildbo}, the inclusion of grey body factors greatly increases sparsity. The higher the particles' spin, the higher the influence of grey body factors. For example, for scalars the numerical values nearly agree with the semi-analytic ones (note particularly $\eta_{\text{peak, }E,E}$ in this regard), while gravitons' sparsities increase two orders of magnitude. Our semi-analytic approach using several broad-stroke approximations, indeed is too conservative, as it was its intention.

\subsubsection{Reissner--Nordström Space-Time}
For the case of a charged, non-rotating black hole we can easily create a direct link with the Schwarzschild case. As described in section~\ref{sec:genBH}, the Reissner--Nordström solution has two horizons, located at $r^\text{H}_\pm$. The outer one, $r_+^\text{H}$ gives the area $A$ which, thanks to the spherical symmetry, just reads
\begin{equation}
	A = 4\pi r_+^\text{H}.
\end{equation}
Thermal wavelength depending on Hawking temperature, Hawking temperature depending on surface gravity, we now require knowledge of the surface gravity $\kappa$ in the Reissner--Nordström metric. This is
\begin{subequations}
	\begin{align}
	 	\kappa &= \frac{r_+^\text{H} - r_-^\text{H}}{2(r_+^\text{H})^2},\\
	 	&= \kappa_\text{Schwarzschild} \frac{r_+^\text{H} - r_-^\text{H}}{r_+^\text{H}}.
	\end{align}
\end{subequations}
The quantity we need to transfer to the Reissner--Nordström case is $\lambda_{\text{thermal}}^2/A$, which is proportional to $1/\kappa^2$. Thus we are able to see that adding charge to a non-rotating black hole further increases its sparsity:
\begin{equation}
	\eta_\text{Reissner--Nordström} = \eta_\text{Schwarzschild}\frac{(r_+^\text{H})^2}{(r_+^\text{H} - r_-^\text{H})^2} > \eta_\text{Schwarzschild}.
\end{equation}
In making the inequality a strict inequality we tacitly assume that an uncharged Reissner--Nordström metric ($Q=0$) should not be considered a Reissner--Nordström metric any more, but rather simply be referred to as a Schwarzschild metric. Any non-zero charge will immediately result in the strict inequality, as $r_+^\text{H} >r_+^\text{H} - r_-^\text{H}$.

\subsubsection{Dirty Black Hole Space-Times}
If we consider the slightly more general case of dirty black holes, let us demand, like in their introduction in section~\ref{sec:dirty}, that the NEC holds in the radial direction, and that the WEC is fulfilled at the horizon. Let us compare the dirty black hole's sparsity with that of a Schwarzschild black hole of the same horizon radius $r_\text{H}$. From section~\ref{sec:dirty} we know that
\begin{equation}
	\kappa_\text{dirty} = \kappa_\text{Schwarzschild} \times \exp\kl{-\Phi(r_\text{H})} \kl{1- \frac{8\pi G \rho(r_\text{H}) r_\text{H}^2}{c^4}},\label{eq:dirtykappa}
\end{equation}
and thus, that for the temperature the same kind of equality will hold. Thanks to the amount of symmetry in this class of black holes, we can reduce the comparison to a closer look at
\begin{equation}
	\frac{\lambda_{\text{thermal}}^2}{A}.
\end{equation}
By construction, we have $A_\text{Schwarzschild} = A_\text{dirty}$. Together with equation~\eqref{eq:dirtykappa} this then implies
\begin{equation}
	\eta_\text{dirty} = \eta_{\text{\,Schwarzschild}} \times \kl{\frac{ e^{\Phi(r_\text{H})} }{ 1-8\pi G \rho(r_\text{H}) r_\text{H}^2/c^4}}^2,
\end{equation}
for any of our sparsity definitions. Given our demands (NEC in radial direction, WEC), this then implies
\begin{equation}
	\eta_\text{dirty} \geq \eta_{\text{\,Schwarzschild}}.
\end{equation}

\subsubsection{Results Sans Superradiance}
Excluding superradiance we see that the sparsity of black holes is bounded \emph{below} by the sparsity of the Schwarzschild black hole. Given our results regarding the inclusion of particle mass, also particle mass will only further increase the sparsity, as the results transfer straight to the curved space-time analysis, though the precise notion of the effective area would have to be carefully chosen, given the diverging results for $c_\text{eff}$ in the low velocity regime. These results were also independent of the particles spin, as long as we neglect backreaction which would add spin to the black hole and thus open up the regime of superradiance. Similarly, as long as we are not dealing with a charged black hole, a particle's charge will not change these results.

\subsection{Including Superradiance} \label{sec:superradiance}
If the space-time allows superradiance, it means that --- for some reason --- there is a chemical potential $\mu$ present, which can have significant influence on the occupation number $\langle n\rangle_\omega$ of bosonic particles. The most natural occurrences for chemical potentials in our context are charge and angular momentum. The former will result in a changed Hawking emission from charged black holes, the latter changes the emission of particles with spin from rotating black holes. Let us recall the change induced in the bosonic occupation number:
\begin{equation}
	\langle n\rangle_\omega = \ed{\exp\{ (\hbar\omega)/\kB T_\text{H}\} - 1 } \to \langle n\rangle_\omega = \ed{\exp\{ (\hbar\omega-\mu)/\kB T_\text{H}\} - 1 }.
\end{equation}
For positive chemical potentials, this expression will diverge if $\hbar \omega = \mu$, and become even negative if $\omega$ is less than $\mu/\hbar$. In the situations we are concerned with, this negative occupation number will occur concurrently with a negative grey body factor $T_{N}(\omega)$. Here, $N$ is a collective index listing all dependencies of the grey body factors besides frequency. We will mostly be concerned with the case of dependence on the spin $s$, and angular momentum $\ell$ and $m$. 

Unless one introduces interactions, superradiance occurs only for bosons \cite{SuperradianceFlux,LNPSuperradiance,SuperradiancePenroseProcess}; the Penrose process, however, can be seen as a particle analogue of the wave phenomenon of superradiance, see \cite{SuperradiancePenroseProcess}. As both Hawking effect and superradiance can be phrased in terms of quantum field theory and radiation linked to the vacua appearing there, some definitions lump the two notions together. We find this not particularly helpful. For example, the Penrose process requires only an ergo-region (which can exist independently of a horizon, though unstably so), while the Hawking effect can only occur for a space-time with (some kind of) horizon \cite{EssIness}. We will therefore reserve the name \enquote{Hawking radiation} to only those modes which are outside the superradiant regime.

It is vital to understand that our approach so far, mostly excluding grey body factors, is ill-suited to deal with superradiance: The occupation number becomes negative, and without the also negative grey body factors no sensible notion of radiation is possible. Our semi-analytic approach of approximating the radiation as Planckian is therefore severely cast into doubt. In the following we shall try to alleviate some of these shortcomings: First with some fairly general comments on superradiance in the context of charged black holes (concretely, the static case of Reissner--Nordström, so we do not run the risk of conflating two kinds of superradiance), mostly concerned with its physical irrelevance, and why it is more useful to concern oneself with our second case study, that of the Kerr space-time and superradiance due to angular momentum.

\subsubsection{Reissner--Nordström Black Holes Revisited} 

The chemical potential in the Reissner--Nordström space-time is due to the charge of the black hole and reads
\begin{equation}
	\mu = q V_\text{H},
\end{equation}
where $q$ is the charge of the emitted quantum, and $V_\text{H}$ is the electric potential at the horizon. Since the only charged particles are actually massive, it makes sense to include for the time being also the particles mass. We thus take the frequency $\omega$ in the Bose--Einstein distribution not proportional to $k$, but to equal $\sqrt{k^2 c^2 \hbar^2 + m^2 c^4}$. If the particle mass is now small enough such that
\begin{equation}
	mc^2 < q V_\text{H},
\end{equation}
then we have the possibility of superradiance. One can now define a generalised notion of an ergoregion $E$ to handle this case, by defining it to be
\begin{equation}
	E \defi \{ x=(t,r,\theta,\phi) \in M | mc^2 < q V(r) \}.
\end{equation}
In counterpoint to the more frequently encountered ergoregions in rotating space-times this ergoregion would actually be \emph{particle-dependent}. It depends both on rest mass and charge of the particle. Apart from this, however, it is a perfectly well-behaved region of space-time and the corresponding ergosurface would be given by those points where the radial coordinate $r$ satisfies
\begin{equation}
	mc^2 = q V(r).
\end{equation}
Astrophysically, however, a charged black hole tends to neutralise too quickly to be of relevance --- and this is the main point of this short introduction or interlude: They neutralise fast, both through the accretion of plasma, as well as through the Schwinger effect close to the horizon.

\subsubsection{Superradiance in the Kerr Metric} 

As with our previous example of the Reissner--Nordström space-time for \emph{uncharged} particles (that is \emph{excluding} superradiance), we will begin our analysis with a careful look at the physical quantities that appeared in the Schwarzschild results. The reason for this is two-fold: One, we want to see how our approach can be taken to this new space-time --- again, for the moment ignoring the issue of superradiance. Two, it will provide the opportunity to introduce some of the quantities that will be of importance in the later discussion that \emph{does} include superradiance.

The point is that all sparsities in the non-superradiant cases, and even earlier, in the flat space-time examples, where given in terms of $\lambda_{\text{thermal}}^2/A_\text{H}$. Let us have a look at the corresponding quantities in the case of the Kerr geometry.\footnote{Strictly speaking, the following works verbatim in Kerr--Newman; since we consider massless particles, we rule out all physical (known) particles with charge, and we thus would not add another layer of superradiance we need to consciously ignore for a moment.} Similar to what we did for the emission of uncharged particles in the Reissner--Nordström space-time in section~\ref{sec:nonsuper}, we can bring both inner and outer horizon radii, $r_\pm$, to bear. This gives us, see also section~\ref{sec:Kerr},
\begin{align}
	A_\text{H} &= 4\pi(r_+^2 + a^2),\label{eq:KerrA2}\\
	\kappa &= \frac{r_+ - r_-}{2(r_+^2+a^2)},\label{eq:Kerrkappa2}\\
	\kB T_\text{H} &= \frac{\hbar c \kappa}{2\pi}.\label{eq:KerrT2}
\end{align}
As long as our semi-classical approximation holds we can then see that
\begin{equation}
	\kl{\kappa A_\text{H}}_\text{Kerr} = \underbrace{\kl{\kappa A_\text{H}}_\text{Schwarzschild}}_{=\pi} \times \frac{(r_+-r_-)^2}{r^2_+ + a^2}.
\end{equation}
This in turn allows us to give an estimate for the sparsity (of any type) in the realm of validity of the black body and semi-classical approximation:
\begin{align}
	\eta_\text{Kerr} &= \eta_{\text{\,Schwarzschild}}\times \frac{(r_+-r_-)^2}{(r^2_+ + a^2)^2}.
\end{align}
So in our original approximation, sparsity would increase yet again by the addition of angular momentum or charge (or any other similar change).\footnote{While it may seem tempting to also have a closer look at the Taub--NUT solution presented in section~\ref{sec:unphysical}, the concept of surface gravity is much more complicated in these cases, and either way, this solution seems tenuous at best from a physics point of view  \cite{TaubNUTsurfacegravity1,TaubNUTsurfacegravity2}. (Mostly these doubts, and the trouble with calculating meaningful surface gravities, is related to the presence of closed time-like curves.)}

This notwithstanding, the Kerr black hole \emph{does} feature superradiance and it hence requires further consideration. Hod used a less restrictive definition of Hawking radiation (compared to the one given above) in \cite{HodRotating} to achieve $\eta \approx O(1)$ for near-extremal Kerr black holes.\footnote{In particular, this is stated explicitly in equation~(9) of \cite{HodRotating}.} In the extremal limit $\kappa\to 0$ (and thus $a/M\nearrow 1$), the grey body factors $T_{s\ell m}(\omega)$ can be approximated as 
\begin{equation}
	T_{s \ell m}(\omega) \approx C_{\ell s} \kle{A_\text{H}\omega\kl{\omega-m \Omega_\text{H}}}^{2\ell +1}.
\end{equation}
Reminding ourselves of the bosonic occupation number in this case,
\begin{equation}
	\langle n\rangle_\omega = \ed{\exp\{ (\hbar\omega-\hbar m\Omega_{\text{H}})/\kB T_\text{H}\} - 1 },
\end{equation}
we see that, as promised, both grey body factors and occupation number change sign at the same frequency, $\omega = m\Omega_{\text{H}}$. This means that in the frequency range $[0,m\Omega_{\text{H}}]$ the spectrum is not well described by our semi-classical, Planckian approximation. It is also well-known that superradiance dominates over the Hawking flux in this range, see \cite{PageThesis,Page1,Page2,Page3}. We thus propose a different procedure (to that of \cite{HodRotating}) to talk about sparsity in this frequency range. In order to achieve this, let us make use of our earlier definition of superradiance which separated it from the Hawking effect. By doing so, we can employ the earlier mentioned (and as of now unused) property~\eqref{eq:binnedsum}, the reciprocal summability of $\eta_{\text{binned, }i}$ for different radiation channels $i$ --- we simply consider Hawking radiation and superradiance different channels.

Slightly modifying the analysis of Page (as cited above), we can take this to translate to
\begin{equation}
	\ed{\eta_{\text{binned}}} = 2\pi \sum_{\ell m} \int T_{s \ell m}(\omega) \langle n\rangle_\omega \frac{\dif \omega}{\omega}.\label{eq:sup2pi}
\end{equation}
This becomes in the near-extremal limit
\begin{equation}
	\ed{\eta_{\text{binned}}} \approx \kl{A_\text{H} \Omega_{\text{H}}^2}^{2\ell+1} \sum_{\ell m} m^2 C_{\ell s} \int \frac{\kle{x(x-1)}^{2\ell+1}}{e^{\epsilon(x-1)}-1} \frac{\dif x}{x},
\end{equation}
where $x \defi \omega/(m\Omega_{\text{H}})$ and $\epsilon \defi \hbar m \Omega_{\text{H}}/(\kB T_\text{H})$. We also absorbed numerical constants both times in the grey body factor (either $T_{s \ell m}$ or $C_{\ell s}$), with the exception of a single factor of $2\pi$ in equation~\eqref{eq:sup2pi} inherent in the definition of sparsity, see equation~\eqref{eq:defbinned}. Note that $\epsilon$ will be very large in the near-extremal limit. This, $\epsilon \gg 1$, is the pivotal inequality to both agree with previous results, but --- contrary to Hod's conclusion --- retain sparsity of the \emph{Hawking radiation}. As the lowest angular momentum available, $\ell=m=s$, will have to overcome the least angular momentum barrier in the relevant wave equations (Teukolsky equation, Regge--Wheeler, Zerilli, \dots), the corresponding mode will dominate, and we can further approximate the sparsity as
\begin{equation}
	\ed{\eta_{\text{binned}}} \approx s^2\kl{A_\text{H} \Omega_{\text{H}}^2}^{2s+1} C_{s s} \int_{0}^{\infty} \frac{\kle{x(x-1)}^{2s+1}}{e^{\epsilon(x-1)}-1} \frac{\dif x}{x}.
\end{equation}
In principle, we could have kept the sum, and considered emission of each angular momentum mode as a separate radiation channel. But the final result would not be changed by this, due to the dominance of the inequality $\epsilon \gg 1$ in the ensuing analysis. Now let us apply the reciprocal additivity of $\eta_{\text{binned}}$:
\begin{align}
	&\ed{\eta_{\text{binned}}} \approx \underbrace{s^2\kl{A_\text{H} \Omega_{\text{H}}^2}^{2s+1} C_{s s} \int_{0}^{1} \frac{\kle{x(x-1)}^{2s+1}}{e^{\epsilon(x-1)}-1} \frac{\dif x}{x}}_{\ifed\ed{ \eta_{\text{superradiant}}}} \nonumber\\&\hphantom{\ed{\eta_{\text{binned}}} \approx} \qquad + \underbrace{s^2\kl{A_\text{H} \Omega_{\text{H}}^2}^{2s+1} C_{s s} \int_{1}^{\infty} \frac{\kle{x(x-1)}^{2s+1}}{e^{\epsilon(x-1)}-1} \frac{\dif x}{x}}_{\ifed \ed{\eta_{\text{Hawking}}}}.
\end{align}
It is here that integration techniques seen in the case of massive particles prove useful again --- though this time resulting in modified Bessel functions of the first kind. Let us start by looking at $\eta_{\text{superradiant}}$ first, and apply again the trick of performing a geometric series expansion.\footnote{Whose convergence, thanks to the presence of $\epsilon$ will be fantastic this time.} The subsequent separation and substitution into a form amenable to relation~\eqref{eq:Icool} give
\begin{align}
	&\ed{\eta_{\text{superradiant}}} = \sum_{n=0}^{\infty} s^2\kl{A_\text{H} \Omega_{\text{H}}^2}^{2s+1} C_{s s} e^{(n+1)\epsilon} \int_{0}^{1} \kle{x(x-1)}^{2s+1}e^{-(n+1)\epsilon x} \frac{\dif x}{x},\\
	&\qquad = \sum_{n=0}^{\infty} s^3\kl{A_\text{H} \Omega_{\text{H}}^2}^{2s+1} C_{s s} e^{(n+1)\epsilon} \frac{\sqrt{\pi} e^{-(n+1)\epsilon/2}}{\kl{(n+1)\epsilon}^{2s+3/2}}\Gamma(2s)\nonumber\\
	&\qquad \hphantom{=}\times \kle{(n+1)\epsilon I_{2s-1/2}((n+1)\epsilon/2) + \kl{(n+1)\epsilon-4(2s+1/2)} I_{2s+1/2}((n+1)\epsilon/2)}.
\end{align}
Given $\epsilon\gg 1$ we now apply our knowledge of the asymptotic behaviour of the modified Bessel functions of the first kind, see equation~\eqref{eq:asymptoticI}, to get
\begin{align}
	&\ed{\eta_{\text{superradiant}}} \sim \sum_{n,k}^{\infty} s^3\kl{A_\text{H} \Omega_{\text{H}}^2}^{2s+1} C_{s s} e^{(n+1)\epsilon} \frac{\sqrt{\pi} e^{-(n+1)\epsilon/2}}{\kl{(n+1)\epsilon}^{2s+3/2}}\Gamma(2s)\nonumber\\
	&\times \frac{e^{(n+1)\epsilon/2}}{\sqrt{\pi (n+1)\epsilon}}\kle{(n+1)\epsilon \frac{a_k(2s-1/2)}{\kl{(n+1)\epsilon/2}^k} + \kl{(n+1)\epsilon - 4(2s+\ed{2})}\frac{a_k(2s+1/2)}{\kl{(n+1)\epsilon/2}^k}}.
\end{align}
Note that the exponentials with argument $\pm(n+1)\epsilon/2$ cancel, leaving an overall $\exp((n+1)\epsilon)$. For the purpose of calculating $\eta_{\text{superradiant}}$ it is now sufficient to limit attention to the first term of the asymptotic expansion where $k=0$: All occurrences of $\epsilon$ will be dominated by the overall, remaining exponential. Hence for $\epsilon\to \infty$, $1/\eta_{\text{superradiant}}\to 0$ and $\eta_{\text{superradiant}} \to\infty$.

Now let us examine $\eta_{\text{Hawking}}$ more closely. Proceeding similarly, we get that
\begin{align}
	&\ed{\eta_{\text{Hawking}}} = \sum_{n=0}^{\infty} s^3\kl{A_\text{H} \Omega_{\text{H}}^2}^{2s+1} C_{s s} e^{(n+1)\epsilon} \frac{e^{-(n+1)\epsilon/2}}{\sqrt{\pi}\kl{(n+1)\epsilon}^{2s+3/2}}\nonumber\\
	&\times\kle{\kl{4\kl{2s+\ed{2}}-(n+1)\epsilon} K_{-2s-1/2}\kl{\frac{(n+1)\epsilon}{2}} + (n+1)\epsilon K_{-2s-1/2}\kl{\frac{(n+1)\epsilon}{2}}}.
\end{align}
The appearance of modified Bessel functions of the second kind now vastly changes the asymptotic behaviour, see equation~\eqref{eq:asymptoticK}:
\begin{align}
	&\ed{\eta_{\text{Hawking}}} \sim \sum_{n=0}^{\infty} s^3\kl{A_\text{H} \Omega_{\text{H}}^2}^{2s+1} C_{s s} e^{(n+1)\epsilon} \frac{e^{-(n+1)\epsilon/2}}{\kl{(n+1)\epsilon}^{2s+5/2}}\nonumber\\
	&\hphantom{\sim} \times\kle{\kl{4\kl{2s+\ed{2}}-(n+1)\epsilon} \exp((n+1)\epsilon/2) + (n+1)\epsilon \exp((n+1)\epsilon/2)}.
\end{align}
This time, \emph{all} exponentials cancel. We are left with negative powers of $\epsilon$, and thus for very large $\epsilon$, $\eta_{\text{Hawking}}$ will diverge. This means that the actual Hawking radiation becomes absolutely negligible, even compared to our earlier results. This is not just an effect of the presence of superradiance dominating the radiation process. The Hawking process itself becomes sparser in the extremal limit, concurrently with that. To summarise:
\begin{equation}
	\eta_{\text{Hawking}} = O\kl{\epsilon^{2s+5/2}} \ggg 1,
\end{equation}
while
\begin{equation}
	\eta_{\text{superradiant}} = O\kl{\epsilon^{2s+1} e^{-\epsilon}} \lll 1.
\end{equation}
This is (matters of definition aside) a very different result to that obtained in \cite{HodRotating}.

\subsection{Summary}
We see that in all situations we looked at, the sparsity of Hawking radiation is a fundamental feature of Hawking radiation. While our analysis strictly speaking is a Planckian, geometric optics approximation, it agrees with analyses coming from more quantum mechanical points of view. Sparsity thus proves to be a surprisingly stable aspect of the astrophysical Hawking effect. The only mild challenge is the incorporation of superradiance --- which is easily overcome once one realises that it makes sense to separate these processes as two different physical effects rather than lumping them together.

\FloatBarrier
\section[An Example from D+1-dimensional General Relativity]{An Example from D+1-Dimensional Extensions of General Relativity} \label{sec:ndim}
One natural extension of general relativity is that to higher dimensions than the traditional 4 space-time dimensions.\footnote{One want for higher dimensions usually stems from particle physics models going beyond the standard model of particle physics. The crossing of high energy particle physics and cosmology, however, leads quite naturally to applications of higher dimensions already in general relativity itself.} In the presence of exact black hole solutions for these higher dimensional versions of general relativity it is equally natural to search for an extension of the sparsity results of the previous section~\ref{sec:sparsity}. In this section, we shall calculate the sparsity of a particular solution to the $D+1$-dimensional Einstein equations exactly, and contrast these results to an approximate analysis done in \cite{HodNdim}.

As the solution we will analyse will be spherically symmetric, let us fix our coordinates first. As our time coordinate does not appear in the fluxes considered in section~\ref{sec:sparsity}, we only need to think about angular coordinates in $D-1$ dimensions (one spatial dimension already covered by the corresponding radial coordinate). These spherical polar coordinates are such that $\varphi_{D-1}\in[0,2\pi)$, $\varphi_i \in [0,\pi)$, if $i \in \{2,\dots,D-2\}$, and $\varphi_{1} \in [0,\frac{\pi}{2})$. Our radial coordinate shall be named $k$, as the radial coordinate appears mostly in connection with the wave vector $\mathbf{k}$. The volume form $\dif V$ takes the form 
\begin{equation}\label{eq:ndimVol}
	\dif^{D} k = k^{D-1} \sin^{D-2} \varphi_1 \dots \sin \varphi_{D-2} \sin^0 \varphi_{D-1} \dif k \dif \varphi_1 \dif\varphi_{D-1}.
\end{equation}

The uniqueness theorems for black holes hold (without further assumptions) only in 3+1 dimensions \cite{Papantonopoulos2009,NumBlackSaturns,QGHolo}. While there are several solutions and solution families known, at the time of writing this might not have been done exhaustingly. Nonetheless, there exists a fairly straightforward generalization of the Schwarzschild metric which can be found simply from generalizing to $D$ dimensions the method employed for finding the spherically symmetric solution in 3+1 dimensions. This gives the Tangherlini metric, named after its finder Frank R. Tangherlini \cite{Tangherlini}, and is given by the following line element:
\begin{equation}\label{eq:Tangherlini}
	\dif s^2 = -\kl{1-\kl{\frac{r_\text{H}}{r}}^{D-2}}\dif t^2 + \kl{1-\kl{\frac{r_\text{H}}{r}}^{D-2}}^{-1}\dif r^2 + r^2 \dif \Omega_{D-1}^2,
\end{equation}
where we follow the notation of \cite{FrolovZelnikov2011}, apart from our counting of dimensions: We count space dimensions, \cite{FrolovZelnikov2011} counts space-time dimensions. The horizon radius $r_\text{H}$ is now given as
\begin{equation}\label{eq:rhndim}
	r_\text{H} = \sqrt[D-2]{\frac{8 \Gamma(\frac{D}{2})GM}{(D-1)\pi^{\nicefrac{(D-2)}{2}}}}.
\end{equation}
The surface area of the horizon becomes
\begin{equation}
	A_{\text{H}} = 2 \frac{\pi^{D/2}}{\Gamma(D/2)} r_{\text{H}}^{D-1},
\end{equation}
while the Hawking temperature is
\begin{equation}
	\frac{D-2}{4\pi r_\text{H}}\frac{\hbar c}{\kB}.
\end{equation}

\subsection{The Number Flux}
First, let us integrate the number flux, which in $D+1$ dimensions takes the form
\begin{equation}
	\dif \Upgamma_n = \frac{g}{(2\pi)^{D}}\frac{c k^{D-1} \cos\varphi_1 \sin^{D-2}\varphi_1 \dots \sin\varphi_{D-2} \dif k \dif \varphi_1 \dots \dif \varphi_{D-1}\dif A}{\exp({\frac{\hbar c k }{ {\kB T}}}-\mu)+s}.\label{eq:ndGamma}
\end{equation}
Here, we already included different particle species and possible chemical potentials --- however, given the lack of black hole uniqueness theorems, it is somewhat doubtful how much physical relevance can be given to these chemical potentials.\footnote{Note that, compared to the previous discussion in four space-time dimensions in section~\ref{sec:sparseflat}, we included the physically important factors of $1/(\kB T)$ in $\mu$ for more legible expressions.} Their inclusion is more a testament to the prowess of the exact analytic results than to their physical applicability.\footnote{If we put aside the fact that we \emph{are} already discussing black holes in dimensions other than 3+1.} It is important to note that the projection onto the sphere, responsible for the appearance of the cosine, is necessary for connecting this $D+1$-dimensional case back to standard (flat-space) results from thermodynamics.

Inserting the limits for our $D$-dimensional spherical polar coordinates, we can then progress to the total number flux
\begin{align}
	\Upgamma_n = \frac{gc}{(2\pi)^{D}} &\underbrace{\int_{0}^{\infty}\frac{k^{D-1} \dif k}{\exp({\frac{\hbar c k }{ \kB T}}-\mu) + s}}_{\ifed \int_k}\int_{0}^{A}\dif A \int_0^{2\pi} \dif \varphi_{D-1} \nonumber\\
	&\times\underbrace{\int_{0}^{\pi}\dif \varphi_{D-2} \sin \varphi_{D-2} \cdots \int_{0}^{\pi}\dif \varphi_2 \sin^{D-3}\varphi_2 \int_{0}^{\frac{\pi}{ 2}} \dif \varphi_1 \cos\varphi_1 \sin^{D-2}\varphi_1}_{\ifed \int_{\text{ang}}},\label{eq:fullint}\\
	= \frac{g c A}{(2\pi)^{D-1}}&\int_k \int_\text{ang}.\label{eq:fullintshort}
\end{align}
As these integrals are sufficiently involved to become cumbersome when done at once, we defined $\int_{\text{ang}}$ and $\int_k$ to \enquote{divide and conquer} them. The simple integral (involving a change of variables, then recognizing the polylogarithm) is $\int_k$:
\begin{equation}
	\int_k = \kl{\frac{\kB T }{ \hbar c}}^{D} (D-1)! {\frac{\Li{D}(-s e^{\mu})}{ (-s)}}.\label{eq:intk}
\end{equation}
For $\int_{\text{ang}}$ we first evaluate the integral over $\varphi_1$:
\begin{equation}
	\int_{0}^{\frac{\pi}{2}} \cos \varphi_1 \sin^{D-2} \varphi_1 \dif \varphi_1 = \int_{0}^{1} \sin^{D-2} \varphi_1 \dif \sin \varphi_1 = \ed{D-1}.
\end{equation}
The remaining integrals from $0$ to $\pi$ can be evaluated or looked up in integral tables\footnote{The result can be checked with, for example, \cite{GradshteynRyzhik1980}, formula {3.621.5}, p.~369. It easily follows from setting $\nu=1$. Similarly for \cite{Olver2010a}, p.142, formula 5.12.2.}. The last step involves recognizing a telescope product. It is important to note that this will not be an area of a hypersphere -- the cosine from the scalar product prevents this from simplifying to that case. This last comment explains quantitative differences of our (still to be gained) results compared to \cite{KantiNdimBH,HodNdim}. It bears repeating that the omission of this cosine will lead to disagreement with standard results in thermodynamics\footnote{Specifically the Stefan--Boltzmann law and related results.} when transitioning back to $3+1$ dimensions.
\begin{subequations}
	\begin{align}
		\int_\text{ang} =& \ed{D-1}\prod_{i=2}^{D-2} \int_0^\pi \dif \varphi_i \sin^{D-1-i}\varphi_i ,\\
		&= \ed{D-1} \prod_{i=2}^{D-2}\sqrt{\pi}\frac{\Gamma(\ed{2}(D-i))}{\Gamma(\ed{2}(D-i+1))} ,\\
		&= \frac{\sqrt{\pi}^{D-3}}{D-1}\prod_{i=2}^{D-2}\frac{\Gamma(\ed{2}(D-i))}{\Gamma(\ed{2}(D-i+1))},\\
		&= \frac{\sqrt{\pi}^{D-3}}{D-1} \frac{\cancel{\Gamma(\ed{2}(D-2))}}{\Gamma(\ed{2}(D-1))} \frac{\cancel{\Gamma(\ed{2}(D-3))}}{\cancel{\Gamma(\ed{2}(D-2))}} \cdots \frac{\cancel{\Gamma(\frac{3}{2})}}{\cancel{\Gamma(2))}} \frac{\Gamma(1)}{\cancel{\Gamma(\frac{3}{2})}},\\
		&= \ed{D-1}\frac{\sqrt{\pi}^{D-3}}{\Gamma(\ed{2}(D-1))}. 	
	\end{align}
\end{subequations}

Collecting the results, we get:
\begin{equation}
	\Upgamma_n = \frac{g c A}{(2\pi)^{D-1}}\frac{\sqrt{\pi}^{D-3}}{\Gamma(\ed{2}(D-1))}\kl{\frac{\kB T }{ \hbar c}}^{D} (D-2)! {\frac{\Li{D}(-s e^{\mu})}{ (-s)}}.\label{eq:Gammafinal}
\end{equation}
It is instructive to investigate already here, how this compares to the results from the paper~\cite{HawkFlux1}, concretely formula (3.4): For bosons ($s=-1$) we have after setting $D=3$ and $\mu=0$: 
\begin{subequations}
	\begin{align}
		\Upgamma_n =& \underbrace{\frac{g c A}{(2\pi)^{D-1}}}_{=\frac{gcA}{(2\pi)^2}}\underbrace{\frac{\sqrt{\pi}^{D-3}}{\Gamma(\ed{2}(D-1))}}_{=1}\underbrace{\kl{\frac{\kB T }{ \hbar c}}^{D}}_{=\kl{\frac{\kB T }{ \hbar c}}^{3}} \underbrace{(D-2)!}_{=1} \underbrace{{\frac{\Li{D}(-s e^{\mu})}{ (-s)}}}_{= \Li{3}(1) = \zeta(3)},\\
		=& \frac{g\zeta(3)}{4\pi^2}\frac{\kB^3 T^3}{\hbar^3 c^2}A. \label{eq:GammaCheckBos}
	\end{align}
\end{subequations}
We see: (3.4) of \cite{HawkFlux1} agrees with \eqref{eq:GammaCheckBos}, as well as with the corresponding equation~\eqref{eq:3.4} of this thesis.

\subsection{The Frequencies}
Having calculated the number flux, we can continue with deriving the corresponding $D+1$-dimensional expressions for the frequencies used to define all sparsities with the exception of $\eta_\text{binned}$. We shall calculate three types of frequencies: Peak frequencies for both the number flux spectrum and the energy flux spectrum, and the average frequency of the number flux. The only influence on the peak frequencies is a trivial change regarding the way the Lambert W-function appears. The simple result is that the peak frequencies change to
\begin{align}
	\omega_{\text{peak, }E,E} =&  c \kl{\frac{\kB T}{\hbar c}}(D+W(D s e^{\mu-D+2})),\label{eq:omegapE}\\
	\omega_{\text{peak, }E,n} =&  c \kl{\frac{\kB T}{\hbar c}}(D-1+W((D-1)s e^{\mu-D+3}))\label{eq:omegapn}.
\end{align}
For the average frequency,
\begin{equation}
	\omega_{\text{average, }E,n} = \frac{\int ck \kl{\nicefrac{\dif \Upgamma_{n}}{\dif k}} \dif k}{\int \kl{\nicefrac{\dif \Upgamma_{n}}{\dif k}} \dif k}.
\end{equation}
Any angular integrals drop out, being the same in numerator and denominator. Therefore:
\begin{equation}
	\omega_{\text{average, }E,n} = \frac{\int_{0}^{\infty} \dif k\frac{c k^{D}}{\exp({\frac{\hbar c k }{ {\kB T}}}-\mu)+s}}{\int_{0}^{\infty} \dif k\frac{c k^{D-1}}{\exp({\frac{\hbar c k }{ {\kB T}}}-\mu)+s}} = \frac{D \Li{D+1}(-s e^{\mu})}{\Li{D}(-s e^{\mu})}\frac{\kB T}{\hbar}.\label{eq:omEav1}
\end{equation}
This gives in four space-time dimensions the correct results. For example, setting $s=-1$ (that is, considering bosons) and $\mu= 0$, the argument of the polylogarithms turns to $1$ and we can use $\Li{D+1}(1) = \zeta(D+1)$. In full:
\begin{equation}
	\omega_{\text{bosons, 4D, avg., }E,n} = \frac{3 \cdot \nicefrac{\pi^4}{90}}{\zeta(3)}\frac{\kB T}{\hbar} = \frac{\pi^4}{30\zeta(3)}\frac{\kB T}{\hbar}.\label{eq:omEav4D}
\end{equation}
This obviously agrees with equation~\eqref{eq:omEav4D1}.

\subsection{D+1-Dimensional Sparsities Fallen Flat}\label{sec:sparsendimflat}
$\eta_{\text{peak, }E,n}, \eta_{\text{peak, }E,E}$, and $\eta_{\text{average, }E, n}$ are easily obtained by just putting together the results of the previous subsections. In order to later see how the results behave with regards to varying $D+1$, we need to write out any variable depending on $D+1$.

Putting together our results for $\Upgamma_{n}$ and the frequencies (remembering the additional $\nicefrac{1}{2\pi}$ from our conservative estimation of $\tau_\text{loc}$), we get:
\begin{subequations}
\begin{align}
	\eta_{\text{average, }E, n} = & \frac{\frac{D}{2\pi}\frac{\Li{D+1}(-s e^{\mu})}{\Li{D}(-s e^{\mu})}\frac{\kB T}{\hbar}}{\frac{g c A}{(2\pi)^{D-1}}\frac{\sqrt{\pi}^{D-3}}{\Gamma(\ed{2}(D-1))}\kl{\frac{\kB T }{ \hbar c}}^{D} (D-2)! {\frac{\Li{D}(-s e^{\mu})}{ (-s)}}},\\
	= & \frac{D\; 2^{D-2} \pi^{\nicefrac{(D-1)}{2}}\Gamma(\frac{D-1}{2})}{(D-2)!} \frac{\frac{\Li{D+1}(-s e^{\mu})}{(-s)}}{\kl{\frac{\Li{D}(-s e^{\mu})}{(-s)}}^2} \ed{g A} \kl{\frac{\hbar c}{\kB T}}^{D-1},\\
	= & \frac{D}{2\pi (D-2)!} \frac{\Gamma(\frac{D-1}{2})}{\pi^{\nicefrac{(D-3)}{2}}}\frac{\frac{\Li{D+1}(-s e^{\mu})}{(-s)}}{\kl{\frac{\Li{D}(-s e^{\mu})}{(-s)}}^2} \frac{\lambda_\text{thermal}^{D-1}}{g A} .
\end{align}
\end{subequations}
Note that the term ${\sqrt{\pi}^{D-3}}/{\Gamma(\ed{2}(D-1))}$ equals 1 for $D < 4$. We can also see that we get a factor of $A\cdot T^{D-1}$ --- the right combination to cancel contributions from the black hole mass (or the horizon radius).

For the $\eta$'s related to peaks of a spectrum, we only need to forego the appearance of polylogarithms, and the $D$ (or $(D-1)$) in the numerator. Only the $\Li{D}$ in the denominator remains. Finally, introduce the Lambert W-function terms from the peak frequencies themselves. We arrive at:
\begin{subequations}
\begin{align}
	\eta_{\text{peak, }E,E} = & \frac{2^{D-2} \pi^{\nicefrac{(D-1)}{2}}\Gamma(\frac{D-1}{2})}{(D-2)!} \frac{(D+W(D s e^{\mu-D+2}))}{\frac{\Li{D}(-s e^{\mu})}{(-s)}} \ed{g A} \kl{\frac{\hbar c}{\kB T}}^{D-1} ,\\
	= & \frac{1}{2\pi (D-2)!} \frac{\Gamma(\frac{D-1}{2})}{\pi^{\nicefrac{(D-3)}{2}}}\frac{(D+W(D s e^{\mu-D+2}))}{\frac{\Li{D}(-s e^{\mu})}{(-s)}} \frac{\lambda_\text{thermal}^{D-1}}{g A}  ,\\\addtocounter{parentequation}{1}\setcounter{equation}{0}
	\eta_{\text{peak, }E,n} = & \frac{2^{D-2} \pi^{\nicefrac{(D-1)}{2}}\Gamma(\frac{D-1}{2})}{(D-2)!} \frac{(D-1+W((D-1)s e^{\mu-D+3}))}{\frac{\Li{D}(-s e^{\mu})}{(-s)}} \ed{g A} \kl{\frac{\hbar c}{\kB T}}^{D-1} ,\\
	= & \frac{1}{2\pi (D-2)!} \frac{\Gamma(\frac{D-1}{2})}{\pi^{\nicefrac{(D-3)}{2}}}\frac{(D-1+W((D-1)s e^{\mu-D+3}))}{\frac{\Li{D}(-s e^{\mu})}{(-s)}} \frac{\lambda_\text{thermal}^{D-1}}{g A}.
\end{align}
\end{subequations}
Finally, the binned version of sparsity, $\eta_\text{binned}$, has to be calculated separately. Nevertheless, it is straightforward if the appearance of polylogarithms (or their special cases, the Dirichlet eta function, and Riemann zeta function) in previous calculations has been understood.
\begin{subequations}
	\begin{align}
		\eta_\text{binned} = & \ed{\int \frac{2\pi}{ck}\kl{\nicefrac{\dif \Upgamma_n}{\dif k}}\dif k \dif A},\\
		= & \frac{c}{\int_{0}^{\pi}\dif \varphi_{D-2} \sin \varphi_{D-2} \cdots \int_{0}^{\pi}\dif \varphi_2 \sin^{D-3}\varphi_2 \int_{0}^{\frac{\pi}{ 2}} \dif \varphi_1 \cos\varphi_1 \sin^{D-2}\varphi_1},\nonumber\\
		&\times\ed{2\pi \frac{gc}{(2\pi)^{D}} \int_{0}^{\infty}\frac{k^{D-2} \dif k}{\exp({\frac{\hbar c k }{ \kB T}}-\mu) + s}\int_{0}^{A}\dif A \int_0^{2\pi} \dif \varphi_{D-1}},\\
		= & \frac{c}{2\pi \frac{g c A}{(2\pi)^{D-1}}\frac{\sqrt{\pi}^{D-3}}{\Gamma(\ed{2}(D-1))}\kl{\frac{\kB T }{ \hbar c}}^{D-1} \frac{(D-2)!}{D-1} \frac{\Li{D-1}(-s e^{\mu})}{ (-s)}},\\
		= & \frac{(2\pi)^{D-2} \Gamma(\nicefrac{(D-1)}{2}) (D-1)}{\sqrt{\pi}^{D-3} (D-2)!}\frac{1}{\frac{\Li{D-1}(-se^{\mu})}{(-s)}}\ed{gA}\kl{\frac{\hbar c }{\kB T}}^{D-1} ,\\
		= & \frac{\Gamma(\nicefrac{(D-1)}{2}) (D-1)}{2\pi\sqrt{\pi}^{D-3} (D-2)!}\frac{1}{\frac{\Li{D-1}(-se^{\mu})}{(-s)}} \frac{\lambda_\text{thermal}^{D-1}}{gA}.
	\end{align}
\end{subequations}
And this completes the first part of calculating the $D$-dimensional, flat-space sparsities. As has been done before with the frequencies, it is easy to check that these expressions match up nicely with the earlier produced $3+1$-dimensional ones of section~\ref{sec:sparseflat}.

\subsection{Degeneracy Factors}
One additional subtlety, which was hidden from sight in the $3+1$-dimensional context, is the dimensional dependence of the degeneracy factors $g$. In $3+1$ dimensions, the factor $g$ equals $2$ for basically all massless particles, with the exception of a scalar particle, where it equals $1$. In $D+1$ dimensions, this changes --- again, with the exception of the scalar particle, whose degeneracy factor stays $g=1$. For massless spin-1 bosons, the possible degrees of freedom --- the number of polarisation modes --- is the number of transverse modes, $D-1$, see \cite{Zwiebach}, chapter~10.5, or \cite{CardosoBulkHawking}. For massless gravitons, and assuming Einstein's relativity as the theory of gravitation, the possible modes are the transverse traceless modes. Imposing this on the field $h_{ab}$ of the graviton, we see that it (a) has to be symmetric, as it is a (perturbation of the) metric\footnote{The precise wording depends whether we consider weak-field or strong-field situations. Since we are interested in a quantum radiation process, and strong-field GR has not been successfully quantised yet, we will just go with the argument from weak field approximations, where the classical field of a gravitational wave is fully characterised as a linear perturbation of the background metric $g$, that is: $g_{ab}^{\text{full}} = g_{ab} + \epsilon h_{ab} + O(\epsilon^2)$.}, (b) transversality reduces the effective number of rows and columns by $2$, and (c) the demand to be traceless reduces the number of d.o.f. by $1$. (Again, see \cite{Zwiebach} (chapter~10.6) for a more thorough derivation.) All in all, we have that
\begin{equation}
	\mathrm{d.o.f.}(h_{ab}) = g_\text{graviton} = \frac{(D-1)D}{2} - 1 = \frac{(D+1)(D-2)}{2}.\label{eq:ggraviton}
\end{equation}
We will encounter this degeneracy factor further below when comparing the sparsity in the emission of gravitons with previous results in the literature.

For spinors the situation is a bit more complicated and most easily resolved by looking at the little group for a given $D+1$-momentum, here null. If $D=2n-1$ or $D=2n$, then the spinor field will have $g=2^{n-1}$ degrees of freedom, see \cite{WeinbergQFT3}, p.395. The additional factor of $1/2$ comes from the distinction between particles and anti-particles --- which would not be present in the case of Majorana fermions. We will not concern ourselves with other spins, as their physical relevance as fundamental particles will be beyond the standard model of particle physics (at the time of writing). For Rarita--Schwinger fields (spin $3/2$) there are even results suggesting their non-existence on a general curved background space-time, see \cite{BirrellDavies} and \cite{NoGoRaritaSchwinger}.

These considerations change slightly depending on one's subscription to different models of higher dimensional physics. If we suppose to be confined to a \enquote{brane}, the degeneracies would be again the standard $3+1$-dimensional ones for emission in this brane, see \cite{KantiNdimBH2}.

\subsection{Effective Cross-Section}
Just as in the $3+1$-dimensional setting, see section~\ref{sec:3+1cross}, the curved space-time changes the effective cross-section for particle capture. As the $D+1$-dimensional case is less known than the previous one, we shall dwell a bit more on its derivation. Paraphrasing again the derivation in \cite{FrolovZelnikov2011}, this time section~7.10.3, we can deduce expressions for these effective cross-sections \emph{in the geometric optics approximation} needed in the calculations of sparsity if one wants to move beyond the simple flat-space approximations. In the following, we will be limiting our derivation to \emph{massless} particles. As in the Schwarzschild space-time, in the Tangherlini space-time it is always possible to simplify the problem of finding the geodesics of particles and light by restricting it to motion in the equatorial plane. This reduces the metric one has to look at to
\begin{equation}
	\dif s^2  = -\kl{1-\kl{\frac{r_\text{H}}{r}}^{D-2}}\dif t^2 + \kl{1-\kl{\frac{r_\text{H}}{r}}^{D-2}}^{-1} \dif r^2 + r^2 \dif \phi^2.
\end{equation}
This metric has two Killing vector fields: $\partial_\phi$ and $\partial_t$, and two corresponding conserved quantities
\begin{equation}
	E \defi -\kl{1-\kl{\frac{r_\text{H}}{r}}^{D-2}} \frac{\dif t}{\dif \lambda}, \qquad L \defi r^2 \frac{\dif \phi}{\dif \lambda}, 
\end{equation}
where $\lambda$ is the affine parameter of the geodesic --- in the case of a massive particle, this could be chosen to be the eigentime.

Now, take the momentum
\begin{equation}
	p^a = \frac{\dif x^a}{\dif \lambda},
\end{equation}
and use the fact that $p^a p_a = 0$ to be able to write the following equation:
\begin{subequations}
	\begin{align}
		0 &= -\kl{1-\kl{\frac{r_\text{H}}{r}}^{D-2}} \kl{\frac{\dif t}{\dif \lambda}}^2 + \kl{1-\kl{\frac{r_\text{H}}{r}}^{D-2}}^{-1} \kl{\frac{\dif r}{\dif \lambda}}^2 + r^2 \kl{\frac{\dif \phi}{\dif \lambda}}^2,\\
		&= \kl{1-\kl{\frac{r_\text{H}}{r}}^{D-2}}^{-1} E^2 + \kl{1-\kl{\frac{r_\text{H}}{r}}^{D-2}}^{-1} \kl{\frac{\dif r}{\dif \lambda}}^2 + r^{-2} L^2.
	\end{align}
\end{subequations}
Using $E$ one can then read off a differential equation for $r(t)$, which we will further simplify :
\begin{equation}
	\kl{\frac{\dif r}{\dif t}}^2 = \kl{1-\kl{\frac{r_\text{H}}{r}}^{D-2}}^2 - \frac{\kl{1-\kl{\frac{r_\text{H}}{r}}^{D-2}}^3}{E^2}\frac{L^2}{r^2}.
\end{equation}
Non-dimensionalising this by introduction of $\rho\defi \nicefrac{r_\text{H}}{r}$, a rescaled affine parameter $\iota\defi \nicefrac{E \lambda}{r_\text{H}}$, and a dimensionless impact parameter $b \defi \nicefrac{L}{E r_\text{H}}$, one gets the differential equations
\begin{subequations}
	\begin{align}
		\frac{\dif\rho}{\dif \iota} &= \pm \rho^2 \sqrt{1-b\rho^2 (1-\rho^{D-2})},\label{eq:ndimrho} \\ 
		\frac{\dif \phi}{\dif \iota} &= \phantom{+} b \rho^2,\\
		\frac{\dif t}{\dif \iota} &= \phantom{+} (1-\rho^{D-2})^{-1}.
	\end{align}
\end{subequations}

As we are interested in a scattering situation, the trajectory of a scattered photon has to have a radial turning point (just as in $3+1$ dimensions) --- coming from infinity it has to go back to infinity. This means that (a) we only need to consider the radial equation~\eqref{eq:ndimrho} from now on, and (b) there has to be a $\sigma$ such that
\begin{equation}
	\frac{\dif \rho}{\dif \iota} \stackrel{!}{=} 0, \qquad \Longrightarrow \qquad b \stackrel{!}{=} \ed{\rho\sqrt{1-\rho^{D-2}}}.
\end{equation}
Taking a closer look at this expression, it becomes apparent that there exists a minimum where
\begin{equation}
	\frac{\dif b}{\dif \iota} = \frac{D\;\rho^{D-2} -2}{2\rho^2 \sqrt{1-\rho^{D-2}}^3} \stackrel{!}{=} 0.
\end{equation}
The corresponding critical point is at 
\begin{equation}
	\rho_\text{min} = \sqrt[D-2]{\frac{2}{D}}
\end{equation}
with critical value
\begin{equation}
	b_\text{min} = \sqrt[D-2]{\frac{D}{2}}\sqrt{\frac{D}{D-2}}.
\end{equation}

The meaning of the existence of this critical value is the same as in $3+1$ dimensions: Any photon trajectory coming from infinity with an impact parameter less than $b_\text{min}$ will end up being captured by the black hole, thus not taking part in any scattering. As $b_\text{min}$ asymptotes $1$ from above for $D\to \infty$, this results in the effective cross-section being larger than that of a hard sphere of radius $r_\text{H}$. The effective cross-section being given in $D+1$ space-time dimensions as
\begin{equation}\label{eq:n-cross-section}
	\sigma_\text{capture} = \frac{\pi^{\nicefrac{(D-1)}{2}}}{\Gamma(\frac{D+1}{2})}r_\text{H}^{D-1} b_\text{min}^{D-1}
\end{equation}
then evaluates to
\begin{equation}
	\sigma_\text{capture} = \frac{r_\text{H}^{D-1}}{\Gamma(\frac{D+1}{2})}\kl{\frac{D\,\pi}{D-2}}^{\frac{D+1}{2}-1} \kl{\frac{D}{2}}^{\frac{D-1}{D-2}}.
\end{equation}
Comparing this result for the cross-section, it is then possible to read off the correction factor $c_\text{eff}$ between horizon area and effective cross-section:
The only remaining problem with \eqref{eq:n-cross-section} is, that it is derived from an expression involving the $D$-dimensional \textit{volume} of a sphere in contrast to the $D$-dimenional surface $A$ appearing in expressions for the horizon area. Therefore, here comes the derivation of the second equality in \eqref{eq:n-cross-section}. For this, we simply multiply by $\nicefrac{A}{A}$, and insert for the denominator the formula for $A$ as a function of $D$.
\begin{subequations}
	\begin{align}
		\sigma_\text{capture} = &\frac{r{_\text{H}}^{D-1}}{\Gamma(\nicefrac{D+1}{2})}\kl{\frac{D\,\pi}{D-2}}^{\frac{D-1}{2}}\kl{\frac{D}{2}}^{\frac{D-1}{D-2}},\\
		= & \frac{\ed{\Gamma(\nicefrac{D+1}{2})}\kl{\frac{D\,\pi}{D-2}}^{\frac{D-1}{2}}\kl{\frac{D}{2}}^{\frac{D-1}{D-2}}}{2 \frac{\pi^{\frac{D}{2}}}{\Gamma(\frac{D}{2})}}A,\\
		= & \ed{2}\frac{\Gamma(\frac{D}{2})}{\Gamma(\nicefrac{D+1}{2})}\kl{\frac{D\,\pi }{D-2}}^{\frac{D-1}{2}}\kl{\frac{D}{2}}^{\frac{D-1}{D-2}}\pi^{-\frac{D}{2}}A,\\
		= & \underbrace{\ed{2\sqrt{\pi}}\frac{\Gamma(\nicefrac{D}{2})}{\Gamma(\nicefrac{D+1}{2})} \kl{\frac{D}{D-2}}^{\frac{D-1}{2}} \kl{\frac{D}{2}}^{\frac{D-1}{D-2}}}_{\ifed c_\text{eff}} A.
	\end{align}
\end{subequations}
If everything went right, then for $D=3$ the thus newly defined $c_\text{eff}$ should evaluate to $\nicefrac{27}{16}$, which it does:
\begin{subequations}
	\begin{align}
	 	c_\text{eff}\lvert_{D=3} = & \ed{2\sqrt{\pi}}\frac{\Gamma(\nicefrac{3}{2})}{\Gamma(\nicefrac{3+1}{2})} \kl{\frac{3}{3-2}}^{\frac{3-2}{2}} \kl{\frac{3}{2}}^{\frac{3-1}{3-2}},\\
	 	= & \ed{2\sqrt{\pi}}\frac{\sqrt{\pi}}{2\cdot 1} \cdot 3^1 \cdot \kl{\frac{3}{2}}^2 = \frac{3\cdot 9}{4 \cdot 4} = \frac{27}{16} .
	\end{align}
\end{subequations}

\subsection{D+1-Dimensional Sparsity Results for the Tangherlini Metric} 
We are now in a position to give the full results for the Tangherlini metric, with a stand-in for the appropriate degeneracy factors $g(D)$ --- these will change the asymptotic behaviour for large dimensions slightly. We have:
\begin{subequations}\label{eq:DdimCurvedEta}
	\begin{align}
		\eta_{\text{peak, }E,n} &= \frac{2^{2D-1+\ed{D-2}} \pi^{D-\ed{2}} \Gamma\kl{\frac{D-1}{2}}\kl{D-1 + W\kl{s(D-1)e^{-D+1}}}}{g(D)(D-2)^{\frac{D-1}{2}} D^{\frac{D^2-3D}{2(D-2)}}\Gamma\kl{\frac{D}{2}} \Li{D}(-s)/(-s)},\\
		\eta_{\text{peak, }E,E} &= \frac{2^{2D-1+\ed{D-2}} \pi^{D-\ed{2}} \Gamma\kl{\frac{D-1}{2}}\kl{D + W\kl{s D e^{-D}}}}{g(D)(D-2)^{\frac{D-1}{2}} D^{\frac{D^2-3D}{2(D-2)}}\Gamma\kl{\frac{D}{2}} \Li{D}(-s)/(-s)},\\
		\eta_{\text{average, }E,n} &= \frac{2^{2D+\ed{D-2}} \pi^{D-\ed{2}} \Gamma\kl{\frac{D+1}{2}}\Li{D+1}(-s)/(-s)}{g(D)(D-2)^{(D+1)/2} D^{\frac{D-1}{2}-\ed{D-2}}\Gamma\kl{\frac{D}{2}-1} \kl{\Li{D}(-s)/(-s)}^2},\\
		\eta_{\text{binned}} &= \frac{2^{3D-1+\ed{D-2}} \pi^{D-1} \kl{\Gamma\kl{\frac{D+1}{2}}}^2}{g(D)(D-2)^{\frac{D-1}{2}} D^{\frac{D^2-3D}{2(D-2)}}\Gamma\kl{D-1} \Li{D-1}(-s)/(-s)}.
	\end{align}
\end{subequations}
Notice that (independent of $g(D)$, as we know it to be larger than or equal to $1$) we have that
\begin{equation}
	\lim_{D\to \infty} \eta = 0.
\end{equation}
This means that in high dimensions the Hawking flux will behave as classical heat radiation, not as quantum radiation. The exception, again, will be fermions. While the sparsity will still tend towards $0$, the inability to inhabit states more than once still will result in essentially quantum mechanical radiation. Adding particle mass will again change nothing about this --- similar approximations as for four space-time dimensions mean that we still have
\begin{equation}
	\eta_\text{massive} < \eta_\text{massless}.
\end{equation}
However, the precise dimension at which the behaviour of the radiation will change will depend on both the $\eta$ chosen, as well as on the specific particle and its properties, including the degeneracy factor $g(D)$. This particular question was investigated in \cite{HodNdim} for gravitons. Let us therefore compare our results with his. (Both our and Hod's results reproduce earlier results seen from an extension of Page's work to higher dimensions by Cardoso, Carvagli\`{a} and Gualtieri in \cite{CardosoBulkHawking}.)

\subsubsection{Comparison with Previously Published Results} 
Hod considered the sparsity of higher-dimensional Tangherlini black holes in the limit $D\gg 1$ in \cite{HodNdim}, after capturing the physics behind it already earlier in \cite{HodBulkEmission}. We regard some of his approximations to be easily improved, though we agree with his general statement: Black hole radiation \emph{will} become sparse in high-enough dimensions. At what specific dimension this transition happens, however, is dependent on one's choice of sparsity. Also, while $\eta=1$ is the value at which we can separate sparse or quantum ($\eta>1$) from non-sparse or classical ($\eta<1$) situations, we think it likely that $\eta\approx1$ will correspond to a transitional region. This transition will be similar to the classical-to-quantum transition in mesoscopic physics --- though, probably, way less tunable from an experimental point of view.

In order to compare our results with those of Hod, let us quickly review his approximations. The peak frequency of the energy flux is approximated for $D\gg 1$ as
\begin{equation}
	\omega_{\text{peak, }E,E} r_\text{H} \approx \frac{D^2}{4\pi} + O(D),\label{eq:HodMegaApprox}
\end{equation}
simplifying further the earlier published approximation \cite{HodBulkEmission}
\begin{equation}
	\omega_{\text{peak, }E,E} r_\text{H} \approx \frac{D(D-2)}{4\pi} + O(D).
\end{equation}
We find it worthwhile to point out that a mere change to 
\begin{equation}
	\omega_{\text{peak, }E,E} r_\text{H} \approx \frac{D(D-4)+3}{4\pi}
\end{equation}
would already lead to the same asymptotic behaviour as the full expression involving the Lambert-W function.\footnote{As the argument of the Lambert-W function goes to $0$ if $D\to \infty$, its value will also converge to $0$.} The approximation~\eqref{eq:HodMegaApprox} actually has a different asymptotic behaviour and diverges faster. Either way, Hod uses this approximation~\eqref{eq:HodMegaApprox} to arrive at an asymptotic sparsity for large dimension $D$:
\begin{equation}
	\eta_\text{Hod} \approx \frac{e}{8\pi^2}\kl{\frac{4\pi}{D}}^{D+1}.\label{eq:etaHod}
\end{equation}

To compare this with our results, let us rewrite equations~\eqref{eq:DdimCurvedEta} with the value of $g_\text{graviton}(D)$ from equation~\eqref{eq:ggraviton}:
\begin{subequations}
	\begin{align}
		\eta_{\text{peak, }E,n} &= \frac{2^{2D+\ed{D-2}} \pi^{D-\ed{2}} \Gamma\kl{\frac{D-1}{2}}\kl{D-1 + W\kl{s(D-1)e^{-D+1}}}}{(D+1)(D-2)^{\frac{D+1}{2}} D^{\frac{D^2-3D}{2(D-2)}}\Gamma\kl{\frac{D}{2}} \Li{D}(-s)/(-s)},\\
		\eta_{\text{peak, }E,E} &= \frac{2^{2D+\ed{D-2}} \pi^{D-\ed{2}} \Gamma\kl{\frac{D-1}{2}}\kl{D + W\kl{s D e^{-D}}}}{(D+1)(D-2)^{\frac{D+1}{2}} D^{\frac{D^2-3D}{2(D-2)}}\Gamma\kl{\frac{D}{2}} \Li{D}(-s)/(-s)},\\
		\eta_{\text{average, }E,n} &= \frac{2^{2D+1+\ed{D-2}} \pi^{D-\ed{2}} \Gamma\kl{\frac{D+1}{2}}\Li{D+1}(-s)/(-s)}{(D+1)(D-2)^{(D+3)/2} D^{\frac{D-1}{2}-\ed{D-2}}\Gamma\kl{\frac{D}{2}-1} \kl{\Li{D}(-s)/(-s)}^2},\\
		\eta_{\text{binned}} &= \frac{2^{3D+\ed{D-2}} \pi^{D-1} \kl{\Gamma\kl{\frac{D+1}{2}}}^2}{(D+1)(D-2)^{\frac{D+1}{2}} D^{\frac{D^2-3D}{2(D-2)}}\Gamma\kl{D-1} \Li{D-1}(-s)/(-s)}.
	\end{align}
\end{subequations}

In figure~\ref{fig:ndim}, we plotted our dimension-dependent sparsities for gravitons in the geometric optics and black body approximation and compare these with Hod's approximated sparsity, equation~\eqref{eq:etaHod}. As to be expected (based on its origin as an approximation), for low dimensions the difference is large. Nevertheless, $\eta_{\text{Hod}}$ captures the rough range of sparsity and non-sparsity, while naturally differing on the exact values.

The numerical results including grey body factors of \cite{HodNdim} suggest a transition to the non-sparse regime at around $D\approx 11$. It is interesting to note that our black body approximation agrees roughly with this results. For the different sparsities we have that $\eta(D) = 1$ at approximately the following values:\footnote{These values can be found using a simple Newton--Raphson method.}
\begin{center}
	\begin{tabular}{||c|cccc||}
		\hline\hline
		& $\eta_{\text{peak, }E,n}$ & $\eta_{\text{peak, }E,E}$ & $\eta_{\text{average, }E,n}$ & $\eta_{\text{binned}}$\\\hline
		$D$ & $\num{11.1709}$ & $\num{11.2716}$ & $\num{11.2714}$ & $\num{11.1704}$\\
		\hline\hline
	\end{tabular}
\end{center}
\noindent This seems to suggest that at least in this regard, the black body approximation becomes better once one increases $D$. Given the additional complexity of black hole space-times in $D$ dimensions, this result needs to be taken with a grain of salt.

Also note, that it is known that the decomposition of Hawking radiation changes with dimension. This is usually shown using an analysis based on grey body factors (something our approach cannot achieve\footnote{Though, one could in principle extend the analysis of Gray and Visser in \cite{SparsityNumerical} to encompass the Kodama--Ishibashi master equation for non-rotating black holes in higher dimensions. This would nonetheless just add to the numerical techniques employed to calculate the relevant grey body factors, not to new results entirely.}), see \cite{KantiNdimBH,CardosoBulkHawking,KantiNdimBH2}. It is worthwhile to have a look at one single, particular sparsity and compare the different particle species' sparsities. The result can be seen in figure~\ref{fig:species}. We can see that the first particle to go below $\eta=1$ is the graviton. This matches the previous numerical results, despite complete neglect of grey body factors. Without numerical results from the literature, this result itself would be of little weight, as we have seen in the inclusion of the numerics done by \cite{SparsityNumerical} for sparsity calculations that gravitons are much sparser once grey body factors are included. Armed with just our pictorial result, there would be no guarantee that a similar defect would not happen for other values of the dimension $D$. The next particles to pass this bound are fermions, but as we know that there is no classical limit to fermions in the sense it exists for bosons (see section~\ref{sec:geoquantum} and \cite{Duncan}), the more important feature is when which boson species passes the threshold $\eta=1$.

\subsection{Summary}
Again, we could see that sparsity reproduces and captures previously gained results also in the setting of general $D$ space dimensions. This includes both the qualitative change\footnote{This change indeed \emph{is} qualitative: It signifies a change from quantum mechanical to classical radiation gas. The absence of sparsity thus certainly is in disagreement with the statement made on page~344 of \cite{FrolovZelnikov2011}, that Hawking radiation is qualitatively the same in higher dimensions!} of the behaviour of emitted black hole radiation, as well as some tentative indications (on our side, not on the side of numerical, previous results) of the different behaviour of different particle types. The biggest complication in transferring the sparsity machinery to $D+1$ dimensions is mostly of algebraic nature, as most integrals involved become quite cumbersome. Yet from a technical point of view, the calculations are the same. Theoretically, one could even try to extend results to other space-times besides Tangherlini, or to include more complications considered in the previous sections, but from our personal point of view there is little to be gained from this without a specific question in mind, first.

\begin{figure}
	\centering
\includegraphics[width=\textwidth]{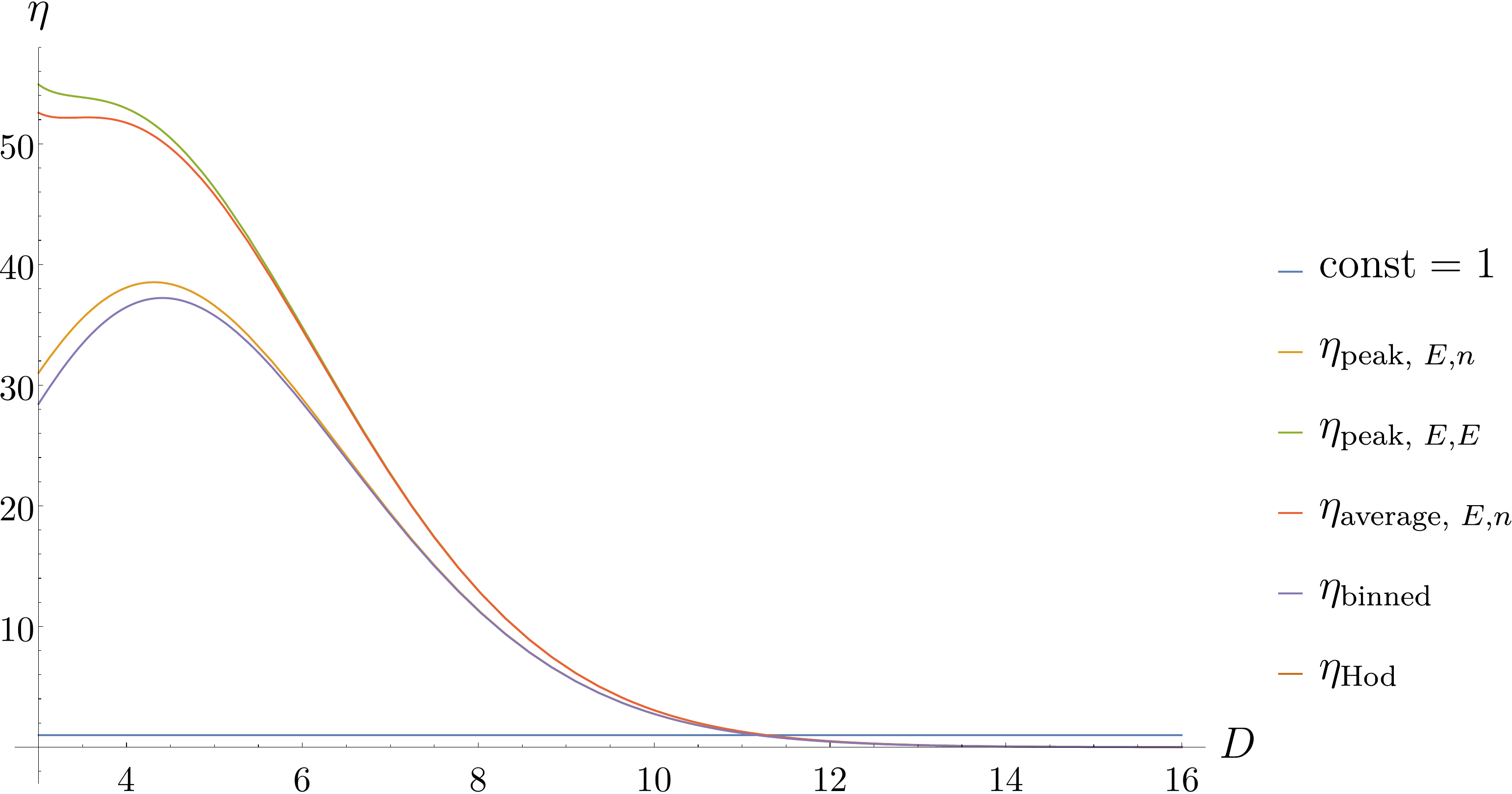}
\caption[Plot of Dependence of Sparsity of Gravitons in Dimension D]{A comparison of various $\eta$ for massless gravitons in $D$ space dimensions. The constant $1$ indicates the transition sparse to non-sparse.}
\label{fig:ndim}
\end{figure}
\begin{figure}
\centering
\includegraphics[width=\textwidth]{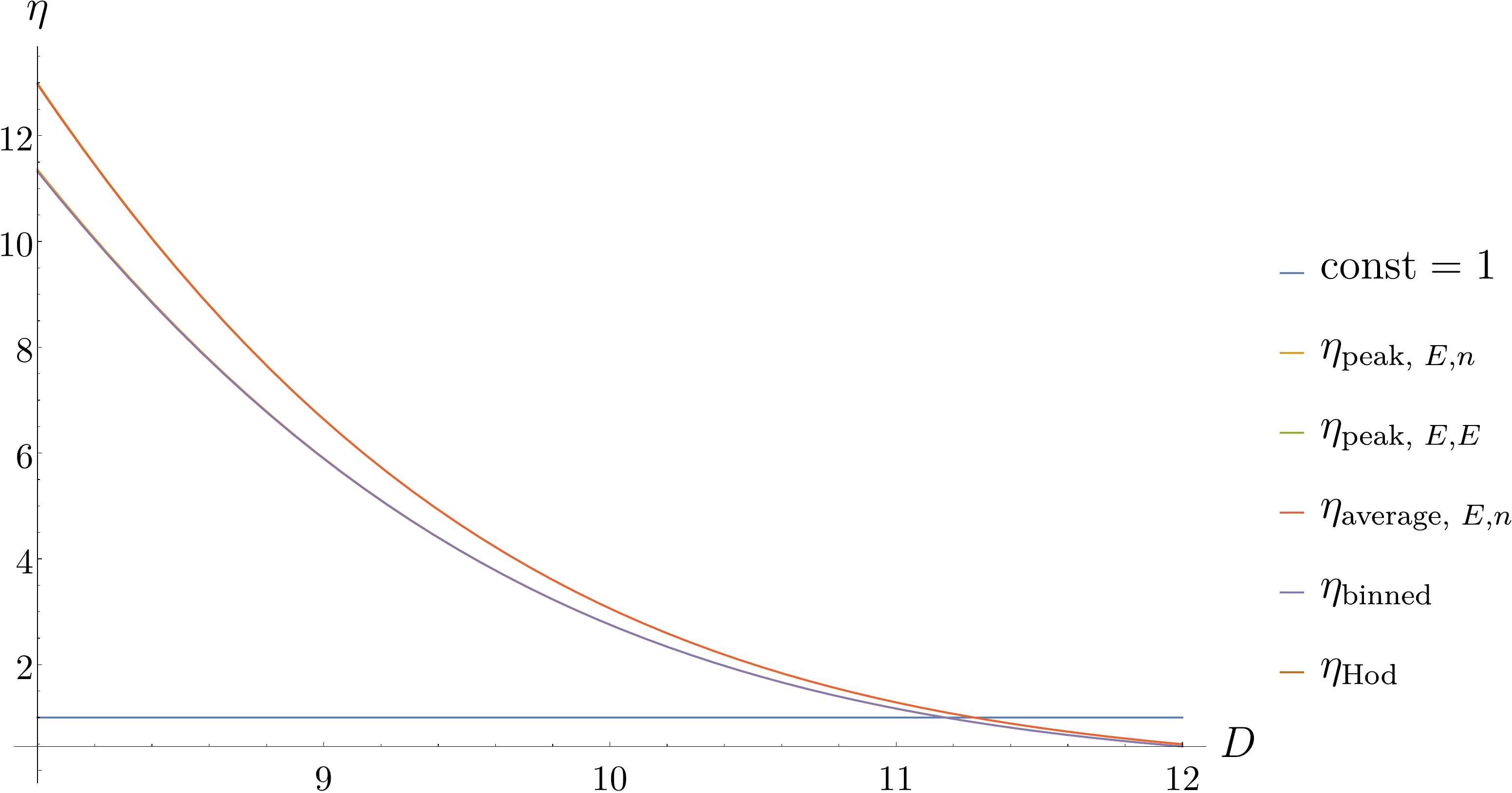}
\caption[Plot of Sparse-Nonsparse Transition for Gravitons in Dimension D]{A comparison of various $\eta$ for massless gravitons in $D$ space dimensions. The constant $1$ indicates the transition sparse to non-sparse. Here we zoomed in on the region where these transitions happen.}
\label{fig:ndimzoom}
\end{figure}

\begin{figure}
	\includegraphics[width=\textwidth]{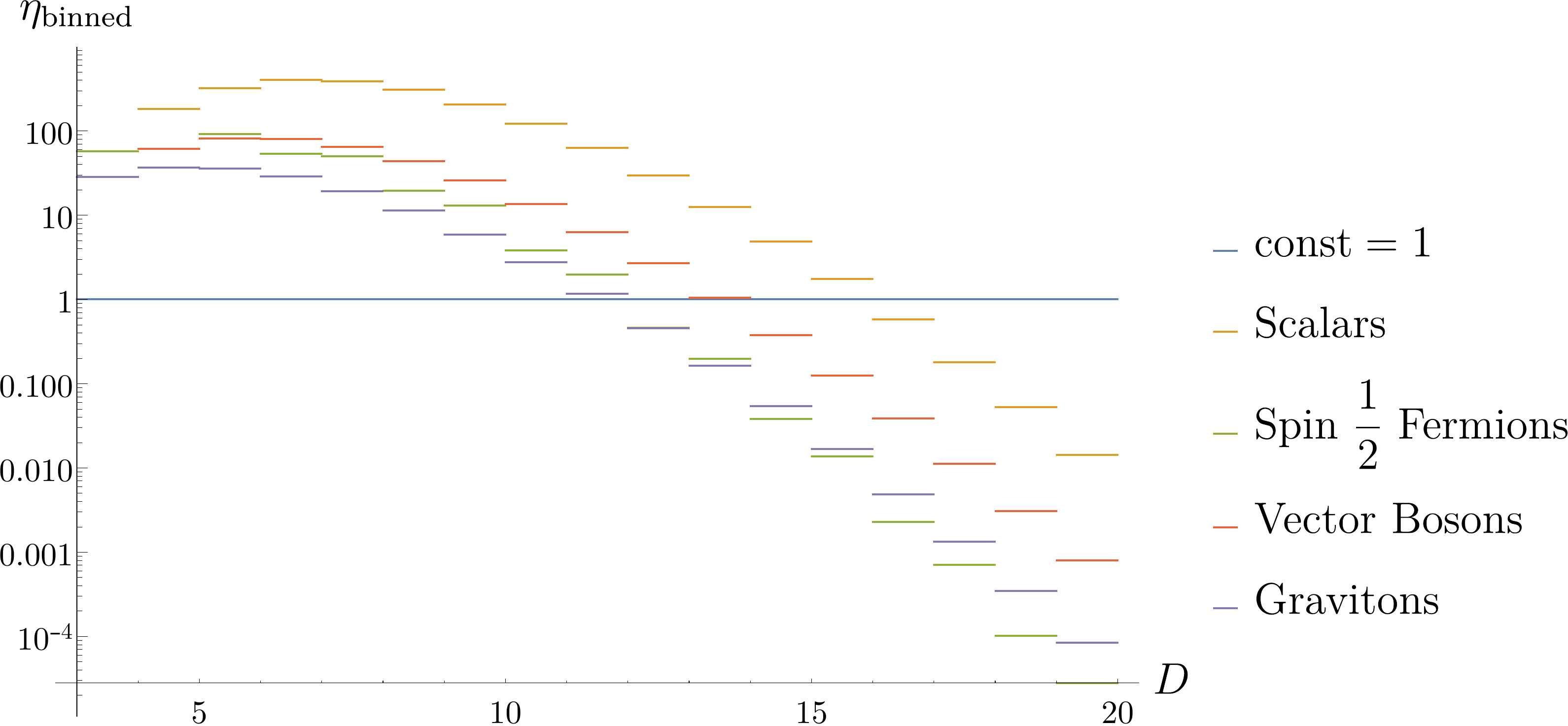}
	\caption[Plot of Dependence of Binned Sparsity of Different Particles in Dimension D]{The binned sparsity $\eta_{\text{binned}}$ for different (massless) particle species.}
	\label{fig:species}
\end{figure}

\FloatBarrier
\section{Sparsity and Analogues}\label{sec:sparseEM} 
After the discussion of the previous sections it should have become clear that the sparsity of Hawking radiation, hence its non-classicality, are characteristic features of this radiation. Despite this, it is reasonably rare that this feature is carefully scrutinised. (Some of the occasional exceptions known to us have been cited in the appropriate places.) Given the inaccessibility of astrophysical instances of the Hawking effect, the related experimental questions will have to be checked in the analogue space-time framework (for the time being) --- here sparsity or non-classicality are just one of many lines of inquiry one could follow up on in the analogy. The present section intends to explain why this is \emph{far} from easy to achieve. Much of this discussion will focus on the particular issues encountered in the context of the algebraic analogue space-time encountered in the largest part of chapter~\ref{ch:analogues}.

In order to see what the additional difficulties are, let us take a careful look at the core quantities involved in defining notions of sparsity. All of these quantities are, before even stating \emph{what} they are, defined in the context of the space-time which can give rise to the Hawking effect in the first place. In the analogue space-time scheme, this is the analogue space-time itself. This space-time, as has been stressed repeatedly, is not necessarily directly accessible to experiment. Any quantity defined in the analogue space-time first has to be pulled back (mathematically and figuratively) into the laboratory space-time. And this is where the issues arise. At the core of the Hawking effect are $3+1$ splits to single out (a class of) observers which measure particle numbers. Without this ADM split, no Fock space picture is available. Already in the (analogue) space-time itself, this split will break Lorentz invariance. Any pull back to the laboratory space-time, naturally, will only further dismantle earlier notions of nice transformation properties due to cartographic distortions. The slicing the experimenter will use in his personal, laboratory Minkowski frame will not necessarily align with the slicing his analogue space-time is concerned with. This can quickly be seen by looking at the time coordinates of the laboratory coordinates and the effective coordinates for our bespoke space-time mimics of section~\ref{sec:bespoke}. The simplest case is, yet again, Schwarzschild space-time, where the proper time of the analogue space-time will be non-trivially connected to the proper time of the laboratory, in a way that will depend on the laboratory coordinates. Every point in the analogue space-time on the table will therefore experience a different lapse of time, compared to what it would experience within the analogue space-time. This already is a severe result: Our whole discussion of sparsity was coined in terms of time scales. As the time scales were all given in terms of asymptotic quantities, the hope might be that asymptotic flatness of the analogy is enough to guarantee no changes to the sparsity. 

Even that, however, is not as simple: As we have seen in section~\ref{sec:HawkingLab}, already differences in scaling will influence the temperature, thus the frequencies, thus the time scales. This will be further exacerbated if the laboratory coordinates have to be chosen less adapted to the analogue space-time's coordinates than a simple scaling of spatial coordinates.

And at this point we have not even begun to look more closely at the number flux $\dif \Upgamma_n$:
\begin{equation}\label{eq:dGammaTheSecond}
	\dif \Upgamma_n = \frac{g}{(2\pi)^3} \frac{c (\hat{k}\cdot \hat{n})}{\exp(\hbar c k / \kB T) - 1} \dif^3 \vec{k}\, \dif A.
\end{equation}
Many of the quantities involved behave very differently under a rescaling of the spatial coordinates: $A$ quadratically, $k$ inversely proportional, $T$, as we have seen in section~\ref{sec:HawkingLab}, inversely to the power of $9/2$. Also the natural constants have to change: They all are \emph{effective} natural constants valid only in the analogue space-time itself. The physically relevant, real natural constants are the usual, but the ones appearing in our quantity~\eqref{eq:dGammaTheSecond} will depend on the scaling --- and more generally on the mapping of laboratory to effective coordinates in a much more complicated way. 	

Since the laboratory will have to contend itself with finite slabs of material performing the role of an analogue space-time, this finiteness also has to be taken into account. In and of itself, this is still of little concern: Since the Hawking temperature is conformally invariant (see, again, \cite{JacobsonKang93}), a conformal compactification will not change the temperature in the analogue space-time. For sparsity, this is less obvious, especially once one includes grey body factors in the analysis.

The case for analytic analogue space-times is slightly less problematic: Here, the issues described in section~\ref{sec:PDEana} will be absent, as the analogy is based completely on a PDE already available to the laboratory. Even then, this may require the mimicked space-time in specific representations. For example, as the foundation of the analogue space-time encountered in a fluid dynamics context often requires a Gordon form of the metric \cite{lrrAnalogue,VorticityAnalogues,TowardsKerrGordonForm}, not all coordinate representations of a given metric are amenable to this framework. Note, however, that the analytic \emph{analogy} presented in section~\ref{sec:refractive} has a different, new difficulty. Since this analogy is not geometric in nature, understanding the Hawking effect is more difficult. It might be possible to rephrase it as a $1$$+$$1$-dimensional space-time, but it will require careful thought if the higher-dimensional origins of the ODEs used may not lead to residual effects preventing this.

To summarise this state of affairs, we can see that the question if sparsity of an analogue system is measurable will require careful and meticulous case-by-case calculations. This will \emph{not} be simple. Certainly, general statements could be made by ample use of pull-backs and push-forwards,\footnote{This is what has been done in \cite{CovOptMet1,CovOptMet2,CovOptMet3,TrafoOpticsCartographDistort,CovOptMet4} in the purely electromagnetic context from the point of view of transformation optics. To extend this to the case of sparsity one would have to also introduce the correspondingly pulled-back and pushed-forward notions of the Hawking effect.} but from a practical point of view this would just be a formalised manual for the calculation to be performed. Many of the steps are likely to require numerical treatment --- especially given that already now the inclusion of grey body factors requires numerical methods (something unlikely to change before the theory of Heun functions has been brought to equal fruition as the theory of hypergeometric functions is currently at).

\chapter{Conclusion}\label{ch:finale}
\epigraph{\enquote{Hear the cruel no-answer\\
	Until blood drips down\\
	Beat your head against the wall of it}}{Ikky\={u} S\={o}jun, translation Stephen Berg, \\\emph{Crow with no Mouth}}
Let us conclude with a short summary, followed by a list of possible future avenues of inquiry.

\section{Summary and Discussion}

One main goal of this thesis was to describe more closely an under-appreciated aspect of the Hawking effect: Its sparsity. For this, we started with an analysis of sparsity in astrophysical black holes, based on our work in \cite{HawkFlux1}. Since astrophysical black holes in general, and their Hawking radiation in particular, are, however, incredibly hard to reach experimentally or even observationally, the analogue space-time framework became an immediate and pressingly needed extension for \enquote{sparsity} beyond its astrophysical origin. A substantial amount of research for this thesis went into preparing this extension by investigating analogue models, specifically, electromagnetic analogues. As this work is mostly (apart from section~\ref{sec:HawkingLab}) concerned with classical considerations, it made sense to place this extensive chapter in front of the sparsity chapter --- thus we could easily draw from it when discussing the (so far failed) attempts of bringing the notion of sparsity into the realm of analogue space-times.

Despite this shortcoming, the notion of sparsity in general is nevertheless incredibly useful. As we have seen in chapter~\ref{ch:sparsity}, most extensions of the simplest physical, underlying model (flat space-time, massless, spinless, uncharged particles emitted, \dots) increase the sparsity. It is only with the inclusion of superradiance or going to (very) higher-dimensional extensions of general relativity, that sparsity could be decreased far enough to talk of Hawking radiation being a classical radiation process \cite{HodBulkEmission,HodRotating,HodNdim}. As we discussed in \ref{sec:superradiance}, we believe it more sensible to exclude superradiation from the Hawking effect, just as it seems at the present stage rather hypothetical (though still standard fare for a theorist!) to consider extra-dimensions. This shows that the astrophysical notion of sparsity for all \enquote{practical} matters of current experimental and observational knowledge in the semi-classical regime is a stable result. Note also the complete cancellation of physical constants: Sparsity is a \emph{dimensionless} figure of merit, independent of one's choice of system of units.

Our geometric optics approach can also be seen as adding to the pool of previous similar results from a new viewpoint: The early results \cite{Page1,Page2,Page3}, as well as more recent quantum theoretic reimaginations of the Hawking effect \cite{KieferDecoherenceHawkingRad,BHradNonclass}, show the sparsity from a quantum (field) theoretic point of view. The results of sparsity gained in these works (though not yet named such) are then extended in both a more quantum direction \cite{BekensteinMukhanovQuantumBH}, as well as a more classical rephrasing (the present work). This further reinforces the importance of sparsity (or non-classicality) of the Hawking effect.

More recent results trying to incorporate quantum gravity effects indicate $\eta$ becoming less than $1$ in late stages of the evaporation process. While \cite{SparsityBackreaction} concludes with a regime where (formally) $\eta<0$, thus a point where radiation should stop and one is left with a remnant, analyses based on general uncertainty principles (GUP) do not imply a black hole remnant \cite{SparsityAna,OngGUPSparsity}. Both approaches agree on the formation of a regime of non-sparse radiation close to the Planck mass. The corrections considered in \cite{SparsityBackreaction} and in the GUP approach \cite{SparsityAna,OngGUPSparsity} can be considered a phenomenological descriptor of quantum gravity effects. As such it falls slightly outside of the area of concern to us --- we are concerned with semi-classical gravity, \emph{i.e.}, quantum field theory in curved space-times (and most of the time even that only approximately, if not in a mutilated way, see the comments regarding the geometric optics approximation in section~\ref{sec:geoquantum}). Also Hod's results in \cite{HodNdim} (and our confirmation thereof) for $D+1$ dimensional relativity can be seen as inspired by model building beyond the standard model (of both particle physics and cosmology). It seems to us rather surprising that an effect becomes classical in exactly that Planck regime which is most infamous for its trouble with quantisation. It is noteworthy that the earlier calculation by Bekenstein and Mukhanov can be seen as another instance of such a phenomenological quantum gravity approach to sparsity/\allowbreak non-classicality of Hawking radiation --- and does agree with our results, even though its considerations are firmly placed on grounds going beyond CSTQFT. It will be interesting to see what future research into this will reveal. This can be put into the pointed question: How generic is the \emph{classicality} of the Hawking flux in the Planck regime, of all energy regimes, in theories of quantum gravity (effects)?

\section{Possible Extensions}\label{sec:extensions}
The work presented in this thesis can easily be extended in a number of directions. We shall list some of these here:
\begin{itemize}
	\item As mentioned at the end of section~\ref{sec:conscond}, with the transformation laws between reference frames of appendix~\ref{sec:frames} in place, it is interesting to have a look at metrics which in their most common forms have non-vanishing magneto-electric effects (and thus are non-trivial to engineer), and transform the corresponding constitutive tensor into that of the natural rest-frame. As the pointwise existence of this rest-frame is guaranteed (unless we would have a material with superluminal fundamental interactions, which would beg a slew of much more urgent physical questions\dots), it would be interesting if the constitutive matrices in that particular rest-frame are easier to engineer than the ones derived from common expressions for a given metric.
	\item As already noted in a footnote, it might be possible to extend Perlick's asymptotic analysis for a mathematically rigorous derivation of ray optics in curved space-times using our formalism. Thus one could circumvent the need to go to the natural rest frame of a given medium.
	\item While less of relevance for actual (table-top) experiments, it might be worthwhile to have a look at what happens if one took the bespoke meta-material methods described in section~\ref{sec:bespoke} beyond the Minkowski background. A post-Minkowskian approximation, for example, might be interesting for identifying potential analogue system in astrophysical situations. However, we do not think this approach likely to produce observable effects, nor being amenable to non-numeric methods.
	\item With some hard work, it might even be possible to then further extend the previous point to the post-Newtonian (PN) formalism. This \emph{may} have experimental relevance even in tabletop laboratory scenarios.
	\item We already discussed in section~\ref{sec:PDEana} the need to repeat its analysis in the case of general mappings $x_\text{eff}(x_\text{lab})$, \emph{i.e.} of cartographic distortions. For the experimentalist, this is important preliminary work still left to be done. It certainly will be interesting to see, how much of the analysis presented in \cite{CovOptMet1,CovOptMet2,CovOptMet3,TrafoOpticsCartographDistort} can be repurposed for this.
	\item As mentioned in section~\ref{sec:refractive}, there exists a different approach to one-dimensional refractive index profiles related to Heun functions, described in \cite{HeunRefIndex}. There, the example of a Heun equation transformed to a Helmholtz equation for plasma media is considered. They find the (physical) requirement that
	\begin{equation}
		0< \lim\limits_{z\to\pm\infty}n^2(z) < \infty,
	\end{equation}
	where $z$ is the vertical component of a Cartesian coordinate system,\footnote{This means that the stratification considered there is in the $z$-coordinate.} which results in two of the exponential parameters having to pick up an imaginary part. The similarity to the complex coefficients appearing in the standard form of the Teukolsky equation (and more generally, the spin-weighted spheroidal equation) is obvious. It is certainly interesting, if the intermediate step through Detweiler's work \cite{DetweilerPotential} can be skipped using this more general analysis, or whether it would have to be modified, as \cite{HeunRefIndex} was specifically looking for refractive index profiles associated with resonances. As resonances could be the right means to get the high refractive indices needed close to the horizon in our analogy, we estimate that little changes will be needed.
	\item It is also possible to follow section~\ref{sec:refractive} orthogonally to the previous point: Instead of trying to bypass Detweiler's work, one might see how many other space-times and their wave-equations can be rephrased using a purely real potential. As all Petrov type~$D$ space-times admit separable wave equations, it is certainly interesting how many type~$D$ space-times admit potentials that can be written out explicitly along these lines. Similarly, one can ask which space-times beyond type~$D$ with separable wave equations admit this analysis. For those that do one could then perform a similar derivation of refractive index profiles as presented in this thesis as a foundation for analogies to space-time experiments.
	\item Likewise, we could change the particle under consideration in section~\ref{sec:refractive}: Instead of focussing on \enquote{photons in curved space-times mimicked by photons in an optical medium}, one could equally well turn this into \enquote{gravitons in curved space-times mimicked by photons in an optical medium}, or scalars, or massless fermions.
	\item Picking up where one of the last items left off, it would be interesting to start looking for materials admitting refractive index profiles like those arising out of our analogy of section~\ref{sec:refractive}; with materials available one could then even take these \enquote{space-times} to the laboratory. As this is one of the big goals of the analogue space-time program in the first place, this would certainly be a satisfying conclusion of this particular project. This would also provide reason for a closer look at possibly available refractive index profiles to systematically search for experimentally realisable profiles.
	\item Using the separation of the Teukolsky (master) equation, or even the Kodama--Ishibashi equation, it might be worthwhile to look at further numerical evaluations of sparsity. Ideally this approach would also immediately include possible superradiance; a thorough numerical look at the binned sparsity $\eta_{\text{binned}}$ seems particularly fruitful in this regard.
	\item So far not mentioned was the subfield of CSTQFT studying detector models to get a better handle on the observer-dependent nature of a state's particle content. It might be possible to marry this line of inquiry with the notion of sparsity and turn sparsity from a \enquote{mere} figure of merit into a truly operational notion.
	\item Ultimately, it will be important to actually calculate the measured sparsity of analogue systems. As described in section~\ref{sec:sparseEM}, one has to specify for this:
	\begin{itemize}
		\item \begin{sloppypar}The analogue space-time model (fluid, Bose--Einstein condensate, optical medium, \dots),\end{sloppypar}
		\item Both effective and laboratory coordinates,
		\item Both effective and laboratory space-times,
		\item How the coordinates are related,
		\item The quantisation procedure in the analogue space-time resulting in its Hawking effect.
	\end{itemize}
	This promises to be an important, but arduous case-by-case analysis, and thus basically an open-ended task.
\end{itemize}

We hope that the amount of possible future research topics arising out of the present thesis piques people's interest. The field of analogue gravity holds fascinating questions from a wide range of fields, applied, as well as fundamental. Likewise a diverse set of skills can be found of use here, making both learning and collaborating a joy. We sure wish that some of the excitement could be communicated!
 	
 	\appendix
 	\renewcommand{\thechapter}{\Alph{chapter}}
 	\renewcommand{\thesection}{\Alph{chapter}.\arabic{section}}
 	\renewcommand{\thesubsection}{\Alph{chapter}.\arabic{section}.\arabic{subsection}}
\chapter{Some Background Material from Differential Geometry}\label{ch:DG}
\epigraph{\enquote{And as imagination bodies forth\\ 
	The forms of things unknown, the poet's pen\\ 
	Turns them to shapes and gives to airy nothing\\ 
	A local habitation and a name.} }{William Shakespeare, \emph{A Midsummer Night's Dream}}

This appendix will collect differential geometric results that are of particular use for the development of our electromagnetic analogue space-times. As much of the calculational convenience enjoyed in our discussion of electromagnetism and the resulting analogue space-times relies on the conformal invariance of electrodynamics, the first part of this appendix will collect some results concerning conformal transformations. These results in place, we will then discuss orthogonal decompositions with respect to an observer of four-velocity $V^a$. Specific attention will be paid to two forms and tensors of the kind the constitutive tensor in chapter~\ref{ch:analogues} takes. 

 	\section{Conformal Equivalence and Conformal Transformations} \label{sec:conformal}
 	If a metric $g$ is related to a second metric $\tilde{g}$ through the relation
 	\begin{equation}
	 	\tilde{g} = \Omega^2 g,
 	\end{equation}
 	for some function $\Omega(x)$, one says that $g$ and $\tilde{g}$ are conformally equivalent, or conformally related to each other, and $\tilde{g}$ is related to $g$ through a conformal transformation (also called Weyl transformation). This transformation corresponds to a local rescaling of the original metric $g$. We call $g$ the physical metric, and $\tilde{g}$ is called the unphysical metric, though strictly speaking their roles can be reversed. This corresponds to the distinction between a physical manifold $(M,g)$ and a conformally transformed, unphysical manifold $(M,\tilde{g})$. A subset of conformal transformations which arises as the pull-back of the metric from a diffeomorphism is also called a conformal isometry. If two metrics are conformally related, their causal structure is the same, as angles are the same --- which in the case of Lorentzian geometry also implies that space-like, light-like, or time-like vectors will stay space-like, light-like, or time-like, respectively, under a conformal transformation. Some care has to be taken regarding zeroes of $\Omega(x)$, but for the sake of brevity (as appropriate for a mere appendix), we shall omit these technical details and refer to \cite{Kroon} for this. Our presentation shall follow mostly appendix~D of \cite{Wald}.
 	
 	As the individual Levi--Civita connections of $g$ and $\tilde{g}$ will be different, quantities related to the connections will involve non-trivial appearances of $\Omega(x)$ and its derivatives. Two connections (on the same manifold) always differ by a tensor $Q^a{}_{bc}$, and in the present context one gets for an arbitrary one-form $\omega_a$
 	\begin{equation}
 		\nabla_a \omega_b - \tilde{\nabla}_a \omega_b = Q^c{}_{ab} \omega_c = \kl{2 \delta^c{}_{(a}\nabla_{b)}\ln\Omega - g_{ab}g^{cd}\nabla_d \ln\Omega}\omega_c.
 	\end{equation}
 	As we are assuming vanishing torsion, $Q$ is symmetric in the lower two indices.
 	
 	Since we are mostly unconcerned with questions of general relativity's dynamics, we shall refrain from giving a summary on how precisely the physical geometric quantities are associated to geometric quantities of the unphysical metric.
 	
 	When considering additional fields (tensorial, spinorial, densities, \dots) which follow certain equations in the physical metric, the natural question (in the context of conformal transformations) is if these equations change under the conformal transformation, and if so, how. Was the original field $\Psi$ a solution of these equations on the physical manifold $(M,g)$, and there is a real number $s$ such that $\tilde{\Psi} = \Omega^s \Psi$ is a solution to the corresponding equation on the unphysical manifold $(M,\tilde{g})$, we call the field, its equations, and the associated physical theory conformally invariant. The number $s$ is called the conformal weight.
 	
 	The field whose conformal transformation behaviour is of interest to us in the main text, is the field strength tensor of Maxwellian electrodynamics, both microscopic and macroscopic. Following the presentation in \cite{Wald}, and thus highlighting the importance of four space-time dimensions to our search for analogue space-times, we shall for the time being, and only within this section, consider $D$+1-dimensional electrodynamics. The corresponding Maxwell equations remain unchanged to those introduced in section~\ref{sec:CovEMana}:
 	\begin{subequations}
 		\begin{align}
 			\nabla_a F^{ab} &= J^b, \label{eq:AppInHomMaxwell}\\
 			\nabla_{[a} F_{bc]} &= 0.\label{eq:AppHomMaxwell}
 		\end{align}
 	\end{subequations}
 	Thus, suppose that $F$ has conformal weight $s$, and $J$ conformal weight $t$. Let us start by looking at the homogeneous Maxwell equation~\eqref{eq:AppHomMaxwell} and its properties under a conformal transformation:
 	\begin{subequations}
	 	\begin{align}
		 	\tilde{\nabla}_{[a} \Omega^s F_{bc]} &= \kl{{\nabla}_{[a}\Omega^s}F_{bc]} + \Omega^s \kl{\tilde{\nabla}_{[a}F_{bc]}},\\
		 	&= s \Omega^{s-1} \Omega_{[,a}F_{bc]} + \Omega^s \nabla_{[a} F_{bc]}.\label{eq:confinvhom}
	 	\end{align}
	\end{subequations}
	Note that we could make use of the symmetry of $Q$ to exchange $\tilde{\nabla}$ for $\nabla$ in the second term on the right hand side. From equation~\eqref{eq:confinvhom} we see that the weight $s$ has to vanish if we want the equations to be conformally invariant. This alone, however, is not enough, as we also have to check the conformal invariance of the inhomogeneous equation~\eqref{eq:AppInHomMaxwell}. As we are using the inhomogeneous Maxwell equation with the twice-contravariant field strength tensor $F^{ab}$, it acquires the conformal weight of the inverse metric twice, resulting in a weight $s-4$:
	\begin{subequations}
		\begin{align}
			\Omega^t J^b &= \tilde{\nabla}_a \Omega^{s-4} F^{ab},\\
			&= (s-4)\Omega^{s-5} \Omega_{,a} F^{ab} + \Omega^{s-4} \tilde{\nabla}_a F^{ab},\\
			&= (s-4)\Omega^{s-5} \Omega_{,a} F^{ab} + \Omega^{s-4} \kl{\nabla_a F^{ab} + Q^a{}_{ad} F^{db} + Q^b{}_{ad} F^{ad}},\\
			&= (s-4)\Omega^{s-5} \Omega_{,a} F^{ab} + \Omega^{s-4} \kl{\nabla_a F^{ab} + \kl{2 \delta^a{}_{(d}\nabla_{a)}\ln\Omega - g_{da}g^{ae}\nabla_e \ln\Omega} F^{db}},\\
			&= \Omega^{s-4} \kl{(s-4)\frac{\Omega_{,a}}{\Omega} F^{ab} + \nabla_a F^{ab} + \kl{\Omega^{-1} \Omega_{,d} + (D+1)\Omega^{-1} \Omega_{,d} - \Omega^{-1} \Omega_{,d}}F^{db}},\\
			&= (s+D-3)\Omega^{s-5} \Omega_{,a} F^{ab} + \Omega^{s-4}\nabla_a F^{ab},\\
			&= (D-3) \Omega^{-5} \Omega_{,a} F^{ab} + \Omega^{-4}\nabla_a F^{ab}.
		\end{align}
	\end{subequations}
	In the last step, we made use of our previously-gained knowledge that $s$ has to be $0$ for the homogeneous equations to be conformally invariant. From this we can gain two pieces of information: For one, the first term on the right hand side only vanishes in four space-time dimensions, \emph{i.e.}, $D=3$. For the other, $J^b$ would have to have a conformal weight of $t=-4$. The same result can be gained from looking at $J$'s divergence, which in the physical space-time was $0$:
	\begin{subequations}
		\begin{align}
			\tilde{\nabla}_a \Omega^t J^a &= t \Omega^{t-1} \Omega_{,a} J^a + \Omega^t \kl{\nabla_a J^a + Q^a{}_{ba} J^b},\\
			&= (t+D+1) \Omega^{t-1} \Omega_{,a} J^a.
		\end{align}
	\end{subequations}
	But without looking at the inhomogeneous Maxwell equation, this divergence alone would not be enough to answer the question of conformal invariance, as this equation alone can be fulfilled in any given space-time dimension.

	There remains one further variation to be examined: The case of fully covariant \emph{macroscopic} electrodynamics as described in chapter~\ref{ch:analogues}. As it is only the inhomogeneous equation,
	\begin{equation}
		J^b = \nabla_a Z^{abcd} F_{cd},
	\end{equation}
	that changes, only its transformation behaviour has to be checked. For this purpose let us introduce a third conformal weight, $u$ for $Z$, such that $\tilde{Z}^{abcd} = \Omega^u Z^{abcd}$:
	\begin{subequations}
		\begin{align}
			\Omega^t J^b =& \tilde{\nabla} \Omega^{s+u} Z^{abcd} F_{cd},\\
			=& (s+u) \Omega^{s+u-1} \Omega_{,a} Z^{abcd} F_{cd} + \Omega^{s+u} \left( \nabla_a Z^{abcd} F_{cd} + Q^a{}_{ae} Z^{ebcd} F_{cd}+Q^b{}_{ae}Z^{aecd}F_{cd}\right.\nonumber\\
			& \left. + Q^c{}_{ae} Z^{abed} F_{cd} + Q^d{}_{ae} Z^{abce} F_{cd} - Q^e{}_{ac}Z^{abcd} F_{ed} - Q^e{}_{ad}Z^{abcd} F_{ce}\right),\\
			=& (s+u+D+1) \Omega^{s+u-1} \Omega_{,a} Z^{abcd} F_{cd} + \Omega^{s+u} \nabla_a Z^{abcd} F_{cd}.
		\end{align}
	\end{subequations}
	Together with the results of the unchanged homogeneous equations, this results in $s=0, t=-D-1$, and $u=-D-1$. This means that macroscopic electrodynamics allows for conformal invariance in any space-time dimension, as long as the constitutive tensor has the right conformal weight. It is worthwhile to compare this to the case of the constitutive tensor of a vacuum space-time,
	\begin{equation}
		Z^{abcd}_\text{vacuum} = \ed{2}\kl{g^{ac} g^{bd} - g^{ad}g^{bc}}.
	\end{equation}
	In this case, the conformal weight is fixed to be $-4$, which again also fixes the space-time dimension in which the corresponding theory can be conformally invariant.
	
 	Similar to our omission of the behaviour of geometric quantities under conformal transformations, we shall also not present the transformation properties of the energy-momentum tensor of a given field (conformally invariant or not) as, again, in the main text the energy-momentum tensor makes no appearance. For this, we refer again to \cite{Wald} and \cite{Kroon}.
 	
 	Lastly, we mention that there are three distinct occurrences, though often simultaneously, of conformal invariance in the thesis: (1) The above described conformal invariance of electrodynamics. (2) The conformal invariance of light cone structures, also known as the conformal invariance of the causal structure. Many analogue models are constructed in a way that cannot distinguish between more than causal structures --- but this is less a feature of the underlying theory (as it is in electrodynamics) and more a matter of how the metric is derived: The assumption of the wave equation under consideration being described by null curves in an effective metric by the very question it asks can only look at null curves. Space-like or time-like curves do not come in --- even if the underlying physics of the analogue might be very different depending on the conformal factor. (Again, note the difference to our electromagnetic analogues in four space-time dimensions, where the underlying theory is actually conformally invariant!) (3) The conformal invariance of the Hawking temperature itself, as described in \cite{JacobsonKang93}. As the (current) holy grail of analogue space-times is the experimental observation of analogues of Hawking temperature and similar quantum processes, its conformal invariance is, given (2), certainly a boon. However, this invariance will usually be broken when transitioning to the laboratory coordinates, see section~\ref{sec:HawkingLab}.
 	
 	\section{Orthogonal Decomposition}\label{sec:ortho}
 	While, strictly speaking, any discussion of (Maurer--Cartan) frame fields (n-beins, vielbeins, \dots) in the context of Lorentzian geometry is such an orthogonal decomposition, this goes hand in hand with the introduction of a slightly different notation. In our notation, tetrad or triad components have a hat, to indicate this. Nevertheless, tensorial methods using (abstract) index notation are possible, and this is our focus. There are names other than \enquote{orthogonal decomposition} for this process (\emph{e.g.}, Bel decomposition, and depending on one's taste one might even call it an ADM or Kaluza--Klein decomposition,
 	), but \enquote{orthogonal decomposition} best describes the geometric meaning of this procedure: Given an observer with four-velocity $V$, decompose a given tensor $A$ in parts orthogonal and parallel to this four-velocity. At least locally, this will always be possible, as one can find a space-like hyper-surface orthogonal to $V$. Hence, the idea is to find projection operators onto the direction of $V$ and orthogonal to it. The key identity to do this is the realisation that
 	\begin{equation}
	 	\delta^a{}_b = \underbrace{-V^a V_b}_{\ifed t^a{}_b} + \underbrace{g^a{}_b + V^a V_b}_{\ifed h^a{}_b}.
 	\end{equation}
 	Here, $t^a{}_b$ is the time-like projection operator and $h^a{}_b$ the space-like one. As $t^a{}_b$ is a projection onto a one-dimensional subspace, just looking at a contraction of the corresponding index with $V_b$ has the same information and can serve as a substitute for \enquote{time-like projection}. We will make use of this equivalence and abandon the time-like projection operator in most places in favour of the contraction with the four-velocity.
 	
 	With this identity at hand, one then rewrites and collects all the different, possible terms arising of combinations of $h$ and $t$:
 	\begin{subequations}\label{eq:orthdecom}
 		\begin{align}
		 	A^{a_1 \dots a_n}{}_{b_1\dots b_m} =& \delta^{a_1}{}_{c_1} \cdots \delta^{a_n}{}_{c_n} \delta_{b_1}{}^{d_1}\cdots \delta_{b_m}{}^{d_m} A^{c_1 \dots c_n}{}_{d_1 \dots d_m,}\\
		 	=&\kl{t^{a_1}{}_{c_1} + h^{a_1}{}_{c_1}}\cdots \kl{t^{a_1}{}_{c_1} + h^{a_n}{}_{c_n}} \kl{t_{b_1}{}^{d_1} + h_{b_1}{}^{d_1}}\cdots\nonumber\\
		 	&\cdots \kl{t_{b_m}{}^{d_m} + h_{b_m}{}^{d_m}}A^{c_1 \dots c_n}{}_{d_1 \dots d_m},\\
		 	=& t^{a_1}{}_{c_1} \cdots t^{a_n}{}_{c_n}t_{b_1}{}^{d_1} \cdots t_{b_m}{}^{d_m}A^{c_1 \dots c_n}{}_{d_1 \dots d_m}+\nonumber\\
		 	& h^{a_1}{}_{c_1} \cdots t^{a_n}{}_{c_n}t_{b_1}{}^{d_1} \cdots t_{b_m}{}^{d_m}A^{c_1 \dots c_n}{}_{d_1 \dots d_m}+\nonumber\\
		 	& t^{a_1}{}_{c_1}h^{a_2}{}_{c_2}t^{a_3}{}_{c_3} \cdots t^{a_n}{}_{c_n}t_{b_1}{}^{d_1} \cdots t_{b_m}{}^{d_m}A^{c_1 \dots c_n}{}_{d_1 \dots d_m}+\dots\nonumber\\
		 	&h^{a_1}{}_{c_1} \cdots h^{a_n}{}_{c_n}h_{b_1}{}^{d_1} \cdots h_{b_m}{}^{d_m}A^{c_1 \dots c_n}{}_{d_1 \dots d_m}.
	 	\end{align}
 	\end{subequations}
 	While this is the general idea behind the orthogonal decomposition, usually less straightforward ways of arriving at the decomposition are more advisable, as they tend to reduce the amount of combinatorics needed by facilitating properties of the tensor $A$. Likewise, we shall refrain from proving statements about this general orthogonal decomposition as we shall only be concerned with two specific types of tensors. In the rest of this appendix, we shall consider this procedure for two particular cases of tensors: (a) The orthogonal decomposition of a two-form (like the field strength tensor), and (b) the orthogonal decomposition of a tensor $Z^{abcd}$ with the symmetries $Z^{abcd} = Z^{cdab} = -Z^{bacd}$ (as the Riemann tensor, area metrics, or the constitutive tensor of electromagnetism \cite{AreaMetrClass,HehlKieferAreaMetrics}). Case (b) is usually considered in the context of the Riemann tensor which additionally fulfils $R^{[abcd]}=0$, however, we shall not make this additional assumption.
 	
 	Furthermore, it is worthwhile to introduce the language of a tensor being orthogonal\footnote{As we are using the orthogonal decomposition only in the context of four space-time dimensions, we shall frequently say \enquote{four-orthogonal}. This both emphasises the dimensional dependence of the results as well as helps if more than one notion of orthogonality is used. For example, in the discussion of the Doran coordinates of the Kerr metric in section~\ref{sec:quasicart} quantities both four- and three-orthogonal to some vector are encountered.} to a vector (in practice this is almost always a four-velocity describing an observer). A tensor $T^{a_1 \dots}{}_{b_1\dots}$ is considered orthogonal to the vector $V^a$ in index $c$, if the contraction of the tensor $T$ with $V$ in that index vanishes. The tensor is simply called \enquote{orthogonal to $V$} if it is orthogonal to $V$ in all indices.
 
 	We should, at this point, also remind ourselves of the definition of the Hodge-star which, with the help of a metric, turns $s$-forms on an $n$-dimensional manifold into $(n-s)$-forms via
 	\begin{equation}
 		(\ast T)_{a_1 \dots a_{n-s}} = \ed{s!} \epsilon_{a_1\dots a_{n-s}}{}^{b_1\dots b_s} F_{b_1\dots b_s}.
 	\end{equation}
 	
 	The orthogonal decomposition of two-forms and its physical application goes back all the way to the early days of special relativistic formulations of electrodynamics, see for example \cite{PauliRelativity}. For other tensors, the literature is less easy to navigate: For example, the results we present in section~\ref{sec:4tensor}, \emph{do} have a long history --- going back to at least Bel \cite{Bel1,BelSem,BelTrans}, or Matte \cite{MatteDecom}, and it has been frequently employed both in a named way \cite{DynLawSuEGR,GravEMagAna}, and nameless way \cite{BalNi}. Nevertheless, it is safe to call it sufficiently ill-known to warrant an extensive exposition in this appendix. Before we commence with this, we set the stage by going through the analysis for two-forms.
 	
 	\subsection{Orthogonal Decomposition of a Two-Form}\label{sec:twoform}
 	As a first example, we shall develop the orthogonal decomposition of a two-form, following the discussion of \cite{GourSR}, pp.~83. This, as already alluded to, also gives an example of a less straightforward derivation, while providing a different kind of bookkeeping. We shall proof:
	\begin{thm}\label{thm:Fdecom}
 		For any two-form $F_{ab}$ in four dimensional space-time and any four-velocity $V^a$, there exist two vector fields $E^a$ and $B^a$, both orthogonal to $V^a$, such that
 		\begin{equation}\label{eq:Fdecom}
	 		F_{ab} = V_a E_b - V_b E_a + \epsilon_{abcd} V^c B^d.
 		\end{equation}
 	\end{thm}
 
 	Before starting the proof, let us first capture some observations: As the naming already suggest, this result is of obvious relevance to electrodynamics and the identification of the field strength tensor as the object collecting the information of electric and magnetics fields in a relativistically sensible way. This is usually written in matrix form in the following way:
 	\begin{equation}\label{eq:Fmatrix}
	 	\kl{F_{ab}}_{a,b\in \{0,1,2,3\}} = \begin{pmatrix}
	 		0 & E_1 & E_2 & E_3\\
	 		-E_1 & 0 & B_3 & -B_2\\
	 		-E_2 & -B_3 & 0 & B_1\\
	 		-E_3 & B_2 & -B_1 & 0
	 	\end{pmatrix}.
 	\end{equation}
 	While this equation can be derived from equation~\eqref{eq:Fdecom}, it is also immediately obvious that the observer dependence is much more apparent in \eqref{eq:Fdecom}, just as the theorem contains additional geometric information. Also note the importance of specifying the space-time dimension: Only in space-time dimension four do the degrees of freedom of a two-form, $\frac{{(4-1)\times 4}}{ 2 }= 6$, match the degrees of freedom of two vectors four-orthogonal to a third, $2\times (4-1) = 6$. This is one instance where it become obvious that the particular tensors used in the orthogonal decomposition, \emph{i.e.}, the objects a given observer would consider in a non-relativistic context, depend on the dimension of the space-time. As a side-effect, the discussion of electromagnetic analogues as done in section~\ref{sec:CovEMana} would change substantially were one to look for higher dimensional analogues. And this even before noting the even more obvious fact that fitting the four-dimensional laboratory situation would also complicate the bi-metric nature of the endeavour, as for a four-dimensional analogue both the analogue and the laboratory will \enquote{live in the same dimension}.
 	
 	What can also be deduced --- and observed in equation~\eqref{eq:Fmatrix} --- is that due to the antisymmetry of $F$, applying the time-projection operator $t^a{}_b$ twice to $F$ (or equivalently just contracting twice with $V$) will give a vanishing result. This partial result remains true in any space-time dimension.
 
 	\begin{proof}
 		First, note that indeed equation~\eqref{eq:Fdecom} defines a two-form. Then set $q_a\defi F_{ab}V^b$. By the antisymmetry of $F$, $q$ is automatically four-orthogonal to $V$. The idea now is to compare this one-form $q$ with the contributions of $B$ and $E$ (more precisely: The by metric duality associated one-forms $B^\flat$ and $E^\flat$) in equation~\eqref{eq:Fdecom}. 
 		
 		Next, define the two-form $G_{ab} \defi F_{ab} - V_a q_b + q_a V_b$ which is obviously four-orthogonal to $V$. That is, $G_{ab}$ is exactly the part of $F_{ab}$ that would not be captured by $q_a$.
 		
 		Choosing a triad $e_{\hat{i}}$, $\hat{i}\in \{1,2,3\}$, for the space-like hyper-surface orthogonal to $V$, define three numbers $b^i$ as
 		\begin{equation}
 			b^1 \defi G_{ab}e^a_{\hat{2}} e^b_{\hat{3}}, \quad b^2 \defi G_{ab}e^a_{\hat{3}} e^b_{\hat{1}}, \quad b^3 \defi G_{ab}e^a_{\hat{1}} e^b_{\hat{2}},
 		\end{equation}
 		and collect these numbers in the vector $b^a \defi b^i e^a_{\hat{i}}$. By definition of the triad, this vector will be orthogonal to $V$. If two vectors $W_A^a$ and $W_B^a$ are orthogonal to $V$, we can expand these in our chosen triad as $W^a_{A,B} = W^{\hat{i}}_{A,B}\, e_{\hat{i}}^a$ and examine their contraction with $G_{ab}$:
 		\begin{subequations}
			\begin{align}
				G_{ab} W_A^a W_B^b &= G_{ab} W^{\hat{i}}_{A}\, e_{\hat{i}}^a W^{\hat{j}}_{B}\, e_{\hat{j}}^b,\\
				&= \begin{vmatrix}
	 			b^{\hat{1}} & W_A^{\hat{1}} & W_B^{\hat{1}}\\
	 			b^{\hat{2}} & W_A^{\hat{2}} & W_B^{\hat{2}}\\
	 			b^{\hat{3}} & W_A^{\hat{3}} & W_B^{\hat{3}}
				\end{vmatrix},\\
				&= \epsilon_{abcd} V^a b^b W_A^c W_B^d. \label{eq:Gidentified}
	 		\end{align}
 		\end{subequations}
 		In the last step we used the fact that the three-dimensional Levi-Civita symbol is related (by definition) to the four-dimensional one by contracting the latter with the four-velocity of a given observer. Note that this last step, equation~\eqref{eq:Gidentified}, is of the form required for equation~\eqref{eq:Fdecom} --- thus the only remaining part of the proof of the theorem is to show the uniqueness of $q$ and $b$ whence we can identify $q$ with $E$ and $b$ with $B$. Thus far, we at least can say that $F$ \emph{can} be re-written in the form
 		\begin{equation}\label{eq:Ffact}
	 		F_{ab} = V_a q_b - V_b q_a + \epsilon_{abcd} V^c b^d.
 		\end{equation}
 		
 		For $q$ we simply observe from the previous equation~\eqref{eq:Ffact} the fact that for any vector $W^a$
 		\begin{subequations}
 			\begin{align}
	 			F_{ab} W^a V^b &= V_a W^a \underbrace{V^b q_b}_{=0} - q_a W^a \underbrace{V^b V_b}_{=-1} + \underbrace{\epsilon_{abcd} V^c b^d W^a V^b}_{=0},\\
	 			&= q_a W^a.
 			\end{align}
 		\end{subequations}
 		This establishes uniqueness of $q$ and we can identify $q=E^\flat$.
 		
 		Finally, restricting equation~\eqref{eq:Ffact} to the space-like hyper-surface orthogonal to $V$ by contracting it with two arbitrary vectors $W_{A,B}$ four-orthogonal to $V$:
 		\begin{subequations}
 			\begin{align}
 			F_{ab} W_A^a W_B^b &= \underbrace{V_a W^a_A}_{=0} W^b_B E_b - E_a W_A^a \underbrace{W_B^b V_b}_{=0} + \epsilon_{abcd} V^c b^d W_A^a W_B^b,\\
 			&= \epsilon_{abcd} V^c b^d W_A^a W_B^b.
 			\end{align}
 		\end{subequations}
 		It follows that any restriction to the hyper-surface orthogonal to $V$ contains all information regarding $b$. As $\epsilon_{abcd} V^c$ is a non-degenerate three-form on this hyper-surface, we can then deduce the uniqueness of $q$ and thus take it to be $B^\flat$. This completes the proof.
 	\end{proof}
 
 	Now is a good opportunity to show the possibility of proving theorem~\ref{thm:Fdecom} using the general orthogonal decomposition formula~\ref{eq:Fdecom}. As any uniqueness proof would either rely on the proof of the uniqueness of the \emph{general} decomposition formula (which we did not give), or reduce to one reminiscent of the one just delivered, this proof will only be the \enquote{existence part}.
 	\begin{proof}
	 	We start by applying the general formula~\eqref{eq:Fdecom} to $F_{ab}$:
	 	\begin{subequations}
	 		\begin{align}
		 		F_{ab} =& \delta_a{}^c \delta_b{}^d F_{cd},\\
		 		=& (t_a{}^c + h_a{}^c) (t_b{}^d + h_b{}^d) F_{cd},\\
		 		=& t_a{}^c t_b{}^d F_{cd} + h_a{}^c h_b{}^d F_{cd} + (t_a{}^c h_b{}^d + h_a{}^c t_b{}^d) F_{cd},\\
		 		=& \underbrace{+V_a{V}^c V_b{V}^d F_{cd}}_{=0} + h_a{}^c h_b{}^d F_{cd} - V_a{V}^c (g_b{}^d + V_b V^d) F_{cd} - h_a{}^c V_b{V}^d F_{cd},\\
		 		=& h_a{}^c h_b{}^d F_{cd} + (g_b{}^d V_a{V}^c + V_b V^d V_a{V}^c) F_{dc} - h_a{}^c V_b{V}^d F_{cd},\\
		 		=& h_a{}^c h_b{}^d F_{cd} + (h_b{}^c V_a - h_a{}^c V_b) F_{cd} {V}^d,\label{eq:Fbeforetetrad}\\
	 		\end{align}
	 	\end{subequations}
 	where in the last step we relabelled indices.
 	
 	From here on out it is easiest to change to a tetrad $e^a_{\hat{a}}$ with $V^a = e_{\hat{0}}^a$. Using this definition and the orthogonality of the vierbeins among each other, the only non-vanishing components of $h^{\hat{a}}{}_{\hat{b}}$, the spatial projection operator, will be $h^{\hat{i}}{}_{\hat{j}}$. We can then identify the first term in equation~\eqref{eq:Fbeforetetrad} as $F_{\hat{i}\hat{j}}$ --- hence a three-dimensional two-form, thus describable with a three-vector. Similarly, the second term involves $F_{\hat{i}\hat{0}}$ --- again something that can be rewritten as a three-vector. Without going into the details, the actual identification with a three-vector for the first term would involve a Hodge--star giving precisely the wanted Levi-Civita symbol for the last term in equation~\eqref{eq:Fdecom}. The second term already has the required structure. Reverting the introduction of the tetrad then gives equation~\eqref{eq:Fdecom}.
 	\end{proof}
 
 	It is worthwhile to note at this point, and easy to show using equation~\eqref{eq:Fdecom}, that $E$ and $B$ change the role when one looks at the Hodge-dual $\ast F$ of $F$ and its orthogonal decomposition. We will encounter a similar interconnection between the tensors appearing in the orthogonal decomposition in the next section, too.
 	
 	\subsection{Orthogonal Decomposition of Certain Fourth Rank Tensors}\label{sec:4tensor} 
 	In this section, we want to analyse the orthogonal decomposition of a fourth-rank tensor $Z^{abcd}$ such that $Z^{abcd} = Z^{cdab} = -Z^{bacd}$. With the orthogonal decomposition of a two-form in place, this analysis is made much more accessible. To see how this comes about it is worthwhile to have a closer look at the degrees of freedom of the $Z$. As explained in section~\ref{sec:CovEMana}, in four space-time dimensions, $Z$ has 21 d.o.f. and can be interpreted as a two-vector-valued symmetric matrix $Z^{AB}$, such that
 	\begin{equation}
 		Z^{abcd} = Y^{ab}_A Z^{AB} Y_B^{cd}.
 	\end{equation}
 	Again, we shall label the four-velocity of the observer with respect to whom we are performing the orthogonal decomposition with $V$. Using the result of theorem~\ref{thm:Fdecom} from the previous section, we can then use this decomposition of two-forms to have a go at the decomposition of $Z$. Remembering that there are six two-vectors we will have six $E_A$ and $B_A$ for their decomposition\footnote{The naming of $E_A$ and $B_A$ is chosen such that their role in the corresponding version of equation~\eqref{eq:Fdecom} is clear; this is not to mean that they are six electric or magnetic fields!}. Inserting this, we get
 	\begin{subequations}
 		\begin{align}
 			Z^{abcd} =& (V^a E^b_A -V^b E_A^a + \epsilon^{abef} V_e B^A_f) Z^{AB} (V^c E^d_B -V^d E_B^c + \epsilon^{cdgh} V_g B^B_h),\\
 			=& V^a E^b_A Z^{AB} V^c E^d_B - V^a E^b_A Z^{AB} V^d E_B^c + V^a E^b_A Z^{AB} \epsilon^{cdgh} V_g B^B_h\nonumber\\
 			& -V^b E_A^a Z^{AB} V^c E^d_B + V^b E_A^a Z^{AB} V^d E_B^c - V^b E_A^a Z^{AB} \epsilon^{cdgh} V_g B^B_h\nonumber\\
 			& +\epsilon^{abef} V_e B^A_f Z^{AB} V^c E^d_B - \epsilon^{abef} V_e B^A_f Z^{AB} V^d E_B^c + \epsilon^{abef} V_e B^A_f Z^{AB} \epsilon^{cdgh} V_g B^B_h.
 		\end{align}
 	\end{subequations}
 	Now it is worthwhile to introduce the following four shorthand notations:
 	\begin{center}
 		\begin{subequations}\label{eq:defW}
 			\begin{minipage}{.45\textwidth}
 				\begin{align}
 				W_\epsilon^{ab} &\defi E_A^a Z^{AB} E_B^b,\hphantom{blabla}\\
 				\addtocounter{equation}{+1}[W_\zeta^T]^{ab} & = B_A^a Z^{AB} B_B^b,
 				\end{align}
 			\end{minipage}\hfill
 			\begin{minipage}{.45\textwidth}
 				\begin{align}
 				\addtocounter{equation}{-2}W_\zeta^{ab} &\defi E_A^a Z^{AB} B^b_B,\hphantom{bladiebla}\\
 				\addtocounter{equation}{1}W_\mu^{ab} &\defi B_A^a Z^{AB} B^b_B.
 				\end{align}
 			\end{minipage}
 		\end{subequations}
 	\end{center}
 	Using the symmetry of $Z^{AB}$ and collecting terms, one then arives at
 	\begin{align}
	 	Z^{abcd} =& V^b V^d W_\epsilon^{ac} + V^a V^c W_\epsilon^{bd} - V^a V^d W_\epsilon^{bc} - V^b V^c W_\epsilon^{ad}\nonumber\\
	 	&\;+V^f (\eps^{ab}{}_{ef} W_\mu^{eg} \eps^{cd}{}_{gh}) V^h\; +\kl{W_\zeta^{ag} V^b - W_\zeta^{bg} V^a} \eps^{cd}{}_{gh} V^h\nonumber\\
	 	& \;+V^f \eps^{ab}{}_{ef} \kl{[W_\zeta^T]^{ec} V^d - [W_\zeta^T]^{ed} V^c}.\label{eq:ZBel1}	 	
 	\end{align}
 	Finally, the further definitions
 	\begin{equation}
	 	-2 W_\epsilon = \epsilon, \quad 2 W_\zeta = \zeta, 2 W_\mu = \mu^{-1}
 	\end{equation}
 	provide the link to the use of this decomposition in macroscopic electrodynamics, as it is done in this thesis in section~\ref{sec:CovEMana}. With some patience and index algebra it is then possible to rephrase equation~\eqref{eq:ZBel1} as:
 	\begin{align}
	 	Z^{abcd} = &\ed{2}\kl{V^a V^d \epsilon^{bc} + V^b V^c \epsilon^{ad} -V^b V^d \epsilon^{ac} - V^a V^c \epsilon^{bd} }\nonumber\\
	 	&+ \ed{8} \eps^{ab}{}_{ef} \eps^{cd}{}_{gh}\kl{V^f \mi^{eg} V^h + V^e \mi^{fh} V^g - V^e \mi^{fg} V^h - V^f \mi^{eh} V^g}\nonumber\\
	 	&+ \ed{4} \eps^{ab}{}_{ef} \kl{\zeta^{fc}V^d V^e + \zeta^{ed} V^c V^f - \zeta^{ec}V^d V^f - \zeta^{fd} V^c V^e}\nonumber\\
	 	&+ \ed{4} \eps^{cd}{}_{gh} \kl{\zt^{bg} V^a V^h  + \zt^{ah} V^b V^g   - \zt^{ag} V^b V^h  - \zt^{bh} V^a V^g }.\label{eq:ZBel2}
 	\end{align}
 	Having arrived at a first version of the orthogonal decomposition of $Z$ as used in electrodynamics, it is useful to say a few words on the uniqueness of this result: The uniqueness of the orthogonal decomposition in this particular case is the question of the uniqueness of the matrices $\eps$, $\mu^{-1}$, and $\zeta$. This question is quick to answer: Since they were defined in equations~\eqref{eq:defW} (up to one sign and factors of $2$ later introduced in definition) in terms of the \emph{unique} decomposition of the six $F_A$ into six $E_A$ and six $B_A$, this uniqueness carries through. A simple counting of d.o.f. of $\eps$, $\mu^{-1}$, and $\zeta$ also helps demonstrating this: Both $\eps$ and $\mu^{-1}$ are symmetric $3\times 3$ matrices, which is easiest seen by either switching to a tetrad with $V$ as $e_0$, or by noting the effective reduction to a $3\times 3$ matrix from their orthogonality to $V$. The symmetry is obvious from their definition. $\zeta$, on the other hand, has only the orthogonality going for it, thus is an unconstrained $3\times 3$ matrix. We end up with a total of $6+6$ d.o.f. from $\eps$ and $\mu^{-1}$, and a further $9$ d.o.f. from $\zeta$. In total, this makes $21$, as needed to fully represent the 21 degrees of freedom of $Z$.
 	
 	Now for some further observations regarding the orthogonal decomposition of $Z$. First, observe that
 	\begin{equation}
 		\epsilon_{cd}{}^{af} \epsilon^{ebcd} V_e V_f = -2 (g^{ab}+V^a V^b) = -2h^{ab},\label{eq:projection}
 	\end{equation}
 	and second that
 	\begin{equation}
 		g^{b_1 c_1}\cdots g^{b_n c_n} \eps_{c_1\dots c_n} \eps^{a_1\dots a_n} = -n! g^{b_1 c_1}\cdots g^{b_n c_n} \delta^{a_1}{}_{[c_1}\cdots\delta^{a_n}{}_{c_n]}.
 	\end{equation}
 	The first fact's usefulness can be already anticipated by the appearance of $h^{ab}$ in both~\eqref{eq:projection} and the earlier, very general equations~\eqref{eq:orthdecom}. With these observations it is now possible to rephrase the orthogonal decomposition of $Z$ yet again as:
 	\begin{align}
	 	Z^{abcd} = &\frac{1}{2} \left( V^d V^a \epsilon^{bc} -V^c V^a \epsilon^{bd} + V^c V^b \epsilon^{ad} - V^d V^b \epsilon^{ac} +h^{ad} \left[ \mu^{-1} \right]^{cb} - h^{ac} \left[ \mu^{-1} \right]^{db}\right.\nonumber\\
	 	& + h^{bc} \left[ \mu^{-1} \right]^{ad} - h^{bd} \left[ \mu^{-1} \right]^{ac} + (h^{bd} h^{ac} - h^{bc} h^{ad}) \left[ \mu^{-1} \right]^e{}_e \nonumber\\
	 	&+\left. \eps^{fabe}(V^d \zeta_e{}^c - V^c \zeta_e{}^d)V_f + \eps^{fcde}(V^b \zeta_e{}^a - V^a \zeta_e{}^b)V_f\right).\label{eq:ZBel3}
 	\end{align}
 	In order to move from equation~\eqref{eq:ZBel2} to equation~\eqref{eq:ZBel3}, it is useful to work from both sides as non-trivial cancellations of terms need to be reinserted. The permittivity needs no work, so let us quickly go through the process for permeability, as this is the hardest part. The magneto-electric tensor can be worked out in a similar manner. It will prove useful to forget about the naming of indices contracted with a four-velocity $V$ or $\mi$. Contractions with the former will be written with a stand-in index $\bullet$ and contractions with the latter with $\circ$. That has the benefit that we can easily find full contractions of $V$ without having to rename indices. The full contractions then simplify to $V_\bullet V_\bullet g^{\bullet\bullet} = -1$.
 	Furthermore, make use of the definition
 	\begin{subequations}
 	\begin{align}
	 	g^{a_1\dots a_n:b_1\dots b_n} &\defi g^{b_1 c_1}\cdots g^{b_n c_n} \eps_{c_1\dots c_n} \eps^{a_1\dots a_n},\label{eq:permut}\\
	 	&= -\delta^{a_1\dots a_n}{}_{c_1\dots c_n}g^{b_1 c_1}\cdots g^{b_n c_n},\\
	 	&= -n! g^{b_1 c_1}\cdots g^{b_n c_n} \delta^{a_1}{}_{[c_1}\cdots\delta^{a_n}{}_{c_n]}.
 	\end{align}
 	\end{subequations}
 	Now, what we shall prove is that
 	\begin{align}
	 	h^{ad} \left[ \mu^{-1} \right]^{cb} - h^{ac} \left[ \mu^{-1} \right]^{db} + h^{bc} \left[ \mu^{-1} \right]^{ad} - h^{bd} \left[ \mu^{-1} \right]^{ac} + (h^{bd} h^{ac} - h^{bc} h^{ad}) \left[ \mu^{-1} \right]^\circ{}_\circ\nonumber\\
	 	= \eps^{ab}{}_{e\bullet} \eps^{cd}{}_{g\bullet} V^\bullet V^\bullet g^{\circ e} g^{\circ g} \mi_{\circ\circ}.\label{eq:tbs}
 	\end{align}
 	The first step is to make tremendous use of \eqref{eq:projection} to get:
 	\begin{subequations}
	 	\begin{align}
		 	&h^{ad} \left[ \mu^{-1} \right]^{cb} - h^{ac} \left[ \mu^{-1} \right]^{db} + h^{bc} \left[ \mu^{-1} \right]^{ad} - h^{bd} \left[ \mu^{-1} \right]^{ac} + (h^{bd} h^{ac} - h^{bc} h^{ad}) \left[ \mu^{-1} \right]^\circ{}_\circ\\
		 	=&V_\bullet V_\bullet \mi_{\circ\circ} \left( (g^{\bullet\bullet} g^{ca} - g^{\bullet a} g^{\bullet c})g^{d\circ} g^{b\circ} - (g^{\bullet\bullet}g^{da} - g^{\bullet a} g^{d\bullet})g^{c\circ}g^{b\circ}  \right.\nonumber\\
		 	&- (g^{\bullet\bullet}g^{cb} - g^{\bullet b} g^{c\bullet})g^{a\circ} g^{d\circ} + (g^{\bullet\bullet}g^{db} - g^{\bullet b}g^{d\bullet})g^{a\circ}g^{c\circ}\nonumber\\
		 	&+\left.\textcolor{formulared}{\kle{ (g^{\bullet\bullet}g^{db} - g^{\bullet b}g^{d\bullet})(g^{\bullet\bullet}g^{ca}-g^{\bullet a}g^{\bullet c}) - (g^{\bullet\bullet}g^{cb} - g^{\bullet b} g^{c\bullet})( g^{\bullet\bullet}g^{da} - g^{\bullet a}g^{d\bullet} ) }}V_\bullet V_\bullet g^{\circ\circ}\right),\\
		 	=&V_\bullet V_\bullet \mi_{\circ\circ} \left(\textcolor{formulablue}{g^{\bullet\bullet} g^{ca}g^{d\circ} g^{b\circ} - g^{\bullet a} g^{\bullet c}g^{d\circ} g^{b\circ} - g^{\bullet\bullet}g^{da}g^{c\circ}g^{b\circ} + g^{\bullet a} g^{d\bullet}g^{c\circ}g^{b\circ}}\right.\nonumber\\
		 	&\textcolor{formulablue}{- g^{\bullet\bullet}g^{cb}g^{a\circ} g^{d\circ} + g^{\bullet b} g^{c\bullet}g^{a\circ} g^{d\circ} + g^{\bullet\bullet}g^{db}g^{a\circ}g^{c\circ} - g^{\bullet b}g^{d\bullet}g^{a\circ}g^{c\circ}}\nonumber\\
		 	&\left.+\textcolor{formulared}{\kle{ g^{db} g^{\bullet a} g^{\bullet c} - g^{\bullet\bullet} g^{db} g^{ca} + g^{\bullet b} g^{d\bullet} g^{ca}+g^{\bullet\bullet} g^{cb} g^{da} - g^{cb} g^{\bullet a} g^{d\bullet}+g^{c\bullet} g^{\bullet b} g^{da} }} g^{\circ\circ}\right).
	 	\end{align}
	\end{subequations}
	 	We have to explicitly write out the permutations in \eqref{eq:permut}. In \textcolor{formulagreen}{green} we will mark all the cross-terms of $\bullet$ and $\circ$ --- thanks to $\mi$ being orthogonal to $V$, these will all vanish. In \textcolor{formulared}{red} we will mark the six pieces involving the trace from the previous calculation, the remainder of the previous calculation will be \textcolor{formulablue}{blue}. For this matching we do not look at the final expression for $Z$ in \eqref{eq:ZBel3} --- rather we will look at the expression involving $\mi$ in the first step of the proof of \eqref{eq:ZBel3}:
	 \begin{subequations}
	 	\begin{align}
		 	&\eps^{ab}{}_{e\bullet}\eps^{cd}{}_{g\bullet} V^\bullet V^\bullet g^{\circ e} g^{\circ g} \mi_{\circ\circ} = g^{ab\circ\bullet:cd\circ\bullet} V_\bullet V_\bullet \mi_{\circ\circ},\\
		 	=&-\frac{4!}{4!}V_\bullet V_\bullet \mi_{\circ\circ}\left[ \textcolor{formulared}{g^{ac}g^{bd} g^{\circ\circ}g^{\bullet\bullet}} + \textcolor{formulagreen}{g^{ac} g^{b\circ} g^{\circ\bullet} g^{\bullet d} + g^{ac} g^{b\bullet} g^{\circ d} g^{\bullet \circ} + g^{ad}g^{bc} g^{\circ\bullet}g^{\bullet \circ}} \right.\nonumber\\
		 	&+ \textcolor{formulablue}{g^{ad} g^{b\circ} g^{\circ c} g^{\bullet\bullet}} +\textcolor{formulared}{g^{ad} g^{b\bullet} g^{\circ\circ} g^{\bullet c}} +\textcolor{formulablue}{g^{a\circ} g^{bc} g^{\circ d} g^{\bullet\bullet}} +\textcolor{formulagreen}{g^{a\circ} g^{bd} g^{\circ\bullet} g^{\bullet c}}\nonumber\\
		 	&+\textcolor{formulablue}{g^{a\circ} g^{b\bullet} g^{\circ c} g^{\bullet d}} + \textcolor{formulared}{g^{a\bullet} g^{bc} g^{\circ\circ} g^{\bullet d}} + \textcolor{formulagreen}{g^{a\bullet} g^{bd} g^{\circ c} g^{\bullet \circ}} + \textcolor{formulablue}{g^{a\bullet} g^{b\circ} g^{\circ d} g^{\bullet c}}\nonumber\\
		 	&\textcolor{formulagreen}{-g^{ac} g^{bd} g^{\circ\bullet} g^{\bullet \circ}} \textcolor{formulablue}{- g^{ac} g^{b\circ} g^{\circ d} g^{\bullet\bullet}} \textcolor{formulared}{-g^{ac} g^{b\bullet} g^{\circ\circ} g^{\bullet d} - g^{ad}g^{bc}g^{\circ\circ} g^{\bullet\bullet}}\nonumber\\
		 	&\textcolor{formulagreen}{-g^{ad} g^{b\circ} g^{\circ\bullet} g^{\bullet c} - g^{ad} g^{b\bullet} g^{\circ c} g^{\bullet \circ} - g^{a\circ} g^{bc} g^{\circ\bullet} g^{\bullet d}} \textcolor{formulablue}{- g^{a\circ} g^{bd} g^{\circ c} g^{\bullet\bullet}}\nonumber\\
		 	&\left.\textcolor{formulablue}{-g^{a\circ} g^{b\bullet} g^{\circ d} g^{\bullet c}} \textcolor{formulagreen}{-g^{a\bullet} g^{bc} g^{\circ d} g^{\bullet \circ}} \textcolor{formulared}{-g^{a\bullet} g^{bd} g^{\circ\circ} g^{\bullet c}} \textcolor{formulablue}{- g^{a\bullet} g^{b\circ} g^{\circ c} g^{\bullet d}}\right].
	 	\end{align}
 	\end{subequations}
 	Thus, we have shown \eqref{eq:tbs}. Together with a similar procedure for $\zeta$, we arrive at equation~\eqref{eq:ZBel3}. Since both \eqref{eq:ZBel2} and \eqref{eq:ZBel3} can be found in the literature, it is useful to have these means to switch from one decomposition to the other.
 	
 	Now it is time to come back to the notion of Hodge-duality mentioned in the previous section and its connection to the three matrices we identified as the components of the orthogonal decomposition of $Z$. However, it is obvious that the original notion of Hodge-duality does not apply --- $Z$ simply is not an $n$-form (nor even an $n$-vector). For this, it is necessary to introduce the notion of \enquote{left-}, \enquote{right-}, and \enquote{double-dual}. As we are restricting ourselves to four space-time dimensions in the current discussion, we shall restrict these definitions similarly. Let us summarise the definitions immediately together with their application in the current context:
 	\begin{subequations}\label{eq:WandDuals}
 		\begin{align}
 		W_\epsilon^{bd} &&= V_a V_c Z^{abcd},\\
 		W_\mu^{bd} &= V_a V_c (\ed{2} \epsilon^{ab}{}_{ef} Z^{efgh} \ed{2} \epsilon_{gh}{}^{cd} ) &\ifed V_a V_c \kl{\ast Z \ast}^{abcd},\\
 		[W_\zeta^T]^{bd} &= V_a V_c (\ed{2} \epsilon^{ab}{}_{ef} Z^{efcd}) &\ifed  V_a V_c \kl{\ast Z}^{abcd},\\
 		W_\zeta^{bd} &= V_a V_c (Z^{abgh} \ed{2} \epsilon_{gh}{}^{cd})&\ifed V_a V_c \kl{Z\ast}^{abcd}.
 		\end{align}
 	\end{subequations}
 	This (four-dimensional) definition of duals adapted to tensors of the kind of $Z$ can be found, for example, in \cite{MTW} and \cite{ExactSolsEinstein} applied in the context of the curvature tensors. However, as the curvature tensors have additional symmetry, much of the discussion of the orthogonal discussion of a curvature tensor needs to be re-examined carefully before applying (and copying) it to that of, for example, a \emph{general} constitutive tensor $Z$. The above formulas highlight also the similarity to $F$ and $\ast F$ --- applying the various duality operations switches the positions of the various classical constitutive tensors (permittivity, permeability, and magneto-electric tensor) inside the fourth rank tensor $Z$.
 	
 	What becomes especially apparent and obvious in this derivation is the mixing of components of $Z$ under a change of observer. For example, a medium at rest having no magneto-electric effect will exhibit a magneto-electric effect when observed in a laboratory moving with respect to this medium. While the transformation behaviour of electromagnetic quantities \emph{can} be derived in a $3+1$-dimensional formalism (or a $6\times6$-formalism, see section~\ref{sec:6dim}), this certainly is done in the fastest way by resorting to a fully covariant calculation. The connection to the other formalisms is then best made at the very first and very last steps.
 
 	Lastly, let us stress a caveat regarding the naming of the components arising from the orthogonal decomposition of $Z$. The most frequent encounter of an orthogonal decomposition of a tensor $Z^{abcd}$ with the aforementioned symmetries is in regard to the Riemann curvature tensor. The result is that, ironically, the names given to the three independent matrices $W_\epsilon$, $W_\zeta$, and $W_\mu$ encountered in this decomposition \emph{in the GR community} are very misleading in the context of macroscopic electrodynamics: In GR, the decomposition --- under the name of Bel decomposition --- of the Riemann tensor is used to find dynamical analogies between the Einstein equations on the one hand, and the Maxwell equations on the other hand. In our case, the role of the Bel decomposition is purely kinematical and entirely in the realm of electromagnetism itself. For example, what goes under the name of \enquote{electric tensors} in \cite{ChoBru} corresponds to both the permittivity and the permeability tensors, while the \enquote{magnetic tensors} here are the magneto-electric tensor and its transpose. Beware.
 	
 	\subsection{Changing Reference Frames}\label{sec:frames}
 	With the orthogonal decomposition in place, it is possible to describe the method to give the permittivity $\epsilon^{ab}_W$, permeability $\mi^{ab}_W$ and magneto-electric tensor $\zeta^{ab}_W$ as seen by an observer of four-velocity $W$ in terms of the permittivity $\epsilon^{ab}_V$, permeability $\mi^{ab}_V$ and magneto-electric tensor $\zeta^{ab}_V$ as seen by another observer of four-velocity $V$. While this question seems mostly physical in nature, we think it natural to insert it in this appendix for two reasons: Firstly, it \emph{can} be rephrased in geometric terms as expressing the orthogonal decomposition of a tensor by one foliation in terms of the orthogonal decomposition of that same tensor by another foliation.\footnote{Having said that, we will strongly rely on physical language in this subsection.} Secondly, the resulting expressions will make heavy use of the notation used in the previous paragraphs --- keeping the discussions spatially close will thus help the reader.
 	
 	Choose any four-velocity $V^a$ and an arbitrary, not necessarily symmetric matrix $q^{ab}$ four-orthogonal to it:
 	\begin{equation}
 		q^{ab}V_b = V_b \, q^{ba} = 0.
 	\end{equation}
 	Let us then define the following fourth-rank tensor
 	\begin{equation}
 		Q^{abcd} \defi V^a V^d  q^{bc} + V^b V^c  q^{ad} - V^b V^d  q^{ac} - V^a V^c  q^{bd}.
 	\end{equation}
 	Furthermore, let us use this tensor $ Q^{abcd}$ to define four more tensors by setting $q$ equal to one of the four \enquote{constitutive matrices} $\epsilon_V^{ab}$, $[\mu^{-1}_V]^{ab}$, $\zeta_V^{ab}$, and its transpose $[\zeta^T_V]^{ab}$ as measured with respect to four-velocity $V^a$:
 	
 	\begin{center}
 		\begin{subequations}
 			\begin{minipage}{.45\textwidth}
 				\begin{align}
 				E^{abcd}_V &\defi  Q^{abcd}_{ q\rightarrow\epsilon_V},\hphantom{bladiebla}\\
 				\addtocounter{equation}{+1}A^{abcd}_V &\defi  Q^{abcd}_{ q\rightarrow\zeta_V},
 				\end{align}
 			\end{minipage}\hfill
 			\begin{minipage}{.45\textwidth}
 				\begin{align}
 				\addtocounter{equation}{-2}M^{abcd}_V &\defi  Q^{abcd}_{ q\rightarrow\mu^{-1}_V},\hphantom{bladie}\\
 				\addtocounter{equation}{1}(A^T_V)^{abcd} &\defi  Q^{abcd}_{ q\rightarrow\zeta^{T}_V}.
 				\end{align}
 			\end{minipage}
 		\end{subequations}
 	\end{center}
 	
 	If we now compare this with the Bel-decomposed expression for the constitutive tensor $Z^{abcd}$ in equation~\eqref{eq:ZBel2}, we see that we can rewrite equation~\eqref{eq:ZBel2} in terms of these four tensors in the following way:
 	\begin{equation}
 		Z^{abcd} = \ed{2}\kl{E_V + (\ast M_V\ast) + (\ast A_V) + (A^T_V \ast)}^{abcd}.
 	\end{equation}
 	While the right-hand side is implicitly dependent on the previously chosen four-velocity $V^a$, the left-hand side is general and independent of it. Note that this is precisely the form of equation~\eqref{eq:ZBel2}, as was to be expected when comparing the previous slightly more abstract discussion with that preceding equation~\eqref{eq:ZBel2}. This, then, enables us to give deceptively simple expressions for how to calculate the \enquote{constitutive matrices} $\epsilon^{ab}_W$, $[\mu^{-1}_W]^{ab}$, and $\zeta_W^{ab}$ as seen by a different observer with four-velocity $W^a$ 
 	
 	\begin{flushright}
 		\begin{subequations}\label{eq:consmatW}
 			\begin{minipage}{.45\textwidth}
 				\begin{align}
	 				\epsilon_W^{ab} &= -2 Z^{dacb} W_d W_c,\hphantom{blad}\label{eq:epsW}\\
	 				\addtocounter{equation}{+1}\zeta_W^{ab} &= \hphantom{-}2 (\ast Z)^{dacb} W_d W_c,\label{eq:zetaW}
 				\end{align}
 			\end{minipage}\hfill
 			\begin{minipage}{.45\textwidth}
 				\begin{align}
	 				\addtocounter{equation}{-2}[\mu^{-1}_W]^{ab} &= 2(\ast Z \ast)^{dacb} W_d W_c,~\label{eq:muW}\\
	 				\addtocounter{equation}{1}  \zt^{ab} &= 2(Z \ast)^{dacb} W_d W_c.
 				\end{align}
 			\end{minipage}
 		\end{subequations}
 	\end{flushright}
 	
 	At this point it is now possible to give $\epsilon^{ab}_W$, $[\mu^{-1}_W]^{ab}$, and $\zeta_W^{ab}$ in full generality in terms of $\epsilon_V^{ab}$, $[\mu^{-1}_V]^{ab}$, $\zeta_V^{ab}$, $V^a$, and $W^a$. The resulting expressions are stigmatized by being unilluminatingly and excessively complicated. Nevertheless, in special cases this will be much less of a problem. Also, the existence of closed-form expressions will prove useful when working numerically in this formalism. We shall therefore brave the indices and give these expressions now:
 	\begin{subequations}
 		\begin{alignat}{1}
	 		\epsilon^{bd}_W \!= &\frac{W_a W_c}{4} \eps^{ab}{}_{ef} \eps^{cd}{}_{gh}\!\kl{V^e [\mu^{-1}_V]^{fg} V^h + V^f [\mu^{-1}_V]^{eh} V^g-V^f [\mu^{-1}_V]^{eg} V^h - V^e [\mu^{-1}_V]^{fh} V^g }\nonumber\\
	 		&-\kl{(V\cdot W) \kl{V^d \epsilon_V^{bc} + V^b \epsilon_V^{dc}}W_c -V^b V^d W_a \epsilon_V^{ac} W_c - (V\cdot W)^2 \epsilon_V^{bd} }\nonumber\\
	 		&- \frac{W_a}{2} \eps^{ab}{}_{ef} \kl{(V\cdot W)\kl{\zeta_V^{ed} V^f - \zeta_V^{fd} V^e} + \kl{\zeta_V^{fc} V^e - \zeta_V^{ec}V^f} W_c V^d}\nonumber\\
	 		&- \frac{W_c}{2} \eps^{cd}{}_{gh} \kl{(V\cdot W)\kl{[\zeta^{\dagger}_V]^{bg} V^h - [\zeta^{\dagger}_V]^{bh} V^g}  + W_a \kl{[\zeta^{\dagger}_V]^{ah} V^g   - [\zeta^{\dagger}_V]^{ag}V^h} V^b},\label{eq:epsilonVW}\\
	 		\![\mu^{-1}_W]^{ij} \!=& \ed{2} \kl{\!W_k [\mu^{-1}_V]^{kl} W_l V^i V^j + (V\cdot W)^2 [\mu^{-1}_V]^{ij} - (V\cdot W)\kl{[\mu^{-1}_V]^{il}V^j + [\mu^{-1}_V]^{jl}V^i}W_l\!}\nonumber\\
	 		& + \ed{4} \eps^{ki}{}_{ab} \eps^{lj}{}_{cd} W_k W_l \kl{\epsilon_V^{bc} V^a V^d + \epsilon^{ad}_V V^b V^c - \epsilon^{ac}_V V^b V^d - \epsilon^{bd}_V V^a V^c}\nonumber\\
	 		& + \ed{2} \eps^{lj}{}_{cd} W_l \kl{(V\cdot W)\kl{\zeta_V^{ic} V^d - \zeta_V^{id} V^c} + W_k\kl{\zeta_V^{kd} V^c - \zeta_V^{kc} V^d}V^i}\nonumber\\
	 		& + \ed{2} \eps^{ki}{}_{ab} W_k \kl{(V\cdot W)\kl{[\zeta^{\dagger}_V]^{aj} V^b - [\zeta^{\dagger}_V]^{bj} V^a} + \kl{[\zeta^{\dagger}_V]^{bl} - [\zeta^{\dagger}_V]^{al}}W_l V^j},\label{eq:muVW}\\
	 		\!\zeta_W^{ld} \!= & \ed{2} \eps^{cd}{}_{gh} W_c \!\kl{\!W_k\kl{[\mu^{-1}_V]^{kg} V^h - [\mu^{-1}_V]^{kh} V^g}V^l + (V\cdot W)\kl{[\mu^{-1}_V]^{lh} V^g - [\mu^{-1}_V]^{lg} V^h}\!}\nonumber\\
	 		& + \ed{2} \eps^{kl}{}_{ab} W_k \kl{\kl{\epsilon_V^{bs}V^a - \epsilon_V^{ac}V^b}V^d W_c + (V\cdot W)\kl{\epsilon_V^{ad}V^b - \epsilon_V^{bd}V^a}} \nonumber\\
	 		& - \kl{W_k \zeta_V^{kc} W_c V^lV^d - (V\cdot W)\kl{\zeta^{cd}_V V^l + \zeta_V^{lc}V^d}W_c + (V\cdot W)^2 \zeta_V^{ld}}\nonumber\\
	 		& + \ed{4} \eps^{kl}{}_{ab} \eps^{cd}{}_{gh} W_kW_c \kl{[\zeta^{\dagger}_V]^{ah}V^b V^g + [\zeta^{\dagger}_V]^{bg}V^a V^h - [\zeta^{\dagger}_V]^{ag}V^bV^h - [\zeta^{\dagger}_V]^{bh}V^aV^g},\\
	 		\![\zeta^{\dagger}_W]^{bi} \!= & \ed{2} \eps^{ab}{}_{ef} W_a \! \kl{\!\kl{[\mu^{-1}_V]^{ej} V^f - [\mu^{-1}_V]^{fj} V^e}W_j V^i + (V\cdot W) \kl{[\mu^{-1}_V]^{fi}V^e - [\mu^{-1}_V]^{ei} V^f}\!}\nonumber\\
	 		& +\ed{2} \eps^{ji}{}_{cd}W_j \kl{W_a \kl{\epsilon_V^{ad} V^c - \epsilon_V^{ac} V^d}V^b + (V\cdot W) \kl{\epsilon_V^{bc}V^d - \epsilon_V^{bd}V^c}}\nonumber\\
	 		& +\ed{4} \eps^{ab}{}_{ef} \eps^{ji}{}_{cd} W_j W_a \kl{\zeta_V^{fc}V^d V^e + \zeta_V^{ed}V^cV^f-\zeta_V^{ec}V^dV^f - \zeta_V^{fd}V^c V^e}\nonumber\\
	 		& -\kl{W_a [\zeta^{\dagger}_V]^{aj} W_j V^b V^i - (V\cdot W) \kl{[\zeta^{\dagger}_V]^{bj} V^i + [\zeta^{\dagger}_V]^{ji} V^b}W_j + (V\cdot W)^2 [\zeta^{\dagger}_V]^{bi}}.
 		\end{alignat}
 	\end{subequations}
 	In section~\ref{sec:iso} we give an explicit example on how to use this. Specifically, we look at an isotropic medium in motion and regain the well known magneto-electric effect of moving media \cite{Post,ODell,TrafoOptics2}, of which the Fresnel--Fizeau effect is a special case \cite{Jackson,LanLifVIII}.
 	
\chapter{A Smorgasbord of Special Functions}\label{ch:Special}
	\epigraph{\enquote{She found her way into all the deserted rooms, where no one ever set foot, and she did not lose her way in underground passages, dark pits, and cellar vaults. The secret passages of the fortress and the secret paths of the forest---she knew them all now.}}{Astrid~Lindgren,~translation~Patricia~Crompton, \emph{Ronia, the robber's daughter}} 
 	
 	This appendix will summarise some properties of a variety of special functions used and encountered in the course of this thesis. We aim for brevity, and therefore only include properties of immediate relevance. For properties of the functions going beyond those employed in the main text we refer to the literature cited here. The only possible exception to this will be the description of the Heun functions, as we imagine these to be the special functions (of those listed here, at least) least seen in the wild, yet also that discussion will be comparatively brief and not exhaustive.
 	
 	\section{Solutions of Linear Second Order Ordinary Differential Equations}\label{sec:2ndODE}
 	Many problems of physics can be related to questions of linear second order ODEs and their solutions. In general, these can be written as
 	\begin{equation}
	 	y''(z) + p_1(z) y'(z) + p_2(z) y = 0.
 	\end{equation}
 	For classifying these differential equations, it is useful to distinguish between so-called \enquote{ordinary}, \enquote{regular singular}, and \enquote{irregular singular} points. \enquote{Regular points} are those $z\in\mathbb{C}$ that have a neighbourhood within which both $p_1$ and $p_2$ are analytic. A point $c$ is singular if either $p_1$ or $p_2$ have a singularity at $c$. A point is a \enquote{regular singular point} if there are two functions $P_1$ and $P_2$ analytic at $c$, such that the singular behaviour at $c$ can be captured in the form
 	\begin{subequations}
 		\begin{align}
	 		p_2(z) &= \frac{P_2(z)}{(z-c)^2},\\
	 		p_1(z) &= \frac{P_1(z)}{z-c}.
 		\end{align}
 	\end{subequations}
 	If a singular point is not regular, it is an \enquote{irregular singular point}.\footnote{These notions \emph{can} be generalised to higher order ODEs --- we shall not do this, as this appendix is only concerned with functions arising from second order ODEs. For the more general approach, see \cite{SlavyanovLay00}.} If a linear, second order, ordinary differential equation has at most two singular points, any solution is not too complicated, see \cite{KristenssonODE}. (There, those functions are called \enquote{elementary}.) Much of the realm of special function theory is concerned with those equations having more than two singular points. Any Möbius transformation in the plane, that is, a transformation
 	\begin{equation}
	 	\tilde{z} = \frac{az+b}{cz + d},
 	\end{equation}
 	where $ad-bc\neq0$, leaves the number of the singular points intact. This is helpful in classifying linear ODEs. If all points on the Riemann sphere (that is, the one-point compactification of $\mathbb{C}$) are either regular points of the ODE or regular singularities, the equation is called \enquote{of Fuchsian type}. If two singular points are merged, this process is called \enquote{confluence}. For the behaviour of the resulting singular points, see \cite{SlavyanovLay00}.
 	
 	\subsection{Modified Bessel Functions}
 	The modified Bessel functions are the solutions to the second order differential equation
 	\begin{equation}
	 	z^2 y''(z) + z y'(z) -(z^2+\alpha^2) y(z) = 0.
 	\end{equation}
 	According to the classification scheme, this is an example of a second order ODE with two singular points, one of which is irregular. The equation is related to the confluent hypergeometric ODE \cite{KristenssonODE}, and by a particularly simple Möbius transformation (multiplication by $i$) with the \enquote{normal} Bessel equation (which we will not be using).
 	
 	As a second order ODE, this differential equation will have two linearly independent solutions, denoted by $I_\alpha(z)$ and $K_\alpha(z)$. The first, $I_\alpha$ is called the modified Bessel function of the first kind, the second, $K_\alpha$ is called the modified Bessel function of the second kind. Note that some identities involving either $I_\alpha$ or $K_\alpha$ need from a technical point of view a limit $\lim_{\alpha'\to \alpha}$ if $\alpha \in \mathbb{Z}$ \cite{BesselK}. 
 	
 	When evaluating superradiance, we will make use of modified Bessel functions of the first kind. For this we make use of the asymptotic expansion for fixed index $\alpha$ as $z\to\infty$ (as our argument is strictly real and positive, we do not need to worry about the necessary condition on the phase of $z$, it will be fulfilled in our application):
 	\begin{subequations}
 		\begin{align}
	 		I_\alpha(z) &\sim \frac{e^z}{\sqrt{2\pi z}}\sum_{k=0}^{\infty} (-1)^k \frac{a_k(\alpha)}{z^k}.\\
	 		&\sim \frac{e^z}{\sqrt{2\pi z}}.\label{eq:asymptoticI}
 		\end{align}
 	\end{subequations}
 	The coefficient $a_k(\alpha)$ is contained in the so-called Hankel expansions, see \cite{Olver2010a}. Their precise value is of little relevance to us, but for completeness' sake reads:
 	\begin{subequations}
 		\begin{align}
 		a_0(\alpha) &= 1,\\
 		a_k(\alpha) &= \frac{\kl{4\alpha^2-1^2}\kl{4\alpha^2-3^2}\cdots\kl{4\alpha^2-(2k-1)^2}}{k! 8^k}.
 		\end{align}
 	\end{subequations}
 	They fulfil furthermore the integral identity
 	\begin{equation}
 	I_\alpha(z) = \frac{\kl{z/2}^\alpha}{\sqrt{\pi}\Gamma(\alpha+1/2)} \int_{0}^{\pi} e^{\pm z \cos(t)} \sin^{2\alpha}(t) \dif t.\label{eq:Icool}
 	\end{equation}
 	
 	In our attempt to find analytic expressions for sparsity including particle rest masses for the emitted particles, see section~\ref{sec:masses}, we encounter identities related to the modified Bessel functions of the second kind. These are also known as: Modified Bessel functions of the third kind, Basset functions, modified Hankel functions, and Macdonald functions. For sources see to the standard references \cite{GradshteynRyzhik1980,Olver2010a}, and we refer to \cite{BesselK} for corrections to \cite{GradshteynRyzhik1980}. A digital repository of \cite{Olver2010a} can be found at \url{https://dlmf.nist.gov}, the \enquote{Digital Library of Mathematical Functions}. Specifically, we only make use of the following three identities:
 	\begin{align}
	 	\frac{K_\alpha(z)}{z} &= \ed{2\alpha} \kl{K_{\alpha+1}(z) - K_{\alpha-1}(z)},\label{eq:Krecursion}\\
	 	K_\alpha(z) &= \int_{0}^{\infty} \exp\kl{-z\cosh(x)} \cosh(\alpha x) \dif x,\label{eq:Kint}\\
	 	\ed{\sqrt{\pi}} \kl{\frac{2}{z}}^\alpha \Gamma(\alpha+1/2) K_\alpha(z) &= \int_{0}^{\infty} \exp\kl{-z\cosh(x)} \sinh^{2\alpha}(x) \dif x.\label{eq:Kool}
 	\end{align}
 	The first is a recursion relation, the second an integral representation, and the third a integral identity. Many more are possible, for example, recursion relations involving a derivative on the LHS. Again, we refer to the literature for this. The properties given here are the ones we use. Furthermore, under (for us) similar conditions as for $I_\alpha(z)$, we also have the asymptotic behaviour for fixed $\alpha$ and $z\to\infty$,
 	\begin{equation}
	 	K_\alpha(z) \sim \sqrt{\frac{\pi}{2z}}e^{-z}.\label{eq:asymptoticK}
 	\end{equation}

 	\subsection{Heun Functions}\label{sec:Heun}
 	In the above-described classification scheme for second order differential equations \cite{KristenssonODE}, the Heun differential equation \cite{HeunOrig} is the differential equation with four regular singular points, located at $0$, $1$, $c$, and $\infty$. Its solution is the equally-named Heun function. Variants are related to different confluence processes, more on this below. As the archetype and starting point for the discussion of these further, related differential equations (which are the ones that arise in the context of black hole physics) is the (general) Heun equation, we shall start with it. In its reduced form \cite{KovacicAlgo} it reads
 	\begin{align}\label{eq:HeunReduced}
 	y''(z) + &\left[\frac{\frac{\alpha\beta}{2}+\frac{\alpha\gamma}{2c}-\frac{\delta\eta h}{c}}{z} + \frac{\frac{\alpha}{2}\kl{1-\frac{\alpha}{2}}}{z^2} + \frac{\frac{\beta\gamma}{2(c-1)} - \frac{\alpha\beta}{2} +\frac{\delta\eta(h-1)}{c-1}}{z-1} \right.\nonumber\\
 	&\left.  + \frac{\frac{\beta}{2}\kl{1-\frac{\beta}{2}}}{(z-1)^2} +\frac{\frac{\delta\eta(c-h)}{c(c-1)} - \frac{\alpha\gamma}{2c} - \frac{\beta\gamma}{2(c-1)}}{z-c} +  \frac{\frac{\gamma}{2}\kl{1-\frac{\gamma}{2}}}{(z-c)^2}\right]y(z) = 0,
 	\end{align}
 	where
 	\begin{equation}\label{eq:HeunParamConstraint}
	 	\alpha + \beta + \gamma - \delta - \eta = 1.
 	\end{equation}
 	The parameter $h$ is called the accessory or auxiliary parameter, $c$ is the singularity parameter, and the remaining parameters are the exponential parameters \cite{RonveauxHeun}. Apart from the constraint~\eqref{eq:HeunParamConstraint}, the exponential parameters can take on any complex value. If \eqref{eq:HeunParamConstraint} does not hold, the singular point at $\infty$ will not be a regular point. Furthermore, the accessory parameter is an arbitrary complex number $h\in \mathbb{C}$, while $c\in \mathbb{C}\setminus\{0,1\}$. Due to $h$'s arbitrariness, it can be redefined to include the pre-factor going along with it in equation~\eqref{eq:HeunReduced} --- as is done in several of the references given here. 
 	
 	Another, for the discussion in section~\ref{sec:refractive} important version of Heun's differential equation is (one of) the canonical\footnote{Both the version given after this footnote and one where the equation has been brought into a form involving only polynomial coefficients by appropriate multiplication are known as \enquote{canonical}.} (Fuchsian) form(s):
 	\begin{equation}\label{eq:HeunCanonical}
	 	u''(z) + \kle{\frac{\alpha}{z} + \frac{\beta}{z-1} + \frac{\gamma}{z-c}}u'(z) + \frac{\delta\eta (z-h)}{z(z-1)(z-c)} u(z) = 0.
 	\end{equation}
 	Other standard forms of the Heun equation can be found in \cite{SlavyanovLay00}.
 	
 	While for the hypergeometric differential equation there are two different confluence processes \cite{KristenssonODE}, for the Heun equation there are a total of four. The following table gives the confluence processes, the name of the corresponding differential equation and the coefficient of the zeroth order term $r(z)$ of the second order ODE
 	\begin{equation}
 		y''(z) + r(z) y(z) = 0,
 	\end{equation}
 	which is the form first presented here for the Heun equation itself, equation~\eqref{eq:HeunReduced}, as well as the form employed in the services of section~\ref{sec:refractive}. Again, we use the notation of \cite{KovacicAlgo}, mildly adapted.
 	\begin{align}
		&\emph{Confluent:} & c&\to\infty & -\frac{\alpha^2}{4} + \frac{1-2\eta}{2z} - \frac{1-2(\delta+\eta)}{2(z-1)} + \frac{1-\beta^2}{4z^2} + \frac{1-\gamma^2}{4(z-1)^2}.\\
		&\emph{Biconfluent:} & \left.\begin{array}{l}c\\1 \end{array}\right\} &\to \infty & \gamma - z^2-\beta z - \frac{\beta^2}{4} - \frac{\delta}{2z}+\frac{1-\alpha^2}{4z^2}.\\
		&\emph{Double Confluent:} & \begin{array}{l}
		c\\1
		\end{array}&\!\begin{array}{l}
		\to\infty\\\to 0
		\end{array} & \frac{\gamma}{z} + \frac{\delta}{z^2} + \frac{\beta}{z^3} - \frac{\alpha^2}{4z^4} - \frac{\alpha^2}{4}.\\
		&\emph{Triconfluent:} & \left.\begin{array}{l}
		c\\1\\0
	\end{array}\right\}&\to\infty & \alpha - \frac{\gamma^2}{4} +\beta z - \frac{3}{2}\gamma z^2 - \frac{9}{4}z^4.
	\end{align}
	Both the radial and the azimuthal part of the Teukolsky equation can be rewritten as (single) confluent Heun differential equation, the difference being that in the limit of extremal rotation the radial one will become a double confluent Heun equation while the azimuthal one does not change type \cite{SlavyanovLay00}. Both are variations of the oblate spheroidal equation \cite{MeixnerSchaefke}.
	
	For our applications, the particular behaviour of the solutions of the equations is of less importance. As we are only looking at the form of differential equations to read off refractive indices in section~\ref{sec:refractive}, this will suffice. For information on solutions, we refer to \cite{MeixnerSchaefke,ConnectionProblemLinearODE,RonveauxHeun,FroNo98,ConnProbDoubleHeun,SlavyanovLay00,Maier192,KerrMixingSphSperoi}. The two references \cite{ConnectionProblemLinearODE,ConnProbDoubleHeun} are especially relevant to analyses of the two-point connection problem of the solutions: Given expansions of the solutions around two different singular points, how can these two descriptions of the solutions be linked? While not discussed in this thesis, this issue is intimately related to the question of how to relate the behaviour of a wave solution (\emph{i.e.}, a solution of the Teukolsky equation) close to the horizon to that seen at spatial infinity. For physicists this situation is relevant as it occurs in scattering processes. It is noteworthy that the general two-point connection problem for the Heun equation is still unsolved, unlike the case for the general hypergeometric equation. References \cite{QNMLRR,EVandEFspheroidal,KerrMixingSphSperoi} are related to Heun functions through quasi-normal modes of black holes (and the corresponding ring-down processes of perturbations of black hole space-times), as they again are directly related to the Teukolsky equation(s).
	
	While other aspects of the Heun equation are considerably simpler, its solutions are still involved enough that only fairly recent applications of computer algebra in \cite{Maier192} found and corrected errors going back to Heun's original analysis \cite{HeunOrig}. If explicit solutions and their notation (especially those related to Kummer's 24 solutions of the Gauss hypergeometric equation or the analogue of Riemann P-symbols for Heun's equation) are used, it is advisable to use \cite{Maier192} as a litmus test.

 	Many other second order ordinary differential equations and their corresponding solutions appear as special cases of the Heun equation and its counterparts after a confluence process. A partial list of these is: The hypergeometric equation \cite{KristenssonODE} (and thus all equations related to the hypergeometric equation), the Heine equation, the Wangerin equation, the Lamé equation, the spheroidal and spin-weighted spheroidal equations, the Whittaker--Ince equation, and the Mathieu equation \cite{RonveauxHeun}.

 	\section{Complete Elliptic Integrals of the Second Kind}\label{sec:elliptic}
 	While they only have a very brief appearance in section~\ref{sec:unphysical}, it is worthwhile to also remind ourselves of elliptic integrals of Legendre form, specifically those of second kind.\footnote{Strictly speaking, they could equally well have been introduced in section~\ref{sec:2ndODE}, as they fulfil a special case of the hypergeometric differential equation --- in our case $(1-k^2)\kl{k \mathrm{EllipticE}'(k)}' = k\mathrm{EllipticE}(k)$ --- but their rather distinct occurrence in the main part somewhat vindicates a separate mentioning.} The incomplete elliptic integral of second kind is defined as
 	\begin{equation}
 		\mathrm{EllipticE}(\phi;k) \defi \int_{0}^{\phi} \sqrt{1- k^2\sin^2\theta}\dif \theta. 
 	\end{equation}
 	The complete elliptic integral of second kind is gained from this by letting the upper bound equal $\nicefrac{\pi}{2}$:
 	\begin{equation}
	 	\mathrm{EllipticE}(k) \defi \int_{0}^{\frac{\pi}{2}} \sqrt{1- k^2\sin^2\theta}\dif \theta. 
 	\end{equation}
 	This integral can also be related to the circumference $C$ of an ellipse of semi-major axis $a$ and semi-minor axis $b$ as
 	\begin{equation}
 		C = 4a\mathrm{EllipticE}(\sqrt{1-b^2/a^2}),\label{eq:ellipticellipse}
 	\end{equation}
 	where the argument of the complete elliptic integral of second kind is exactly the eccentricity of the ellipse.
 	
 	It would certainly be possible to also list the elliptic integrals of first and third kind, both complete and incomplete --- as they, however, do not appear in the main text, unlike those of second kind, we shall refrain from it. For the main text it proves useful to be aware of the following identity \cite{Olver2010a}:
 	\begin{equation}
 		\mathrm{EllipticE}(m\pi\pm\phi;k) = 2m\mathrm{EllipticE}(k)\pm \mathrm{EllipticE}(\phi;k),\label{eq:ellipticaddition}
 	\end{equation}	
 	where $m$ is an integer.
 	
 	\section{The Lambert W-Function}\label{sec:Lambert} 
 	The Lambert W-function $W(x)$ is defined as the inverse function of
 	\begin{equation}
	 	f(x) = x\, e^x,
 	\end{equation}
 	thus
 	\begin{equation}
 		x = W(x)\, e^{W(x)}.
 	\end{equation}
 	It has countably many branches, usually denoted by $W_k(x)$, with $k\in \mathbb{N}$. The principal branch we shall denote with $W(x)$, which fulfils $W(x)\geq -1$. The second branch, $W_1(x)$, is $\leq -1$. The two branches meet at $x=-\nicefrac{1}{e}$. $W(x)$ has domain $[-\nicefrac{1}{e},\infty)$, while $W_1(x)$ has domain $[-\nicefrac{1}{e},0)$. This actually has some relevance when looking at bosons: For negative enough chemical potentials the argument of the Lambert-W function can in these circumstances encounter the branch point --- a sure sign of trouble for the model.

 	\section{Riemann Zeta Function, Polylogarithms and Related Functions}
 	Here we shall give a quick introduction to the definition of this interrelated set of special functions that occurs naturally and frequently in the calculation of sparsities, as seen in chapter~\ref{ch:sparsity}. The first three functions are basically just special cases of the fourth one, each closely linked to the thermal spectrum of particles of different spin (bosons, fermions, and classical --- \emph{i.e.}, \enquote{spinless} --- particles). Unifying these special cases with the notion of polylogarithms makes it fairly easy to capture all kinds of particles \emph{and} including possible contributions from chemical potentials of any kind at the same time.
 	
 	\subsection{The Gamma Function}
 	The gamma function is defined as
 	\begin{equation}
	 	\Gamma(z) = \int_{0}^{\infty} x^{z-1} e^{-x} \dif x.
 	\end{equation}
 	It has the nice property of extending the factorial to all complex numbers except the non-positive integers. The integral definition above is convergent for positive real parts of $z$. The Bohr--Mollerup theorem \cite{Koenigsberger1} can be used to characterise the gamma function as the unique function $\Gamma$ that fulfils the following three properties:
	\begin{itemize}
		\item $\Gamma(1) = 1$,
		\item For all $x>0$, $\Gamma(x+1) = x\Gamma(x)$.
		\item $\Gamma$ is logarithmically convex, \emph{i.e.}, $\log(\Gamma(x))$ is convex.
	\end{itemize}
	In particular, we have that
	\begin{equation}
		\Gamma(n+1) = n!
	\end{equation}
	for natural numbers $n$. If $m$ is a negative integer, the gamma function has a pole of order $m$ at $\Gamma(m)$. At $\Gamma(0)$ it is logarithmically divergent. Lastly, let us give a value that will be encountered frequently and is of great use:
	\begin{equation}
		\Gamma\kl{\ed{2}} = \sqrt{\pi}.
	\end{equation}
 	
 	\subsection{The Riemann Zeta Function}\label{sec:Riemann}
 	The Riemann zeta function $\zeta(z)$ is the analytic continuation of the Dirichlet series
 	\begin{equation}
	 	\zeta(z) = \sum_{n=1}^{\infty} \ed{n^z}
 	\end{equation} 
 	to the complex plane. For the series to be convergent, the real part of $z$ has to be larger than 1, which is the reason for including the analytic continuation in the definition. Equivalently, the Riemann zeta function can be written down and defined as the integral
 	\begin{equation}\label{eq:zetaint}
	 	\zeta(z) = \ed{\Gamma(z)}\int_{0}^{\infty} \frac{x^{z-1}}{e^x-1}\dif x,
 	\end{equation}
 	which is the form that is encountered in chapter~\ref{ch:sparsity}. There it appears when integrating the differential number/energy fluxes of bosonic particles. More a curio than of immediate use to this thesis\footnote{Unless one starts digging in the cited quantum field theory literature where at least the Euler--Mascheroni constant $\gamma$ appears.} is the fact that
 	\begin{equation}
	 	\ln\Gamma(1+z) = -\gamma z + \sum_{k=2}^{\infty}\frac{\zeta(k)}{k}(-z)^k,
 	\end{equation}
 	with convergence radius $1$, and where $\gamma$ is the Euler--Mascheroni constant.

 	\subsection{The Dirichlet Eta Function}\label{sec:Dirichlet}
 	The Dirichlet eta function is analogously defined to the Riemann zeta function. In this case, for positive real part of the argument $z$, it is defined as the analytic continuation to the complex plane of the series
 	\begin{equation}
	 	\eta(z) = \sum_{n=1}^{\infty} \frac{(-1)^{n-1}}{n^z}.
 	\end{equation}
 	By simple comparison with the case of the Riemann zeta function it can already be guessed (as it can be proved!) that an equivalent definition through the following integral is available:
 	\begin{equation}
	 	\eta(z) = \ed{\Gamma(z)}\int_{0}^{\infty} \frac{x^{z-1}}{e^x + 1}\dif x.
 	\end{equation}
 	
 	\subsection{Polylogarithms}\label{sec:polylogs}
 	The polylogarithms $\Li{r}(z)$ are defined as the analytic continuation of the series
 	\begin{equation}\label{eq:DefLi}
 		\Li{r}(z) \defi \sum_{n=1}^{\infty} \frac{z^n}{n^r}.
 	\end{equation}
 	It is also known as Jonquière's function and sometimes denoted $\phi(z,r)$ \cite{Olver2010a}. For us of interest is their relation to the Bose--Einstein and Fermi--Dirac distributions through a corresponding integral representation,
 	\begin{equation}
	 	\mp \Li{r+1}(\mp e^{y}) = \ed{\Gamma(r+1)} \int_{0}^{\infty} \frac{x^{r}}{e^{x-y} \pm 1}\dif x.
 	\end{equation}
 	Furthermore, while $\Li{r}(0)$ vanishes (as a quick look at the definition~\eqref{eq:DefLi} shows), the limit
 	\begin{equation}\label{eq:LiLimit}
 		\lim\limits_{z\to 0} \frac{\Li{r}(z e^{y})}{z} = e^{y}
 	\end{equation}
 	exists.
 	
 	As we are, in the context of our sparsity calculations, also interested in integrals of the kind
 	\begin{equation}
 		\int_0^\infty \frac{x^k}{e^{x-y} + 0}\dif x,
 	\end{equation}
 	the previous limit is helpful to realise that these integrals (yielding the gamma function) can indeed be collected in one expression using the polylogarithm. Thus, the concrete correspondence takes the form
 	\begin{equation}
 		\frac{\Li{k+1}(-s e^y)}{-s} = \ed{\Gamma(k+1)}\int_0^\infty \frac{x^k}{e^{x-y} + s}\dif x.
 	\end{equation}
 	In the language of, for example, \cite{Olver2010a} the cases $s=1$ and $s=-1$ are called \enquote{complete Fermi--Dirac} integral and \enquote{complete Bose--Einstein} integral, respectively. The naming scheme follows obviously the (to physicists) equally obvious link of these integrals to the above-mentioned distributions of energy levels of a non-interacting gas of fermions or bosons in thermal equilibrium with an environment (with which energy and particle exchange is allowed). For the case $s=0$ one could therefore introduce the parlance \enquote{complete Boltzmann integral} --- though given the trivial nature of that particular integral, a similar name for it seems hardly justified, given that it already is well-known as the Gamma function. Comparison with subsections~\ref{sec:Dirichlet} and \ref{sec:Riemann} yields then the connection between polylogarithms and the Dirichlet eta function or Riemann zeta function, respectively.

\chapter{Abstracts of Published Articles}
All publications are listed in chronological order.
\section{Articles}
\begin{center}
	{\sffamily \bfseries \Large The Hawking cascade from a black hole is extremely sparse}\\
	\vspace{.3\baselineskip}
	{\sffamily\large Finnian Gray, Sebastian Schuster, Alexander Van-Brunt, Matt Visser}\\
	\vspace{.3\baselineskip}
	{\sffamily\large Classical and Quantum Gravity \textbf{33}(11), 115003 (2016)}\\\vspace{.3\baselineskip}
	\begin{tabular}{llc}
		DOI: &  \href{https://doi.org/10.1088/0264-9381/33/11/115003}{10.1088/0264-9381/33/11/115003}&\cite{HawkFlux1} \\
		arXiv: & \href{http://arxiv.org/abs/1506.03975}{1506.03975} &[gr-qc]
	\end{tabular}
\end{center}
\textbf{Abstract:} The average time between emission of subsequent quanta in the Hawking process is extremely large. While this sparsity result has been known for a long time, it is neither well-known, nor do (semi-)analytic results currently exist, the prior focus being placed on numerical results. We define several ways of quantifying this sparsity, and starting from a black body approximation for the Schwarzschild case, we derive analytic expressions for a lower bound on this average time. We also check the validity of the results in presence of greybody factors by numerical analysis. Furthermore, we show how to separate the super-radiance in the low-frequency regime from the genuine Hawking effect itself. This enables us to extend the previous lower bounds to Reissner--Nordström, Kerr and dirty black holes in addition to different particle species. Lastly, we want to draw attention to some of the physical consequences of this under-appreciated fact of the Hawking process.

\begin{center}
	{\sffamily \bfseries \Large \enquote{Twisted} black holes are unphysical}\\
	\vspace{.3\baselineskip}
	{\sffamily\large Finnian Gray, Jessica Santiago, Sebastian Schuster, Matt Visser}\\
	\vspace{.3\baselineskip}
	{\sffamily\large Modern Physics Letters \textbf{A32}(18), 1771001 (2017)}\\\vspace{.3\baselineskip}
	\begin{tabular}{llc}
		DOI: &  \href{https://doi.org/10.1142/S0217732317710018}{10.1142/S0217732317710018}&\cite{TwistedUnphysical} \\
		arXiv: & \href{http://arxiv.org/abs/1610.06135}{1610.06135} &[gr-qc]
	\end{tabular}
\end{center}
\textbf{Abstract:} So-called \enquote{twisted} black holes were recently proposed by [H. Zhang, \href{http://arxiv.org/abs/1609.09721}{arXiv:\allowbreak 1609.09721}], and were further considered by [S. Chen and J. Jing, \href{http://arxiv.org/abs/1610.00886}{arXiv:1610.00886}]. More recently, they were severely criticized by [Y. C. Ong, J. Cosmol. Astropart. Phys. 1701, 001 (2017)]. While these spacetimes are certainly Ricci-flat, and so mathematically satisfy the vacuum Einstein equations, they are also merely minor variants on Taub--NUT spacetimes. Consequently, they exhibit several unphysical features that make them quite unreasonable as realistic astrophysical objects. Specifically, these \enquote{twisted} black holes are not (globally) asymptotically flat. Furthermore, they contain closed time-like curves that are not hidden behind any event horizon --- the most obvious of these closed time-like curves are small azimuthal circles around the rotation axis, but the effect is more general. The entire region outside the horizon is infested with closed time-like curves.

\begin{center}
	{\sffamily \bfseries \Large Effective metrics and a fully covariant description of constitutive tensors in electrodynamics}\\
	\vspace{.3\baselineskip}
	{\sffamily\large Sebastian Schuster, Matt Visser}\\
	\vspace{.3\baselineskip}
	{\sffamily\large Physical Review \textbf{D96}(12), 124019 (2017)}\\\vspace{.3\baselineskip}
	\begin{tabular}{llc}
		DOI: &  \href{https://doi.org/10.1103/PhysRevD.96.124019}{10.1103/PhysRevD.96.124019}&\cite{CovEffMetrics} \\
		arXiv: & \href{http://arxiv.org/abs/1706.06280}{1706.06280} &[gr-qc]
	\end{tabular}
\end{center}
\textbf{Abstract:} Using electromagnetism to study analogue space-times is tantamount to considering consistency conditions for when a given (meta-)material would provide an analogue space-time model or --- vice versa --- characterizing which given metric could be modelled with a (meta-)material. 
While the consistency conditions themselves are by now well known and studied, the form the metric takes once they are satisfied is not. This question is mostly easily answered by keeping the formalisms of the two research fields here in contact as close to each other as possible. While fully covariant formulations of the electrodynamics of media have been around for a long while, they are usually abandoned for (3+1)- or 6-dimensional formalisms. Here we shall use the fully unified and fully covariant approach. This enables us even to generalize the consistency conditions for the existence of an effective metric to arbitrary background metrics beyond flat space-time electrodynamics. We also show how the familiar matrices for permittivity $\epsilon$, permeability $\mu^{-1}$, and magneto-electric effects $\zeta$ can be seen as the three independent pieces of the Bel decomposition for the constitutive tensor $Z^{abcd}$, i.e., the components of an orthogonal decomposition with respect to a given observer with four-velocity $V^a$. 
Finally, we shall use the Moore--Penrose pseudo-inverse and the closely related pseudo-determinant to then gain the desired reconstruction of the effective metric in terms of the permittivity tensor $\epsilon^{ab}$, the permeability tensor $\mi^{ab}$, and the magneto-electric tensor $\zeta^{ab}$, as an explicit function $\g(\epsilon,\mu^{-1},\zeta)$.	

\begin{center}
	{\sffamily \bfseries \Large Bespoke analogue space-times: Meta-material mimics}\\
	\vspace{.3\baselineskip}
	{\sffamily\large Sebastian Schuster, Matt Visser}\\
	\vspace{.3\baselineskip}
	{\sffamily\large General Relativity and Gravitation \textbf{50}(6). 55 (2018)}\\\vspace{.3\baselineskip}
	\begin{tabular}{llc}
		DOI: &  \href{https://doi.org/10.1007/s10714-018-2376-2}{10.1007/s10714-018-2376-2}&\cite{Bespoke1} \\
		arXiv: & \href{http://arxiv.org/abs/1801.05549}{1801.05549} &[gr-qc]
	\end{tabular}
\end{center}
\textbf{Abstract:} Modern meta-materials allow one to construct electromagnetic media with almost arbitrary bespoke permittivity, permeability, and magneto-electric tensors. If (and only if) the permittivity, permeability, and magneto-electric tensors satisfy certain stringent compatibility conditions, can the meta-material be fully described (at the wave optics level) in terms of an effective Lorentzian metric—an analogue spacetime. We shall consider some of the standard black-hole spacetimes of primary interest in general relativity, in various coordinate systems, and determine the equivalent meta-material susceptibility tensors in a laboratory setting. In static black hole spacetimes (Schwarzschild and the like) certain eigenvalues of the susceptibility tensors will be seen to diverge on the horizon. In stationary black hole spacetimes (Kerr and the like) certain eigenvalues of the susceptibility tensors will be seen to diverge on the ergo-surface.

\section{Conference Proceedings}
\begin{center}
	{\sffamily \bfseries \Large Sparsity of the Hawking flux}\\
	\vspace{.3\baselineskip}
	{\sffamily\large Finnian Gray, Sebastian Schuster, Alexander Van-Brunt, Matt Visser}\\
	\vspace{.3\baselineskip}
	{\sffamily\large 14\textsuperscript{th} Marcel Grossmann Meeting on Recent Developments in Theoretical and Experimental General Relativity, Astrophysics, and Relativistic Field Theories (MG14) (2015)}\\\vspace{.3\baselineskip}
	\begin{tabular}{llc}
		DOI: &  \href{https://doi.org/10.1142/9789813226609_0175}{10.1142/9789813226609\_0175}&\cite{HawkFlux2} \\
		Conference: & \href{http://inspirehep.net/record/1339307}{C15-07-12} &\\
		arXiv: & \href{http://arxiv.org/abs/1512.05809}{1512.05809} &[gr-qc]
	\end{tabular}
\end{center}
\textbf{Abstract:} It is (or should be) well-known that the Hawking flux that reaches spatial infinity is extremely sparse, and extremely thin, with the Hawking quanta, one-by-one, slowly dribbling out of the black hole. The typical time between quanta reaching infinity is much larger than the timescale set by the energy of the quanta. Among other things, this means that the Hawking evaporation of a black hole should be viewed as a sequential cascade of 2-body decays.

\section{Preprints}
\begin{center}
	{\sffamily \bfseries \Large Boyer--Lindquist space-times and beyond: Meta-material analogues}\\
	\vspace{.3\baselineskip}
	{\sffamily\large Sebastian Schuster, Matt Visser}\\
	\vspace{.3\baselineskip}
	\begin{tabular}{llcc}
		arXiv: & \href{http://arxiv.org/abs/1802.09807}{1802.09807} &[gr-qc]&\cite{Bespoke2}
	\end{tabular}
\end{center}
\textbf{Abstract:} Physically reasonable stationary axisymmetric spacetimes can (under very mild technical conditions) be put into Boyer--Lindquist form. Unfortunately a metric presented in Boyer--Lindquist form is not well-adapted to the \enquote{quasi-Cartesian} meta-material analysis we developed in our previous article on \enquote{bespoke analogue spacetimes} (\href{http://arxiv.org/abs/1801.05549}{arXiv:1801.05549} [gr-qc]). In the current article we first focus specifically on spacetime metrics presented in Boyer--Lindquist form, and determine the equivalent meta-material susceptibility tensors in a laboratory setting. We then turn to analyzing generic stationary spacetimes, again determining the equivalent meta-material susceptibility tensors. While the background laboratory metric is always taken to be Riemann-flat, we now allow for arbitrary curvilinear coordinate systems. Finally, we reconsider static spherically symmetric spacetimes, but now in general spherical polar rather than quasi-Cartesian coordinates. The article provides a set of general tools for mimicking various interesting spacetimes by using non-trivial susceptibility tensors in general laboratory settings.
\clearpage
\begin{center}
	{\sffamily \bfseries \Large Electromagnetic analogue space-times, analytically and algebraically}\\
	\vspace{.3\baselineskip}
	{\sffamily\large Sebastian Schuster, Matt Visser}\\
	\vspace{.3\baselineskip}
	\begin{tabular}{llcc}
		arXiv: & \href{http://arxiv.org/abs/1808.07987}{1808.07987} &[gr-qc]&\cite{EMAnaloguesAlgAna}
	\end{tabular}
\end{center}
\textbf{Abstract:} While quantum field theory could more aptly be called the \enquote{quantum field framework} --- as it encompasses a vast variety of varying concepts and theories --- in comparison, relativity, both special and general, is more commonly portrayed as less of a \enquote{general framework}. Viewed from this perspective, the paradigm of \enquote{analogue space-times} is to promote the specific theory of general relativity (Einstein gravity) to a framework which covers relativistic phenomena at large. Ultimately, this then also gives rise to new proposals for experiments in the laboratory, as it allows one to move general features of the \enquote{relativistic framework} from general relativity to entirely new fields. This allows experiments looking into analogies of currently unobservable phenomena of general relativity proper. The only requirement for this to work is the presence of a notion of an upper limit for propagation speeds in this new field. Systems of such a kind abound in physics, as all hyperbolic wave equations fulfil this requirement. 

Consequently, models for analogue space-times can be found aplenty. We shall demonstrate this here in two separate analogue space-time models, both taken from electrodynamics in continuous media. First of all, one can distinguish between analytic analogue models (where the analogy is based on some specific hyperbolic differential equation), on the one hand, and algebraic models (where the analogy is fashioned from the more or less explicit appearance of a metric tensor), on the other hand. Yet this distinction is more than just a matter of taste: The analogue space-time model's nature will also determine which physical concepts from general relativity can be taken easily into an experimental context. Examples of this will the main aim of this paper, and the Hawking effect in one of the two models considered the example of most immediate experimental interest.
 	
 	\backmatter
 	\phantomsection
 	\addcontentsline{toc}{chapter}{Bibliography}
 	\printbibliography
 \end{document}